# Mean-field and cumulant approaches to modelling organic polariton physics

Piper Fowler-Wright

A thesis completed under the supervision of J. M. Keeling and B. W. Lovett
in requirement for the degree of Doctor of Philosophy in Theoretical Physics
at the University of St Andrews

University of
St Andrews

February 2024

## Candidate's Declaration

I, Piper Fowler-Wright, do hereby certify that this thesis, submitted for the degree of PhD, which is approximately 46,000 words in length, has been written by me, and that it is the record of work carried out by me, or principally by myself in collaboration with others as acknowledged, and that it has not been submitted in any previous application for any degree. I confirm that any appendices included in my thesis contain only material permitted by the 'Assessment of Postgraduate Research Students' policy. I was admitted as a research student at the University of St Andrews in August 2020. I received funding from an organisation or institution and have acknowledged the funder(s) in the full text of my thesis.

Date 26/2/24 Signature of candidate Piper Fowler-Wright

## Supervisor's Declaration

I hereby certify that the candidate has fulfilled the conditions of the Resolution and Regulations appropriate for the degree of PhD in the University of St Andrews and that the candidate is qualified to submit this thesis in application for that degree. I confirm that any appendices included in the thesis contain only material permitted by the 'Assessment of Postgraduate Research Students' policy.

Date 26/2/24 Signature of supervisor

## Permission for publication

In submitting this thesis to the University of St Andrews we understand that we are giving permission for it to be made available for use in accordance with the regulations of the University Library for the time being in force, subject to any copyright vested in the work not being affected thereby. We also understand, unless exempt by an award of an embargo as requested below, that the title and the abstract will be published, and that a copy of the work may be made and supplied to any bona fide library or research worker, that this thesis will be electronically accessible for personal or research use and that the library has the right to migrate this thesis into new electronic forms as required to ensure continued access to the thesis.

I, Piper Fowler-Wright, have obtained, or am in the process of obtaining, third-party copyright permissions that are required or have requested the appropriate embargo below.

The following is an agreed request by candidate and supervisor regarding the publication of this thesis:

Printed copy
No embargo on print copy.

Electronic copy
No embargo on electronic copy.

Date 26/2/24 Signature of candidate Piper Fowler-Wright

Date 26/2/24 Signature of supervisor

## Underpinning Research Data or Digital Outputs

## Candidate's declaration

I, Piper Fowler-Wright, understand that by declaring that I have original research data or digital outputs, I should make every effort in meeting the University's and research funders' requirements on the deposit and sharing of research data or research digital outputs.

Date 26/2/24 Signature of candidate Piper Fowler-Wright

## Permission for publication of underpinning research data or digital outputs

We understand that for any original research data or digital outputs which are deposited, we are giving permission for them to be made available for use in accordance with the requirements of the University and research funders, for the time being in force. We also understand that the title and the description will be published, and that the underpinning research data or digital outputs will be electronically accessible for use in accordance with the license specified at the point of deposit, unless exempt by award of an embargo as requested below. The following is an agreed request by candidate and supervisor regarding the publication of underpinning research data or digital outputs: No embargo on underpinning research data or digital outputs.

Date 26/2/24 Signature of candidate Piper Fowler-Wright

Date 26/2/24 Signature of supervisor

**mens sana in corpore sano**

*Summarise your PhD in one picture or less.*

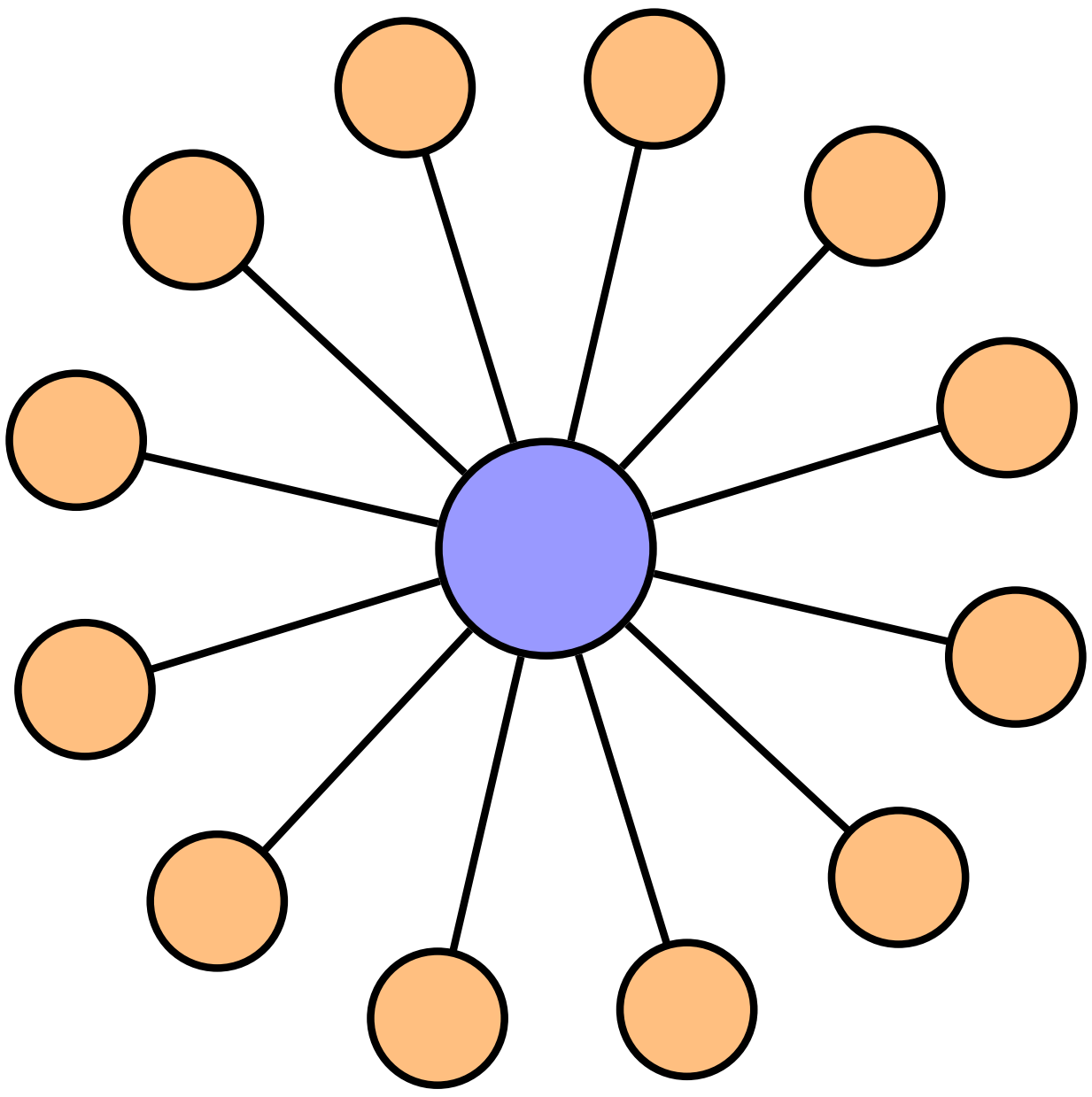

## Acknowledgments

It has been a great privilege to have been supervised by Brendon Lovett and Jonathan Keeling. Their expertise and unrelenting good natures have made the PhD not only an immensely rewarding experience, but also an enjoyable one. That experience was enhanced by the other students in the Keeling/Lovett groups I have had the pleasure of working with during the course. I would also like to acknowledge external collaborators I had the opportunity to work with and learn from: Peter Kirton, Erik Gauger, Joel Yuen-Zhou, Federico Carollo and Igor Lesanovsky.

Lastly, my thanks to family, and friends (which is largely to say those I ride my bicycle with). To a sound mind in a healthy body, each of you have contributed.

## Funding

This work was supported by the Engineering and Physical Sciences Research Council under grant number EP/T518062/1.

## Research data access statement

Research data underpinning this thesis are available at

https://doi.org/10.17630/27aa0850-9bd6-442b-abf7-49434c061464 (Chapter 4)
https://doi.org/10.17630/e1f1b609-f324-4b7c-9b7a-9cd4655b8b5d (Chapter 5)
https://doi.org/10.17630/8b395bc4-6d39-4c10-bd99-fc07cb8d8ebc (Chapter 6)

# Abstract

In this thesis we develop methods for many-body open quantum systems and apply them to systems of organic polaritons. The methods employ a mean-field approach to reduce the dimensionality of large-scale problems. Initially assuming the absence of correlations in the many-body state, this approach is built upon in two ways.

First, we show how the mean-field approximation can be combined with matrix product operator methods to efficiently simulate the non-Markovian dynamics of a many-body system with strong coupling to multiple environments. We apply this method to calculate the threshold and photoluminescence for a realistic model of an organic laser.

Second, we extend the mean-field description by systematically including higher-order correlations via cumulant expansions of the Heisenberg equations of motion. We investigate the validity and convergence properties of these expansions, both with respect to expansion order and system size, for many-body systems with many-to-one network structures. We then show how the cumulant expansions may be used to calculate spatially resolved dynamics of organic polaritons. This enables a study of organic polariton transport in which we observe reversible conversion to dark exciton states and sub-group-velocity propagation.

The methods established in this work offer versatile tools for analysing large, many-body open quantum systems and investigating finite-size effects. Their application reveals the intricate dynamics of organic polaritons resulting from the interplay of strong light-matter coupling and vibrational effects.

# Contents

# List of Figures

# Part I

# Background

# Chapter 1

# Introduction ↑TOC

The sooner we start the younger we will be when we finish.

Yulii Shikhmurzaev

## Contents

## 1.1 Synopsis ↑

No experiment is completely free from noise, no practical device operates in perfect isolation from its environment. Real systems are *open* to external influences. This applies all the way down to the smallest scales in the design and operation of quantum devices. Whilst physically small, such devices are not limited in action to only a few particles. Indeed, it is often in the aggregation of hundreds, thousands or millions of particles that useful collective quantum behaviours emerge. In order to harness these behaviours, and to describe many of the complex physical, biological and chemical systems in the world around us, it is necessary to consider systems of many interacting particles subject to external influences. It is the challenge of solving *realistic* models for such systems that this thesis is to address, in providing *methods for many-body open quantum systems*.

The study of many-body physics has a long and plentiful history [1–3], with powerful approximations and other methods of dimensional reduction designed to handle extended systems. Over recent decades a set of standard approaches to describe open, *few-body* quantum systems has also been established [4]. The intersection between many-body and open systems, however, delineates a frontier only recently made accessible by the advancement of numerical methods and hardware, and the application of sophisticated techniques from other fields such as tensor network methods and machine learning. Our first aim is to contribute to this rapidly evolving field of many-body open quantum systems.

The second pursuit of this work is the study of organic polaritons. These are systems of light interacting with molecular matter [5] that continue to attract great interest for their potential use in a large number of emerging technologies, ranging from ultra-small lasing devices [6] to novel computing architectures [7, 8] and even chemical catalysis [9, 10]. The rich photophysics of organic molecules that affords such varied applications makes the task of modelling these systems, which may comprise many, e.g., $10^6$ [11] molecules, a challenging one: in addition to collective coupling to light, each molecule has a local vibrational environment which affects the dynamics. The methods for many-body open systems we introduce are suited to this task and thus provide the opportunity to advance efforts in establishing realistic models of organic polaritons, to both guide the implementation of technologies using these systems and inspire the development of new ones.

## 1.2 Outline

The thesis comprises two parts. The first part provides background material on both the physical systems of interest and the methods to be developed. The second part includes the development of those methods and their application to the physical systems, and discussion of the results.

In more detail, Chapter 2 first introduces the class of many-body open quantum systems and broadly the problem to be addressed. Particular attention is given to many-to-one networks and other high connectivity structures our methods are well suited for. Second in this chapter is an introduction to exciton-polaritons, their occurrence in organic materials, and their role in organic lasing and transport. An essential discussion here is how the vibrational physics of molecules may be included in models of light-matter interaction. Chapter 3 follows with three methodological developments: the mathematical framework for modelling of open quantum systems, the use of tensor network methods to calculate exact open system dynamics, and the mean-field and cumulant expansion approaches central to our work.

The first part is hence largely introductory 'known' material. Its purpose is to supply the reader with the knowledge required to follow our lines of investigation as well as keep the thesis as self-contained as possible. With that in mind, researchers of organic polaritons and open quantum systems may wish to skip portions of those sections (2.2 & 3.1) and any others they find familiar. The exception to this advice would be the sections on mean-field theory and cumulant expansions (3.3 & 3.4), since there is considerable variation of the meaning of these terms between different fields of study. Our presentation of these approaches may be new and, moreover, underpins the major part of the thesis.

The second and main part of this work separates into three chapters:

- Chapter 4: a new approach using matrix product operator methods in conjunction with mean-field theory to simulate many-body systems with strong coupling to multiple environments, and its application to a realistic model of an organic laser
- Chapter 5: an investigation of the validity and convergence behaviour of mean-field theory and cumulant expansions for central spin models, with consequences for the use of these methods in central spin and central boson problems
- Chapter 6: the calculation of spatially resolved dynamics for a model of transport in organic materials, to capture the unexplained transport mechanism and below-group-velocity propagation observed in these materials.

These projects will be motivated in both the background material and at the start of the respective chapter. Chapters 4 and 6 directly address applications of polariton lasing and transport, respectively. Chapter 5 on the other hand is more theoretically oriented, and considers a central spin model. However, the results have broad implications for studies using mean-field theory and cumulant expansions, including those of organic polaritons.

Along with the analysis and discussion of results, comments on method applicability and avenues for further investigation will be included in each of Chapters 4 to 6. Additional questions for future work will be presented alongside a closing summary in Chapter 7.

Chapters 4 to 6 represent work conducted at the University of St Andrews during the course of study for the PhD in Theoretical Physics from Autumn 2020 to late 2023. The work of Chapters 4 and 5 resulted in publications [12, 13] in peer reviewed journals. Chapter 6 represents the latest work on an open problem that may be the subject of a future publication.

## 1.3 Document information ↵

To make navigating the writing as straightforward as possible, each chapter includes at its start its own table of contents, in addition to the main contents list above. At the very end of the document there is a general index as well as one for authors named in the text (usually for historical context).

Those reading the electronic version of this document have access to the modern wonder of hyperlinks. These are coloured blue if they point to a citation or external URL, otherwise they are red. At the end of the header of a section or subsection there is a small return (↵) symbol which takes you to the top of the parent chapter or section, respectively. The return arrow at the end of chapter headings instead takes you to the main contents list. Did I say how good hyperlinks were?

Corrections for typos and other errors are welcomed, as are more general discussions. Up-to-date contact details should be found on my webpage `https://phf23.user.srcf.net`. I'll try to maintain a list of any erratum and a corrected version of the thesis there.

This document was typeset with LaTeX 2$_\varepsilon$ using the `XCharter` font. Diagrams and figures not from external sources were created using PGF/TikZ as well as the Seaborn Python data visualisation library. The bibliography is formatted in a fairly well-modified version of the APS style from `biblatex-phys`.

## 1.4 Citing the thesis ↵

Where possible please refer to the archived version on the St Andrews Research Repository,

> P. Fowler-Wright, *Mean-field and cumulant approaches to modelling organic polariton physics*, PhD thesis (University of St Andrews, 2024) and link the identifier `https://doi.org/10.17630/sta/872`.

# Chapter 2

# Physical Systems ↑TOC

Someone should be studying the whole system, however crudely that has to be done, because no gluing together of partial studies of a complex nonlinear system can give a good idea of the behavior of the whole.

Murray Gell-Mann

## Contents

The content of this thesis described in the previous chapter can be summarised as i. the development of methods for many-body open quantum systems, and ii. the application of theses methods to systems of organic polaritons. In this chapter we provide physical motivations and experimental context for both objectives.

For the first (i.) we explain, in Section 2.1, what a many-body open quantum system is and the challenges faced in developing *realistic* models of such systems. We also describe the many-to-one and many-to-many network structures underlying the systems studied in this thesis, and for which our methods are well suited.

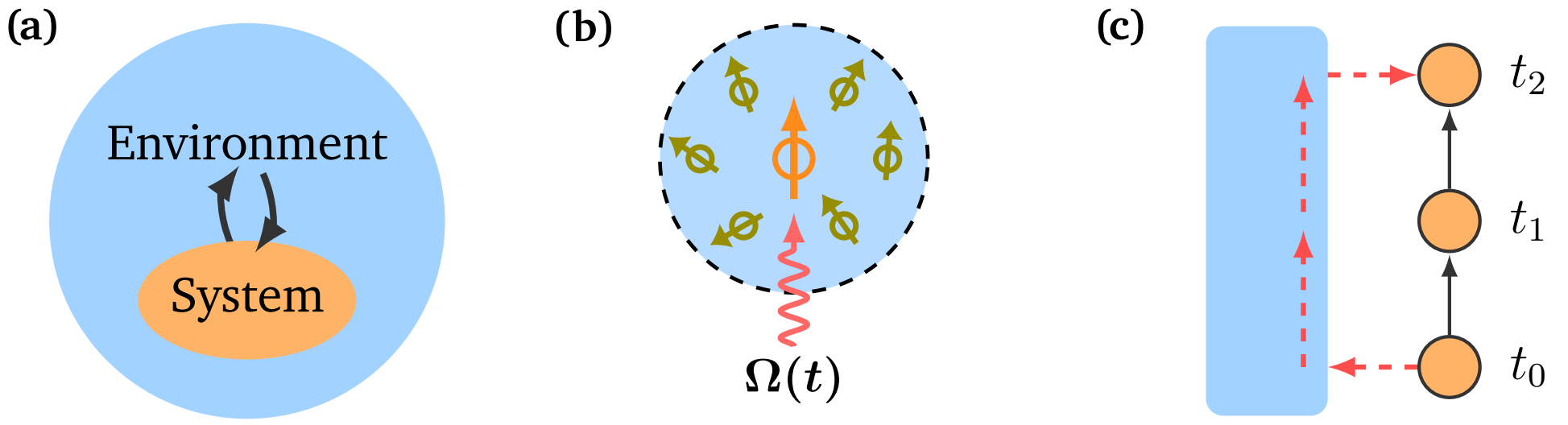

**Figure 2.1:** (a) Schematic of an open quantum system. We are directly interested in the system, but cannot ignore its interaction with the environment. (b) Electron spin (orange) interacting with nuclear spins (green) in its local environment. In this example the electron is also subject to an external drive $\boldsymbol{\Omega}(t)$. (c) Memory (non-Markovian) effects occur when information from the interaction of the system with the environment at $t_0$ influences the interaction at a later time $t_2$. Time increases vertically upwards in this sketch. ⤴TOF

For the second (ii.), in Section 2.2 we explain the nature of electronic excitations in organic materials, how these may couple to light in optical microcavities and nanostructures, and the influence of vibrational degrees of freedom in the dynamics. We also introduce the particular topics of organic polariton lasing and transport studied in Chapters 4 and 6.

## 2.1 Open many-body systems ⤴

### 2.1.1 Realistic models ⤴

Often when modelling physical situations one considers a system to be closed[1]: an insulated flask for the chemical reaction, a sealed working reservoir of water. Like the frictionless pulley or the inextensible rope, this is an idealisation: heat escapes from the flask, water evaporates from the reservoir. A good approximation of practical use for many macroscopic systems, but ultimately still an approximation. Any *real* system interacts, to a certain extent, with its surrounding environment. The effect of this interaction is not always negligible. This is unmistakably true in the domain of quantum mechanics: when working on the smallest scales even a minute thermal fluctuation or mechanical vibration can irrevocably change the trajectory of a carefully prepared system. Hence one is forced to face the reality of *open quantum systems*.

It is not necessary to go *ad absurdum* to include the entire universe and its interactions in the model of the quantum system. Instead, in all practical cases a larger region including the system of interest and its immediate environment may be considered closed[2]. A schematic for this arrangement is shown in Fig. 2.1a.

So far we have described the scenario of an uncontrolled, ambient environment. Yet this is not the only way in which a system may be open. In many experiments and technological applications it is necessary to apply an external drive or heating (e.g., using a laser) to the system in order to balance losses to the surroundings or initiate desired behaviour. Such influences must be accounted for in addition to effects from the local environment. The study of *driven-dissipative* systems is an immensely active field for the rich set of phenomena they can exhibit. The polariton systems in this thesis fall within this category, although we mainly consider the simple case of incoherent, time-independent drive. A key facet of these systems is that they are non-equilibrium, that is, have steady states determined not by a simple thermodynamic distribution but the *balance* of driving (gain) and dissipative (loss) processes.

[1] We refer to what may be termed *isolated* in classical thermodynamics [14]: no exchange of heat, work or matter.

[2] Whether the combined system plus environment is closed in the technical (isolated) sense may or may not be relevant. External reservoirs for example must be sustained, e.g., by heating, so as to remain at equilibrium despite the transfer of energy or particles to the system.

In Chapter 3 we set out the operational framework in which a practical description of an open system may be obtained. Here we discuss conceptually the difficulties inherent in doing so.

The first basic challenge is the *size* of the environment. This refers to the dimension of the *Hilbert space* of the environment rather that its physical extent—which may or may not be large. Typically, an environment comprises a vast number of of degrees of freedom that cannot be independently controlled. For example, an electron spin in a GaAs semiconductor quantum dot [15] may interact with $\sim 10^5$ nuclear spins in the surrounding lattice (Fig. 2.1b). This number precludes a complete microscopic description of the environment or generally any 'first principles' approach.

A common aim in the theory of open quantum systems is obtaining an effective description of the open system, accounting for the influence of the environment, in terms of a small set of degrees of freedom—often those of the system alone. In fact, the problem of the electron spin coupled to many nuclear spins is one we show, in Chapter 5, can be reduced to that of the electron interacting with one or a small number of representative nuclear spins.

A second main difficulty is the possibility of *non-Markovian* behaviour: disturbances of a structured environment by the system may come back to act on the system at a later time (Fig. 2.1b). Memory of previous interactions is retained by the environment and, even if one manages to obtain a description in terms of the system degrees of freedom, the information in that description grows with time as the *history* of system states must be known in order to calculate the dynamics at any one instant. The opposite memoryless or Markovian case, where only the *current* state of the system is required to calculate the dynamics, is normally limited to weakly coupled or unstructured environments [4]. For many physical systems of interest [15–21], this is not the case, and non-Markovian effects manifest in experimentally relevant regimes (see Ref. [22] for a review). This includes, for example, the electron-nuclear spin system in quantum dots where non-Markovian effects may be important to determine decoherence times [15, 23–25].

The previous two difficulties are each compounded in cases where the system itself comprises a large number of interacting degrees of freedom, i.e., is many-body. For then even an effective description of the system degrees of freedom—obtained under a Markovian approximation or otherwise—may still present a large and intractable problem. Systems of organic polaritons are within this class. Electronic excitations of molecules in an optical cavity, for example, interact collectively with light forming a many-body system, whilst each molecule also has vibrational degrees of freedom coupled to the local excitation dynamics[3]. The coupling is not necessarily weak, nor is the number of molecules necessarily small. Standard approaches of open quantum system theory cannot handle such complexity.

The above are exactly the type of problems we wish to address, as per the mission statement of methods for realistic models of many-body open quantum systems. Considering the *many-body* allows for a rich set of emergent behaviour [1, 3], i.e., behaviour arising from collective interaction that may not be observed in, or simply predicted from, the individual parts of the system. Further considering the *open* accommodates the fact that real systems interact with their surroundings and may, in addition, be subject to external drive or other experimental interventions.

[3]Note here one has not only a many-body system coupled to a single environment, but an interacting system with many open parts.

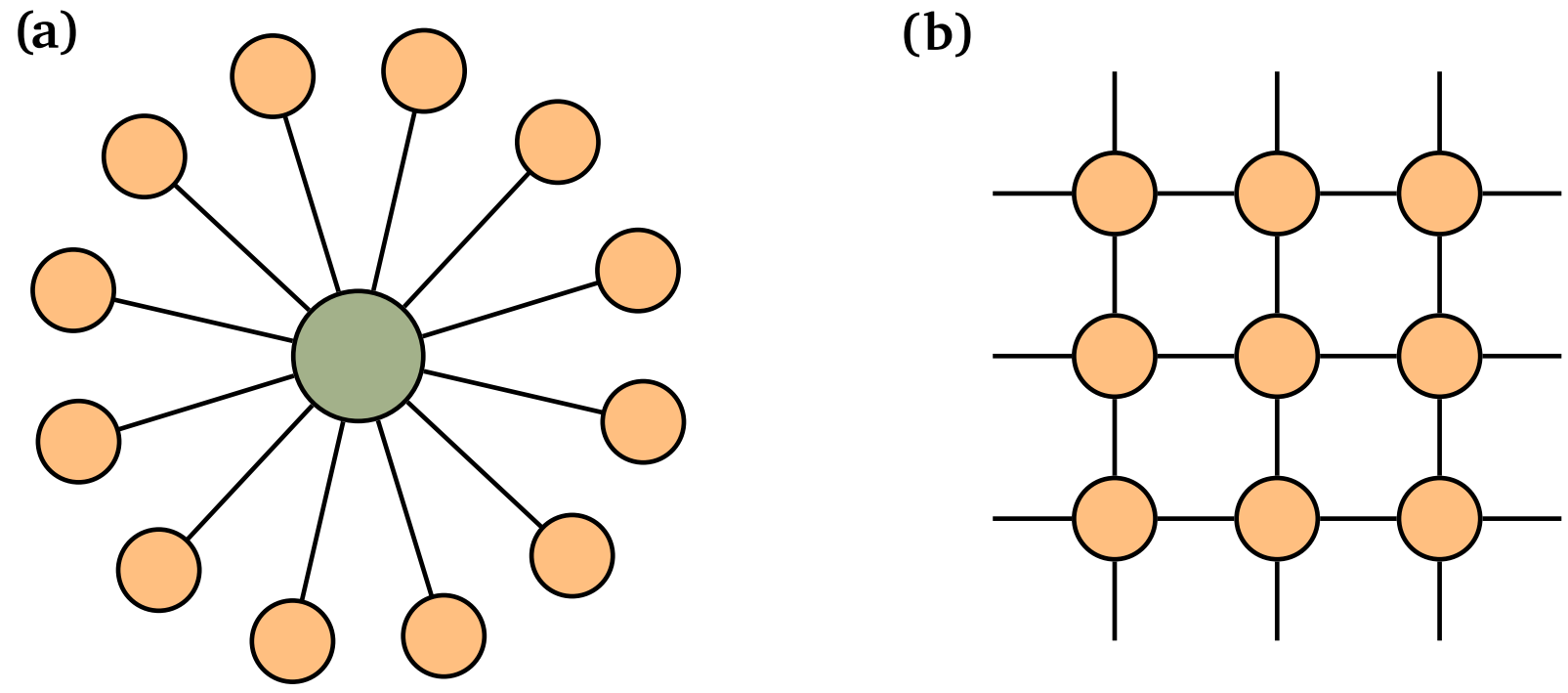

**Figure 2.2:** (a) Many-to-one network. $N$ satellites sites (orange) couple to a common central site (green). The connectivity of the central site grows with $N$. (b) The square lattice in two dimensions has a fixed connectivity: each site has four neighbours regardless of how large the system (i.e. $N$) is. These are closed many-body systems—no environments have been included in this figure. ↩TOF

### 2.1.2 Many-to-one models ↩

Networks in which one site couples non-locally to many satellite sites occur in a wide range of many-body open quantum systems. For example, models where a driven electronic spin interacts with a bath of nuclear spins (as in GaAs mentioned above) are relevant to nuclear magnetic resonance spectroscopy [26–29], quantum sensing [30–32] and quantum information processing [33–39]. The network structure is also ubiquitous in quantum optics where it defines the interaction of a single emitter with many electromagnetic modes [40], or equally a single mode with an ensemble of emitters [41]. The latter includes single-mode models of polariton systems discussed further below. In many such cases, the large number of satellite sites precludes exact calculations, especially when considering open system dynamics, e.g., the electron-nuclear system is both driven by microwave laser and subject to environmental noise [27]. Consequently there is great need for methods capable of handling large, driven-dissipative systems with many-to-one connectivity.

A key property of the many-to-one network, seen in Fig. 2.2a, is its high connectivity: whilst each satellite couples to the central site only, the central site couples to *every* satellite. It is for this reason the network can describe not only systems such as an electron interacting with a finite–albeit very large–number of adjacent nuclear spins, but also models where interactions are *non-local* and so not limited in range or number, such as the case between an ensemble of emitters and a delocalised photon mode. This contrasts conventional lattice models of condensed matter physics [2, 42], where connectivity is a small, finite number fixed by the dimension $d$ of space ($d = 2$ in Fig. 2.2b). As we explain in Chapter 3, this difference is the essential reason mean-field approaches may be expected to be effective for many-to-one and other high connectivity models, whereas they break down for lattice models at small $d$.

In Chapter 4 a model of organic lasing based on the Tavis-Cummings Hamiltonian is defined. This describes a many-to-one network where the central degree of freedom is a boson, also known as a central boson model. The class of central *spin* models are later introduced in Chapter 5. Finally, to study polariton transport, Chapter 6 goes beyond the many-to-one network with a emitter-cavity model including multiple photon modes, each coupling to all emitters: a many-to-many model. A similar multimode extension to the Tavis-Cummings Hamiltonian is considered earlier to determine momentum dependent spectra.

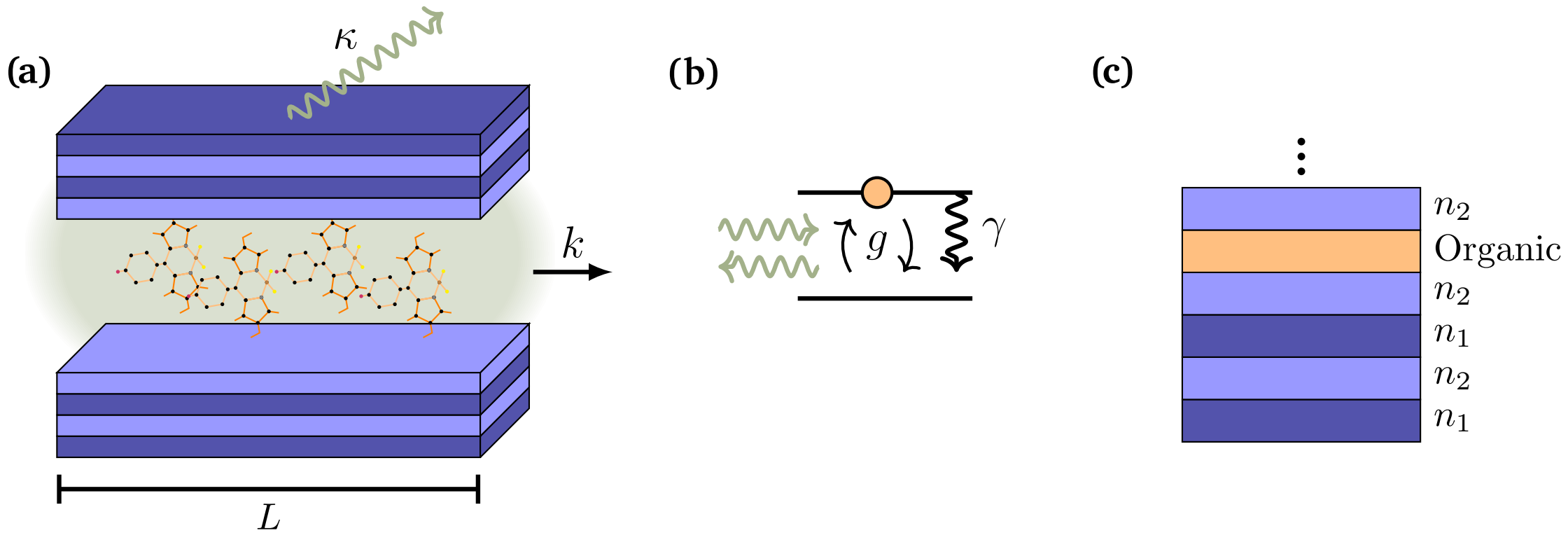

**Figure 2.3:** Microcavity polaritons. (a) An ensemble of organic molecules inside a planar cavity with in-plane momentum $k$. Photon loss occurs at rate $\kappa$ due to imperfect mirrors. (b) Each molecule has an electronic transition (two-level system) that interacts with the cavity field at strength $g$ and is subject to dissipation $\gamma$. (c) The mirrors comprise layers of dielectrics of alternate refractive index called Distributed Bragg Reflectors (DBRs). A film or matrix containing the organic molecules is deposited between the DBRs. ↑TOF

## 2.2 Organic polaritons

### 2.2.1 Exciton-polaritons

Exciton-polaritons are hybrid light-matter excitations that result from the strong coupling between electronic excitations and electromagnetic radiation [5]. We are interested in their formation with organic matter, i.e., carbon-based compounds, for the flexibility and range of favourable properties organic molecules can convey.

A prototypical system is that of organic *microcavity* polaritons. Refer to Fig. 2.3a which depicts a collection of molecules—organic emitters—placed within an optical cavity. Each molecule has an electronic transition near-resonant with one or more cavity modes, enabling interaction between electronic excitations of the molecules and photons confined to the cavity. The significance of the emitters being molecules rather than atoms, say, is that molecules have rovibrational degrees of freedom. These may strongly couple to the electronic state and so significantly affect the exciton-polariton dynamics [43]. In the models below we refer to high frequency vibrational modes, but similar considerations could be given to low frequency rotational or torsional modes of the molecules [5].

Setting the vibrational physics aside for the time being, strong *light-matter* coupling is achieved when the rate of coherent exchange between the electronic and photonic degrees of freedom exceeds those of decay and decoherence processes from either part [5]. This is equivalent to the resolution of new normal modes of the system, the polaritons, in the spectrum. Exciton-polaritons therefore are superpositions of excitons and photons with properties between that of light and matter: they may travel ballistically with a small—light-like—effective mass, but also interact and scatter. Since all microcavity polaritons have in common the form of light, differences between organic and *in*organic microcavity polaritons arises from the differing nature of electronic excitations in the two types of systems, as we now discuss.

### 2.2.2 Organic and inorganic excitons

Excitons are electronic excitations in crystalline and molecular matter comprising a negatively charged electron and a positively charged 'hole' bound together by Coulomb attraction [44]. The key difference between their occurrence in organic and inorganic matter is in their spatial extent. For organics, excitons are typically sharply localised on the atomic scale—a tightly bound pairing that moves via interstitial hopping. These are called *Frenkel* excitons and may be found in molecular crystals, aggregates, polymers and biological molecules such as proteins and chromophores [44]. The opposite limit, where the excitons span many, e.g., hundreds [44] of lattice sites, is realised in inorganic materials. These are *Mott-Wannier* excitons; delocalised electron-hole pairs that can diffuse throughout the lattice.

In both cases, excitons carry energy and momentum but no charge, and may travel slowly through the material, scattering with each other as well as any non-electronic excitations present such as phonons, the quanta of lattice vibrations. As they can be created by any process that forms electron-hole pairs, a possible life-cycle for an exciton is: creation via the absorption of light, propagation, and then recombination to emit light. It is precisely when this cycle becomes rapid enough—at strong light-matter coupling—such that the coherent exchange of energy between the emitters and photons exceeds losses, that new quasiparticles, the polaritons, are effectively formed.

Crucially, the smaller radius of Frenkel excitons corresponds to larger *binding energies* $E_b$, of the order of $1$ eV, compared to $\sim 10$ meV in inorganic semiconductors for example [45]. This conveys stability up to far higher temperatures (a basic requirement is that $E_b$ must exceed the thermal energy $k_B T$; note $k_B T = 26$ meV at $T = 300$ K). Further, excitons in organic materials have large optical dipole moments, i.e., intrinsic coupling to light, and may be prepared at high densities [45, 46]. So while conventional[4] inorganic semiconductors such as GaAs [51] and CdTe [52, 53] require cryogenic temperatures to achieve strong light-matter coupling, in organic systems it may be attained at room temperatures [45]. This remains true despite the *shorter* photon lifetimes often found in organic microcavities and other confinement schemes [43]. Room temperature operation is a main attraction of organic materials for practical applications of polaritonic devices.

### 2.2.3 Vibrations and disorder

Having discussed the nature of electronic excitations, there are two principal features of systems of organic polaritons that need to be considered. The first, mentioned above, is significant interaction of the electronic systems with vibrational degrees of freedom, that is, *vibronic coupling*. We introduced discrete (high-frequency) intramolecular vibrational modes of a molecule above, but quite generally in molecular matter there is coupling to a broad spectrum of modes arising from a number of possible sources in the local environment, e.g., conferred by a host matrix. Critically, this environment may be structured and beyond weak coupling treatments, making the task of determining their effect on the dynamics a difficult one. Note in these cases we treat the environment of each site as independent. The problem of exchange between different environments is potentially interesting, but challenging[5]. Further below we explain how our approach to modelling the two types of vibrational environment—discrete and continuous—is different.

[4]Wide-band semi conductors, e.g., GaN [47], ZnO [48], with large $E_b$ and optical dipoles do permit room temperature strong coupling and lasing. However, fabricating microcavities with these materials requires complex techniques and tends to result in large inhomogeneities, which may limit their practical use [49]. See Ref. [50] for a review.

[5]Another complication that we do not consider is that in molecular assemblies such as aggregates many different molecules may be involved in the electronic transition [50].

| Process | Timescale |
|---|---|
| Light-matter coupling | 1-5 fs |
| Polariton lifetime | 100-200 fs |
| Vibronic coupling | 10 fs |
| Vibronic relaxation | 1000 fs |

**Table 2.1:** Illustrative timescales for organic microcavity polaritons [5, 43]. Here we used $\sim 1/\Omega[\text{fs}]$ as the timescale associated with frequency $\Omega$ in (inverse) femtoseconds ($2\pi/\Omega[\text{fs}]$ is also a common choice [10]). Later we refer to frequencies in units of electronvolts (i.e., state $\hbar\Omega$ with $\hbar = 1$) and so for reference note $\Omega[\text{fs}] = 10^{-15} \cdot 10^{-3} \cdot (e/\hbar) \cdot \Omega[\text{meV}]$.

The timescales over which coherent and incoherent vibrational processes act vary, but may be comparable to that set by the light-matter coupling [54]. Some typical values are provided in Table 2.1. The interplay between system and environment timescales in organic materials results in varied dynamical regimes, and so modelling requirements. In particular in the following chapter we see how it—and generally strong coupling to a structured environment—may give rise to non-Markovian dynamics as introduced above.

The second feature of organic samples is the disorder they exhibit. This may include variation in the electronic transition frequency as well as structural disorder [54]. We show below how both may be captured by inhomogeneities of the microscopic Hamiltonian. In this thesis we do not consider any explicit[6] forms of disorder, instead choosing to focus on the vibrational physics. We note homogeneous models can provide many useful predictions robust to disorder [43]. Further, all the methods we develop can in principle accommodate disordered Hamiltonians, and this may be the target of future work.

Both vibrational and disorder properties are highly dependent on the particular organic compound and sample preparation. For example, for molecules embedded in a host matrix, the local vibrational environment is determined by the choice of the host material in conjunction with the concentration of active molecules [55]. The presence of disorder similarly depends on sample quality and the chosen fabrication technique. Besides these variations, the sheer variety of organic materials that may be used in polariton experiments presents a wealth of possible properties and behaviours that may be selected to match the requirements of an application. Our aim, through simple models, is to capture the common and essential physics of these systems.

### 2.2.4 Models of microcavity polaritons

We now consider in further detail a model of microcavity polaritons. The aspects we introduce readily extend to describe other systems of organic polaritons.

As was shown in Fig. 2.3, a typical setup is a planar (2-dimensional) microcavity with mirrors consisting of a series of dielectrics of alternative refractive index called Distributed Bragg Reflectors (DBRs) [56]. This provides a highly efficient confinement mechanism for the light which interacts with organic material deposited between the DBR layers.

First we describe the energies of the *uncoupled* light-matter system where 'matter' refers to the electronic part. The photon dispersion $\omega_k$ for the planar cavity is a function of the in-plane momentum $k$ of the form [57]

$$\omega_k = \sqrt{\omega_c^2 + k^2 c^2}, \tag{2.2.1}$$

[6] We show in Chapters 4 and 6 how vibronic coupling confers *dynamical disorder* and hence scope for behaviours otherwise thought to require disorder in the light-matter Hamiltonian.

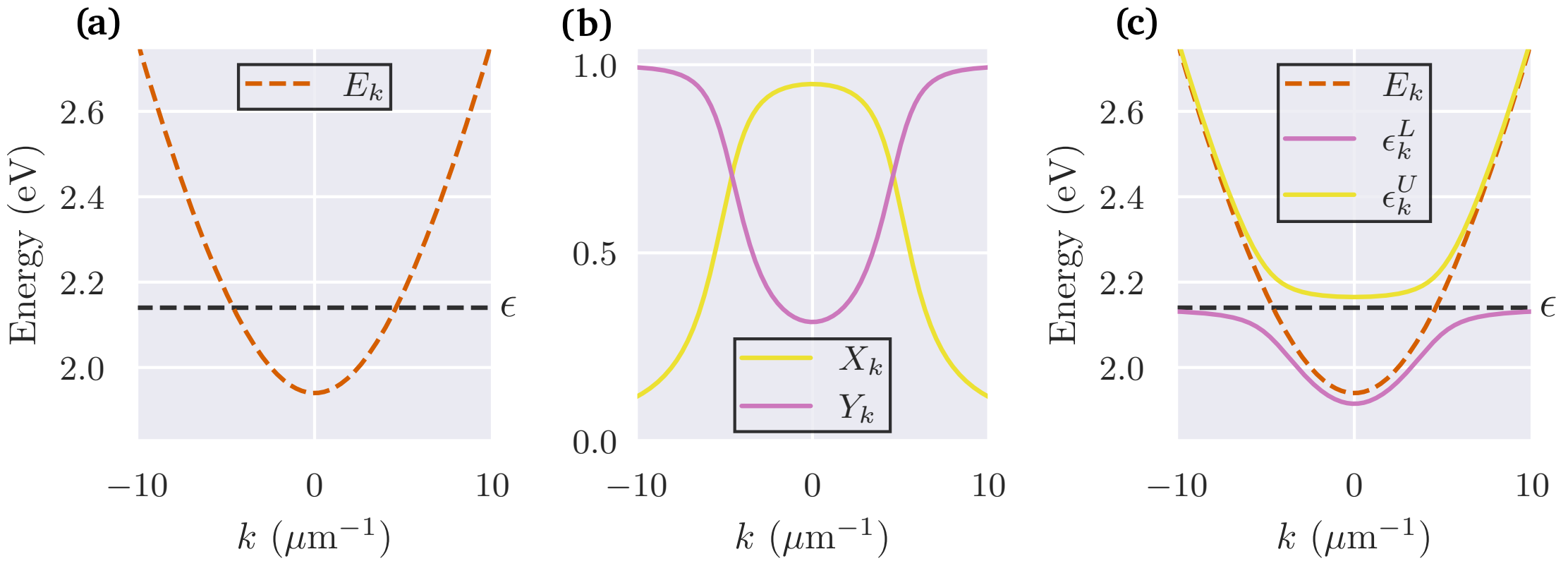

**Figure 2.4:** (a) Cavity (red) and exciton energies for the uncoupled system as functions of the in-plane momentum $k$. A small negative detuning $\omega_c - \epsilon = -0.2$ eV is present in this example. (b) Hopfield coefficients for an example multimode Tavis-Cummings model. These are real and satisfy $X_k^2 + Y_k^2 = 1$. The ratio $X_k^2/Y_k^2$ controls the (inverse) ratio of light to matter for the (upper) lower polariton. (c) Upper (U) and lower (L) polariton energies for the multimode model. An avoided crossing occurs where the bare cavity and exciton energies (dashed lines) intersect. The separation of the polariton branches is determined by the collective coupling $g\sqrt{N}$. (Parameter values and derivations of the curves in (b), (c) are given Section 6.3.2.) ↑TOF

where $\omega_c$ is the minimum frequency and $k$ is quantised due to the confinement (we generally indicate discrete variables using subscripts). If $L$ is the effective length[7] of the cavity then

$$k = 0, \pm\frac{2\pi}{L}, \pm\frac{4\pi}{L}, \pm\frac{6\pi}{L}, \dots \tag{2.2.2}$$

describe photons of increasing momentum and energy $\hbar\omega_k$. While the number of modes is in principle infinite, only a subset will be near resonant with the exciton energy and so relevant to the dynamics. Consequently in practice it is sufficient to consider a finite number of modes up to a maximum magnitude $k_{\text{max.}}$. In the simplest case, which we initially consider below, only a single mode couples to the electronic system.

For small in-plane momentum $k$, $\omega_k$ is quadratic in $k$. This allows for the assignment of an effective mass to the cavity photons, in analogy with the quadratic dispersion of massive free particles:

$$E_k = \hbar\omega_k = \hbar\omega_c + \underbrace{\frac{\hbar c^2 k^2}{2\omega_c}}_{=\hbar^2 k^2/2m_{\text{phot}}} + \dots \;\Rightarrow\; m_{\text{phot}} = \frac{\hbar\omega_c}{c^2}. \tag{2.2.3}$$

Here and in Eq. (2.2.1) we wrote the speed of light in vacuum $c$; more generally $\tilde{c} = c/n_r$ where $n_r$ is the refractive index of the intracavity material.

The dispersion Eq. (2.2.1) is plotted in Fig. 2.4a together with the exciton energy $\epsilon$. The latter is effectively constant since the exciton mass is orders of magnitude larger than that of the photon. In other words, excitons are dispersionless on this scale. Without any interactions, the system eigenstates have the photon modes in number states and, independently, each molecule in its ground or excited state.

[7]The width (distance between mirrors) also imposes a mode structure [57], but we consider the single mode close to the exciton energy.

When coupling between light and matter is switched on, new normal modes arise. The standard model used to describe the combined light-matter system in the simple case of a single mode cavity ($k = 0$ in Eq. (2.2.1)) is the Tavis-Cummings Hamiltonian [58],

$$H = \omega_c a^\dagger a + \sum_{n=1}^{N} \left[ \frac{\epsilon}{2} \sigma_n^z + g \left( a \sigma_n^+ + a^\dagger \sigma_n^- \right) \right] \qquad (\hbar = 1)\,. \tag{2.2.4}$$

Here $a$ is a annihilation operator for the bosonic photon mode and $\sigma_n^\alpha$ a Pauli matrix operating on the electronic site of the $n^{\text{th}}$ molecule, which is modelled as a two-level system, i.e., a spin-1/2 particle. The light-matter coupling strength is parametrised by $g$. As indicated, we set $\hbar = 1$ so that $\omega_c$, $\epsilon$ and $g$ have units of energy. We do this throughout the thesis and tend to work with energies in electronvolts—reminders will be given.

The eigenstates of the coupled system may be obtained by noting $H$ conserves the total excitation number. Indeed, $a^\dagger a$ and $\sum_n \sigma_n^z$ effectively count the number of photons and excitons, respectively, whilst $a\sigma_n^+$ corresponds to the loss of a photon and a gain of an electronic excitation, and vice versa for $a^\dagger \sigma_n^-$. Thus the model may be solved for a fixed excitation number $n_{\text{ex.}}$. This is readily done for small values of $n_{\text{ex.}}$ or large systems with low excitation densities in various ways. One we find instructive is to recognise Eq. (2.2.4) may be written in terms of the collective spin operations $J^z = \sum_n \sigma_n^z/2$ and $J^\pm = \sum_n \sigma_n^\pm$:

$$H = \omega_c a^\dagger a + \epsilon J^z + g \left( a J^+ + a^\dagger J^- \right). \tag{2.2.5}$$

It is now possible to use a bosonic representation, obtained from the transform (after Holstein-Primakoff [59]),

$$J^z \to -\frac{N}{2} + b^\dagger b, \quad J^+ \to b^\dagger \sqrt{N - b^\dagger b}, \tag{2.2.6}$$

such that $b^\dagger$ acts as a creation operator for deviations from the fully polarised state with no electronic excitations [60]. Then, for $N \gg \langle b^\dagger b \rangle$, i.e., low excitation densities, $J^+ \approx \sqrt{N} b^\dagger$, $J^- \approx \sqrt{N} b$, and

$$H \sim \omega_c a^\dagger a + \epsilon b^\dagger b + g\sqrt{N} \left( a b^\dagger + a^\dagger b \right), \tag{2.2.7}$$

where a constant offset $-N\epsilon/2$ was ignored. The remaining eigenproblem is exactly solvable, being that of the $2 \times 2$ matrix

$$\begin{pmatrix} \epsilon & g\sqrt{N} \\ g\sqrt{N} & \omega_c \end{pmatrix}. \tag{2.2.8}$$

In Appendix A.1 we show that this provides the eigenstates $|U\rangle = U^\dagger |0\rangle$, $|L\rangle = L^\dagger |0\rangle$, where $|0\rangle$ is the vacuum state (no photons or excitons), and the operators

$$U^\dagger = X \frac{1}{\sqrt{N}} \sum_{n=1}^{N} \sigma_n^+ + Y a^\dagger, \tag{2.2.9}$$

$$L^\dagger = -Y \frac{1}{\sqrt{N}} \sum_{n=1}^{N} \sigma_n^+ + X a^\dagger, \tag{2.2.10}$$

create superpositions of photons and delocalised molecular excitations, i.e., the polaritons. Note in particular the symmetric combination of molecular excitations $\frac{1}{\sqrt{N}} \sum_{n=1}^{N} \sigma_n^+ |0\rangle$ is identified as the *bright* state, since it is the part of matter that actually couples to light.

The states $|U\rangle$, $|L\rangle$ define the *upper* and *lower* polariton, respectively. The optical-material character of the polaritons is given by $X$ and $Y$, known as Hopfield coefficients, where

$$X = \frac{1}{\sqrt{2}}\sqrt{1 + \frac{\epsilon - \omega_c}{2\sqrt{(\epsilon-\omega_c)^2 + 4g^2N}}}, \quad Y = \frac{1}{\sqrt{2}}\sqrt{1 - \frac{\epsilon - \omega_c}{2\sqrt{(\epsilon-\omega_c)^2 + 4g^2N}}}. \tag{2.2.11}$$

The energies of the polaritons are

$$\epsilon^{U/L} = \frac{1}{2}\left[\epsilon + \omega_c \pm \sqrt{(\epsilon-\omega_c)^2 + 4g^2N}\right]. \tag{2.2.12}$$

These energies are centred about the average of $\epsilon$ and $\omega_c$ and, at resonance $\epsilon = \omega_c$, differ as $\pm g\sqrt{N}$. The quantity $\Omega = 2g\sqrt{N}$ is identified as the vacuum *Rabi splitting*. This is one of two key parameters to determine strong coupling in polariton experiments, the other being the linewidth due to the finite polariton lifetime $\tau_p$. Many authors (including us in Chapter 4) hence write the light-matter coupling as $(\Omega/2\sqrt{N})$ rather than $g$.

That the Rabi splitting $\Omega$ may be many hundreds of meV for organic materials whilst $1/\tau_p \sim 5$ meV [5] is the statement of strong coupling. Note the role of the *collective* coupling $g\sqrt{N}$: compared to a single molecule the Rabi splitting for an ensemble is enhanced by a factor of $\sqrt{N}$, although, as we discuss below, $g$ normally scales inversely with $\sqrt{N}$ such that $\Omega$ remains constant with system size. We also see that for the Tavis-Cummings model it is really only the detuning $\Delta = \epsilon - \omega_c$ that is relevant for the physics. This is generally true for models describing strong, but not ultrastrong, light-matter coupling (see Section 2.2.11).

An additional point to be taken from the above derivation is that it relied on a representation in terms of collective spin operators. This is a widely used strategy for many-to-one models. However, it cannot be applied in the case of environments or processes that act on the many-body sites individually. In the problems we consider, such processes will always be present.

In Chapter 6 we extend the calculation above to a multimode model such that the Hopfield coefficients and polariton energies become, through the $k$-dependent cavity dispersion $\omega_c \to \omega_k$, functions of in-plane momentum: $X_k, Y_k$ and $\epsilon_k^{U/L}$. We show the result in Figs. 2.4b and 2.4c. The spectrum, Fig. 2.4b, demonstrates the upper and lower polariton branches separated by $\sim \Omega$. In a real system, with dissipative processes, this splitting must be larger than the linewidth $\sim 1/\tau_p$ in order for the polaritons to be resolved. On the other hand, as $\Omega \to 0$ the avoided crossings between branches vanish and the uncoupled dispersions $\omega_k$, $\epsilon$ are recovered. In Chapter 4 we develop a method that can provide spectral information of a real, lossy system, including an exact description of the relevant vibrational physics not present in the Tavis-Cummings model.

### 2.2.5 Models of organic polaritons

We now outline how discrete and continuous vibrational structure may be added to the model[8].

First, when one has discrete intramolecular modes, these may be included directly in the system Hamiltonian. In the simplest case of a single harmonic mode of frequency $\omega_\nu$ for each molecule,

$$H = \omega_c a^\dagger a + \sum_{n=1}^{N}\left[\frac{\epsilon}{2}\sigma_n^z + g\left(a\sigma_n^+ + a^\dagger\sigma_n^-\right)\right] + \sum_{n=1}^{N}\omega_\nu\left[b_n^\dagger b_n + \sqrt{S}\left(b_n^\dagger + b_n\right)\sigma_n^z\right]. \tag{2.2.13}$$

[8] In addition to the Holstein-Tavis-Cumming model for discrete modes described below, an approach often used to fit experimental spectra is to solve a coupled harmonic oscillator Hamiltonian where the leading vibrational transitions are included as separate species. See Ref. [61] for an example.

Here $b_n$ is a bosonic annihilation operator for the mode of the $n^{\text{th}}$ molecule and $S$—known as the Huang-Rhys parameter—characterises the exciton-vibration coupling strength.

Equation (2.2.13) is called the Dicke-Holstein or Holstein-Tavis-Cummings (HTC) model. Despite its simplicity, it is able to capture a wide range of vibrational physics, and has proven effective in modelling many phenomena of organic polariton system [9, 10, 43, 62–69]. Note in this model the system operator $\sigma_i^z$ coupling to the vibrational mode is diagonal in the local electronic basis. As we discuss in the next chapter, this affects molecular dephasing, which is the loss of quantum coherence of the electronic state.

For coupling to a continuum of vibrational modes we instead take an open systems approach where the vibrational degrees of freedom are treated as an environment for each the molecules. Specifically, we consider a bath of harmonic oscillators (bosonic operators $b_j$) with a diagonal coupling to the electronic state,

$$H_E^{(n)} = \sum_j \left[ \nu_j b_j^\dagger b_j + \left( \xi_j b_j + \bar{\xi}_j b_j^\dagger \right) \sigma_n^z \right], \quad n = 1, \dots, N. \tag{2.2.14}$$

This is a very common model for open quantum systems which we discuss in detail in the following chapter. Here we note the coupling to the bath may be characterised by a continuous function $J(\nu)$ known as the spectral density, $J(\nu) = \sum_j |\xi|^2 \delta(\nu - \nu_j)$, and that the possibility of strong system-environment coupling prevents standard, Markovian treatments of the environment. The last observation leads us to consider (Section 3.2) tensor network methods that can provide an efficient means to the exact dynamics of the system in such cases.

Whilst in Chapters 4 and 6 we work with the two types of vibrational environment separately, they can in principle be combined. That is, one can have a model with coupling to both a low frequency continuum and one or more discrete vibrational modes included in the system Hamiltonian.

### 2.2.6 Dark exciton states

Above we determined two eigenstates of the Tavis-Cummings Hamiltonian, the polaritons, as superpositions of photons and the bright excitonic state $\frac{1}{\sqrt{N}} \sum_{n=1}^{N} \sigma_n^+ |0\rangle$. The full problem, for $N$ molecules plus one photon, has $N + 1$ degrees of freedom. Hence there are a further $N - 1$ eigenstates that do not involve light. These *dark* exciton states are degenerate with energy $\epsilon$ and orthogonal to the bright state. A basis that is commonly chosen are the delocalised plane waves

$$\frac{1}{\sqrt{N}} \sum_{n=1}^{N} e^{i2\pi kn/N} \sigma_n^+ |0\rangle, \quad k \in [1, N-1]. \tag{2.2.15}$$

At first, one may dismiss the dark states as irrelevant to strong light-matter coupling. Indeed, they do not gain any population under the unitary evolution of $H$ in Eq. (2.2.4). However, a real system has disorder and scattering, and these serve to mix the dark states into the dynamics, that is, they become optically active. In this case the basis states Eq. (2.2.15) become semi-localised over different sites of the lattice [70]. The overwhelming density of states of dark excitons means that they may be expected to contribute to, if not dominate, dynamical processes within a microcavity [71, 72].

The obvious challenge presented by dark states is that studies of microcavity polaritons are foremost *optical* experiments. Hence the dark states cannot be directly observed, and their role in many dynamical processes remains poorly understood. We discuss their potential role in mediating polariton transport and chemistry below. We also explain how they are essential to experiments of polariton lasing in providing a reservoir of excitations.

Solving the Tavis-Cummings model in the presence of disorder is more challenging; see Refs. [70, 73–78] for recent developments. We also note Ref. [79] which shows how small amounts of disorder can lead to an *enhancement* of the vacuum Rabi splitting $\Omega$. As commented above, we do not include explicit forms of disorder (energetic or inhomogeneous couplings) in our models, but dephasing from the vibrational environment does result in the mixing of dark states into the dynamics.

### 2.2.7 Experiments and observables

Before discussing potential applications of organic polaritons we need to connect our modelling to what actually occurs in experiment, and explain the relevant physical observables.

The essential feature not included in any *Hamiltonian* model such as the Tavis-Cummings model Eq. (2.2.4) is the non-equilibrium, dissipative nature of organic polariton systems: the rate of thermalisation[9] is not sufficiently fast for a thermal state to be reached before significant losses occur [5]. The continuous operation of any polariton device hence requires a supply of energy—normally via laser pumping [81]—in order to balance losses. The correct starting point for modelling organic polariton systems therefore is in fact a *driven-dissipative* model, such as the driven-dissipative Tavis-Cummings model, where both pump and loss are included in addition to the Hamiltonian evolution. In the next chapter we explain how this may be done via the master equation for the open system dynamics.

This brings us to discuss what is actually measured in an organic polariton experiment. As mentioned above, all observations are fundamentally tied to the *optical* response of the system, since it is only the light emitted from the cavity that one can record. The non-equilibrium character means both the spectrum of possible excitations—the density of states $\varrho$—*and* their occupation $n$ must be determined. These combine to give the photoluminescence $\mathcal{L} = \varrho n$, which is the actual light emitted from the system.

First, the polariton spectra (dispersions in Fig. 2.4b) are normally determined in preliminary reflectivity or transmissivity studies. Here the reflection $R$ and transmission $T$ of the cavity subject to probe radiation—but otherwise unpumped—relate to the absorption $A$ by the cavity as $R+T+A=1$. For a high quality cavity ($\kappa \to 0$), $A$ corresponds directly to the spectral weight $\varrho$, although the general relation for a lossy cavity is not so straightforward, as we discuss in Appendix A.2.

Measurements are often made using various forms of *angle-resolved* spectroscopy where the intensity of light is recorded at different angles of emission $\theta$ from the cavity normal. Since the parallel component of the electric field is continuous across each point of the DBR dielectric boundaries [82], the in-plane momentum $k$ of escaping light is conserved and $k = k_0 \sin\theta$ where $k_0$ is the wavevector in free space (Fig. 2.5a).

Figure 2.6 includes results from a study [55] using the organic molecular dye BODIPY-Br. In the first part of this study, reflectivity measurements, shown in Fig. 2.6a, were made to tune the thickness of the organic film used. Here white light was shone on a microcavity mounted on a rotation stage such that the reflectivity as a function of viewing angle could be recorded. The dips in reflectivity at each angle, indicating absorption, were then plotted to reveal the lower and upper polariton branches—Fig. 2.6b. The second part, shown in Figs. 2.6c and 2.6d, was a photoluminescence experiment. These figures reveal emission from the lower polariton is strongly dependent on the cavity detuning $\Delta = \omega_c - \epsilon$. It is dependence such as this that we intend to capture with our theoretical models.

We do not discuss experimental methods in further detail. An overview of the principles is given in the review article Ref. [5], and a survey of some of the main measurement setups

[9]Establishing the microscopic processes responsible for thermalisation in systems of organic polaritons is a difficult problem. See Ref. [80] for a recent discussion.

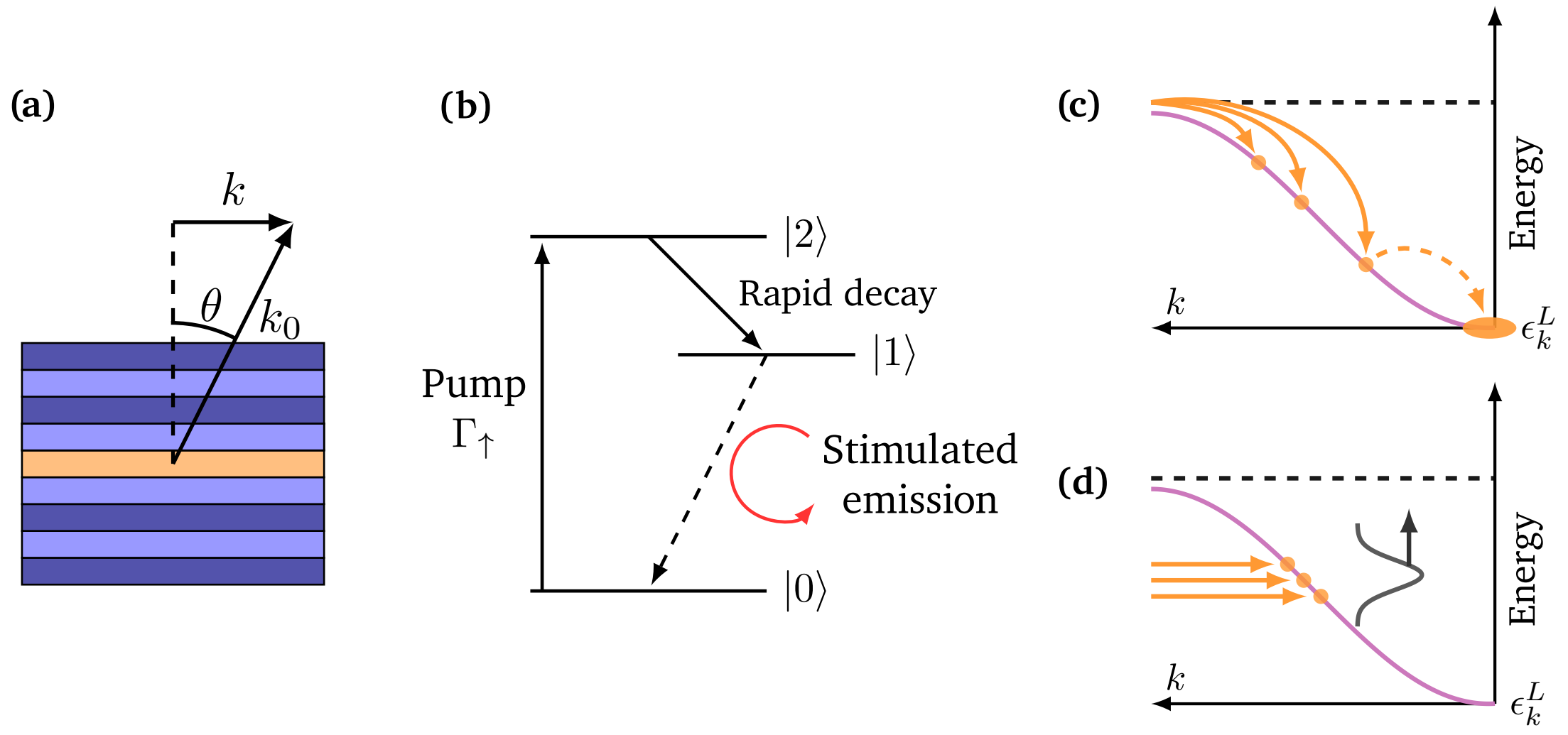

**Figure 2.5:** (a) The in-plane momentum $k$ is determined from the angle of emission $\theta$ and wavevector in free space $k_0$ as $k = k_0 \sin\theta$ [5]. (b) 3-level scheme for a photon laser discussed in Section 2.2.8. Strong pumping to a short-lived excited state $|2\rangle$ creates population inversion between excited state $|1\rangle$ and ground state $|0\rangle$, allowing for lasing operation between these states. (c) In polariton lasing experiments under non-resonant pumping, the system is pumped at high energies. This creates an exciton population that relaxes via scattering to populate the lower polariton branch. The polaritons may subsequently condense to the ground state [72, 83]. (d) Some transport experiments [71, 84, 85] instead use *resonant* pumping where a short laser pulse creates a Gaussian wavepacket by directly exciting a polariton branch [85, 86]. ↑TOF

and their operation may be found in the book [56]. Separate to these references, we wish to explain that all of the above spectroscopic quantities (absorption, photoluminescence etc.) are fundamentally related to certain response functions of the system, the photon's Green's functions. These functions provide the critical link between theory and experimental observations, as we demonstrate in Chapter 4. The characterisation of equilibrium and non-equilibrium many-body systems by Green's functions is a vast topic; we refer the reader to texts [2, 87, 88] where detailed treatment can be found. Based on these we discuss in Appendix A.2 key results relevant to the calculation of the spectroscopic variables, including the spectral weight and photoluminescence.

Finally, we make an important point about achieving strong coupling in microcavity experiments. The expression $\Omega = 2g\sqrt{N}$ might suggest the route to large splitting, and so strong coupling, is to increase the number of molecules $N$. However, increasing $N$ at a fixed molecular density increases the modal volume $V$ of the cavity, which *lowers* the individual couplings to light according to $g \propto 1/\sqrt{V}$ [89]. As a result, the normal situation in fact $g\sqrt{N}$ is a constant function of $N$. In this thesis we investigate behaviour of organic polaritons for large number of molecules, often considering $N \to \infty$ with the collective coupling $g\sqrt{N}$ fixed.

The relevant metric for the potential of a sample for strong coupling then is its dipole *density* [9], a product of the intrinsic coupling strength of the emitters to light and their density. A large part of the this chapter can be summarised by saying organic materials are good candidates for strong coupling due large dipole moments and the possibility of high preparation densities[10].

We next discuss two important applications of organic polariton systems that will be studied in Chapters 4 and 6.

[10]Plasmonic cavities (Section 2.2.10) realise the extreme limit where $V$ is so small strong coupling can be achieved with a few or even a single emitter [90].

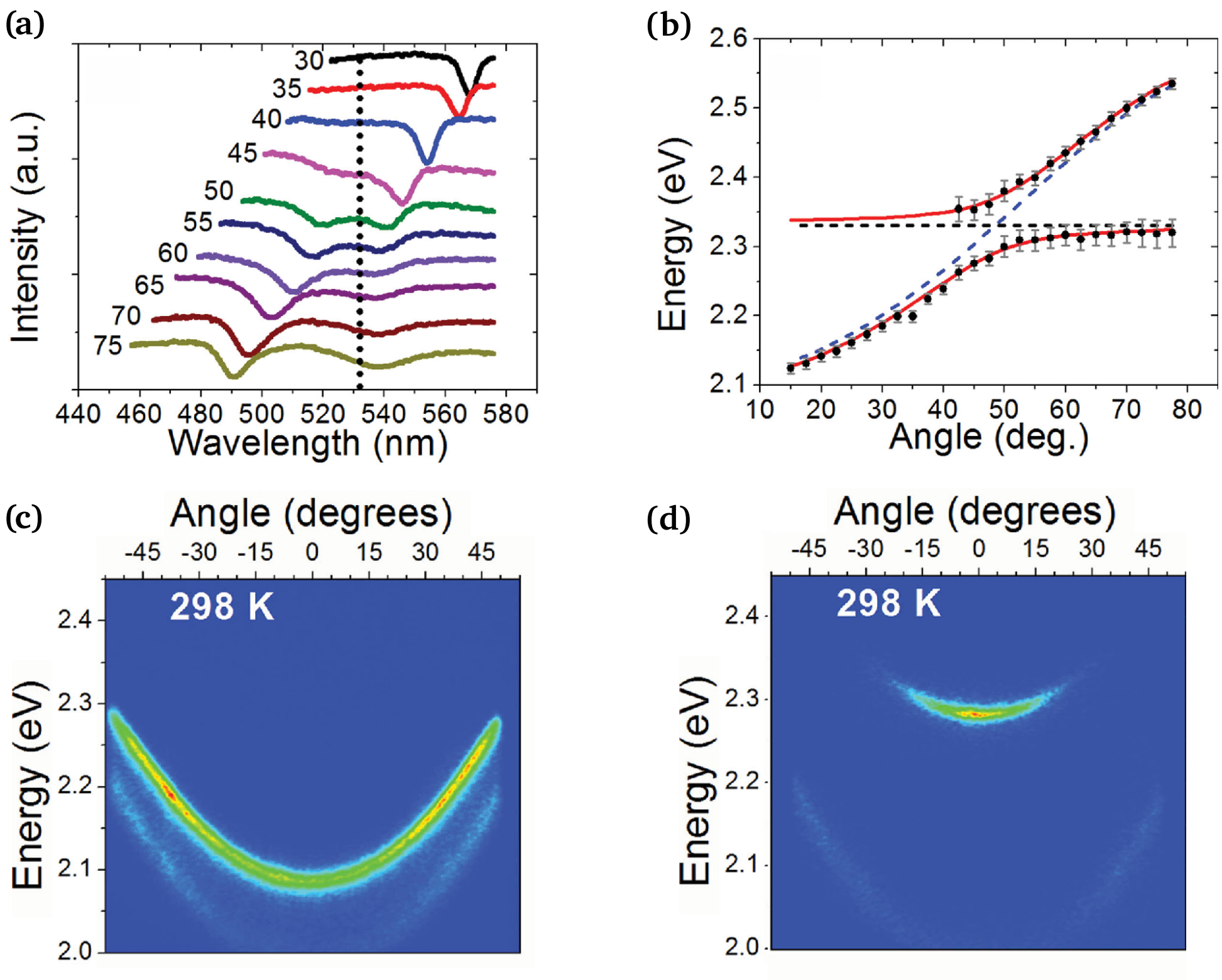

**Figure 2.6:** Strong coupling with the BODIPY-Br molecular dye [55]. (a) Intensity data from a reflectivity experiment with a metal-DBR cavity containing BODIPY-Br in a polystyrene matrix. Light from a tungsten-halogen lamp was focused on the cavity surface and the reflected beam collected at angles indicated for each line. Each dip in reflectivity provides a point on the dispersion (black point in (b)). Below viewing angles of 45 degrees it was not possible to discern a dip for the upper polariton at short wavelengths. The dotted vertical line indicates the exciton energy. (b) Upper and polariton dispersions (red) as a function of viewing angle (i.e., in-plane momentum) from fits of the reflectivity data. (c)-(d) Photoluminescence (PL) measurements for a DBR-DBR cavity at room temperature under non-resonant excitation (see Fig. 2.5c). The PL was measured with cavity detunings of (c) 290 and (d) 106 meV. Scattering was less efficient at larger detunings leading to reduced thermalisation and the intensity begin spread over entire branch (N.B. the plots were individually normalised). In particular, states higher up the lower polariton branch nearer the exciton reservoir remain significantly populated. The weak feature seen at lower energies in both panels is an experimental artefact. Figures reproduced from Ref. [55] with permission from John Wiley and Sons (Copyright 2016 WILEY-VCH Verlag GmbH & Co. KGaA, Weinheim). ↑TOF

### 2.2.8 Organic polariton lasing

Under sufficient pumping, systems of organic polaritons may condense into a coherent or *lasing* state. This phenomena has been demonstrated in a wide range of organic materials, including molecular crystals and aggregates [91–93], polymers [94, 95], dyes [55, 96], and fluorescent proteins [97, 98].

To a first approximation, polaritons at low excitation densities behave as a gas of interacting bosons with a small effective mass [5]. One may then look to a description of polariton condensation phenomena in terms of Bose-Einstein Condensation (BEC) [99]. This is the behaviour of particles obeying Bose-Einstein statistics in thermodynamic equilibrium where, below a critical temperature $T_c$ at which the chemical potential reaches the bottom of the dispersion, macroscopic occupation of the ground state occurs[11] [99]. This description would be appropriate for exciton-polaritons in a high quality microcavity for which the polariton lifetime is long enough for thermalisation to occur [100]. However, we have explained that experiments with organic materials are typically far from this regime and then a description closer to the *non-thermal* limit for coherence, which is conventional lasing, may apply [5].

In a photon laser, a state of population inversion of an electronic transition is obtained, usually by pumping to a short-lived, highly excited state [101] (Fig. 2.5b). If the inversion is sufficient for the net gain from stimulated processes (emission minus absorption) to exceed losses from the system, laser operation initiates whereby one or more[12] photon modes becomes macroscopically occupied and a highly coherent output results. This behaviour may be captured by semi-classical rate equations for gain versus loss in each mode [102] (see Ref. [103] for a similar treatment with organic emitters). Any such equations must realise the fact that the threshold for lasing depends not only on the extent of inversion of the electronic transition, but also on the total linewidth, which must not be too large so as to diminish the gain.

Condensation in systems of organic polaritons lies somewhere between an equilibrium condensate and a photon laser [5, 62, 63, 103]. We generally refer to organic polariton *lasing*, but both 'lasing' and 'condensation' are regularly used in the literature, depending on the intended experiment, the extent of equilibration and author preference.

To discuss a specific example of polariton lasing, first note that in polariton experiments pumping occurs in the form of non-resonant excitation of the dark exciton reservoir above the lower polariton branch [5] (Fig. 2.5c). Relaxation of the excitons via scattering produces a population of lower polaritons which *may* subsequently condense, i.e., lase, if a sufficient level of pumping is sustained. Whether lasing is achieved is readily observed in the photoluminescence (PL) spectrum. Refer to Figs. 2.7a and 2.7b, which show PL from a BODIPY-G1 microcavity system [96] below and above a lasing threshold. The rapid increase of emission and sharp decrease of linewidth (i.e., increased coherence) that occurs with the onset of lasing is demonstrated in Fig. 2.7c.

An additional feature which is often tracked in the development of coherence and condensation in polariton systems is a blueshift of the spectrum. As seen in Fig. 2.7d, the emission moves to higher energies through the transition. For inorganic semiconductor microcavities, blueshift can result from polariton-polariton and polariton-exciton interactions [72]. In organics, the extent of Frenkel excitons limits such interactions and the behaviour likely results from a combination of several effects, including saturation of the optical transitions at higher densities; see Ref. [104] for a complete discussion.

Next, we note the driven-dissipative Tavis-Cummings model contains the basic ingredients to describe lasing. Let $\Gamma_\uparrow$ and $\Gamma_\downarrow$ denote the rate of incoherent pump and decay of each emitter in model, respectively. It can be shown that, as the ratio $\Gamma_\uparrow/\Gamma_\downarrow$ increases, the system transitions from a normal state, with no photons in the cavity, to a lasing one where the number of photons scales with $N$ [41, 105]. This behaviour is analogous to that of conventional lasing under population inversion of the ensemble. What the Tavis-Cummings model lacks, of course, is vibrational physics. In this regard there have been recent studies of lasing in organic systems [62, 63, 103] working either under weak-coupling assumptions or with Holstein-Tavis-Cummings models containing a single vibrational mode. From these studies we highlight several interesting results, in preparation for our own model of polariton lasing in Chapter 4.

[11]This strictly applies to an ideal bosonic gas in $d = 3$ dimensions. For $d = 2$, BEC does not occur at finite temperatures in a uniform gas [99]. However, a separate phase transition (after Berezhinskii–Kosterlitz–Thouless) associated with the proliferation of vortex-antivortex pairs may still produce an ordered state with phase correlations over finite regions [99].

[12]The number of modes that lase depends on the mode spacing (set by the cavity dimensions) and the type of broadening—inhomogeneous or homogeneous—that may allow for the competition between different modes, or not [101].

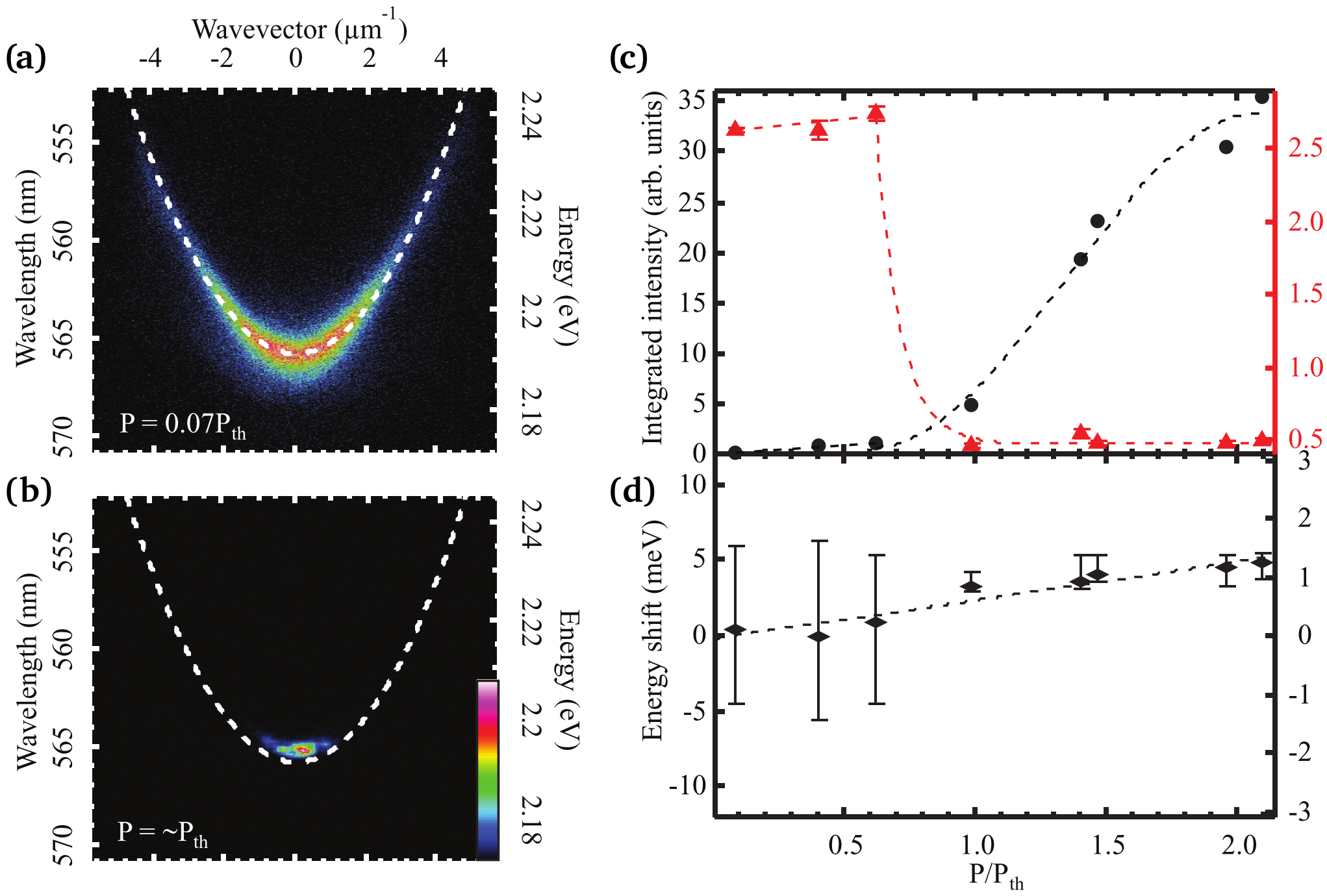

**Figure 2.7:** Polariton lasing for the BODIPY-Br molecular dye [96]. Normalised photoluminescence (PL) (a) below and (b) at a threshold $P_{\text{th}}$ pump intensity for which emission from the lower polariton branch collapsed to the bottom of the dispersion ($P_{\text{th}} \sim 500\ \mu\text{J cm}^{-2}$). The threshold is associated with (c) a sharp reduction in linewidth and rise in output intensity. (d) A blueshift of the spectrum also occurs with increasing pump strength. Figure adapted with permission of John Wiley and Sons from Ref. [96] (Copyright 2017 WILEY-VCH Verlag GmbH & Co. KGaA, Weinheim). ↑TOF

First, lasing in these models does not necessarily require electronic inversion. This is in contrast to the conventional photon laser, where inversion is required for positive gain [101]. We will find lasing without inversion for our model in Chapter 4 too. The possibility of low-threshold lasing is a major attraction of organics in applications, as we discuss shortly below.

Second, Ref. [62] observed re-entrant behaviour: as the pump $\Gamma_{\uparrow}$ was continually increased, lasing switched on, then off, and finally on again. This relied on frequency locking between the polariton and a particular vibrational side-band, so we may not expect to observe the same behaviour in our model which features a continuum of vibrational modes.

Third, Ref. [63] investigated *multimode* organic lasing, and found switching between modes according to system parameters. For our work in Chapter 4, we will not consider this complexity, either working with a single-mode model or a multimode model where condensation is assumed to occur in the $k = 0$ mode. A single $k = 0$ lasing mode was true for the model in Ref. [63] at small or positive detunings.

For further information and the state of the art of organic polariton lasing, we refer to recent review articles [5, 50, 72, 106].

We conclude this section by noting the main attractions of organic materials for lasing devices [107]. Foremost amongst these is room temperature operation. In fact, the threshold of organic lasers is often nearly independent of temperature [5]. Moreover, this threshold can be low, a crucial factor in the development of low-powered devices.

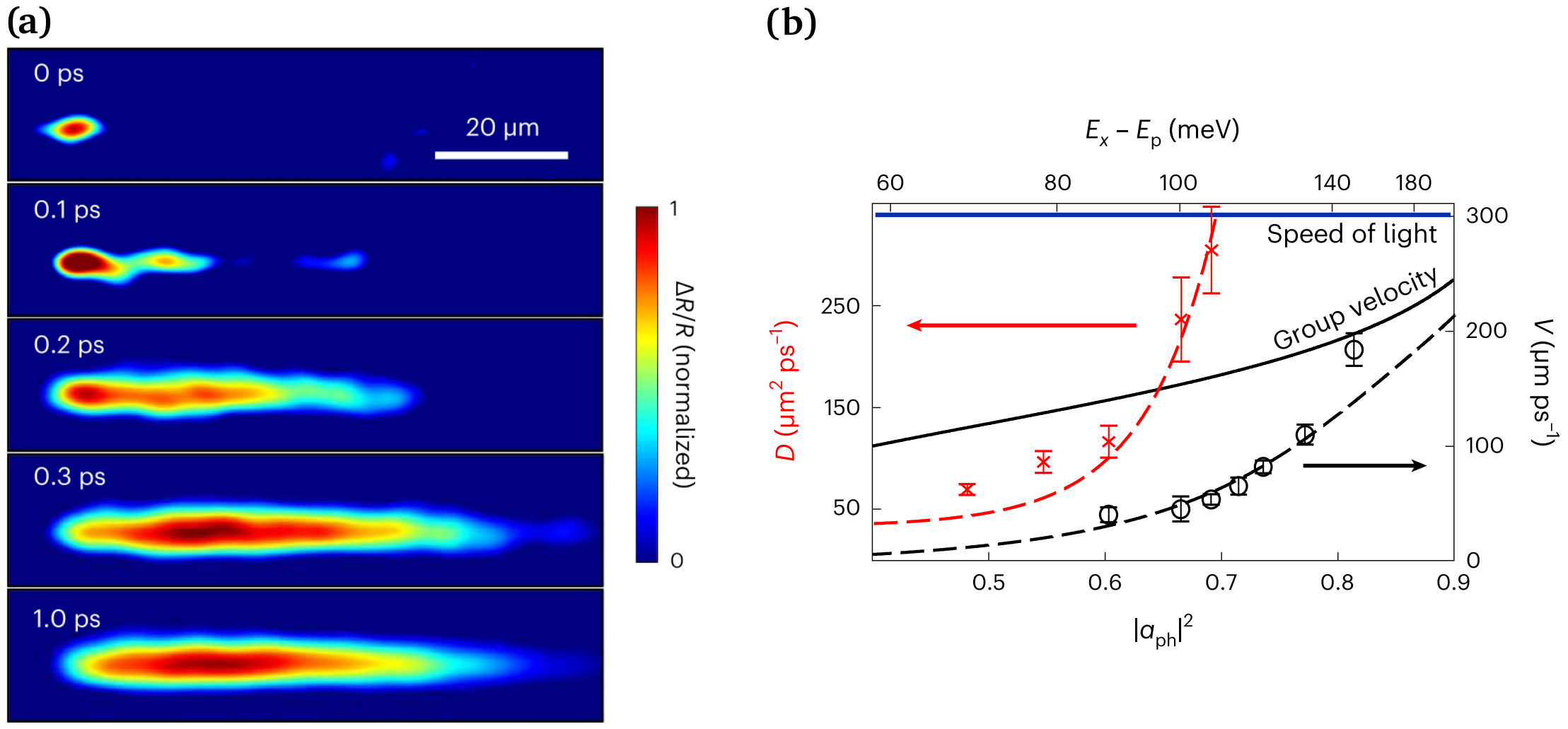

**Figure 2.8:** Spatially resolved dynamics of organic polaritons [116]. (a) Snapshots of time-resolved microscopy showing the expansion of a polariton cloud following non-resonant excitation at a photonic fraction of 82%. (b) Diffusive (red crosses) and ballistic (black circles) transport was observed dependent on the photonic fraction $|\alpha_{\text{phot}}|^2$. The ballistic transport occurs noticeably below the polariton group velocity. The dashed lines show the prediction of a kinetic model [116] for reversible scattering to dark exciton states which we examine in Chapter 6. Figures reproduced with permission from Springer Nature. ⮥TOF

Next, the flexibility of carbon chemistry and the vast number of organic compounds that might be considered. Combined with the ease and cost-effectiveness of fabricating organic films for microcavities [50], there is opportunity to design organic devices with properties tailored to the specific demands of an application. In addition, organic lasing is not limited to the microcavity setup. Another realisation we consider below are *nano*-plasmonic cavities which hold promise ultra-small lasing devices [6, 108, 109].

The properties of organic systems described in the preceding sections that allow for the possibility of room-temperature, low-threshold, and tunable devices also make determining the optimum conditions for lasing a challenging one. This is because one must take into account the effect of the vibronic coupling for each molecule. As explained above, this coupling is generally strong and so beyond weak-coupling or other simplified treatments. This motivates our work in Chapter 4 where we show a method capable of handling the complex vibrational density of states for many molecules in a realistic model of an organic laser.

### 2.2.9 Organic polariton transport ⮥

A second promising application of organic polaritons is polariton-mediated transport. While exciton transport is generally limited by the localised nature of Frenkel excitons, due to their optical character—namely their small effective mass—organic polaritons may travel with far greater efficiency [72]. Polariton enhanced transport in organic materials has the potential to be harnessed for high-speed, long-range transport of energy in optoelectronic devices and quantum circuitry [8, 72, 110–115].

Recently, the development of ultrafast microscopy techniques has allowed for spatially-resolved imaging of organic polaritons on femtosecond timescales [71, 84, 85, 116–118]. Data from a recent experiment [116] is included in Fig. 2.8. These and similar results have triggered great interest in understanding the mechanism of polariton transport in real systems with dynamical or static disorder [71, 72, 78, 85, 86, 119–121].

In the ultrafast microscopy studies, an initially localised polariton is prepared using non-resonant or resonant excitation (Fig. 2.3c), and the subsequent expansion of the excitation density, a polariton cloud, is observed. Commonly, distinct diffusive (slow) and ballistic (rapid) transports regimes are found, depending on the character—photonic versus excitonic—of the excitation. Critically, the ballistic transport, which is relevant to potential applications, has consistently been found to occur at speeds below the polariton group velocities $v_g$.

It is the last observation in particular that has led to much discussion regarding the nature of organic polariton transport. Specifically, the role dark states may have [71, 86, 116, 119, 122–124]. Dark states are stationary (or slowly diffusing), so conversion to and from these states could serve to slow the polaritons that would otherwise propagate at $v_g$. In Chapter 6 we look to investigate this possibility. There we explain in more detail the observations of the transport experiment [116] shown in Fig. 2.8, and develop a method to calculate spatially resolved dynamics of transport in a model based on perovskite materials [85, 125, 126].

For a recent review on studies of transport in organic materials, see Ref. [72].

### 2.2.10 Other systems of organic polaritons

Although exciton-polaritons in a planar microcavity are a mainstay for organic polariton studies, organic polaritons are supported by many other structures and are not limited to electromagnetic radiation in the form of visible light or even to electronic excitations. We point out two realisations of organic polaritons relevant to the applications of lasing (plasmonic nanocavities) and transport (Bloch surface wave polaritons), as well as third (vibrational polaritons) of interest not related to our work.

#### Plasmonic nanocavities

Near the surface of metals collective excitations of conduction electrons gives rise to electromagnetic modes called surface plasmons[13] [127, 128]. These may be harnessed in nanostructures where the surface modes of two metals placed closed together hybridise to provide confinement of light on the nanoscale [109]. The exceptionally small modal volume allows for strong light-matter coupling with a few or even a single quantum emitter trapped in the gap between metals [90].

The most robust realisation, which has been used to form sub-nanometer gaps [129, 130], is the nanoparticle-on-mirror structure [109, 131]. Figure 2.9a shows an example where a gold (Au) nanoparticle traps several organic emitters above a metallic film. Organic polariton lasing in this system has recently been studied [6] for the potential development of ultra-small lasing devices. We point out this example since, as we explain in the next chapter, the cumulant expansion methods we develop in this thesis are useful not only for describing many-body systems of $N \gg 1$ particles, but may also be used to study complex systems of small and intermediary sizes.

Besides nanolasers, plasmonic nanocavities are promising for other cutting edge nanoscale technologies such as biosensors [132] and molecular junctions [132]. We refer to a recent review article [109] of this new field for further information.

[13] Surface plasmon *polaritons* is a more accurate term, since they form under resonant interaction between light and free electrons [127]. But we already have enough uses of 'polariton'!

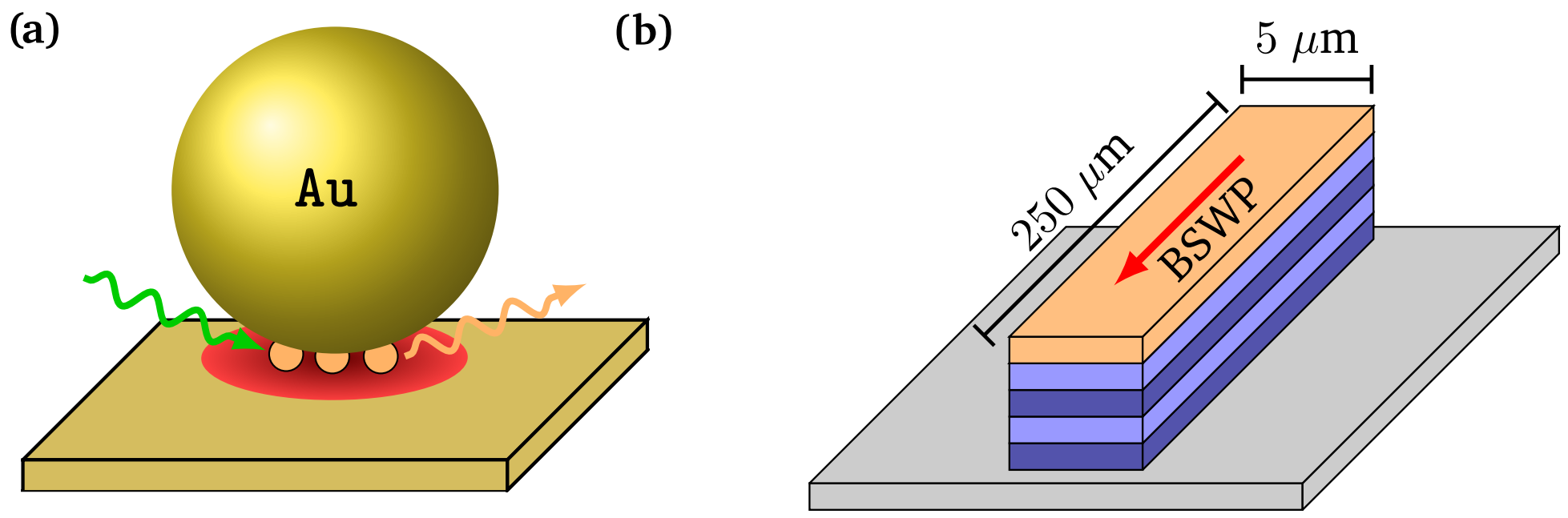

**Figure 2.9:** (a) Nanoparticle-on-mirror structure [109, 129, 130]. A gold (Au) nanoparticle traps several molecular emitters above a metallic film, here also gold. This geometry has been used to construct plasmonic nanocavities with sub-nanometer spacings [129, 130] and has applications ranging from biosensing to ultra-small lasing devices [6, 109]. (b) Polariton waveguide for guided transport by block surface wave polaritons (BSWPs) [113]. The device comprises a thin film of organic material deposited on a DBR stripe. ↑TOF

### Bloch surface wave polaritons

Another set of electromagnetic surface modes that may be used for strong coupling with excitons in organic or inorganic materials are those supported on the surface of periodic multilayered dielectric structures [133] such as DBRs. These are called Bloch surface waves [133] and exciton-polaritons formed under strong coupling to these modes Bloch Surface Wave polaritons (BSWPs). The strong confinement and electric field enhancement presented by the surface modes make BSWPs well suited to applications requiring guided low-loss, long-range energy transport [118, 134]. The advantages of organic materials in ease of fabrication, tunability and room temperature operation all apply here.

Figure 2.9b shows an organic BSWP device [113] that was recently used to achieve guided transport over distances as long as $60\ \mu$m at room temperature. For a second example, and the connection to our own study of organic polariton transport, the ultrafast imaging experiment in Ref. [116], whose results we discuss further in Chapter 6, investigated BSWPs formed by strong coupling with an organic semiconductor. We note however the model we develop is not limited to a particular polariton dispersion. In fact, in Chapter 6 we use a dispersion for a planar micro-cavity [57] in line with another recent work [85] on transport in halide perovskites.

### Vibrational polaritons

A separate class of organic polaritons are those formed from the hybridisation of optical cavity modes and molecular *vibrations* [135, 136]. Collective vibrational strong coupling (VSC) can be reached at room temperatures with Rabi splittings between vibrational polaritons also scaling with $\sqrt{N}$ [135, 136]. We wish to distinguish this type of polaritons from the exciton-polaritons we consider (both may be referred to as 'molecular polaritons' in the literature [137]), but also point out the rapidly developing field of vibropolaritonic chemistry. This explores the potential for modified chemical kinetics under VSC [138]. We note in particular that the potential importance of dark states in this context has also been discussed [138–140]. For further information, see reviews [141–144] of vibropolaritonic chemistry and more generally the modification of photo-physics and chemical properties not limited to VSC [114, 145–147].

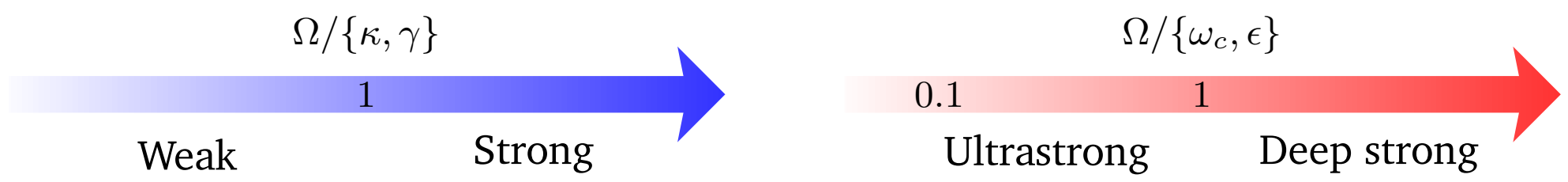

**Figure 2.10:** Regimes of light-matter interaction. The systems we consider in this thesis are in the regime of strong light-matter coupling where the Rabi splitting $\Omega$ (collective coupling $2g\sqrt{N}$) exceeds cavity and emitter linewidths. They are not however in the ultrastrong (or further deep strong) regime where $\Omega$ is a significant fraction of (or exceeds) the bare system energies $\omega_c$, $\epsilon$ [148, 149]. Note the scales shown are technically separate, so it is possible to envision a system where the light-matter coupling is both weak ($\Omega < \kappa, \gamma$) and ultrastrong ($\Omega \gtrsim 0.1\omega_c, \epsilon$), for example [150]. TOF

Hereafter 'polariton', without further qualification, should be taken to refer to exciton-polaritons in organic materials not necessarily limited to a particular form of electromagnetic radiation or cavity structure. Similarly, 'vibrational strong coupling' will only be used to describe the coupling of vibrations with electronic degrees of freedom, not light. That being said, for our studies in Chapters 4 and 6, we will consider specific organic materials in microcavities (BODIPY-Br and halide perovskites, respectively), in mind of deriving realistic models.

### 2.2.11 Ultrastrong light-matter coupling

The Tavis-Cummings Hamiltonian Eq. (2.2.4) provides a rich set of behaviours sufficient to describe light-matter interactions in the strong coupling regime, where the collective light-matter coupling (Rabi splitting) $\Omega = 2g\sqrt{N}$ is larger than losses $\kappa$, $\gamma$, but less than the system energies $\omega_c$, $\epsilon$ ($\hbar = 1$). However, experiments (e.g., [151–154]) have been able to achieve even higher light-matter coupling strengths, with splittings $\Omega \sim 1$ eV comparable to the system energies. In this regime of *ultrastrong coupling* (Fig. 2.10), additional anti-resonant (or 'counter-rotating') terms $a^\dagger\sigma_n^+$, $a\sigma_n^-$, must be considered in the light-matter interaction [155]:

$$H = \omega_c a^\dagger a + \sum_{n=1}^{N}\left[\frac{\epsilon}{2}\sigma_n^z + g\left(a\sigma_n^+ + a^\dagger\sigma_n^-\right) + g\left(a\sigma_n^- + a^\dagger\sigma_n^+\right)\right] \tag{2.2.16}$$

$$= \omega_c a^\dagger a + \sum_{n=1}^{N}\left[\frac{\epsilon}{2}\sigma_n^z + g\left(a + a^\dagger\right)\sigma_n^x\right]. \tag{2.2.17}$$

This is known as the Dicke model [156, 157]. It presents a more general model of light-matter interaction that reduces to the Tavis-Cummings model outwith the ultrastrong coupling regime [155]. This is because in the interaction picture (see Section 3.1), the operators $a(t)$, $a^\dagger(t)$, oscillate as $e^{i\omega_c t}$, $e^{-i\omega_c t}$ respectively, and similarly $\sigma_n^\pm(t)$ as $e^{\mp i\epsilon t}$. Hence the anti-resonant terms oscillate rapidly with $e^{\pm i(\omega_c+\epsilon)t}$, at optical frequencies of the order of $1$ eV$\sim 10^{15}$ Hz. The contribution of these terms is then effectively an average over a very large number of cycles and, provided the light-matter coupling strength is not too large [158], may be neglected compared to contributions from the resonant terms, which vary relatively slowly according to the detuning $\omega_c - \epsilon$.

The neglect of anti-resonant terms from the light-matter interaction is referred to as a *rotating-wave approximation* (RWA). In this thesis we will always assume a regime where the RWA provides a good approximation, given this will be sufficient capture the physics in the polariton lasing and transport experiments discussed above. Our methods can readily accommodate the anti-resonant terms, however.

The effects of ultrastrong coupling in light-matter interactions for systems of organic molecules have been considered elsewhere [46, 62, 65, 150, 159, 160]. See Refs. [149, 161] for recent reviews.

# Chapter 3

# Methods ↑TOC

> The self-referential nature of the exercise is thus apparent: the average state of the system is both an explanatory variable and the variable itself to be explained.
>
> Fernando Vega-Redondo

## Contents

We now establish the methodological groundwork for the thesis. As explained in the previous chapter, the methods we develop are to be applied to realistic models of open quantum systems. Specifically, many-body ones (realistic systems are not limited to those of a few parts!).

Section 3.1 starts with the foundations: the theory of open quantum systems. This includes the basic mathematics and notations that will be used throughout the thesis. Building from this, in Section 3.2 we explain how tensor network methods, specifically the time-evolving matrix product operator (TEMPO) method [162], may be used to calculate exact open quantum system dynamics. This material is relevant to the method we develop in Chapter 4. Next, Section 3.3 introduces mean-field theory, which provides a means to address many-body problems. It is no exaggeration to say that mean-field theory underlies all work of this thesis. We then go beyond mean-field theory in Section 3.4 by introducing cumulant expansions of the Heisenberg equations of motion. These expansions will be further explored and put to use in Chapters 5 and 6.

The text here is intended to be accessible to researchers in any field, assuming at most familiarity with an undergraduate level of quantum mechanics—Schrödinger's equation and Dirac notation—and related concepts from linear algebra, e.g., Hilbert space and tensor product, although we recap many of the main features or otherwise provide references for additional reading (Ref. [163] contains a good primer on linear algebra and quantum mechanics).

## 3.1 Theory of open systems

### 3.1.1 Problem to be addressed

We firstly discuss the problem of open quantum systems and the mathematical framework developed to address it. Our treatment is introductory, aiming to cover the theoretical concepts and mathematics required to understand the approaches developed in this thesis. A primary target is the description of open quantum system dynamics in terms of Heisenberg equations of motion, since these provide an operational starting point for the main part of our methods. We cover additional aspects of relevance in detail: correlations in multipartite systems, harmonic environments and the Markovian master equation. For further information we refer to established textbooks on quantum mechanics [163, 164] and open systems [4] as well as review and tutorial-style articles [22, 165–167].

As discussed in Chapter 2, the physical picture of an open quantum system is a (typically) small subset of a larger interacting quantum system that is of experimental interest. The mathematical division is of the total Hilbert space $\mathcal{H}$ into a product $\mathcal{H} = \mathcal{H}_S \otimes \mathcal{H}_E$ where $\mathcal{H}_S$ is the system of interest and $\mathcal{H}_E$ its environment. The practical statement of this division, which defines the problem to be solved, is of total Hamiltonian $H$,

$$H(t) = H_S(t) + H_E(t) + H_{SE}(t), \tag{3.1.1}$$

where $H_S = H_S \otimes I_E$ acts non-trivially only on the system degrees of freedom, $H_E = I_S \otimes H_E$ on the environment, and $H_{SE}$ acts on both, hence describing an interaction. For brevity we will omit tensor products with identity operators where the behaviour of an operator can be reasonably inferred from notation (so $H_S$ acts only the system, $\sigma_1^x$ only on the first spin, etc.). The division or *factorisation* of operators, states and structures between Hilbert spaces will be a key motif in our work, so pay attention.

In Eq. (3.1.1) we allowed for the possibility of explicit time-dependence, $H = H(t)$, in which case the total system is not closed in the strict sense since its energy is changing. For example, the entire system may be subject to a periodic drive. The term *isolated* [4] may be reserved specifically for the case when $H$ is time-independent and energy is conserved, but for our purposes the distinction will not be important. Similarly, we will be content to use the general term *environment* where one might see *reservoir* to describe an environment with infinite degrees of freedom forming a continuum of energies and *bath* if it is further in a thermal equilibrium state [4].

The problem of open quantum system theory is the following. The environmental degrees of freedom are too vast or complex to handle directly, and moreover not of direct interest. How can an efficient description be obtained for the small subset we are actually interested in, the system, whilst accounting for its interaction with the environment? The complexity of this problem for real systems means that assumptions—or approximations—are required to proceed. Broadly, these occur at the level of describing the environment, i.e., in choosing a model for $\mathcal{H}_E$, and then in treating the system-environment interaction. In order to understand these assumptions we must firstly establish a mathematical framework for open quantum system dynamics using the language of density matrices.

### 3.1.2 Schrödinger dynamics and observables

We start from the Schrödinger formulation of quantum dynamics where unit vectors $|\psi\rangle$ of a Hilbert space $\mathcal{H}$ describe physical states, and evolve according to

$$\partial_t |\psi(t)\rangle = -iH(t) |\psi(t)\rangle , \tag{3.1.2}$$

where we continue to set $\hbar = 1$ and use $\partial_t$ to denote a derivative with respect to time. A solution to this equation may be expressed using a unitary time-evolution operator $U(t, t_0)$,

$$|\psi(t)\rangle = U(t, t_0) |\psi(t_0)\rangle , \tag{3.1.3}$$

such that $U(t, t_0)$ propagates the state from $t_0$ to $t$. For notational simplicity from now on we take the initial time $t_0 = 0$ and write $U(t, 0) = U(t)$. Adjust your stopwatches accordingly.

Differentiating both sides of Eq. (3.1.3),

$$-iH(t) |\psi(t)\rangle = (\partial_t U(t)) |\psi(t_0)\rangle . \tag{3.1.4}$$

Therefore $U$ must itself satisfy

$$\partial_t U(t) = -iH(t)U(t). \tag{3.1.5}$$

If $H$ is time-independent, this (or equally Eq. (3.1.2)) may be directly integrated to obtain $U(t) = e^{-iHt}$ and $|\psi(t)\rangle = e^{-iHt} |\psi(0)\rangle$. For the time-dependent case there is the subtlety that Hamiltonian operators $H(t')$, $H(t'')$ at different times $t' \neq t''$ do not necessarily commute, and then the solution is instead a *time-ordered* exponential[1],

$$U(t, t_0) = \mathcal{T} \exp \left[ -i \int_{t_0}^{t} dt' H(t') \right] , \tag{3.1.6}$$

[1] This ensures operators in the series expansion of $\exp(\ldots)$ occur in the right order. See Appendix B.1 for a derivation.

where the time ordering operator $\mathcal{T}$ is simply an instruction to arrange operators with the earliest times to the *right*:

$$\mathcal{T}H(t'')H(t') = \begin{cases} H(t'')H(t') & t'' > t' \\ H(t')H(t'') & t'' < t'. \end{cases} \tag{3.1.7}$$

Next, *observables* are Hermitian operators $A : \mathcal{H} \to \mathcal{H}$ such that all possible outcomes of a measurement of $A$ are described by its eigenvalues, and the expected outcome for a system in state $|\psi\rangle$ by the inner product

$$\langle A \rangle = \langle \psi | A | \psi \rangle \, . \tag{3.1.8}$$

Here *expectation* reflects the probabilistic nature of quantum theory: unless $|\psi\rangle$ happens to be an eigenstate of $A$, one cannot predict with certainty what value the measurement will reveal, only the probabilities for the different possible outcomes. Put another way, $\langle A \rangle$ is the *average* obtained from many measurements of identical systems prepared in $|\psi\rangle$.

### 3.1.3 Mixed states and density matrices

Knowledge of the state vector $|\psi\rangle$ means knowing all physical information that is possible to determine for a quantum system. The system is said to be in a *pure* state. This is not the typical case. For example, an imperfect experiment will result in the preparation of not a single, precise state $|\psi\rangle$ but rather an uncertain mixture of states $|\psi_i\rangle$ with probabilities $p_i$. This leads to the generalisation of *mixed* states described by a density operator or matrix $\rho$,

$$\rho = \sum_i p_i \, |\psi_i\rangle\langle\psi_i| \, , \tag{3.1.9}$$

where $p_i \geq 0$ are such that $\sum_i p_i = 1$ ($\rho$ is a convex linear combination). This form is motivated by producing the correct average for expectation values of system observables, as is shown below. To describe a real system, a density operator must be Hermitian ($\rho^\dagger = \rho$), of unit trace $\operatorname{Tr} \rho = 1$ and positive semi-definite [164]. The description of a pure state is obtained when a single $p_i = 1$ and all others are zero, $\rho = |\psi_i\rangle\langle\psi_i|$, in which case $\operatorname{Tr} \rho^2 = 1$ is also satisfied.

Observe Eq. (3.1.9) is a *classical* mixture of *quantum* states—a dichotomy we will see reflected in the calculation of expectations, and later in the nature of correlations. The density matrix, then, contains both statistical and physical information. Critically, it allows the expected outcome of $A$ to be calculated for the mixture via a trace, $\langle A \rangle = \operatorname{Tr}(A\rho)$. To see this, let $|j\rangle$ denote a complete basis for $\mathcal{H}$ such that $\sum_j |j\rangle\langle j|$ is the identity operator. Then

$$\begin{aligned} \langle A \rangle &= \sum_i p_i \, \langle \psi_i | A | \psi_i \rangle & (3.1.10) \\ &= \sum_{ij} p_i \, \langle \psi_i | j \rangle \langle j | A | \psi_i \rangle & (3.1.11) \\ &= \sum_{ij} p_i \, \langle j | A | \psi_i \rangle \langle \psi_i | j \rangle & (3.1.12) \\ &= \sum_j \, \langle j | A\rho | j \rangle & (3.1.13) \\ &= \operatorname{Tr}(A\rho). & (3.1.14) \end{aligned}$$

Here, in the first line, a classical average over the statistical ensemble (probabilities $p_i$) is combined with a quantum expectation for each state $|\psi_i\rangle$—that's the dichotomy.

A second way in which mixed states arise that is most important to our discussion is in the description of *composite* quantum systems. Suppose the total state $\rho$ of a composite system $\mathcal{H} = \mathcal{H}_S \otimes \mathcal{H}_E$ is known. Then $\rho$ contains complete information of the system *and* the environment. The number of elements of this matrix, $(\text{dim}(\mathcal{H}_S)\text{dim}(\mathcal{H}_E))^2$, make it an impractical—likely impossible—way to keep track of the system degrees of freedom. However, it is possible to extract from $\rho$ a *reduced density matrix* $\rho_S$ containing all pertinent information for the system. To see how this is done, consider the expression for the expectation of an observable $A_S$ on $\mathcal{H}_S$:

$$\text{Tr}(A_S\rho) = \text{Tr}_{S\otimes E}(A_S \otimes I_E\rho). \tag{3.1.15}$$

The trace can be performed in two steps: a trace over the environment, and then over the system. As $A_S$ acts only on $\mathcal{H}_S$, it can be taken out of the first trace,

$$\langle A_S\rangle = \text{Tr}_S\left(A_S\,\text{Tr}_E(\rho)\right) \equiv \text{Tr}_S\left(A_S\rho_S\right), \tag{3.1.16}$$

where $\rho_S = \text{Tr}_E(\rho)$ defines a density operator on $\mathcal{H}_S$ [4].

The reduced system density matrix $\rho_S$ is then of a manageable dimension, $\text{dim}(\mathcal{H}_S)^2$, yet allows the expectation of any observable of $\mathcal{H}_S$, that is, the expected outcome of any possible measurement on the system, to be calculated. It is a clear target in the analysis of an open quantum system.

Crucially, even when the total system is in a pure state, $\rho = |\psi\rangle\langle\psi|$, $\rho_S$ will in general be mixed. As we now discuss, this results from *entanglement* between the system and its environment.

### 3.1.4 Correlations and separability

Assumptions on the extent of correlations are critical in both the standard operation of open quantum system theory and the mean-field approaches we explain below. Fundamentally, a correlation is the failure of two variables to be statistically independent. For state variables, correlations may be either quantum or classical in nature.

Quantum correlations arise in the form of entanglement. This is the observation that the quantum states of different parts of a composite system cannot necessarily be described independently. Consider a pure state $\rho = |\psi\rangle\langle\psi|$ on $\mathcal{H} = \mathcal{H}_S \otimes \mathcal{H}_E$. If $|\psi\rangle$ can be written as a single tensor product, $|\psi\rangle = |\varphi\rangle \otimes |\eta\rangle$ for $|\varphi\rangle \in \mathcal{H}_S$ and $|\eta\rangle \in \mathcal{H}_E$, then it is not entangled and is called *separable*. If this is not possible, $|\psi\rangle$ is entangled.

The qualification *single* is essential here. For if $\{|i\rangle_S\}$ and $\{|j\rangle_E\}$ are bases of $\mathcal{H}_S$ and $\mathcal{H}_E$, respectively, then $|i\rangle_S \otimes |j\rangle_E$ provides a basis of $\mathcal{H}$ and there exist coefficients $c_{ij}$ such that

$$|\psi\rangle = \sum_{ij} c_{ij}\,|i\rangle_S \otimes |j\rangle_E\,. \tag{3.1.17}$$

So $|\psi\rangle$ can always be written as a linear combination of tensor products. Only when this linear combination can be reduced to a single term, meaning $c_{ij} = a_i b_j$ is the outer product of two vectors, is $|\psi\rangle$ separable. The last observation suggests a possible characterisation of entanglement according to the *singular value decomposition* of $c_{ij}$. Since singular value decompositions will be a vital tool in handling tensor networks further below, we cover this characterisation in some detail.

The singular value decomposition (SVD) of a matrix $C \in \mathbb{C}^{m\times n}$ is a factorisation [168]

$$C = U\Sigma V^\dagger, \tag{3.1.18}$$

where $U = [\boldsymbol{u}^{(1)}, \boldsymbol{u}^{(2)}, \ldots, \boldsymbol{u}^{(m)}] \in \mathbb{C}^{m\times m}$ and $V = [\boldsymbol{v}^{(1)}, \boldsymbol{v}^{(2)}, \ldots, \boldsymbol{v}^{(n)}] \in \mathbb{C}^{n\times n}$ are unitary and $\Sigma \in \mathbb{R}^{m\times n}$ is diagonal. The diagonal entries $\sigma_1$, $\sigma_2$,... of $\Sigma$ are non-negative and called the *singular values* of $C$. This decomposition always exists and the number of non-zero singular values equals the rank of $C$ [168].

Equation (3.1.18) can be written using the column vectors $\boldsymbol{u}^{(i)}$ and $\boldsymbol{v}^{(j)}$ from $U$ and $V$ as

$$C = \sum_{r=1}^{\mathrm{rank}(C)} \sigma_r \boldsymbol{u}^{(r)} \otimes \overline{\boldsymbol{v}}^{(r)}, \tag{3.1.19}$$

or, in terms of components,

$$c_{ij} = \sum_{r=1}^{\mathrm{rank}(C)} \sigma_r u_i^{(r)} \overline{v}_j^{(r)}, \tag{3.1.20}$$

where an overline denotes a complex conjugate. We see that, if only one singular value is non-zero ($\mathrm{rank}(C) = 1$), then $C$ is the outer product of two vectors and we have found $|\varphi\rangle := \sigma_1 \sum_i u_i^{(1)} |i\rangle_S$, $|\eta\rangle := \sum_j \overline{v}_j^{(1)} |j\rangle_E$ such that

$$|\psi\rangle = |\varphi\rangle \otimes |\eta\rangle \,. \tag{3.1.21}$$

In the context of quantum information theory, the above decomposition is known as the *Schmidt decomposition* and the number of non-zero singular values the Schmidt number $\alpha_{\text{Sch.}}$ [163]. The latter is hence a measure of entanglement for a bipartite system: if $\alpha_{\text{Sch.}} = 1$ then the state is separable, while $\alpha_{\text{Sch.}} > 1$ implies entanglement. The maximum case, $\alpha_{\text{Sch.}} = \min\{\dim(\mathcal{H}_S), \dim(\mathcal{H}_E)\}$, classifies a maximally entangled state.

Unfortunately this characterisation is not readily extended to describe multipartite ($> 2$) entanglement as relevant to our study of many-body systems. In general, a pure state $|\psi\rangle$ on a Hilbert space $\mathcal{H} = \otimes_{i=1}^N \mathcal{H}_i$ is separable if it can be written as a product state,

$$|\psi\rangle = |\varphi_1\rangle \otimes |\varphi_2\rangle \otimes \ldots \otimes |\varphi_N\rangle = \bigotimes_{i=1}^{N} |\varphi_i\rangle \,. \tag{3.1.22}$$

Otherwise, it is entangled. Many different measures of multipartite entanglement exist such as concurrence and entanglement of formation [169], but we will not have any need for them. Instead, we will simply postulate that only a certain level of entanglement (more specifically, correlations) may exist in the system, and determine the practical consequences of that assumption.

Note the prescription of entanglement is a subjective one in the following sense. Consider a three particle system $\mathcal{H} = \mathcal{H}_1 \otimes \mathcal{H}_2 \otimes \mathcal{H}_3$ where particles 1 and 2 are interacting, and so entangled, whilst particle 3 is entirely decoupled. Then the total state considered across the three particle subspaces is entangled. Yet if you grouped particles 1 and 2 into a single, composite particle, $\mathcal{H}_{12} = \mathcal{H}_1 \otimes \mathcal{H}_2$, the state qualifies as separable (on $\mathcal{H}_{12} \otimes \mathcal{H}_3$). Normally an appropriate division of $\mathcal{H}$ is obvious, e.g., the physical sites of a lattice or the extent of the system vs. environment set by experimental apparatus, but one should ensure this division is clear when making statements regarding entanglement.

For a pure state, the notation of entanglement translates simply to the language of density matrices: $\rho = |\psi\rangle\langle\psi|$ is separable if it can be written as a product

$$\rho = |\varphi\rangle\langle\varphi| \otimes |\eta\rangle\langle\eta| = \rho_1 \otimes \rho_2. \tag{3.1.23}$$

We note that in this case the reduced density matrices $\rho_S = \mathrm{Tr}_E[\rho] = \rho_1$ and $\rho_E = \mathrm{Tr}_S[\rho] = \rho_2$ are themselves pure, and can be used to construct the total state exactly:

$$\rho = \rho_S \otimes \rho_E = \mathrm{Tr}_E[\rho] \otimes \mathrm{Tr}_S[\rho]. \tag{3.1.24}$$

On the other hand, if there is entanglement between $S$ and $E$ then the reduced density matrices will be mixed and their product does not faithfully recreate the total state.

For mixed states, the question of separability is nuanced due to the possibility of *classical correlations*. Strictly, a general density matrix $\rho$ for $\mathcal{H} = \bigotimes_{i=1}^{N} \mathcal{H}_i$ is separable if [169]

$$\rho = \sum_k p_k \bigotimes_{i=1}^{N} \rho_i^{(k)}, \quad p_k \geq 0, \quad \sum_k p_k = 1. \tag{3.1.25}$$

In other words, $\rho$ is separable if it can be written as a probability distribution over uncorrelated product states. In this case, there are no *quantum* correlations (entanglement). However, if more than one $p_k > 0$ then there exists *classical* correlations. This is simply because the expectation of two operators $\langle A_1 A_2 \rangle$ does not reduce to a product $\langle A_1 \rangle \langle A_2 \rangle$ of expectations. Explicitly,

$$\langle A_1 A_2 \rangle = \sum_k p_k \langle A_1 \rangle^{(k)} \langle A_2 \rangle^{(k)}, \quad \langle A_i \rangle^{(k)} = \mathrm{Tr}_i[A_i \rho_i^{(k)}], \tag{3.1.26}$$

which is not generally equal to the product

$$\langle A_1 \rangle \langle A_2 \rangle = \left( \sum_k p_k \langle A_1 \rangle^{(k)} \right) \left( \sum_j p_j \langle A_2 \rangle^{(j)} \right), \tag{3.1.27}$$

unless only one of the $p_k = 1$ and the rest are zero.

These ideas of separability are crucial for our use of mean-field theory in Section 3.3. There we assert a lack of correlations of any kind, quantum or classical, i.e., a single product state

$$\rho = \bigotimes_{i=1}^{N} \rho_i, \tag{3.1.28}$$

in which case $\rho$ may be called *simply separable*.

> *Aside.* While testing for entanglement of a pure state is straightforward (are the reduced density matrices pure?), as similarly is testing for simple separability (is the state a product of the reduced density matrices?), the question of whether an arbitrary density matrix $\rho$ is separable according to Eq. (3.1.25), known as the *Quantum Separability Problem*, is NP-hard [170]. For bipartite systems, the Peres–Horodecki criterion [171, 172] provides a necessary condition for $\rho$ to be separable: its partial transpose [169] $\rho^{T_2}$ must have non-negative eigenvalues (for $2 \times 2$ and $2 \times 3$ dimensional spaces it is also a sufficient condition), but other and more general cases are not so straightforward. There is also the interesting problem of constructing the *best separable approximation* [173] of an entangled state. See Refs. [174–176] for a review and recent progress.

### 3.1.5 Liouville-von Neumann equation

Learning the time-dependent reduced system density matrix $\rho_S(t)$ is an ultimate objective in open quantum system theory. From $\rho_S = \mathrm{Tr}_E[\rho]$, its time-evolution follows from that of $\rho(t)$,

$$\partial_t \rho_S(t) = \mathrm{Tr}_E[\partial_t \rho(t)]. \tag{3.1.29}$$

So how does $\rho$ evolve? Return to the definition (3.1.9) of the density operator,

$$\rho(t) = \sum_i p_i \, |\psi_i(t)\rangle\langle\psi_i(t)| \, . \tag{3.1.30}$$

The time-dependence of each $|\psi_i\rangle$ is given by the Schrödinger equation[2]. Differentiating the parts of each product $|\psi_i(t)\rangle\langle\psi_i(t)|$,

$$\partial_t \rho(t) = \sum_i p_i \Big[ \left(\partial_t |\psi_i(t)\rangle\right) \langle\psi_i(t)| + |\psi_i(t)\rangle \left(\partial_t \langle\psi_i(t)|\right) \Big] \tag{3.1.31}$$

$$= \sum_i p_i \Big[ -iH(t) \, |\psi_i(t)\rangle\langle\psi_i(t)| + |\psi_i(t)\rangle\langle\psi_i(t)| \, (+iH(t)) \Big]. \tag{3.1.32}$$

So

$$\partial_t \rho(t) = -i \, [H(t), \rho(t)] \, , \tag{3.1.33}$$

or

$$\partial_t \rho(t) = \mathcal{L}(t)\rho(t), \tag{3.1.34}$$

where $\mathcal{L}(t) = -i[H(t), \cdot]$ is the Liouville operator [4]. As this defines a map between operators, it is called a 'superoperator'.

Equation (3.1.33) is the von Neumann equation for the density operator, also known as the *Liouville*-von Neumann equation. It governs the unitary, i.e., Hamiltonian, evolution of a closed system (recall $\mathcal{H} = \mathcal{H}_S \otimes \mathcal{H}_E$ is considered closed). Analogous to the time-ordered solution to the Schrödinger equation we have the general solution

$$\rho(t) = \mathcal{T} \exp\left( \int_{t_0}^{t} dt' \mathcal{L}(t') \right) \rho(t_0). \tag{3.1.35}$$

So far we have discussed the dynamics of the total density operator $\rho$. We now go about deriving those of the reduced system density matrix $\rho_S$. As $S$ is not closed, Eq. (3.1.33) will be modified to a more general class of dynamical equations including non-unitary evolution: *master equations* (Section 3.1.8).

### 3.1.6 The interaction picture

So far we have considered operators $A$ to be fixed quantities and any time-dependence to be carried by state variables $|\psi(t)\rangle$, $\rho(t)$. This is the Schrödinger picture of quantum mechanics. Recall [177] however one may equally view the observables as evolving in time whilst the states remain static—the Heisenberg picture. The relation between these two viewpoints is observed from the time dependence of an expectation. Again setting the initial time $t_0 = 0$,

$$\langle A \rangle(t) = \mathrm{Tr}(A\rho(t)) \tag{3.1.36}$$

$$= \mathrm{Tr}\big(AU(t)\rho(0)U^\dagger(t)\big) \tag{3.1.37}$$

$$= \mathrm{Tr}\big(U^\dagger(t)AU(t)\rho(0)\big) \tag{3.1.38}$$

$$= \mathrm{Tr}(A_H(t)\rho(0)), \tag{3.1.39}$$

[2]The bra vector $\langle\psi_i(t)|$ obeys the conjugated equation $\partial_t \langle\psi_i(t)| = \left(\partial_t |\psi_i(t)\rangle\right)^\dagger = \langle\psi_i(t)| \, (iH(t)) \quad (H^\dagger = H)$.

where we used the cyclic property of the trace and $A_H(t) = U^\dagger(t) A U(t)$ defines the observable in the Heisenberg picture[3].

A third, intermediate representation is the *interaction picture* (introduced, in fact, by Dirac [178]). Here one divides the total Hamiltonian $H(t)$ into two parts,

$$H(t) = H_0 + H_I(t), \tag{3.1.40}$$

where $H_0$ is a time-independent, non-interacting or 'free' Hamiltonian, and $H_I(t)$ an interaction. For the open system considered above, $H_0 = H_S + H_E$ and $H_I(t) = H_{SE}(t)$ is typical. One then assigns the evolution dictated by $H_0$, $U_0(t) = e^{-iH_0 t}$, to operators of observables in a similar way to before, e.g., by considering $\langle A \rangle(t)$. As with many results in mathematical physics, the working amounts to inserting identities $U_0(t)U_0^\dagger(t)$ in the right place, i.e., multiplying by 1[4]:

$$\begin{aligned}
\langle A \rangle(t) &= \mathrm{Tr}(A\rho(t)) &(3.1.41)\\
&= \mathrm{Tr}\Big(U_0(t)U_0^\dagger(t) A U_0(t) U_0^\dagger(t)\rho(0)\Big) &(3.1.42)\\
&= \mathrm{Tr}\Big(\Big(U_0^\dagger(t) A U_0(t)\Big)\Big(U_0^\dagger(t)\rho(0)U_0(t)\Big)\Big) &(3.1.43)\\
&= \mathrm{Tr}(A_I(t)\rho_I(t)), &(3.1.44)
\end{aligned}$$

where

$$A_I(t) = U_0^\dagger(t) A U_0(t) \tag{3.1.45}$$

and

$$\rho_I(t) = U_0^\dagger(t)\rho(t)U_0(t) = U_0^\dagger(t)U(t)\rho(0)U^\dagger(t)U_0(t) \tag{3.1.46}$$

define the operator $A_I$ and state $\rho_I$ in the interaction picture. From $\partial_t U_0(t) = -iH_0 U_0(t)$,

$$\begin{aligned}
\partial_t A_I(t) &= +iU_0^\dagger(t) H_0 A U_0(t) - iU_0^\dagger(t) A H_0 U_0(t) &(3.1.47)\\
&= -iU_0^\dagger(t)[A, H_0]U_0(t). &(3.1.48)
\end{aligned}$$

The state $\rho_I(t)$ may instead be written $\rho_I(t) = U_I^\dagger(t)\rho(0)U_I(t)$ where $U_I(t) = U_0^\dagger(t)U(t)$ is the propagator for dynamics in the interaction picture. The strategy here is to move the simple, known evolution from $H_0$ to the operators and focus on the state equation of motion for the interacting problem. For, given $U(t) = -iH(t)U(t)$ and $\partial_t U_0(t) = -iH_0 U_0(t)$,

$$\begin{aligned}
\partial_t U_I(t) &= \partial_t \Big(U_0^\dagger(t)U(t)\Big) &(3.1.49)\\
&= +iH_0 U_0^\dagger(t)U(t) - iU_0^\dagger(t)\left(H_0 + H_I(t)\right)U(t) &(3.1.50)\\
&= -iU_0^\dagger(t)H_I(t)U(t), &(3.1.51)
\end{aligned}$$

or

$$\partial_t U_I(t) = -iH_{II}(t)U_I(t), \tag{3.1.52}$$

[3] We assumed $A$ (in the Schrödinger picture) to have no explicit time dependence. If it does, then this must be carried over to the Heisenberg picture, i.e., $A \to A(t)$ in $A_H(t)$.

[4] The other main approach is adding 0 as $x = x + y + (-y)$.

where $H_{II}(t) = U_0^\dagger(t) H_I(t) U_0(t)$ is the interaction Hamiltonian...in the interaction picture. In other words, the state variables in the interaction picture satisfy

$$\partial_t \left|\psi(t)\right\rangle_I = -iH_{II}(t) \left|\psi(t)\right\rangle_I, \quad \partial_t \rho_I(t) = -i \left[H_{II}(t), \rho_I(t)\right], \tag{3.1.53}$$

which is the remaining problem to be solved.

In following sections we will not use an interaction label $I$: if the distinction is important, we will say so in words. Subscripts can then continue to indicate Hilbert spaces. In fact, in most cases we will work exclusively with the Heisenberg equations of motion for expectations of operators which do not care which viewpoint is taken (such as in Eqs. (3.1.36) to (3.1.39))[5]. The interaction picture will nonetheless be a useful tool in the derivation of the master equation governing the reduced system dynamics.

### 3.1.7 Harmonic baths and spin-boson models

While master equations for the reduced system dynamics $\rho_S(t)$ can be derived for many types of microscopic systems [4], we focus on models with environments comprising an infinite number of harmonic oscillators linearly coupled to the system: the *harmonic bath*. This may be referred to as the *Caldeira-Leggett* model after Caldeira and Leggett who were amongst the first to use it to describe solid-state dissipation phenomena in the 1980s [179–184], employing the influence functional method [185] set out by Feynman and Vernon two decades prior. It has since become a most widely used model for quantum open systems [22].

The bosonic modes of a harmonic bath may directly describe photonic and phononic environments [2, 4]. (That's pretty much everything we'd be interested in covered!). For example, an external reservoir of electromagnetic modes coupling to a cavity [155], an impurity interacting with acoustic phonons in a lattice [186], or the local vibrational environment of organic emitters [43]. Yet more generally a many-body environment can be mapped[6] to an *effective* harmonic one with linear coupling to the system [22, 187]—an approximation valid for large environments with coupling to the system spread over many degrees of freedom.

The environment and interaction Hamiltonian of these models take the form

$$H_{IE} = H_E + H_I = \sum_j \left[\nu_j b_j^\dagger b_j + \left(\xi_j b_j + \bar{\xi}_j b_j^\dagger\right) S\right]. \tag{3.1.54}$$

Here $S$ is a system operator that couples the system to each bath mode of frequency $\nu_j$. The strength of this coupling is parametrised by the coefficients $\xi_j$ (complex conjugate $\bar{\xi}_j$). Given the bath modes are infinite in number, in practice the coupling is captured by a continuous function

$$J(\nu) = \sum_j |\xi_j|^2 \delta(\nu - \nu_j), \tag{3.1.55}$$

which defines the spectral density. For a bath of positive frequency modes, $J(\nu) = 0$ when $\nu < 0$.

Spectral densities may be constructed from spectroscopic data [188–191] or given phenomenologically [184, 192]. A common set of functions, suitable to our use, was also introduced by Leggett et al. [184]:

$$J(\nu) = 2\alpha \nu^s \nu_c^{1-s} e^{-(\nu/\nu_c)^2}, \quad s > 0. \tag{3.1.56}$$

[5] We hence use the notation $\langle A(t)\rangle$ instead of $\langle A\rangle(t)$ to avoid confusion when, for example, $\langle A\rangle(t - t_0)$ could be interpreted as either $\langle A\rangle$ evaluated at $(t - t_0)$ or $\langle A\rangle$ (at a fixed time) multiplied by $(t - t_0)$.

[6] For example, an anharmonic bath can be replaced by a harmonic one provided the spectral density is chosen such that the correlation functions (Section 3.1.8) of the baths match. See Ref. [88] for a relevant discussion for organic polaritons.

This describes coupling of the form $\nu^s$ up to some cut-off $\nu_c$, which is appropriate to describe, for example, the low frequency vibrational environments of organic molecules. For that purpose, $s = 1$ is normally chosen, in which case the environment is said to be *Ohmic*, with $s > 1$ and $s < 1$ instead describing super-Ohmic and sub-Ohmic baths, respectively. All types are in use; another example would be the polaron environment of electrons in solids, for which $s = 3$ or $s = 5$ may apply [184].

The cut-off function is often added post hoc to regularise $J$ but may also arise naturally in microscopic derivations [184]. A Gaussian was chosen here, but other exponentials or even a step function may be considered. The dimensionless constant $\alpha$ describes the overall coupling strength of the system to the environment.

In addition to the above parameters, a harmonic bath is characterised by a temperature, $T$, which sets the populations of the modes at thermal equilibrium. In Section 3.1.8 we will see further how $T$, in conjunction with the spectral density, determines the important correlation functions of the bath.

An example of a model including the system Hamiltonian is provided by the class of spin-boson (SB) models [184]. Here the system is a two-level system, and $S = \sigma^z$:

$$H_{SB} = H_S + H_E + H_{IE} \tag{3.1.57}$$

$$= -\frac{\Delta}{2}\sigma^x + \frac{\epsilon}{2}\sigma^z + \sum_j \left[\nu_j b_j^\dagger b_j + \left(\xi_j b_j + \bar{\xi}_j b_j^\dagger\right)\sigma^z\right]. \tag{3.1.58}$$

These models are ubiquitous in many-body physics [2, 22, 192] due to the number of physical systems that are intrinsically two-level or may be effectively described as such. The case $\Delta = 0$ provides an exactly solvable model [2], the *independent* boson model, which is particularly well appreciated in the literature[7].

The diagonal form of the coupling $\sigma^z$ in Eq. (3.1.58)—applicable to many real systems [184]—results in coherence between the two states of the system being lost due to the system-environment interaction. Informally, this can be understood as the interaction 'measuring' the state of the two-level system [193]. It is of little surprise then that the independent boson model is a prototype for environmentally induced decoherence, e.g., in superconducting qubits [194]. This form of coupling will also motivate the pure dephasing models of decoherence we introduce below.

In Section 3.2.3 we show how the TEMPO method [162] can be used to calculate exact dynamics for open quantum systems with coupling to a harmonic bath. In order to understand the challenge posed by this task, and how it is overcome, we move on to the derivation of master equations for the reduced system dynamics.

*Aside.* Students of theoretical physics should not be surprised to find the harmonic oscillator is a linchpin of yet another set of methods: it describes fluctuations around equilibrium of an arbitrary system [164]! Sidney Coleman was right[8].

### 3.1.8 The Markovian master equation

We now discuss a staple of any text on open quantum systems: the derivation of the weak coupling Markovian master equation in Lindblad form. This is by no means the only master equation used to describe open system dynamics [4, 16, 22, 196–199], but it is certainly the most common one. We will use its form directly in Chapters 5 and 6. It also provides a point of reference for the calculation of non-Markovian dynamics using tensor network methods, which we discuss in Section 3.2.3 and implement in Chapter 4. With this in mind the main points of discussion are the assumptions required for the derivation and the interpretation of the final equation. Hence we detail the key steps of the derivation only; see Refs. [4, 167, 200, 201] for additional exposition.

[7] There appears to be an unwritten rule that any paper introducing a new numerical method for open quantum system dynamics should start with a benchmark of the independent boson model.

[8] "The career of a young theoretical physicist consists of treating the harmonic oscillator at ever-increasing levels of abstraction" [195].

We mainly follow the presentation of Ref. [167] and further consider the environment to be a harmonic bath as defined in Section 3.1.7. In the derivation several key physical assumptions are made:

1. The system and environment are initially uncorrelated, $\rho(0) = \rho_S(0) \otimes \rho_E(0)$
2. Weak coupling between the system and environment such that the state of the environment is negligibly changed by the interaction
3. The timescale $\tau_R$ of the dynamics of the open system induced by the bath far exceeds that over which correlations in the environment decay $\tau_c$
4. The timescale of the intrinsic (uncoupled) evolution of the system $\tau_S$ is far smaller than $\tau_R$

Assumption 1 is required practically in the derivation and avoids issues of non-positivity from the outset, i.e., negative eigenvalues appearing in the density matrix that can arise starting from certain correlated states [202–204]. It is consistent with a situation where environment correlations are short lived (assumption 3) or simply an experiment where a system is brought into contact with its environment at $t_0 = 0$. For the harmonic bath we take the initial state to be thermal, meaning

$$\rho_E(0) = \frac{e^{-H_E/T}}{\mathrm{Tr}_E[e^{-H_E/T}]} \qquad (k_B = 1), \tag{3.1.59}$$

with $H_E = \sum_j \nu_k b_j^\dagger b_j$ as in Eq. (3.1.54).

Assumptions 1 through 3 may collectively be referred to as the *Born-Markov approximation* [4]. Together they imply not only that the state of the environment is weakly perturbed by the interaction, but also that this perturbation is not perceptible on the timescales of which the open system evolution is to be determined. As a consequence, as far as the equation of motion for the reduced system density matrix $\rho_S(t)$ is concerned, one may take the separable form $\rho(t) = \rho_S(t) \otimes \rho_E(0)$ to hold for all times.

Note these assumptions do *not* mean the system has no effect on the environment, only that the effect is not resolved in the system dynamics. Put another way, the open system dynamics are coarse-grained at $\tau_R \gg \tau_c$. The result is a *time-local* equation of motion, i.e., one that only depends on the instantaneous state of the system and not previous states. It is in this sense that the equation of motion is Markovian, as discussed in Chapter 2 and shown below.

Finally, assumption 4 allows for a rotating wave approximation (cf. Chapter 2), here known as the *secular approximation*, to be made in the derivation. It is the statement that the intrinsic or 'free' evolution of the system, characterised by the difference in system energies $\tau_S \sim |\omega - \omega'|^{-1}$, is short relative to the timescale over which the open system dynamics are resolved: $\tau_S \ll \tau_R$. Formally, it is required to guarantee complete positivity of the dynamics [4, 205]. While generally crude, it is normally a good approximation for systems of light-matter interaction we consider, where optical frequencies are $\sim 10^{15}$ Hz.

Without the secular approximation, an equation known as the Redfield equation results [16, 206, 207]. This contains more information than the Lindblad master equation, and is also time-local, but does not guarantee positive evolution of the density matrix [4]. This negativity can be anticipated [22, 208], and positive-preserving forms derived from the Redfield equation other than via a secular approximation [196, 198, 209–212]. However, in this thesis we will consider either a regime where the secular approximation is valid, or else one where the weak coupling assumption is not justified and neither Markovian nor Redfield equations apply.

The main steps in the derivation of the Markovian master equation are now summarised. The starting point is the von Neumann equation for the total state $\rho(t)$ on $\mathcal{H} = \mathcal{H}_S \otimes \mathcal{H}_E$ in the interaction picture:

$$\partial_t \rho(t) = -i\delta \left[H_I(t), \rho(t)\right]. \tag{3.1.60}$$

Here we wrote the interaction Hamiltonian as $\delta \cdot H_I(t)$ where $\delta$ is a dimensionless parameter characterising the weak coupling (e.g., $\delta \sim \alpha^{1/2}$ for an Ohmic spectral density $J(\nu) \sim \alpha\nu$). It is useful to write this equation in integral form,

$$\rho(t) = \rho(0) - i\delta \int_0^t ds \left[H_I(s), \rho(s)\right]. \tag{3.1.61}$$

The idea of deriving a Markovian equation for $\rho_S(t) = \mathrm{Tr}_E(\rho)$ is that we want to remove dependence on previous states of the system $\rho_S(s)$ for $s < t$, including anchoring to the initial state preparation. To do so, we take the equation of motion for $\rho(t)$ and repeatedly substitute integral expressions of the form Eq. (3.1.61) into Eq. (3.1.60) (those who read Appendix B.1 may recognise this approach). With one iteration,

$$\partial_t \rho(t) = -i\delta \left[H_I(t), \rho(0)\right] - \delta^2 \int_0^t ds \left[H_I(t), \left[H_I(s), \rho(s)\right]\right]. \tag{3.1.62}$$

It may be shown [167] that for the thermal initial state Eq. (3.1.59), $\mathrm{Tr}_E\left(H_I(t), \rho(0)\right)$ can be assumed to vanish:

$$\partial_t \rho(t) = -\delta^2 \int_0^t ds \left[H_I(t), \left[H_I(s), \rho(s)\right]\right]. \tag{3.1.63}$$

We now repeat the process, but this time integrating from the boundary $s' = t$ to $s' = s$,

$$\rho(s) = \rho(t) - i\delta \int_t^s ds' \left[H_I(s'), \rho(s')\right]. \tag{3.1.64}$$

The result is

$$\partial_t \rho(t) = -\delta^2 \int_0^t ds \left[H_I(t), \left[H_I(s), \rho(t)\right]\right] + O\left(\delta^3\right), \tag{3.1.65}$$

where collecting terms $O(\delta^3)$ realises a Markovian approximation under weak coupling, i.e., in Eq. (3.1.63) the memory kernel is sufficiently short such that we may approximate $\rho(s)$ by $\rho(t)$.

This leads us to apply assumptions 1-3 whereby we neglect the $O(\delta^3)$ terms and substitute the product $\rho(t) \approx \rho_S(t) \otimes \rho_E(0)$. Performing a trace over the environment, we have

$$\partial_t \rho_S(t) = -\int_0^t ds \,\mathrm{Tr}_E \Big( \left[\delta H_I(t), \left[\delta H_I(s), \rho_S(t) \otimes \rho_E(0)\right]\right] \Big). \tag{3.1.66}$$

For the linear interaction $\delta H_I = \sum_j (\xi_j b_j + \bar{\xi}_j b_j^\dagger) S$, the trace yields the bath correlation functions

$$\langle x_j^\dagger(t) x_k(t-s)\rangle = \mathrm{Tr}_E\left(x_j^\dagger(t) x_k(t-s) \rho_E(0)\right), \quad x_j = \xi_j b_j + \bar{\xi}_j b_j^\dagger. \tag{3.1.67}$$

Note $x_j^\dagger = x_j$ for the harmonic bath, but we wrote $x_j^\dagger(t)x_j(t-s)$ for generality.

Under the assumption $\tau_c \ll \tau_R$ that the bath correlations decay faster than the timescale being resolved, the integrand in Eq. (3.1.66) disappears sufficiently quickly over $t-s \gg \tau_c$ to allow the lower limit to be taken to $-\infty$, removing reference to the initial state preparation. After making the substituting $s = t-u$,

$$\partial_t \rho_S(t) = -\delta^2 \int_0^\infty du\, \mathrm{Tr}_E \Big( [H_I(t), [H_I(t-u), \rho_S(t) \otimes \rho_E(0)]] \Big), \tag{3.1.68}$$

which is recognised as a Redfield equation [16, 206, 207]. To obtain a more tractable form, the system operator $S$ coupling to the environment is decomposed in the eigenbasis $\{|\epsilon\rangle\}$ of the system Hamiltonian $H_S$, $S_\omega = \sum_{\epsilon'-\epsilon} |\epsilon\rangle\langle\epsilon|\, S\, |\epsilon'\rangle\langle\epsilon'|$, such that $[H_S, S_\omega] = -\omega S_\omega$ and $\sum_\omega S_\omega = S$. This transforms Eq. (3.1.68) to [4]

$$\partial_t \rho_S(t) = \sum_{\omega\omega'} \sum_{jk} e^{i(\omega'-\omega)} \gamma_{jk}(\omega) \left( S_\omega \rho_S(t) S_{\omega'}^\dagger - S_{\omega'}^\dagger S_\omega \rho_S(t) + \mathrm{H.c.} \right), \tag{3.1.69}$$

where H.c. denotes the Hermitian conjugate and $\gamma_{jk}(\omega)$ are the one-sided Fourier transforms of the bath correlation functions,

$$\gamma_{jk}(\omega) = \int_0^\infty ds\, e^{i\omega s} \langle x_j^\dagger(t) x_k(t-s) \rangle. \tag{3.1.70}$$

The final step is to perform the secular approximation, by negating terms $\omega' \neq \omega$ which are assumed to be rapidly oscillating on the timescale $\tau_R$ of the dynamics to be resolved. The resulting equation may be written [4]

$$\partial_t \rho_S(t) = -i\, [H_S + H_{LS}, \rho_S(t)] + \mathcal{D}(\rho_S(t)), \tag{3.1.71}$$

where the free system Hamiltonian $H_S$ was added to return to the Schrödinger picture, and the Lamb-shift Hamiltonian

$$H_{LS} = \sum_\omega \sum_{jk} \mathrm{Im}\, \gamma_{jk}(\omega) S_\omega^\dagger S_\omega, \tag{3.1.72}$$

provides a unitary contribution. Lastly the dissipator $\mathcal{D}(\rho_S(t))$ is defined by

$$\mathcal{D}(\rho_S(t)) = \sum_\omega \sum_{jk} 2\, \mathrm{Re}\, \gamma_{jk}(\omega) \left( S_\omega \rho_S(t) S_\omega^\dagger - \frac{1}{2} \left\{ S_\omega^\dagger S_\omega, \rho_S(t) \right\} \right). \tag{3.1.73}$$

where $\{\cdot,\cdot\}$ denotes the anti-commutator.

For a harmonic bath in thermal equilibrium, the independence of the modes $x_j$ means $\gamma_{jk}$ is diagonal, hence

$$\mathcal{D}(\rho_S(t)) = \sum_\omega \sum_j 2\, \mathrm{Re}\, \gamma_j(\omega) \left( S_\omega \rho_S(t) S_\omega^\dagger - \frac{1}{2} \left\{ S_\omega^\dagger S_\omega, \rho_S(t) \right\} \right). \tag{3.1.74}$$

The time dependence of the bath operators in the interaction picture ($H_0 = H_S + H_E$) follows simply from the canonical commutation relations $[b_i, b_j^\dagger] = \delta_{ij}$ for the bath creation and annihilation operators:

$$\partial_t b_j(t) = -i U_0^\dagger(t) \left[ b_j, \sum_k \nu_k b_k^\dagger b_k \right] U_0(t) \quad \text{(using Eq. (3.1.48))} \tag{3.1.75}$$

$$= -i\nu_j U_0^\dagger(t) b_j U_0(t) \tag{3.1.76}$$

$$= -i\nu_j b_j(t), \tag{3.1.77}$$

which implies $b_j(t) = b_j e^{-i\nu_j t}$ and $x_k(t) = \xi_j b_j e^{-i\nu_j t} + \overline{\xi}_j b_j^\dagger e^{i\nu_j t}$. Then

$$\langle x_j(t) x_j(t-s)\rangle = |\xi_j|^2 \left( \langle b_j b_j^\dagger \rangle e^{-i\nu_j (t-[t-s])} + \langle b_j^\dagger b_j \rangle e^{i\nu_j (t-[t-s])} \right) \tag{3.1.78}$$

$$= (1 + n_B(\nu_j/T))\, e^{-i\nu_j s} + n_B(\nu_j/T) e^{i\nu_j s}, \tag{3.1.79}$$

where

$$n_B(\nu/T) = \frac{1}{e^{\nu/T} - 1} \equiv \frac{1}{2}\left(\coth\left(\frac{\nu}{2T}\right) - 1\right) \tag{3.1.80}$$

is the Bose-Einstein occupation function for the modes in thermal equilibrium at temperature $T$. The sum of the one-sided transforms $\gamma_j$ may then be written

$$\sum_j \gamma_j(\omega) = \int_0^\infty ds\, e^{i\omega s} \sum_j \frac{|\xi_j|^2}{2} \left[ \left(\coth\left(\frac{\nu_s}{2T}\right) + 1\right) e^{-i\nu_j s} + \left(\coth\left(\frac{\nu_s}{2T}\right) - 1\right) e^{i\nu_j s} \right] \tag{3.1.81}$$

$$= \int_0^\infty ds\, e^{i\omega s} \int_0^\infty d\nu J(\nu) \left[ \coth\left(\frac{\nu}{2T}\right) \cos(\nu s) - i \sin(\nu s) \right] \tag{3.1.82}$$

$$= \int_0^\infty ds\, e^{i\omega s} C(s), \tag{3.1.83}$$

where we used the definition (3.1.56) of the spectral density to rewrite the sum over modes as an integral, and identified the bath correlation or 'autocorrelation' function

$$C(t) = \int_0^\infty d\nu J(\nu) \left[ \coth\left(\frac{\nu}{2T}\right) \cos(\nu t) - i \sin(\nu t) \right]. \tag{3.1.84}$$

$C(t)$ describes the decay of correlations for a given spectral function and temperature. This function frequently occurs when considering molecular spectra, response functions, etc. In Section 3.2 we will see how it controls *non-Markovian* behaviours of an open system as captured by the TEMPO method.

Defining $\Gamma(\omega) = \int_0^\infty ds\, e^{i\omega s} C(s)$, the master equation with the harmonic bath is

$$\partial_t \rho_S(t) = -i[H_S + H_{LS}, \rho_S(t)] + \sum_\omega 2\,\mathrm{Re}\,\Gamma(\omega) \left( S_\omega \rho_S(t) S_\omega^\dagger - \frac{1}{2}\left\{ S_\omega^\dagger S_\omega, \rho_S(t) \right\} \right). \tag{3.1.85}$$

with $H_{LS} = \sum_\omega \mathrm{Im}\,\Gamma(\omega) S_\omega^\dagger S_\omega$.

In Chapter 4 we evaluate the dissipator explicitly for a model with $S = \sigma^z/2$. There we point out a crux of the above analysis: it relies on the eigenstates of $H_S$ being known. While in simple cases these may be available, for large or complex systems such as the many-body systems we consider, this is unlikely.

Equation (3.1.85) is an example of a master equation in Lindblad form, also known as a Gorini–Kossakowski–Sudarshan–Lindblad equation after work by those authors on the generators of quantum dynamical semigroups in the 1970s [213–215]. We now make some general comments regarding master equations of this type relevant to our applications.

1. The dissipator $\mathcal{D}$ in Eq. (3.1.71) involves a sum over certain system operators $S_\omega^\dagger$ that arose from the coupling to the bath. We can generalise this to include contributions from up to $d^2 - 1$ linear independent system operators (excluding the identity), with $d = \dim(\mathcal{H}_S)$:

$$\partial_t \rho = -i[H, \rho] + \sum_i \Gamma_i L[Y_i], \quad L[Y_i] = Y_i \rho Y_i^\dagger - \{Y_i^\dagger Y_i, \rho\}/2. \tag{3.1.86}$$

Here we made several notation changes in line with what is seen in the literature as well as the rest of this thesis: we drop the $S$ subscript and $t$ argument from the system reduced density matrix. It will be assumed that $\rho_S$ is always the target of our calculations and that this is a function of time. Similarly $H$ is taken to mean the system Hamiltonian unless otherwise specified. This may or may not include a Lamb Shift contribution, depending on whether one is interested in the renormalisation of the system energy levels brought about by the coupling.

2. The dissipator $\mathcal{D}$ in Eq. (3.1.71) induces non-unitary, i.e., dissipative dynamics. This should not be surprising: in an open system energy is exchanged, often irreversibly, with the environment. Similarly, each $Y_i$, known as a Lindblad operator, provides a different decay or relaxation mode of the open system, with rate $\Gamma_i$. These may be categorised depending on whether $Y_i$ is a lowering ($y_i$), raising ($y_i^\dagger$) or number conserving ($y_i^\dagger y_i$) operator:

$$\Gamma_i L[y_i] \longrightarrow \text{decay (loss) at rate } \Gamma_i, \tag{3.1.87}$$

$$\Gamma_i L[y_i^\dagger] \longrightarrow \text{gain (drive) at rate } \Gamma_i, \tag{3.1.88}$$

$$\Gamma_i L[y_i^\dagger y_i] \longrightarrow \text{dephasing (decoherence) at rate } \Gamma_i. \tag{3.1.89}$$

Particular examples we will see include (if $a_k$ is a bosonic operator for a photon mode with momentum $k$ and $\sigma^\alpha$ Pauli matrices for a two-level system):

$$\kappa L[a_k] \longrightarrow \text{photon number decay at rate } \kappa, \tag{3.1.90}$$

$$\Gamma_\uparrow L[\sigma^+] \longrightarrow \text{pumping at rate } \Gamma_\uparrow, \tag{3.1.91}$$

$$\Gamma_\uparrow L[\sigma^-] \longrightarrow \text{dissipation at rate } \Gamma_\downarrow, \tag{3.1.92}$$

$$\Gamma_z L[\sigma^z] \longrightarrow \text{dephasing at rate } \Gamma_z. \tag{3.1.93}$$

All of these describe incoherent processes in the sense that no phase information is conveyed.

3. Considering the previous two points allows one to easily design phenomelogical models of open systems. That is, instead of following a particular microscopic derivation, one can simply write down the Hamiltonian evolution plus Lindblad operators describing incoherent processes believed to be physically relevant (the values of the rates $\Gamma_i$ may be inspired by experimental timescales, e.g., of photon or exciton decay). This is particularly useful when the eigenstates of the system cannot be determined, preventing a derivation such as above. The ease of including general loss and gain processes in the quantum evolution is a main reason Markovian master equations are so widely used in the study of open quantum systems, with the caveat of results being limited to the weak coupling or Markovian regime.

4. The assumptions of weak coupling and rapid decay of bath correlations are strong ones [167]. They are known to be applicable for some well known cases [4]. Quantum optics with is weak coupling to an external vacuum reservoir or other non-resonant modes is one such case where in addition the secular approximation may be justified. As such we will always treat cavity loss and non-resonant decay within the Markovian approximation. On the other hand, there are many physical systems where a Markovian description is a poor approximation or fails entirely [22, 166, 216]. Naturally, our interest is in the scope for non-Markovian behaviours in systems of organic polaritons where there is strong coupling to structured vibrational environments in conjunction with strong light-matter coupling (fast system dynamics) [5, 43, 64, 65, 217–221].

We continue the discussion of non-Markovian effects in Section 3.2 where we explain the TEMPO exact numerical method. Before then, we recast the dynamics of the reduced system density operator determined by a Lindblad master equation in the form of equations of motions for the expectations of system operators, and further show how multi-time correlations of those operators may be calculated.

### 3.1.9 Heisenberg equations of motion

Frequently in our applications instead of working with the master equation directly we will derive equations of motion for the expectations of system operators, the *Heisenberg equations of motion*. These provide an equivalent description of the dynamics via $\langle A(t)\rangle = \mathrm{Tr}(A\rho(t))$: the expectations of $d(d+1)/2$ ($d = \dim(\mathcal{H}_S)$) linearly independent operators determine the unique components of a Hermitian operator $\rho$, although fewer are required if you consider $\mathrm{Tr}[\rho] \equiv 1$ (the evolution of the identity operator is trivial) and any other constraints such as symmetries of the Hamiltonian.

For the purpose of calculating the Heisenberg equations, the following result is most useful.

#### A handy Heisenberg shortcut

Given a Lindblad master equation $\partial_t \rho = -i[H,\rho] + \sum_i L[Y_i]$, the equation of motion for an operator $X$ is

$$\partial_t\langle X\rangle = -i\big\langle [X,H]\big\rangle - \frac{1}{2}\sum_i \big\langle [X,Y_i^\dagger]\,Y_i + Y_i^\dagger\,[Y_i,X]\big\rangle. \tag{3.1.94}$$

If $X$ is Hermitian, we further have

$$\partial_t\langle X\rangle = -i\,\langle [X,H]\rangle - \sum_i \mathrm{Re}\,\langle [X,Y_i^\dagger]\,Y_i\rangle\,. \tag{3.1.95}$$

*Proof.* From the master equation in the Schrödinger picture,

$$\begin{aligned}\partial_t\langle X\rangle &= \partial_t\,\mathrm{Tr}\,[X\rho] && (3.1.96)\\ &= \mathrm{Tr}(X\partial_t\rho) && (3.1.97)\\ &= -i\,\mathrm{Tr}(X\,[H,\rho]) + \sum_i \mathrm{Tr}\Big(X\mathcal{L}[Y_i]\Big). && (3.1.98)\end{aligned}$$

Using the cyclic property of the trace,

$$\begin{aligned}\mathrm{Tr}(X\,[H,\rho]) &= \mathrm{Tr}(XH\rho - X\rho H) && (3.1.99)\\ &= \mathrm{Tr}(XH\rho - XH\rho) && (3.1.100)\\ &= \mathrm{Tr}([X,H]\,\rho) && (3.1.101)\\ &= \langle [X,H]\rangle\,. && (3.1.102)\end{aligned}$$

Similarly, from $L[Y_i] = Y_i\rho Y_i^\dagger - \{Y_i^\dagger Y_i,\rho\}/2$,

$$\begin{aligned}\mathrm{Tr}\Big(XL[Y_i]\Big) &= (1/2)\,\mathrm{Tr}\Big(2XY_i\rho Y_i^\dagger - X\,\big(Y_i^\dagger Y_i\rho + \rho Y_i^\dagger Y_i\big)\Big) && (3.1.103)\\ &= (1/2)\,\mathrm{Tr}\Big(Y_i^\dagger XY_i\rho + Y_i^\dagger XY_i\rho - XY_i^\dagger Y_i\rho - Y_i^\dagger Y_i X\rho\Big) && (3.1.104)\\ &= -(1/2)\,\mathrm{Tr}\Big(\Big([X,Y_i^\dagger]\,Y_i + Y_i^\dagger\,[Y_i,X]\Big)\,\rho\Big) && (3.1.105)\\ &= -(1/2)\big\langle [X,Y_i^\dagger]\,Y_i + Y_i^\dagger\,[Y_i,X]\big\rangle. && (3.1.106)\end{aligned}$$

It remains to note if $X$ is Hermitian then $Y_i^\dagger\,[Y_i,X]$ is the Hermitian conjugate of $[X,Y_i^\dagger]\,Y_i$.

#### Solving the Heisenberg equations

Given a master equation of an open system with a single degree of freedom, the Heisenberg equations for single operator expectations will form a closed set. One may then attempt to solve this system of differential equations, either analytically or by numerical integration. Alternatively, one may look for a solution to the steady state by solving the system of equations resulting from setting all derivatives to zero[9].

For many-body systems however the Heisenberg equations for the expectations of operators acting on a single site (subspace) will generally not be sufficient: interaction terms of the Hamiltonian produce expectations involving multiple sites, so the single-site equations do not form a closed set. We discuss how this may be resolved using mean-field theory and higher-order cumulant expansions in Sections 3.3 and 3.4.

### 3.1.10 Multi-time correlations and quantum regression

For Markovian master equations the quantum regression theorem (QRT) is a result [4, 22, 223, 224] that allows one to write down evolution equations for correlation functions of the system, i.e., functions of the form $\langle A_i(t)A_j(t')\rangle$, if the equations for the single-time expectations $\langle A_i(t)\rangle$ are known.

Suppose the Heisenberg equations $\langle A_i(t)\rangle$ form a linear system of first-order differential equations

$$\partial_t\langle A_i(t)\rangle = \sum_j M_{ij}\langle A_j(t)\rangle. \tag{3.1.107}$$

Then

$$\partial_t\langle A_i(t)A_k(t')\rangle = \sum_j M_{ij}\langle A_j(t)A_k(t')\rangle. \tag{3.1.108}$$

This result may be understood by moving evolution operators around under the trace, similar to as has now been done a number of times in this chapter. Assuming $t > t'$,

$$\langle A_i(t)A_k(t')\rangle = \mathrm{Tr}\Big(U^\dagger(t',0)U^\dagger(t,t')A_iU(t,t')\overbrace{U(t',0)U^\dagger(t',0)}^{=1}A_kU(t',0)\rho(0)\Big) \tag{3.1.109}$$

$$= \mathrm{Tr}\big(A_iU(t,t')A_k\rho(t')U^\dagger(t,t')\big) \tag{3.1.110}$$

$$= \mathrm{Tr}(A_i\Phi(t,t')) \tag{3.1.111}$$

where $\Phi(t,t') := U(t,t')(A_k\rho(t'))U^\dagger(t,t')$. If this is regarded as a function of $t$, it should[10] satisfy the same differential equation as $\rho$, e.g., $\partial_t\Phi = -i[H,\Phi]$, and so the equations for the two-time average $\langle A_i(t)A_k(t')\rangle$ follow the same structure as the single-time averages but with a modified initial state $A_k\rho(t')$ [225]. This result may be extended to derive $n$-time correlation functions [22, 226, 227].

While the QRT have proven very effective for quantum optical systems in the Markovian regime [4, 224], it does not apply for non-Markovian dynamics [228–232]: entanglement with the environment means it is not possible to express multi-time correlations using propagators for the reduced system density matrix [233]. Extensions to the QRT can be made based on perturbative expansions of the system-environment coupling [226, 227, 232, 234, 235], but we do not discuss these. In fact, it is has been shown that the QRT may fail even in the Markovian case [230], that is, there are systems for which the Markovian master equation is accurate, but the QRT gives incorrect results.

[9] When there is a difference in timescales between parts of the system, a mixed approach may be effective: eliminate the fast degrees of freedom by setting their derivatives to zero then solve the reduced set of equations of motions for the slow variables. This is known as adiabatic elimination [222].

[10] See Ref. [4] for an actual proof which shows how non-unitary (Markovian) dynamics is included.

We will not use the QRT in our work, but have brought it to attention for two reasons.

First, the matrix product operator method explained in the following section allows for the calculation of arbitrary multi-time correlation functions of the system *without* recourse to a QRT. We make use of this in Chapter 4. Second, the QRT has previously been used [63, 105, 236, 237] in conjunction with the cumulant expansion methods we set out in Section 3.4. For example, the equation of motion of $\langle a^\dagger(t)a(0)\rangle$ may be obtained from that for $\langle a^\dagger(t)\rangle$. There are potential issues when applying cumulant expansions in this context. We discuss this problem in Chapter 7.

## 3.2 Time-evolving matrix product operator method

### 3.2.1 Overview

The time-evolving Matrix Product Operator (TEMPO) method [162, 233] is a numerically exact method for calculating the dynamics of an open quantum system. The essential idea is that the total evolution of an open quantum system can be formulated as a tensor network. This allows standard tensor network techniques [238, 239] to be used to control growth of information in the non-Markovian description of the open system and efficiently propagate the dynamics. In this section we will explain what these techniques are, and how TEMPO harnesses them. The aim is to provide the basis for our extension of the method in Chapter 4 rather than a complete account of the method and its development—there are already three excellent theses on that [233, 240, 241]! We start by discussing the problem of calculating dynamics outwith the Markovian regime.

### 3.2.2 Non-Markovian and exact dynamics

As discussed above, there are many physical systems where a Markovian description is not sufficient [22]. A general method capable of solving non-Markovian dynamics would enable the modelling of these systems. It would also allow for the evaluation of whether proposals for new technologies made using the Markovian master equation, e.g., in quantum information processing [242, 243], would be robust against decoherence and dissipation processes in realistic settings. Here we discuss the challenge of developing such a method.

In the simplest definition[11], Markovian dynamics occur when the evolution of the reduced system density matrix depends only on its current state and not its history. Information may flow from the open system to its environment due to their interaction, but is lost irreversibly. Therefore the environment is memoryless and the amount of information contained in the description of the open system dynamics—hence the computational difficulty—does not grow with time.

On the other hand, for non-Markovian open systems excitations created in the environment by the system may influence the system dynamics at a later time. In other words, there is a backflow of information [227]. To calculate the dynamics one then needs not only knowledge of the current system-environment state but potentially all proceeding states of the total system. The result is a rapid (exponential [245]) growth of the information required to propagate the dynamics that is catastrophic to direct numerical calculations beyond short times. The TEMPO method uses tensor network compression techniques to control this growth of information. This makes possible the calculation of non-Markovian dynamics with memory effects over many hundreds of timesteps.

[11] There exist many different measures of non-Markovianity; see Refs. [166, 216, 227, 244]. A precise characterisation is not important for our discussion.

The description of the dynamics used by TEMPO in fact involves no approximations, making the method *numerically exact*. That is, up to error associated with numerical evaluation, TEMPO provides an exact representation of a quantum system coupled linearly to a harmonic bath[12]. The magnitude of the error can, in principle (at least up to machine precision), be made as small as required by tuning numerical parameters such as the timestep length $\delta t$. In practice, resources are finite and a balance between computational time and accuracy and must be struck. The utility of the TEMPO method is the reality that for many systems of interest sufficient accuracy is obtained with manageable computational costs.

Below we show that the relevant parameters for a TEMPO computation are the timestep size, environment memory length, and a SVD precision. The validity of a calculation is verified by the convergence of the results with respect to these parameters.

### 3.2.3 Tensor networks

Tensor networks are collections of multi-dimensional arrays (tensors) connected together via contractions. These structures have found immense use in quantum simulation [238, 246–248] computation [249–252] and machine learning tasks [253, 254]. Their potential to describe complex quantum networks stems from the natural representation they provide of composite systems where entanglement is restricted in some way [238]. We are interested in how this feature can be harnessed to calculate open quantum system dynamics. We begin by introducing relevant tensor network notations as well as the building blocks used by the TEMPO tensor network: matrix product states and operators.

A tensor is a multidimensional array of complex numbers $C_{i_1 i_2 \ldots i_R}$ where the indices $i_1$, $i_2$,..., $i_R$ range over $d_1$, $d_2$,..., $d_R$ values, respectively. We follow the convention in many-body physics [238] of referring to the number of indices (here $R$) as the *rank* of the tensor. This is opposed to other fields of applied mathematics [255] where *order* refers to the number of indices and *rank* to the tensor equivalent of matrix rank (cf. Section 3.1.4). A rank-2 tensor then is a matrix, a rank-1 tensor a vector and a rank-0 tensor a scalar (single number).

So far we have considered the density operator $\rho$ to be a square $d \times d$ matrix where $d$ is the dimension of the system Hilbert space. For the purpose of calculations involving tensor networks however, the convention is to work in Liouville space where the density matrix is flattened to a vector and so represented by a tensor with a single leg of dimension $d^2$ (Fig. 3.1c).

In a tensor network diagram, a rank-$R$ tensor is represented by a single vertex (node, box, circle, triangle etc.) with $R$ outgoing edges (also known as legs), one for each index. *Contraction* is the operation of summing over all values of repeated indices in a tensor or product of tensors, and is indicated by an edge joining two vertices. Familiar examples include the dot product of vectors, $\boldsymbol{u} \cdot \boldsymbol{v} = \sum_i u_i v_i$, and the matrix product $C_i = \sum_j A_{ij} B_{jk}$. A matrix-*vector* product $U_1 \rho_0$ is shown in Fig. 3.1a. The edges of a tensor may then be referred to as internal, if they connect to another tensor, or external, if they remain free. The diagrammatic representation makes it straightforward to handle calculations involving large networks of tensors that would otherwise be troublesome to write down.

[12] The exactness then is predicated on the harmonic bath model providing an accurate description for the physical system in question.

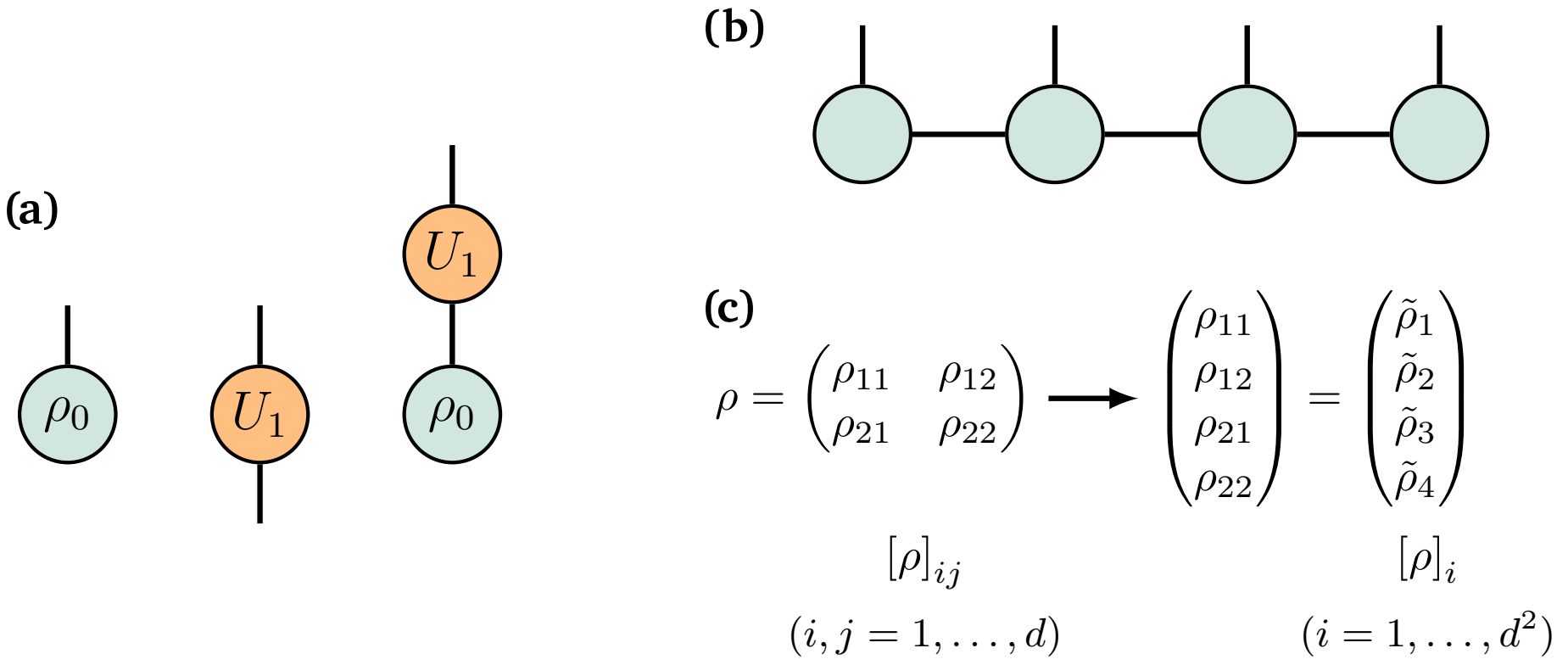

**Figure 3.1:** (a) Tensor network representation of a rank-1 tensor (a vector) $\rho_0$ and a rank-2 tensor (a matrix) $U_1$. Each outgoing edge (leg) represents an index. An internal edge joining two tensors indicates a contraction involving the corresponding indices. (b) Many tensors are joined together (contracted) in certain patterns to construct a tensor network. (c) In Liouville space a $d \times d$ density operator (here $d = 2$) is flattened to a vector of $d^2$ entries. Similarly a superoperator $U$ becomes a $d^2 \times d^2$ matrix. ↑TOF

### Matrix Product States

Matrix product states (MPS) are a particular network structure used to represent composite systems. The idea is to write the total state as a network of many small, low-rank tensors instead of a single high-rank one. Let $|\psi\rangle$ be the state of a many-body system of $N_s$ sites (Hilbert spaces). For simplicity we consider a case where each site is identical, but the following may be applied to more general composite systems. If $|\varphi_i\rangle$ are basis vectors local to each site then a general expression for $|\psi\rangle$ is

$$|\psi\rangle = \sum_{i_1,\ldots,i_{N_s}}^{d} C_{i_1 i_2 \ldots i_{N_s}} |\varphi_{i_1}\rangle \otimes |\varphi_{i_2}\rangle \ldots \otimes |\varphi_{i_{N_S}}\rangle \,, \tag{3.2.1}$$

where $d$ is the local Hilbert space dimension. The $d^{N_s}$ coefficients $C_{i_1 i_2 \ldots i_{N_s}}$ form a single $N_s$-rank tensor, a representation of $|\psi\rangle$ in this basis. An equivalent MPS form may be derived via a series of singular value decompositions (SVDs) [239]. Referring to the tensor diagrams in Fig. 3.2, a possible procedure is:

a) Collect all indices into two groups, dividing the tensor (approximately) in the middle to form a matrix: $C_{i_1 \ldots i_N} \rightarrow \hat{C}_{j_1 j_2}$

b) Perform a SVD, $\hat{C} = U\Sigma V$, and multiply the matrix of singular values into either $U$ or $V^\dagger$. Split up the indices of these matrices to produce two tensors of (approximately) half the rank of $C$.

c) Repeat a)-b) until a chain of rank-3 tensors results with rank-2 tensors at either end.

The result is a decomposition in the form

$$C_{i_1,\ldots i_{N_s}} = \sum_{j_1 \ldots j_{N_s-1}} A_{i_1}^{j_1} A_{i_2 j_1}^{j_2} \ldots A_{i_{N_s-1} j_{N_s-2}}^{j_{N_s-1}} A_{i_{N_s} j_{N_s-1}}, \tag{3.2.2}$$

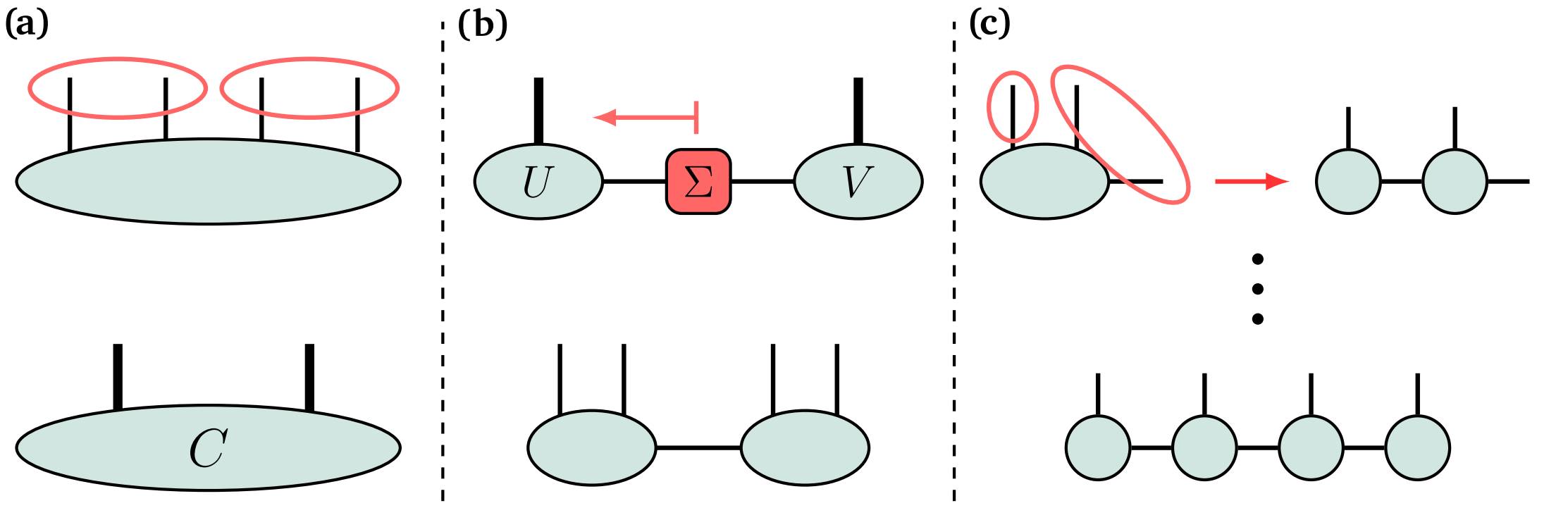

**Figure 3.2:** Example MPS decomposition for a many-body state with $N_s = 4$ sites. (a) The indices are grouped into two large indices (of dimension equal to the product of the dimensions of the combined indices) to give a matrix $C$. (b) A SVD $C = U\Sigma V^\dagger$ is made. The matrix of singular values $\Sigma$ is absorbed into $U$ (or $V^\dagger$) and the large indices split up. (c) Index-grouping, SVD and index-ungrouping is repeated on each of the smaller tensors, finally giving a combination of $N_s = 4$ low-rank tensors, the MPS. ↑TOF

where on the right-hand side the $i$ indices correspond to outgoing edges in Fig. 3.2c and the $j$ indices to internal ones. The dimensions of the internal edges are referred to as *bond dimensions* as opposed to the physical dimension $d$ of each of $i_1, \dots, i_{N_s}$.

The above steps did not detail exactly how the indices were grouped and split up, or how the SVDs were performed. There are different possible choices for these operations, leading to variations in the final MPS. However, these distinctions are not relevant to our discussion. What is important is that so far no information has been lost: Eq. (3.2.2) is an exact representation. As such the maximum bond dimension grows exponentially with $N_s$. So while the above algorithm yields a faithful representation of $|\psi\rangle$, it is not a particularly useful one for large $N_S$. However, suppose during each SVD some of the singular values $\sigma_r$ were close to zero. Then, to a good approximation, those values—and the corresponding columns of $U$, $V$—could be discarded, lowering the bond dimension and providing a more efficient representation.

One might at each step consider keeping only the $\chi$ largest singular values. According to the Schmidt decomposition (Section 3.1.4), this corresponds to discarding weakly entangled contributions to the state, specifically between the two parts divided by each SVD. The truncation introduces error, but this is well controlled by the value of $\chi$[13]. In the limit where $\chi = 1$ is accurate, the many-body state is represented by a simple product state.

The utility of MPSs comes when one considers local Hamiltonians and observables can be formulated as matrix product *operators* (MPOs) comprising low-rank tensors with two outgoing legs. The action of these operators on the state is computed by contracting the sites of the MPO with the MPS, allowing for time evolution and expectations to be calculated efficiently (Fig. 3.3). Other operations such as the norm, trace and derivatives are also straightforward to enact within this framework [239].

In practice, to calculate dynamics one usually begins with a state represented by many low-rank tensors, e.g., a product state, and computes the evolution whilst keeping the bond dimension low by performing sweeps of truncated SVDs along the state. Here one calculates the SVD

[13] An SVD provides the best lower-rank approximation for a matrix [256] in minimising the matrix-norm $||C - \tilde{U}\tilde{\Sigma}\tilde{V}^\dagger||$ with error $\epsilon = \left(\sum_{r=\chi+1}^{\text{rank}(C)} \sigma_r^2\right)^{1/2}$. Here $\sigma_{\chi+1}, \dots \sigma_{\text{rank}(C)}$ are the discarded singular values and $\tilde{U}$, $\tilde{\Sigma}$, $\tilde{V}$ are truncated versions of $U$, $\Sigma$, $V$ according to the removal of these values. In practice, verifying the accuracy of a truncation may be done by repeating the calculation with a larger value of $\chi$ and comparing results.

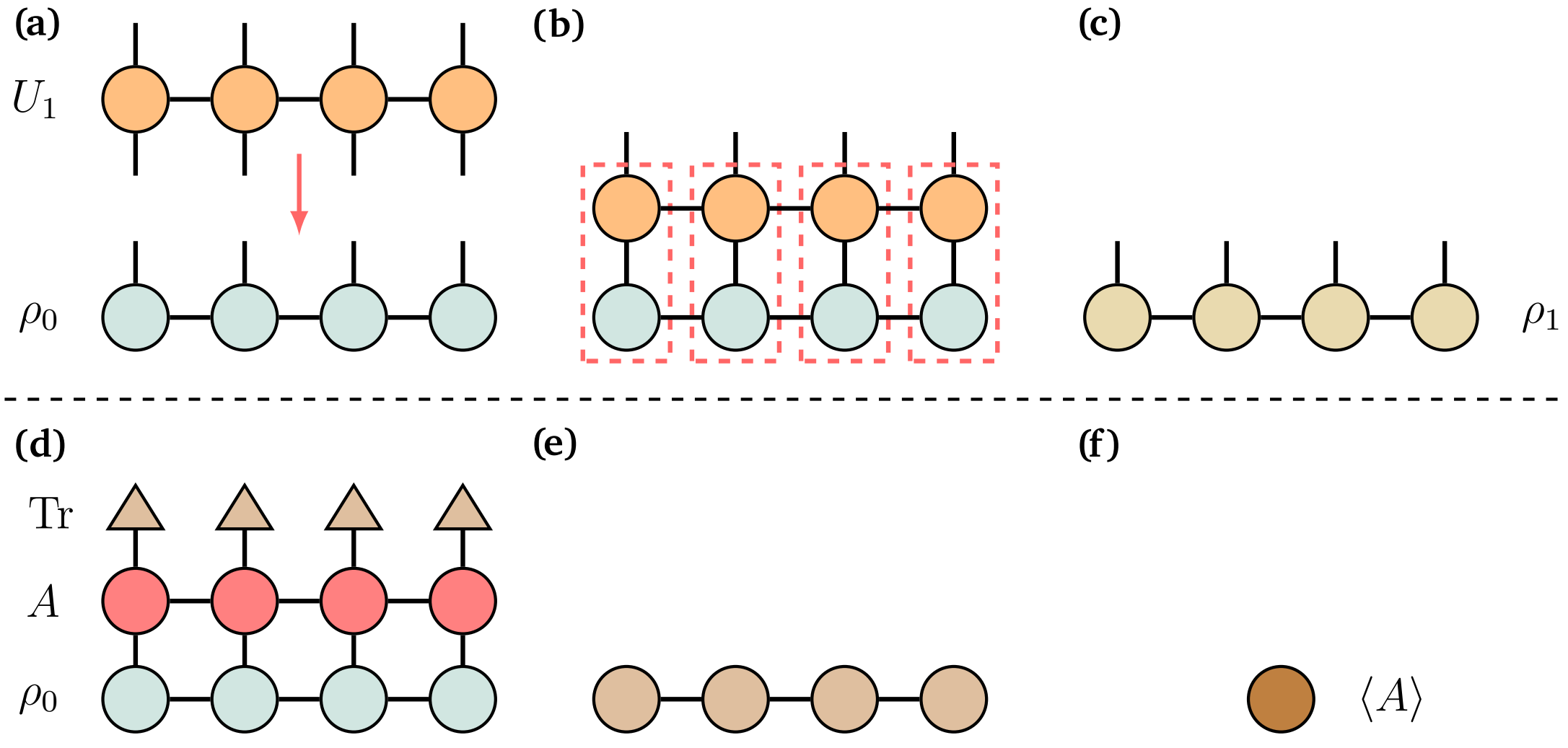

**Figure 3.3:** Matrix product operators (MPOs). (a)–(c) Contracting a MPO $U_1 = U(t_1)$ representing an evolution operator with the state vector $\rho_0 = \rho(0)$ gives the state $\rho(t_1) = U(t_1)\rho(0)$ at time $t_1$. (d)–(f) Calculation of the expectation $\langle A \rangle = \mathrm{Tr}(A\rho_0)$ for the state $\rho_0$. The triangle nodes represent a trace operation. Here the vertical contraction is performed before a horizontal contraction, but any contraction order would give the same result (different orders may be more or less efficient to compute however). Note the result is a scalar and so has no indices (legs). ↑TOF

$U\Sigma V^\dagger$ for the matrix associated with one of the tensors either side of the bond, and then inserts an approximate identity $\tilde{U}\tilde{U}^\dagger \approx I$ between the two tensors, where $\tilde{U}$ is the matrix $U$ with columns corresponding to the discarded singular values removed. This is fundamentally a good strategy for the fact that physically relevant states typically occupy only a small fraction of the entire many-body Hilbert space, namely where entanglement is limited due to the local nature of interactions [238, 257][14].

### 3.2.4 Paths to TEMPO ↑

Many different approaches have been taken to the problem of simulating general, non-Markovian open system dynamics. Numerically exact methods include those using hierarchical equations of motion [259, 260], networks of effective damped harmonic oscillators and representations with Markovian embeddings [261–265], time-dependent density matrix renormalisation group techniques in conjunction with chain mappings [248, 266–272], and quantum trajectories [273]. There are a further array of methods using Monte Carlo and stochastic wave function descriptions [274–278].

The TEMPO method is part of a class of path integral tensor network methods [162, 279–285] based on the Feynman-Vernon influence functional [185] for systems coupled linearly to harmonic environments. In this approach, the environment is integrated out of the full system path integral to yield an influence functional coupling the current evolution of the system to its *history*. In 1995, Makri and Makarov [279, 280] provided a discretisation of the influence functional known as the quasi-adiabatic path integral (QUAPI). Here influence functions $I_k(j, j')$ connect states $j$,$j'$ of a discrete basis across $k$ timesteps. The path integral is then formulated as a propagator for an extension of the density matrix to include states at previous timesteps called the *augmented density tensor* (ADT). While numerically exact, this representation grows exponentially with the environment memory length, i.e., maximum number of influence functions kept in the description.

[14]An MPS with bond dimension $\chi$ encodes what is known as an area law: the entanglement entropy is bounded (in one spatial dimension) by a constant, in this case set by $\chi$ [257]. Typical ground states of quantum many-body systems with local Hamiltonians satisfy such area laws [257, 258].

The key advancement of the TEMPO method is to represent the ADT and its propagator using matrix product states and operators. This allows singular value truncations to control the growth of information and obtain a highly efficient description of non-Markovian dynamics for small systems. We now discuss the key components of the TEMPO tensor network and algorithm. Further details may be found in the methods section of the original TEMPO paper [162], A. Strathearn's thesis [233], and in publications [241, 286–290] concerning additional developments and usage. Information on the path integral formulation of quantum mechanics may be found in standard textbooks [42, 164].

### 3.2.5 TEMPO tensor network

The starting point for the TEMPO method is a system (Hamiltonian $H_S$) coupled linearly to a harmonic bath (see Section 3.1.7):

$$H_{IE} = H_E + H_I = \sum_j \left[ \nu_j b_j^\dagger b_j + \left( \xi_j b_j + \bar{\xi}_j b_j^\dagger \right) S \right]. \tag{3.2.3}$$

The system and environment are assumed to be initially uncorrelated, meaning $\rho_{SE}(0) = \rho(0) \otimes \rho_E(0)$, and the bath in thermal equilibrium such that $\rho_E(0) = e^{-H_E/T} / \operatorname{Tr}_E(e^{-H_E/T})$. Note we continue to denote the reduced system density matrix $\rho$ without a subscript.

The system Hamiltonian $H_S$, coupling operator $S$ and initial state $\rho_S(0)$ are all left to be chosen by the user of the method, as are the spectral density $J(\nu)$ and temperature $T$ characterising the environment. The calculation is performed in a basis $|s\rangle$ in which $S$ is diagonal. While the TEMPO method was originally derived for time independent systems, it has been extended [241] to handle general time-dependent $H_S(t)$, and we consider this generality here. On the other hand, $H_{IE}$ is always time-independent.

To derive the dynamics of the system density matrix $\rho(t)$ given the initial state $\rho_{SE}(0)$, consider the solution to the von Neumann equation for the total evolution,

$$\partial_t \rho_{SE}(t) = \mathcal{L}(t)\rho_{SE}(t) \to \rho_{SE}(t) = \mathcal{T} e^{\int_0^t dt' \mathcal{L}(t')} \rho_{SE}(0) \qquad (\mathcal{L} \equiv -i\,[H, \cdot])\,, \tag{3.2.4}$$

where the total Liouville operator $\mathcal{L}(t) = \mathcal{L}_S(t) + \mathcal{L}_{IE}$ has parts corresponding to $H_S(t)$ and $H_{IE}$. While we have written only a unitary contribution of $H_S(t)$ for $\mathcal{L}_S$, TEMPO can accommodate coupling to additional baths for which the Markovian approximation is justified by the addition of Lindblad operators $L$ to $\mathcal{L}_S$ [162, 291]. We discuss further below how discrete operator interventions describing, e.g., measurements or resets of the system, may also be included in the evolution.

The corresponding solution for $\rho(t)$ is obtained from a trace,

$$\rho(t) = \operatorname{Tr}_E \left( \mathcal{T} e^{\int_0^t dt' \mathcal{L}(t')} \rho(0) \otimes \rho_E(0) \right), \tag{3.2.5}$$

where we substituted the product form of $\rho_{SE}(0)$.

With the goal of evaluating Eq. (3.2.5) via numerical integration, the time evolution is split up into $N$ times steps of length $\delta t = t/N$:

$$\rho(t) = \operatorname{Tr}_E \left( \prod_{n=0}^{N-1} \mathcal{T} e^{\int_{t_n}^{t_{n+1}} dt' \mathcal{L}(t')} \rho(0) \otimes \rho_E(0) \right), \quad t_n = n\delta t. \tag{3.2.6}$$

In order to separate the system and interaction-environment propagators at each timestep, a symmetrised[15] Trotter splitting is applied as

$$\mathcal{T}e^{\int_{t_n}^{t_{n+1}} dt' \mathcal{L}(t')} \equiv \mathcal{T}e^{\int_{t_n}^{t_{n+1}} dt' (\mathcal{L}_S(t') + \mathcal{L}_{IE})} \tag{3.2.7}$$

$$\approx \left(\mathcal{T}e^{\int_{t_n+\delta t/2}^{t_{n+1}} dt' \mathcal{L}_S(t')}\right) e^{\delta t \mathcal{L}_{IE}} \left(\mathcal{T}e^{\int_{t_n}^{t_n+\delta t/2} dt' \mathcal{L}_S(t')}\right) + O(\delta t^3), \tag{3.2.8}$$

where we used that $\mathcal{L}_{IE}$ is time-independent. The full evolution Eq. (3.2.6) becomes

$$\rho(t) = \mathrm{Tr}_E \left[ \left( \prod_{n=0}^{N-1} \mathcal{T}e^{\int_{t_n+\delta t/2}^{t_{n+1}} dt' \mathcal{L}_S(t')} e^{\delta t \mathcal{L}_{IE}} \mathcal{T}e^{\int_{t_n}^{t_n+\delta t/2} dt' \mathcal{L}_S(t')} \right) \rho(0) \otimes \rho_E(0) \right]. \tag{3.2.9}$$

To make the bridge to a path integral, consider the expression for $\rho(t)$ in the eigenbasis $\{|s\rangle\}$ of $S$:

$$\rho(s', s'', t_N) = \mathrm{Tr}_E \left[ \langle s''| \left( \prod_{n=0}^{N-1} \mathcal{T}e^{\int_{t_n+\delta t/2}^{t_{n+1}} dt' \mathcal{L}_S(t')} e^{\delta t \mathcal{L}_{IE}} \mathcal{T}e^{\int_{t_n}^{t_n+\delta t/2} dt' \mathcal{L}_S(t')} \right) \rho_S(0) \otimes \rho_E(0) \, |s'\rangle \right]. \tag{3.2.10}$$

The path integral formulation states that the total amplitude to go from $|s'\rangle$ to $|s''\rangle$ is calculated by summing over all possible intermediary states (i.e. paths) weighted by their probability amplitude. Those familiar with this prescription will recall that in practice it amounts to inserting resolutions of identity $\sum_s |s\rangle\langle s|$ between the propagators of successive timesteps.

In fact, in the TEMPO algorithm two resolutions of identity are made at each timestep, one for each system propagator in Eq. (3.2.8). The calculation is also made in Liouville space, where $\rho_{j_N}$ are the $d^2$ ($d = \dim(\mathcal{H}_S)$) components of the reduced system density matrix. The resolutions of identity then appears as $\sum_{j_i} |j_i\rangle\rangle\langle\langle j_i|$, $\sum_{j_i'} |j_i'\rangle\rangle\langle\langle j_i'|$, where $i = 0, \ldots, N-1$ indicates the timestep and each $|j_i\rangle\rangle$, $|j_i'\rangle\rangle$ is a basis vector of Liouville space formed from the $d^2$ elements $|s\rangle\langle s'|$ (see Ref. [297] for a discussion of this vectorisation).

Crucially, the trace over the environment—given the initial product state involving $\rho_E(0)$—can be performed in the $S$ eigenbasis [279, 280]. This produces a series of 'influence functions' capturing the action of the environment and the main result of Refs. [279, 280], the quasidiabatic propagator path integral. This is formulated for the TEMPO method [162, 233, 241] as the time evolution

$$\rho_{j_N}(t) = \sum_{jj'} \prod_{n=0}^{N-1} \left[ P^{(n+)}_{j_{n+1}j_n'} \left( \prod_{k=0}^{n} I^{(k)}(j_n', j_{n-k}') \right) P^{(n-)}_{j_n' j_n} \right] \rho_{j_0}, \tag{3.2.11}$$

where $\rho_{j_0}$ are the components of the initial system density matrix and the sum runs over all primed $j_0', \ldots, j_{N-1}'$ and unprimed $j_0, \ldots, j_{N-1}$ indices. The $P^{(n\pm)}_{jj'}$ are the components of the system propagators for each half-timestep,

$$P^{(n-)}_{j'j} = \left[ \mathcal{T}e^{\int_{t_n}^{t_n+\delta t/2} dt' \mathcal{L}_S(t')} \right]_{j'j}, \tag{3.2.12}$$

[15] A symmetrised splitting is used for the reason that the non-symmetric version ($e^{A+B} \approx e^A e^B$) has error $O(\delta t^2)$, hence slower convergence under $\delta t \to 0$. Depending on who you ask, the symmetrised splitting may be referred to as a symmetric Trotter splitting [292], Strang splitting [293] or the Trotter-Kato product formula [294], not to be confused with that due to Baker-Campbell-Hausdorff [295] or further Feynman's disentangling technique [296].

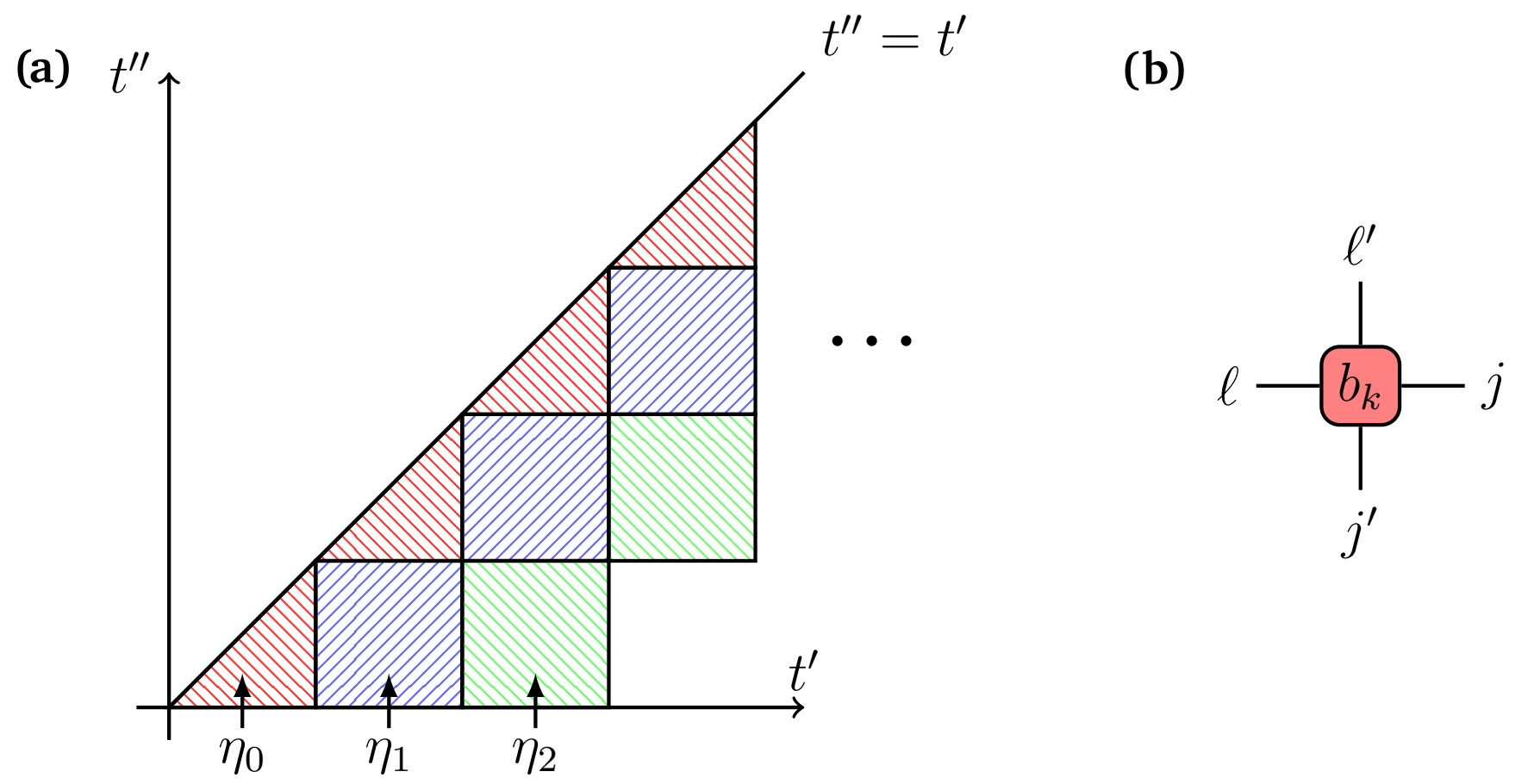

**Figure 3.4:** (a) Domains of integration for the double integrals Eq. (3.2.15). If a finite memory approximation is made, integrals are calculated up to a certain $k_{\max}$ only (here $k_{\max}=2$). (b) Four index bath tensors are constructed by attaching Kronecker deltas to the influence functionals $I^{(k)}_{jj'}$ as per Eq. (3.2.17). ↵TOF

and

$$P^{(n+)}_{jj'} = \left[\mathcal{T} e^{\int_{t_n+\delta t/2}^{t_{n+1}} dt' \mathcal{L}_S(t')}\right]_{jj'}. \tag{3.2.13}$$

The $I^{(k)}$ are the influence functions, defined by

$$I^{(k)}(j,j') = \exp\Big[-\hat{S}^-_j \left(\hat{S}^-_{j'}\,\mathrm{Re}[\eta_k] + i\hat{S}^+_{j'}\,\mathrm{Im}[\eta_k]\right)\Big], \tag{3.2.14}$$

where $\hat{S}^\pm_j$ are the diagonal components of the superoperators $\hat{S}^- := [S,\cdot]$ and $\hat{S}^+ := \{S,\cdot\}$. We comment further on the role of these components in Section 3.2.6 below.

The $\eta_k$ are double integrals of the bath correlation function $C(t)$,

$$\eta_k = \begin{cases} \displaystyle\int_0^{\delta t} dt' \int_0^{t'} dt'' C(t'-t'') & k=0 \\ \displaystyle\int_{k\delta t}^{(k+1)\delta t} dt' \int_0^{\delta t} dt'' C(t'-t'') & k>0, \end{cases} \tag{3.2.15}$$

with (see Section 3.1.7)

$$C(t) = \sum_j \langle x_j^\dagger(t) x_j(t-s)\rangle = \int_0^\infty d\nu J(\nu)\left[\coth\left(\frac{\nu}{2T}\right)\cos(\nu t) - i\sin(\nu t)\right]. \tag{3.2.16}$$

Integrals of time arise in the influence functions due to the continuum limit $\delta \to 0$ being applied to this part (specifically, under the trace of the interaction-environment propagators), in accordance with the result of Feynman and Vernon [185]. To interpret $\eta_k$ for successive $k$, it is helpful to plot the integration region in the $t' - t''$–plane. In Fig. 3.5a we see that $\eta_0$ captures the immediate impulse of the environment on the dynamics, and $\eta_k$ the delayed action over $k$ timesteps that introduces non-Markovianity.

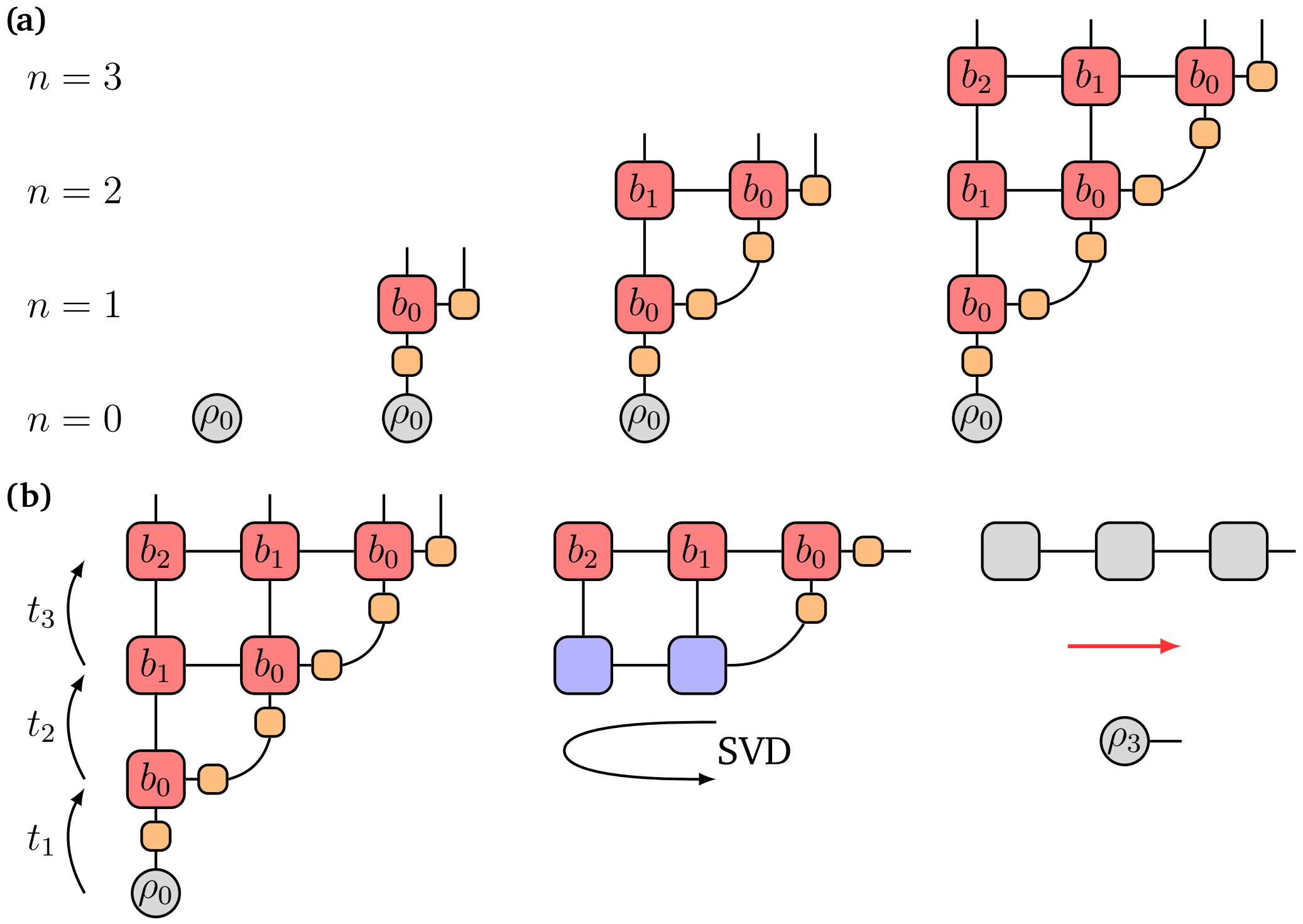

**Figure 3.5:** The TEMPO tensor network. (a) Starting from the initial state $\rho_0$ at each $n > 0$ two half-system propagators $P^{(n)}$ (orange) and one or more bath tensors $b_k$ with $k = 0 \ldots n-1$ are added. The network grows vertically with timestep $t_n = n\delta t$. Control operators (not shown) may also be inserted around the system propagators to allow for experimental interventions and the measurement of multi-time correlations. (b) The dynamics to time $t = t_3$ (say) are calculated by contracting the network vertically (note a trace is performed over the unused legs of the bath tensors). At each stage truncated SVD sweeps are used to compress the representation. The state $\rho_3 = \rho(t_3)$ is calculated by a final horizontal contraction as shown.

↑TOF

Equation (3.2.11) for $\rho_{j_N}(t)$ contains all the pieces to construct the TEMPO tensor network. Referring to this equation and Fig. 3.5a, $n = 0$ describes the initial state. At $n = 1$ there are two system half-propagators $P^{(0\pm)}$, rank-2 tensors drawn as small orange squares in Fig. 3.5a, and a single $k = 0$ influence functional $I^{(0)}$ which defines the red 'bath' tensor $b_0$.

Next, $n = 2$ again has a pair of system propagators surrounding a $k = 0$ bath tensor but. In addition, there is a $k = 1$ bath tensor $b_1$ capturing the second influence function $I^{(1)}$ at this stage. The TEMPO network can be built step-wise in this fashion, with time increasing vertically upwards in Fig. 3.5a. In this diagram we see that the bath tensors appear as rank-3 and rank-4 objects, whereas the influence functions only have two indices. This is achieved by adding one or two Kronecker deltas according to the definition

$$[b_k]_{jj'll'} = I^{(k)}_{jj'}\delta_{lj}\delta_{l'j'}. \tag{3.2.17}$$

As the influence functions depend only on the step difference $k$, each bath tensor $b_k$ only needs to be calculated once for the evolution. For example, the $b_0$ at $n = 1, 2, 3, \ldots$ are all copies of each other.

In order to calculate the dynamics, the tensor network is contracted sequentially upwards as shown in Fig. 3.5b. Crucially, at each step horizontal sweeps of SVDs with truncation are performed to maintain a manageable bond dimension. Here correlations over *time* that are weak are discarded. The array of low-rank tensors at this stage (blue in Fig. 3.5b), encodes the entire history of states and is called the augmented density tensor (ADT). Finally, an additional horizontal contraction is made to produce the system density matrix $\rho(t_n)$ at a given timestep.

The use of SVD compression allows the growth of the ADT with memory step $k$ to be reduced from exponential to linear [233, 241], making memory lengths $k\delta t$ covering hundreds of timesteps possible. For propagation to arbitrary times, one can make use of the physical expectation that the bath correlation function decays to zero over some characteristic time $\tau_c$ [297]. Then a finite memory approximation can be made such that bath tensors $b_k$ are only added up to $k_{\text{max}} = K$, where $K$ is chosen to satisfy $K\delta t \gtrsim \tau_c$.

Observe there is competition between computational parameters $K$ and $\delta t$: one should choose a $\delta t$ small enough for a sufficient resolution of the dynamics, but decreasing $\delta t$ necessitates a larger $K$ to satisfy $K\delta t \gtrsim \tau_c$ and so increases the storage requirements for the ADT.

In Section 3.2.3 we introduced compression via SVD truncations as keeping the $\chi$ largest singular values and so a fixed bond dimension. The TEMPO method actually uses a fixed *precision* $\epsilon_{\text{rel}}$ whilst allowing bond dimension to vary. In this scheme singular values smaller than $\epsilon_{\text{rel}}$ relative to the larger singular value are discarded. This provides more direct control over the precision at the risk of a long or diverging computation time if $\chi$ grows too large to meet the specified accuracy.

After the timestep size $\delta t$ and any finite memory approximation $K\delta t$, $\epsilon_{\text{rel}}$ completes the core set of numerical parameters in a TEMPO calculation for which the convergence of results should be checked.

### 3.2.6 Computational cost and degenerate couplings

As noted above, the MPS representation with singular value truncation reduces the storage requirements for the quasidiabatic path integral from exponential to linear in the bath memory length [233, 298][16]. The main cost are the SVD sweeps, which have an overhead that scales with the third power of the internal bond dimension [245]. In total, the method can be implemented with a computational cost that scales linearly with the memory length but to the third power of both the bond dimension *and* the Liouville space dimension, i.e., $\sim (d^2)^3$ [241]. Hence the cost of a calculation depends strongly on the both the precision $\epsilon_{\text{rel}}$, which controls the bond dimension, and the system size $d$. Nonetheless, the method has proven to be of practical use for many problems involving small systems with a range of bath coupling strengths [162, 286, 303–311].

Due to the unfavourable scaling with system Hilbert space dimension $d$, the method has so far[17] been limited to studies of small, few-body systems. However, we note the form Eq. (3.2.14) of the influence functions allows the cost of increasing $d$ to be mitigated in certain cases. Recall

$$I^{(k)}(j, j') = \exp\left[-\hat{S}_j^- \left(\hat{S}_{j'}^- \operatorname{Re}[\eta_k] + i\hat{S}_{j'}^+ \operatorname{Im}[\eta_k]\right)\right], \tag{3.2.18}$$

where $\hat{S}_j^\pm$ are the diagonal components of the superoperators $\hat{S}^- := [S, \cdot]$, $\hat{S}^+ := \{S, \cdot\}$ in the eigenbasis of the bath coupling operator $S$. These components are equal to the differences and sums of the eigenvalues of $S$, respectively. It was realised [240, 312] that the dimension of the bath tensors used in the TEMPO computation can therefore be reduced if there are degeneracies in

[16] Alternative schemes using different decompositions and path filtering techniques [299–301] have been used to reduce this exponential scaling. Many of these were reviewed alongside the TEMPO algorithm in Ref. [302].

[17] A new development [240, 290] combining TEMPO with time-evolving block decimation allows chains of open quantum systems to be handled. This is not related to approach we take in Chapter 4 for a many-to-one network.

**(a)**

$$S = \sigma^z = \begin{pmatrix} 1 & 0 \\ 0 & -1 \end{pmatrix}$$

| $s_1$ | $s_2$ | $s_1 + s_2$ | $s_1 - s_2$ |
|---|---|---|---|
| 1 | 1 | 2 | 0 |
| 1 | −1 | 0 | 2 |
| −1 | 1 | 0 | −2 |
| −1 | −1 | −2 | 0 |

$$I^{(k)}_{jj'}$$

$$(4 \times 4) \to (3 \times 4)$$

**(b)** $$S = \sigma^z \otimes I_{N_\nu}$$

$$(4N_\nu^2, 4N_\nu^2) \to (3 \times 4)$$

**Figure 3.6:** (a) The eigenvalues for $S = \sigma^z$ coupling a two-level system to the bath have 3 unique differences, so the required dimension of the first coordinate ($j$) of the bath influence functions reduces from $d^2 = 4$ to 3. There are no simultaneous degeneracies in the differences and sums, so the first coordinate remains of dimension 4. (b) If an $N_\nu$-level system couples to the electronic state, but not the environment, i.e., $S = \sigma^z \otimes I_{N_\nu}$, then the eigenvalues of $\sigma^z$ are repeated and the problem reduces to $(4 \times 3)$, for any $N_\nu$. TOF

the differences, or further in the differences *and* sums, of pairs of eigenvalues of $S$ (see Fig. 3.6 for an example). Since operations with these tensors represent the major part of the total calculation, this can provide a vast reduction of computational cost in the case of large degeneracies of this type, e.g., due to repeated eigenvalues or only a subspace of the system being coupled to the bath. In Chapter 4 we discuss an extended Holstein-Tavis-Cummings model as a relevant case.

The degeneracy simplification has been implemented [298] for both the TEMPO method as well as its process tensor formulation, which we now discuss.

### 3.2.7 Process tensor formulation (PT-TEMPO)

So far we have explained the TEMPO method as originally implemented in Ref. [162]. However, there is a second implementation by G. Fux [241, 303] called process tensor TEMPO (PT-TEMPO) that will be useful for our extension of the method in Chapter 4. This follows the realisation, made by Jørgensen and Pollock [281, 313], that the TEMPO network can be reformulated in terms of a an object known as a *process tensor* (PT). Specifically, contracting the bath tensors horizontally as shown in Fig. 3.7a yields a matrix product operator representation of the PT. This can be calculated independently of the system initial state and propagators. Hence it only needs to be computed once for a given environment. A stored PT can be combined with different sets of system propagators or control operations to calculate the system dynamics [244]. As constructing the network of bath tensors represents the bulk of a typical TEMPO calculation, this allows for many evolutions to be calculated at relatively little cost.

PT-TEMPO is particularly well suited to investigate dynamics under the variation of one or more system parameters [303]. The possibility of inserting control operations, i.e., operators between system propagators also enables the efficient calculation of multi-time correlation functions. For example, to compute the two-time correlation function $\langle A_i(t)A_j(t_0)\rangle$, one inserts in the network an operator $A_j$ acting on the state alongside the system propagators at $t = t_0$. The desired correlation function is given by the expectation of $A_i$ in the subsequent dynamics, $\langle A_i(t)\rangle_0 = \langle A_i(t)A_j(t_0)\rangle$ for $t \geq t_0$, where the 0 subscript indicates that an intervention occurred at $t = t_0$.

The process tensor was introduced by Pollock et al. [244] as a mapping from the set of all possible control operations (measurements or other interventions) to output states. We do not discuss the details of this mapping but notes that it provides a powerful *operational* framework

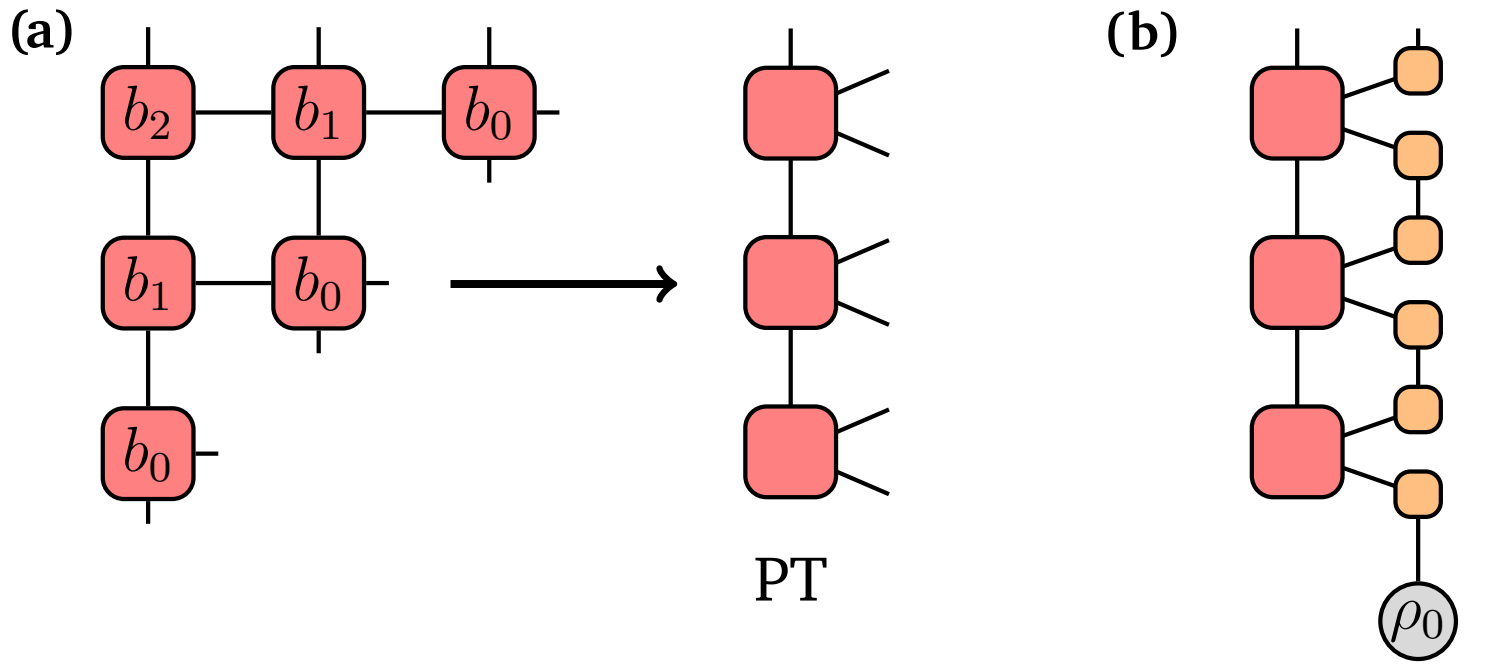

**Figure 3.7:** (a) In the process tensor formulation of TEMPO (PT-TEMPO), the bath influence tensors are computed separately and contracted (horizontally) to form the process tensor (PT). (b) An initial state $\rho_0$ and system propagators (orange) are then sequentially contracted with the PT to calculate system dynamics. As in Fig. 3.5 time increases in the upward vertical direction. ↵TOF

for characterising arbitrary quantum processes, including non-Markovian ones. The MPO representation of the process tensor has also recently been used in the automated compression of environments (ACE) method [314]. This also allows for the computation of non-Markovian open quantum system dynamics, with the feature of being able to accommodate contributions from multiple independent environments that are not necessarily Gaussian.

### 3.2.8 OQuPy Python Package ↵

Open Quantum Systems in Python or *OQuPy* [298] is an open-source Python 3 package incorporating the original TEMPO algorithm [162], PT-TEMPO [303] and extensions [287, 288, 290] of both methods. The extension we develop in Chapter 4 has been included as 'mean-field TEMPO' for both TEMPO and PT-TEMPO implementations. The package is available on the Python Package index as oqupy with API documentation and tutorials available on readthedocs. Contributions to the codebase are welcomed on the OQuPy GitHub responsitory.

### 3.2.9 Many-body opens systems ↵

As discussed in Section 3.2.6, the TEMPO method has so far been limited to study small systems. While leveraging degeneracies in the eigenspectrum of the bath coupling operator may allow for the treatment of certain single or few-site systems of moderate Hilbert space dimension, this cannot provide a solution for the type of many-body system we are interested in, which have multiple open parts. For this purpose we will require an additional strategy: mean-field theory.

The restriction to small system sizes holds true for other exact numerical approaches. Broadly, methods for large many-body systems—open or closed—are limited to specific models. For example, we saw in the previous chapter how many-to-one models involving only collective processes for $N$ satellites can be solved when $N \gg 1$ using a collective spin representation and a Holstein-Primakoff transformation. In the presence of individual processes, e.g. pumping, this approach cannot be used. From the opposite limit, $N$ small, permutation symmetries of the model can be used to calculate dynamics efficiently within the Markovian regime. This can handle individual processes (see Chapter 5), but is limited to solving systems of tens, at most hundreds, of particles. The mean-field approaches we develop on the other hand are widely applicable to large systems.

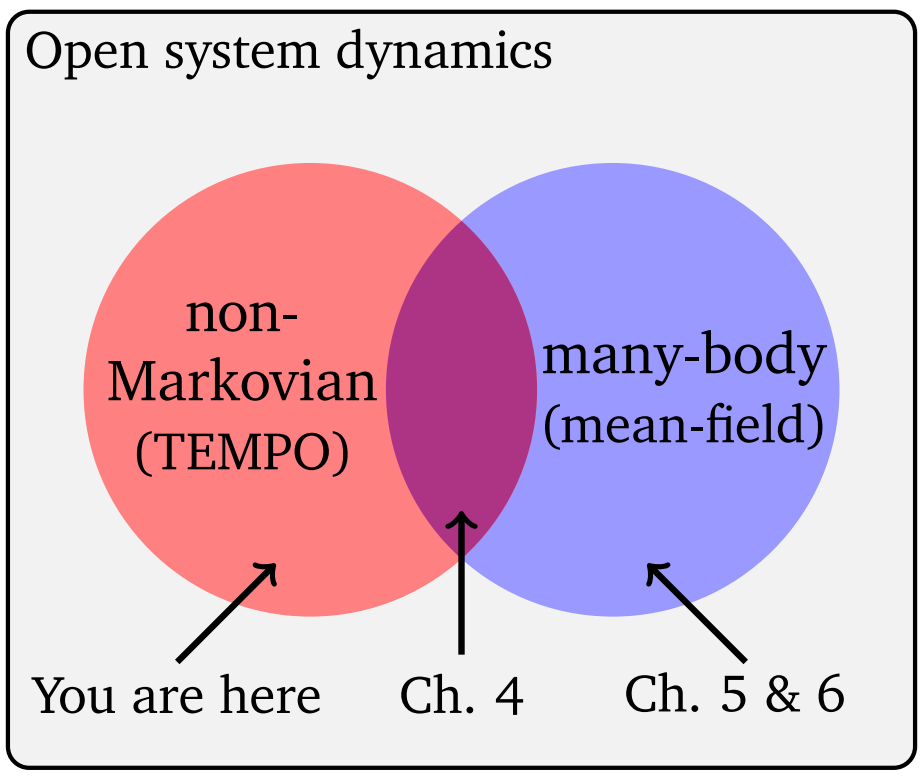

**Figure 3.8:** *More is different, many is difficult.* The TEMPO method discussed in this section calculates exact non-Markovian open system dynamics, whilst mean-field theory (Section 3.3.1) may be used to solve many-body dynamics. In Chapter 4 we combine the two methods to address the difficult intersection: many-body and non-Markovian. Chapters 5 and 6 address many-body models with Markovian master equations. ↑TOF

In particular, in Chapter 4 we explain how mean-field theory (Section 3.3) can be used in conjunction with the TEMPO method to simulate many-body dynamics with strong coupling to multiple environments (Fig. 3.8). Thereafter, in Chapters 5 and 6, we use cumulant expansions to capture finite size effects for the Markovian dynamics of large many-body systems.

## 3.3 Mean-field theory

We now introduce the cumulant expansion approach for models with long-range interactions, starting from mean-field theory and its operation in solving quantum many-body problems. These approaches underpin every method developed in this thesis.

### 3.3.1 Historical introduction

Mean-field theory has a wide-ranging legacy in solving many-body problems in science and mathematics. So varied are its applications that even within physics the term 'mean-field' is interpreted differently according to the field of study. Its origins date back to 1907 with P. Weiss' work on the Curie point for ferromagnetic materials [315]. Inspired by P. Langevin's earlier theory for magnetic properties [316], Weiss hypothesised that every molecule experienced the average action or *mean field* arising from the molecules surrounding it. This earliest incarnation of mean-field theory is perhaps the most widely recognised, being used extensively to study critical behaviour and phases in quantum magnetism as well as other 'molecular field' problems, for example, involving liquid crystals [317] and fluid mixes [318].

A second prominent form of mean-field theory is the saddle point approximation for path integrals in statistical mechanics and field theories [319, 320], stemming from L. Landau's celebrated theory of phase transitions [321, 322]. Mean-field theories have also spread into other disciplines, with diverse applications including protein structure prediction in molecular biology [323], variational Bayesian inference for machine learning tasks [324] and stochastic models of neurological organisation [325].

Our formulation of mean-field theory resembles approaches used in condensed matter physics to handle large systems of interacting fermions or bosons whereby the many-particle wavefunction $\Psi_n(\boldsymbol{x}, t)$ is approximated as a product of single-particle wavefunctions. However, we address open systems and work directly with quantum equations of motion rather than looking to derive effective macroscopic ones [326].

At the fundamental level, mean-field theory prescribes *factorisation* within a probabilistic system. As we describe below, for our purposes this amounts to an assumption on the *structure* of the total many-body state [327–330]. A key tenet of mean-field theory therefore is statistical independence of the parts of a large system. A second is the determination of a *self-consistent* solution, as articulated by F. Vega-Redondo in the quote at the top of this chapter [331]. It stands to reason that mean-field theory is expected to be accurate when many components contribute to the average local field, i.e., for large systems of high connectivity. At the same time, the power of mean-field theory is the reduction of a high dimensional problem to a small one. Hence the approach is useful for handling large, many-body systems that would otherwise be intractable, for example, to direct numerical approaches. In the following we explain our application of mean-field theory to solve quantum many-body dynamics.

### 3.3.2 The mean-field ansatz

Consider an $N$-partite quantum system $\mathcal{H} = \bigotimes_{i=1}^{N} \mathcal{H}_i$. We refer to each part (Hilbert space $\mathcal{H}_i$) as a *site*, whether they represent physical sites of a lattice or not. A mean-field description *may* be applicable for $N$ large, and further *may* become exact as $N \to \infty$. The limit $N \to \infty$ often coincides with the thermodynamic limit [14], meaning the physical extent $L$ of the system increases with $N$ such that the site density $N/L$ remains constant. The $\mathcal{H}_i$ are not necessarily identical, but the greatest simplification comes when many or all of them are.

Mean-field theory is a product-state ansatz for the total many-body state $\rho : \mathcal{H} \to \mathcal{H}$,

$$\rho = \bigotimes_{i=1}^{N} \rho_i. \tag{3.3.1}$$

In other words, there are no correlations—quantum or classical—between different sites. Here $\rho_i$ is the reduced density matrix obtained by tracing over all but subsystem $i$:

$$\rho_i = \mathrm{Tr}_{\otimes j \neq i}\, \rho. \tag{3.3.2}$$

An explicit expression for $\rho$ is not known (otherwise the problem would already be solved!), so we are not intending to construct the $\rho_i$ in this way. Instead, we look to deduce the consequences of ansatz (3.3.1) for the description of the many-body dynamics. This ansatz is general yet operational and, as we show in Section 3.4, readily extended to capture *beyond* mean-field effects.

> *Aside.* We saw in Section 3.2.3 that a MPS with bond dimension $\chi = 1$ describes a product state. One can view the MPS as a variational ansatz that interpolates between the mean-field and exact state as $\chi$ is increased from 1. The cumulant expansion ansatz we develop in the next section also interpolates between the mean-field and exact state with increasing amount of correlations, but in a different way: for $\chi > 1$ the MPS has correlations along the entire chain of sites, i.e., $N$-wise entanglement, whereas in an order-$M$ cumulant expansion entanglement is limited to involve at most $M$ sites. At the same time, the cumulant expansion includes terms for entanglement between every possible combination of sites, beyond that of a MPS.

### 3.3.3 The mean-field approximation

Consider a system undergoing unitary dynamics under $\partial_t \rho = -i[H, \rho]$. From Eq. (3.3.2) an equation of motion for any one of the reduced density matrices is

$$\partial_t \rho_i = -i \operatorname{Tr}_{\otimes j \neq i} [H, \rho] \, . \tag{3.3.3}$$

Ordinarily this is not so useful: the right-hand side involves operators in the full many-body Hilbert space and the trace cannot generally be performed.

This changes with ansatz (3.3.1), however. Then the trace over $j \neq i$ of each term on the right-hand side yields either an expectation $\langle A_j \rangle = \operatorname{Tr}_j(A_j \rho_j)$, if the term contains a non-trivial operator $A_j$ of site $j$, or $\operatorname{Tr}_j(\rho_j) = 1$ in the case that it does not. Evaluating these traces transforms Eq. (3.3.3) from an equation on the total space $\mathcal{H}$ of all sites to one on the single site $\mathcal{H}_i$, albeit with time-dependent coefficients. Those coefficients are the expectations whose Heisenberg equations of motion derive from the same reduced density matrix equation of motion or analogous ones for other types of sites. In total, the result is a reduction of the dimension of the problem[18] for a many-body system with $s$ different types of sites from a product $d_1^{N_1} d_2^{N_2} \dots d_s^{N_s}$ to a sum $d_1 + d_2 + \dots + d_s$, where $d_i$ and $N_i$ are the dimensions and number of each type of site.

In practice, to enact mean-field theory one writes down the Heisenberg equations of motion for single-site operators and factorises expectations of operators from multiple sites into products of single-site expectations, such as $\langle A_1 A_2 \rangle = \langle A_1 \rangle \langle A_2 \rangle$ for operators on $A_1$ and $A_2$ on $\mathcal{H}_1$ and $\mathcal{H}_2$. This prescription follows from the product ansatz (3.3.1) for which

$$\begin{aligned} \langle A_1 A_2 \rangle &= \operatorname{Tr}\left(A_1 A_2 I_3 \dots I_N \rho\right) && (3.3.4) \\ &= \operatorname{Tr}_1\left(A_1 \rho_1\right) \operatorname{Tr}_2(A_2 \rho_2) \operatorname{Tr}_3(I_3 \rho_3) \dots \operatorname{Tr}_N(I_N \rho_N) && (3.3.5) \\ &= \langle A_1 \rangle \langle A_2 \rangle, && (3.3.6) \end{aligned}$$

where we used that each $\rho_i$ is normalised to one.

It is the factorisation of the expectations of operators as above that is normally referred to as the mean-field approximation. It produces a closed system of non-linear equations for the expectations of single-site operators.

### 3.3.4 Example of mean-field theory

A simple example will make the above concepts clear. Note in the following we do not write out individual tensor product symbols $\otimes$; the meaning, e.g., of $\rho_1 \rho_2 = \rho_1 \otimes \rho_2$ should be apparent from the distinct subscripts.

Consider an ensemble of $N$ spins interacting according to

$$H = J \sum_{i,j>i}^{N} \left( \sigma_i^+ \sigma_j^- + \sigma_i^- \sigma_j^+ \right) , \tag{3.3.7}$$

where the sums include the two terms for every pair or sites, and the master equation

$$\partial_t \rho = -i[H, \rho] + \Gamma_\downarrow \sum_{i=1}^{N} L[\sigma_i^-], \tag{3.3.8}$$

[18]This refers to an effective Hilbert space dimension $d_{\text{eff.}}$ for the problem. The number of Heisenberg equations to be solved, in correspondence with the unique components of a density operator (Section 3.1.9), scales as $d_{\text{eff}}^2$.

describing, in addition, individual dissipation for each site.

The choice of a Hamiltonian with non-local, all-to-all interactions is intentional—we only consider models with high (or infinite) connectivity in our work. Otherwise, for lattice models of finite connectivity, the applicability of mean-field theory is highly dependent on the dimension of the embedding space: at low dimensions it is inaccurate—or fails entirely—and instead only becomes accurate beyond a dimension known as the upper critical dimension [42, 320][19]. This follows the intuition that mean-field theory is accurate for models of high connectivity: in higher dimensions each site has more neighbours and so experiences a net effect closer to that of the average field.

At this juncture we point out a key result of our work in Chapter 5 is the *failure* of mean-field theory to capture the $N \to \infty$ limit of a model with unlimited connectivity, contrary to the above rationale. Besides this, mean-field theory is known to be robust for many types of many-to-one and many-to-many models, including the emitter-cavity (Dicke-like) models considered in Chapters 4 and 6. In particular, it can be proven to be exact as $N \to \infty$ [332, 333]. We discuss these and other exact results in detail in Chapter 5.

Returning to the example, all $N$ sites are identical and of dimension 2, so $\dim\mathcal{H} = 2^N$. With the mean-field ansatz (3.3.1), the equation of motion for any $\rho_i$ is

$$\partial_t \rho_i = -iJ\,\mathrm{Tr}_{\otimes j\neq i}\left[\sum_{k,l>k}^{N}\left(\sigma_k^+\sigma_l^- + \mathrm{H.c.}\right), \bigotimes_{r=1}^{N}\rho_r\right] + \Gamma_\downarrow \sum_{k=1}^{N}\mathrm{Tr}_{\otimes j\neq i}\, L[\sigma_k^-], \tag{3.3.9}$$

where H.c. denotes the Hermitian conjugate.

To evaluate the partial traces, first note that any term in the commutator not involving an operator from $\mathcal{H}_i$ vanishes:

$$\begin{aligned}
&\mathrm{Tr}_{\otimes j\neq i}\left(\sigma_k^+\sigma_l^-, \rho_1\ldots\rho_i\ldots\rho_k\ldots\rho_l\ldots\rho_N\right)\\
&= \left(\prod_{j\notin\{i,k,l\}}\underbrace{\mathrm{Tr}_j\,\rho_j}_{=1}\right)\mathrm{Tr}_{k,l}\left(\rho_i\sigma_k^+\rho_k\sigma_l^-\rho_l - \rho_i\rho_k\sigma_k^+\rho_l\sigma_l^-\right) && (3.3.10)\\
&= \left(\rho_i\langle\sigma_k^+\rangle\langle\sigma_l^-\rangle - \rho_i\langle\sigma_k^+\rangle\langle\sigma_l^-\rangle\right) && (3.3.11)\\
&= 0. && (3.3.12)
\end{aligned}$$

Similarly, $\mathrm{Tr}_{j\neq i}\,\mathcal{L}[\sigma_k^-] = 0$ whenever $k \neq i$.

Next, the terms actually involving $\mathcal{H}_i$:

$$\begin{aligned}
&-iJ\left(\sum_{j\neq i}\mathrm{Tr}_j([\sigma_i^+\sigma_j^-, \rho_i\rho_j]) + \sum_{j\neq i}\mathrm{Tr}_j([\sigma_i^-\sigma_j^+, \rho_i\rho_j])\right)\\
&\quad = -iJ(N-1)\Big(\langle\sigma_i^-\rangle[\sigma_i^+, \rho_i] + \overline{\langle\sigma_i^-\rangle}[\sigma_i^-, \rho_i]\Big).
\end{aligned} \tag{3.3.13}$$

Here we used that all sites are identical, so $\langle\sigma_j^-\rangle = \langle\sigma_i^-\rangle$, and $\langle\sigma_j^+\rangle = \overline{\langle\sigma_j^-\rangle}$. The result may be written as $-i[H_i, \rho_i]$ where

$$H_i = J(N-1)\left(\langle\sigma_i^-\rangle\sigma_i^+ + \mathrm{H.c.}\right) \tag{3.3.14}$$

[19]The classical example is the Ising model $H = J\sum_{i=1}^{N}\sigma_i\sigma_{i+1} - B\sum_{i=1}^{N}\sigma_i$. Mean-field theory fails to capture phase behaviour in $d = 1$, is approximate correctly for $d = 2, 3$, and accurate when $d \geq 4$ [42, 320].

may be referred to as the mean-field Hamiltonian. Finally,

$$\Gamma_\downarrow \operatorname{Tr}_{\otimes j\neq i} L[\sigma_i^-] = \Gamma_\downarrow \operatorname{Tr}_{\otimes j\neq i} \left(\rho_i \dots \rho_{i-1} \left(\sigma_i^- \rho_i \sigma_i^+ - \{\sigma_i^+\sigma_i^-, \rho_i\}/2\right) \rho_{i+1} \dots \rho_N\right) \tag{3.3.15}$$

$$= \Gamma_\downarrow \left(\sigma_i^- \rho_i \sigma_i^+ - \{\sigma_i^+\sigma_i^-, \rho_i\}/2\right) \tag{3.3.16}$$

$$= \Gamma_\downarrow L_i[\sigma_i^-], \tag{3.3.17}$$

with $L_i[x] = x\rho_i x^\dagger - \{x^\dagger x, \rho_i\}/2$ being a Lindblad operator for the site $i$ alone.

The resulting equation for $\rho_i$ is

$$\partial_t \rho_i = -i\,[H_i, \rho_i] + \Gamma_\downarrow L_i[\sigma_i^-]. \tag{3.3.18}$$

As this describes the behaviour of every spin, the problem has been reduced from an $N$-body one to that of a single spin. The self-consistency of the approach is also evident in Eq. (3.3.18): the field $\langle\sigma_i^-\rangle$ determined by on-site properties at $i$ also determines the on-site properties at $i$.

From Eq. (3.3.18) it is straightforward to derive the mean-field Heisenberg equations of motion,

$$\partial_t\langle\sigma_i^-\rangle = iJ(N-1)\langle\sigma_i^-\rangle\langle\sigma_i^z\rangle - \frac{\Gamma_\downarrow}{2}\langle\sigma_i^-\rangle \tag{3.3.19}$$

and

$$\partial_t\langle\sigma_i^z\rangle = -\frac{\Gamma_\downarrow}{2}\left(\langle\sigma_i^z\rangle + 1\right). \tag{3.3.20}$$

As stated above, in practice one does not normally need to go via the equation of motion for $\rho_i$. For example, starting from the total master equation,

$$\partial_t\rho = -i[H,\rho] + \Gamma_\downarrow \sum_{i=1}^{N} L[\sigma_i^-], \quad H = J\sum_{i,j>i}^{N}\left(\sigma_i^+\sigma_j^- + \sigma_i^-\sigma_j^+\right), \tag{3.3.21}$$

The full Heisenberg equation of motion for $\langle\sigma_i^-\rangle$ is

$$\partial_t\langle\sigma_i^-\rangle = -iJ\left\langle\left[\sigma_i^-, \sum_{k,l>k}^{N}\left(\sigma_k^+\sigma_l^- + \text{H.c.}\right)\right]\right\rangle - \frac{\Gamma_\downarrow}{2}\sum_{k=1}^{N}\left\langle\left[\sigma_i^-, \sigma_k^+\right]\sigma_k^-\right\rangle. \tag{3.3.22}$$

$$= -iJ\sum_{k\neq i}\left\langle[\sigma_i^-, \sigma_i^+]\sigma_k^-]\right\rangle + \frac{\Gamma_\downarrow}{2}\left\langle\sigma_i^z\sigma_i^-\right\rangle \tag{3.3.23}$$

$$= iJ\sum_{k\neq i}\left\langle\sigma_i^z\sigma_k^-\right\rangle - \frac{\Gamma_\downarrow}{2}\left\langle\sigma_i^-\right\rangle \tag{3.3.24}$$

The mean-field approximation $\langle\sigma_i^z\sigma_k^-\rangle = \langle\sigma_i^z\rangle\langle\sigma_k^-\rangle = \langle\sigma_i^z\rangle\langle\sigma_i^-\rangle$ then gives

$$\partial_t\langle\sigma_i^-\rangle = iJ(N-1)\langle\sigma_i^-\rangle\langle\sigma_i^z\rangle - \frac{\Gamma_\downarrow}{2}\langle\sigma_i^-\rangle \tag{3.3.25}$$

as before. The equation $\partial_t\langle\sigma_i^z\rangle$ is derived in the same way.

It is important to stress that within this scheme factorisation occurs between operators from *different* sites (we applied the mean-field approximation *before* relabeling $\langle\sigma_k^-\rangle = \langle\sigma_i^-\rangle$ above). Indeed, $\sigma_i^z\sigma_i^- = -\sigma_i^-$ implied $\left\langle\sigma_i^z\sigma_i^-\right\rangle = -\langle\sigma_i^-\rangle$ in Eq. (3.3.24), which in general is *not* close to $\langle\sigma_i^z\rangle\langle\sigma_i^-\rangle$.

Factorisation of operators on the same site requires additional approximations. A common assumption in quantum optics is the Gaussianity [334] of a bosonic mode $a$ which, in conjunction with mean-field theory, allows one to approximate $\langle a^\dagger a \sigma_i^z \rangle \approx \langle a^\dagger \rangle \langle a \rangle \langle \sigma_i^z \rangle$ for example. We make such an assumption in Chapter 6, but note that we think it is an interesting and underexplored question as to under which conditions the failure of this approximation, i.e., the breakdown of Gaussianity, may occur. The scope for our methods to address this question is discussed in Chapter 7.

In Chapters 4 and 5 we apply mean-field theory to central boson and spin models. In Chapter 5, and later with the many-to-many model in Chapter 6, we also go beyond mean-field theory with higher-order cumulant expansions. This brings us to the subject of the next section.

## 3.4 Cumulant expansions beyond mean-field theory

Cumulant expansions of the Heisenberg equations of motion provide a systematic approximation scheme in which increasing orders of correlations are included at the cost of growing complexity. The framework for this approach in many-body systems was established in 1962 by R. Kubo who generalised mathematical concepts regarding the cumulants of stochastic variables for applications to problems in quantum and statistical physics.

As with mean-field theory, cumulant expansions have developed a broad set of uses across many disciplines [2, 42, 320, 335, 336]. In particular, we distinguish our implementation from those field theoretic techniques for perturbative expansions of correlation[20] functions [42, 320] as used, for example, to calculate thermodynamic potentials and Green's functions [2, 337, 338]. Those expansions rely on the resummation presented by the cumulant generating function of a random variable $X$ [320],

$$\ln\langle e^X \rangle = \langle X \rangle + \frac{1}{2}\left(\langle X^2 \rangle - \langle X \rangle^2\right) + \ldots, \tag{3.4.1}$$

but are not directly related to the method we describe. Closer to our approach are the cluster expansion[21] techniques [336, 339, 340] developed by M. Kira and S. Koch for quantum optics [336, 340]. These are also expansions of the Heisenberg equations. However, they differ critically in that factorisation is made on the separation of many-particle field operators rather than Hilbert spaces.

### 3.4.1 Second-order ansatz

Our formulation of cumulant expansions follows a natural extension to the mean-field ansatz (3.3.1), which corresponds to a first-order expansion. At second order then, we consider a many-body state including *pairwise* entanglement[22],

$$\rho = \bigotimes_{i=1}^{N} \rho_i + \sum_{(i,j)} \tau_{ij} \bigotimes_{k\notin(i,j)} \rho_k + \sum_{(r,s)\neq(i,j)} \tau_{ij}\tau_{rs} \bigotimes_{\substack{k\notin(i,j)\\ k\notin(r,s)}} \rho_k + \dots \tag{3.4.2}$$

where the sum in the first line runs over all pairs $(i,j)$ of sites exactly once, that in the second line every unique choice for two pairs, and so on. As before, $\rho_i$ is the reduced density matrix for a single site $i$, whilst $\tau_{ij}$ is that for subsystems $i$, $j$ *after* subtracting the simply separable (mean-field) part:

$$\tau_{ij} = \text{Tr}_{\otimes k\neq i,j}\left(\rho - \bigotimes_{r=1}^{N} \rho_r\right). \tag{3.4.3}$$

[20]In these subjects the functions targeted are multipoint correlation functions, e.g., a scalar field $\phi(\boldsymbol{x})$ evaluated at different points in space-time. We instead consider multi*site* correlation functions evaluated at one time (the possibility of multi-time cumulants via the quantum regression theorem is discussed in Chapter 7).

[21]Cumulants may also be referred to as Ursell functions [341] or truncated correlation functions [342] following their use by H. Ursell and others in statistical thermodynamics [343].

With this definition, the trace of a product with $\tau_{ij}$ vanishes if it contains one or fewer non-trivial operators for sites $i$ and $j$:

$$\text{Tr}_{ij}\left(\tau_{ij}\right) = \text{Tr}_{ij}\left(A_i\tau_{ij}\right) = \text{Tr}_{ij}\left(A_j\tau_{ij}\right) = 0. \tag{3.4.4}$$

On the other hand, $\text{Tr}_{ij}\left(A_iA_j\tau_{ij}\right)$ is non-zero and is in fact equal to the joint cumulant $\langle\langle A_iA_j\rangle\rangle$ of $A_i$ and $A_j$, which is defined by [344]

$$\langle\langle A_iA_j\rangle\rangle = \langle A_iA_j\rangle - \langle A_i\rangle\langle A_j\rangle. \tag{3.4.5}$$

This may be seen by calculating the expectation $\langle A_1A_2\rangle$ for Eq. (3.4.2):

$$\langle A_iA_j\rangle = \text{Tr}\left(A_iA_j\rho\right) \tag{3.4.6}$$
$$= \text{Tr}_{ij}\left(A_iA_j\rho_i\rho_j\right) + \text{Tr}_{ij}\left(A_iA_j\tau_{ij}\right) \tag{3.4.7}$$
$$= \langle A_i\rangle\langle A_j\rangle + \text{Tr}_{ij}\left(A_iA_j\tau_{ij}\right) \tag{3.4.8}$$
$$\Rightarrow \text{Tr}_{ij}\left(A_iA_j\tau_{ij}\right) = \langle A_iA_j\rangle - \langle A_i\rangle\langle A_j\rangle. \tag{3.4.9}$$

Note in Eq. (3.4.2) a choice was made to include *all* possible combinations of pairwise entanglement (one pair of entangled particles, two pairs, etc.). This is a reasonable one consistent with the prescription for cumulant expansions we provide below. Naturally, it is not the only possible choice. There may well be problems for which a different form of the ansatz might be more meaningful. For example, one could consider an ansatz including the simple product part and terms in which a single pair of particles is entangled only (first line of Eq. (3.4.2)).

The critical consequence of the second-order ansatz arises in the evaluation of moments (ex-

[22]A closed (i.e., without ellipses ...) expression for this ansatz is provided in Appendix C.1. You may notice an ordering issue arises with the $\tau_{ij}$ terms: the first $\tau_{12}\rho_2\rho_3\rho_4\dots$ is as expected, but already in $\tau_{13}\rho_2\rho_4\dots$ the parts corresponding to the second and third Hilbert (implied by the indices) spaces are in the wrong order. This is simply a failure of notation. Although not necessary for our discussion, for a numerical implementation where $\rho_i$ and $\tau_{ij}$ are matrices, ordering operators should be applied to each tensor product, e.g., $W_{3\leftrightarrow2}(\tau_{13}\rho_2\rho_4\dots)$, to restore the proper order.

pectations) involving operators from *three* sites:

$$\begin{aligned}\langle A_1A_2A_3\rangle &= \langle A_1\rangle\langle A_2\rangle\langle A_3\rangle \\ &\quad + \mathrm{Tr}\left(A_1A_2\tau_{12}A_3\rho_3\rho_4\ldots\right) \\ &\quad + \mathrm{Tr}\left(A_1A_3\tau_{13}\rho_2\rho_4\ldots\right) \\ &\quad + \mathrm{Tr}\left(A_1\rho_1A_2A_3\tau_{23}\rho_4\ldots\right)\end{aligned} \tag{3.4.10}$$

$$\begin{aligned}&= \langle A_1\rangle\langle A_2\rangle\langle A_3\rangle + \langle A_3\rangle\,\mathrm{Tr}_{12}\left(A_1A_2\tau_{12}\right) \\ &\qquad + \langle A_2\rangle\,\mathrm{Tr}_{13}\left(A_1A_3\tau_{13}\right) \\ &\qquad + \langle A_1\rangle\,\mathrm{Tr}_{23}\left(A_2A_3\tau_{23}\right)\end{aligned} \tag{3.4.11}$$

$$\begin{aligned}&= \langle A_1\rangle\langle A_2\rangle\langle A_3\rangle + \langle A_3\rangle\left(\langle A_1A_2\rangle - \langle A_1\rangle\langle A_2\rangle\right) \\ &\qquad + \langle A_2\rangle\left(\langle A_1A_3\rangle - \langle A_1\rangle\langle A_3\rangle\right) \\ &\qquad + \langle A_1\rangle\left(\langle A_2A_3\rangle - \langle A_2\rangle\langle A_3\rangle\right)\end{aligned} \tag{3.4.12}$$

$$\langle A_1A_2A_3\rangle = \langle A_1\rangle\langle A_2A_3\rangle + \langle A_2\rangle\langle A_1A_3\rangle + \langle A_3\rangle\langle A_1A_2\rangle - 2\langle A_1\rangle\langle A_2\rangle\langle A_3\rangle. \tag{3.4.13}$$

Thus moments involving operators from three (or more) sites split up into products of moments involving one or two sites, as expected given the ansatz contained at most pairwise correlations.

We now notice that Eq. (3.4.13) is the approximation for $\langle A_1A_2A_3\rangle$ that would be obtained by setting the third cumulant $\langle\langle A_1A_2A_3\rangle\rangle$ to zero, where

$$\langle\langle A_1A_2A_3\rangle\rangle = \langle A_1A_2A_3\rangle - \langle A_1\rangle\langle A_2A_3\rangle - \langle A_2\rangle\langle A_1A_3\rangle - \langle A_3\rangle\langle A_1A_2\rangle + 2\langle A_1\rangle\langle A_2\rangle\langle A_3\rangle. \tag{3.4.14}$$

What does this mean for the Heisenberg equations of motion? Consider a typical case where the Hamiltonian $H$ involves terms coupling different pairs of sites in the many-body system. If one writes down a set of equations of motion for the expectations of single-site observables $A_i$, then these will not be closed, since they involve moments of two operators brought in via $\sim \langle[A_i, H]\rangle$. The equations for moments of two sites will in turn depend on moments involving three sites, and so on; an infinite hierarchy of equations. A cumulant expansion, ansatz (3.4.2) at second order, is a way of truncating this hierarchy to produce a closed system of equations.

At the level of mean-field theory, one factorises products of operators so that the equations of motion for single-site observables forms a closed set. This can be understood by approximating the second cumulants of operators to zero. For second-order cumulants, one instead sets third-order cumulants to zero, hence factorising moments with operators for three or more sites. The closed set of equations then involves both those for single-site observables and those for all possible two-site moments. In general one *closes the Heisenberg equations of motion at $M^{th}$-order by factorising moments involving $M+1$ sites into nonlinear combinations of lower order cumulants by setting the corresponding $(M+1)^{th}$ cumulant to zero.* This reflects extending ansatz (3.4.2) to include up to $M$-wise entanglement by the introduction of terms with rank-$M$ tensors $\tau_{i_1,\ldots i_M}$.

### 3.4.2 Recipe for a cumulant expansion

**To perform an $M^{\text{th}}$-order cumulant expansion,**

1. Derive Heisenberg equations for operators involving up to $M$ sites
2. Factorise moments involving $M+1$ sites by setting $(M+1)^{\text{th}}$ cumulants to zero

A general formula for the joint cumulant of operators on $M$ sites is provided in Appendix C.2.

Step 2 assumes a linear Hamiltonian that couples pairs of sites. For Hamiltonians with interactions between $k > 2$ sites the equations for $M^{\text{th}}$-order moments will involve of operators for $M + k - 1$ sites. To factorise these, the cumulant expansion can be applied in a nested fashion. For example, if a fourth order moment $\langle A_1 A_2 A_3 A_4 \rangle$ arises in the application of second-order expansion, the moment is first split up according to the prescription $\langle\langle A_1 A_2 A_3 A_4 \rangle\rangle = 0$. Thereafter the third order moments in the result (e.g. $\langle A_1 A_2 A_3 \rangle$) are split up according to setting the third cumulants to zero ($\langle\langle A_1 A_2 A_3 \rangle\rangle = 0$). We show this is consistent with the variational ansatz in Appendix C.1. In other words, an $M^{\text{th}}$-order expansion should mean setting the $K^{\text{th}}$ cumulants to zero for all $K \geq M$.

### 3.4.3 Bosonic degrees of freedom

A particular challenge arises for particles with infinite degrees of freedom such as bosons: the equation of motion for a bosonic operator $a$ will contain expectations involving higher powers of the operator and its conjugate, e.g., in $\langle a^\dagger a \rangle$ and $\langle aa \rangle$, which cannot be written as a linear combination of single powers of $a$ and $a^\dagger$. Thus, even after applying a cumulant expansion between sites an infinite hierarchy of equations is obtained ($\partial_t \langle a^\dagger a \rangle$, $\partial_t \langle aa \rangle$, $\partial_t \langle a^\dagger aa \rangle$, $\ldots$).

To obtain a closed set of equations in this case one may either i. truncate the degree of freedom to a finite number $N_a$ of levels (making it a $N_a$-dimensional Hilbert space) or ii. rely on other approximations such as Gaussianity described above (Section 3.3.1) to allow the splitting up of powers of $a$ between moments. We make use of both of these strategies in Chapter 6 (i. for the vibrational modes $b_n$, ii. for the photon modes $a_k$).

### 3.4.4 Dimension reduction and symmetries

Where mean-field theory ($M = 1$) reduces the problem space for $N$ identical particles from $d^N$ to $d$, an $M^{\text{th}}$-order cumulant expansion reduces it to $d^M$. The actual number of cumulant equations to be solved depends on the nature of interactions in the model. In particular, for Hamiltonians with a given symmetry one only needs to construct a set of *symmetry-preserving* cumulant equations if dynamics from a symmetric initial state is required. This is done by discarding from the cumulant expansion terms that do not respect that symmetry.

Referring to the previous example, Eq. (3.3.7), which is symmetric under rotations $\sigma^\pm \to \sigma^\pm e^{\pm i\theta}$ (U(1) symmetry), the expansion for $\langle \sigma_i^+ \sigma_j^- \sigma_k^z \rangle$ where $i$, $j$, $k$ are distinct would be

$$\langle \sigma_i^+ \sigma_j^- \sigma_k^z \rangle = \langle \sigma_i^+ \sigma_j^- \rangle \langle \sigma_k^z \rangle + \langle \sigma_i^+ \rangle \langle \sigma_j^- \sigma_k^z \rangle + \langle \sigma_i^+ \sigma_k^z \rangle \langle \sigma_j^- \rangle - \langle \sigma_i^+ \rangle \langle \sigma_j^- \rangle \langle \sigma_k^z \rangle \tag{3.4.15}$$

$$= \langle \sigma_i^+ \sigma_j^- \rangle \langle \sigma_i^z \rangle \quad (i \neq j)\,, \tag{3.4.16}$$

since terms which does not conserve the number of excitations vanish, and all sites are identical so we are free to relabel $k \to i$. This significantly reduces the number and complexity of derived equations. Of course, it means the equations cannot be used to determine dynamics starting from an initial state that does not respect the same symmetry—a consideration we will find to be important in Chapter 6.

On a related note, a characteristic feature of mean-field theories is a symmetry breaking transition between a disordered phase in which the order parameter, i.e., the mean field, is zero, to an ordered one in which it is non-zero. This occurs as one or more system parameters are varied. We observe this behaviour in Chapter 4 with the lasing transition in the Tavis-Cummings model ($\langle a \rangle = 0 \to \langle a \rangle > 0$) as molecular pump $\Gamma_\uparrow$ is increased and as well as in Chapter 5 with the analogous transition for the central spin model ($p_0^\uparrow = 0 \to p_0^\uparrow > 0$). A side-effect is that the mean-field descriptions can only describe non-trivial dynamics with symmetry breaking, so are limited in the opposite sense to a symmetry-preserving set of cumulant equations. That is, they may not directly be used to calculate a system whose dynamics is inherently symmetric.

### 3.4.5 Applicability of cumulant expansions

In the previous sections we explained the operation of cumulant expansions but have not commented on their use in solving many-body problems. Extending mean-field theory, the utility of cumulant expansions remains in their ability to reduce a high dimensional many-body problem to an effective few-body one. By including successive orders of correlations higher-order expansions, they may offer a more accurate solution than the mean-field description. In particular, if mean-field theory describes the exact behaviour of a model in the limit that the number of sites $N \to \infty$, increasing orders of cumulant expansions may be expected to provide improved approximations at successively smaller $N$.

The trade-off of higher-order expansion orders is the growth in size and complexity of the system of equations to be solved: typically expansions beyond second or third order are not feasible to derive by hand. Yet for high connectivity models cumulant expansions have repeatedly been found to be effective at low orders of expansion and intermediary $N$ [40, 63, 105, 236, 237, 345–351].

Further to the applicability of cumulant expansions, the Ritsch group at Innsbruck have recently released an open-source Julia framework for computing the cumulant equations of general quantum systems, `QuantumCumulants.jl` [352]. This software is capable of deriving cumulant equations symbolically to any desired order $M$ (at least, for $M$ well beyond $2$ or $3$). The equations can then be solved numerically using the efficient differential equation suite available in Julia. For these reasons cumulant expansions will no doubt continue to be an increasingly popular tool for addressing problems in atomic, molecular and optical physics.

Despite their widespread use, and success in solving many-body problems in quantum optics, there is a lack of rigorous results regarding the validity of cumulant expansions beyond mean-field theory. In Chapter 5 we uncover the inability of these methods to capture the $N \to \infty$ limit of a model for which mean-field behaviour may well be expected. By examining this failure, and further studying the convergence of cumulant expansions for this problem, we look to contribute to the task of establishing firmer theoretical grounds for these methods.

#### Outro

That's all for the background, now we can take off.

# Part II

# Foreground

# Chapter 4

# Many-body non-Markovian dynamics of organic polaritons ↑TOC

When we wish to bring to the knowledge of a person any phenomena or processes of nature, we have the choice of two methods: we may allow the person to observe matters for himself, when instruction comes to an end; or, we may describe to him the phenomena in some way, so as to save him the trouble of personally making anew each experiment.

Ernst Mach

## Contents

In this chapter we develop a method to simulate a model of many molecules with both strong coupling to many vibrational modes and collective coupling to a photon mode. This combines process tensor matrix product operator methods with a mean-field approximation that reduces the dimension of the problem.

We start in Section 4.1 by discussing the precedent for realistic models of organic polariton lasing, and introduce our model based on the BODIPY-Br molecular dye [55]. The new method, combining mean-field evolution with the process tensor time-evolving matrix product operator (PT-TEMPO) method [162, 281, 303], is set out in Section 4.2. The efficiency of the method in solving the model with large numbers of molecules is demonstrated in Section 4.3. Here the steady state under incoherent pumping is extracted to determine the dependence of the threshold for polariton lasing on cavity detuning, light-matter coupling strength and environmental temperature. The basic motivation for these calculations is the problem of determining the optimum conditions for lasing. We explain differences from our results with those of approximate treatments based on weak system-environment or weak light-matter coupling, as well as those of simplified models based on a few discrete vibrational modes (Sections 4.3.3 to 4.3.5).

Next, in Section 4.4, we show how the method can be used to calculate photoluminescence spectra, the actual observable in polariton lasing experiments. For this purpose we extend the model to include multiple photon modes, and study quadratic fluctuations about the mean-field solution by measuring two-time correlations. We benchmark this approach using analytical results for the absorption spectrum at zero in-plane momentum $k$ before presenting full, $k$-dependent spectra (Section 4.4.2). We perform additional analysis of results for the normal state (Section 4.4.3), explaining how the inverse photon Green's functions may be used to track spectral properties as the lasing transition is approached.

Finally, in Section 4.5, we discuss the role of bright and dark excitonic states in the mean-field approach. We show how the bright and dark populations may be calculated under the mean-field approximation, and how these change through the lasing transition. We also demonstrate a spectral feature of the dark states that may be observed at large light-matter coupling strengths. A summary of the chapter and the broader applicability of the method is provided in Section 4.6.

The method and results of this chapter were presented in the publication P. Fowler-Wright et al., Efficient Many-Body Non-Markovian Dynamics of Organic Polaritons, Phys. Rev. Lett. **129**, 173001 (2022) [12].

## 4.1 Model of an organic laser

Chapter 2 introduced the challenge of developing realistic models of organic polariton lasing: one must consider the effect on the dynamics of the vibrational environment of each molecule [43], which is generally structured and beyond weak coupling or Markovian treatments [17, 20, 220, 270, 353–357]. While there have been studies using simplified models with a few vibrational modes [43, 46, 54, 62–64, 66, 358–361], and studies including exact vibrational spectra for a small number of molecules [20, 220], there is a lack of methods capable of accurately handling the vibrational environments of many, e.g., $10^5$ molecules. In this chapter we provide such a method and investigate the consequences for the description of polariton lasing. In particular, we construct a realistic model of an organic laser (Fig. 4.1) with parameters based on BODIPY-Br, an organic molecule which has shown polariton lasing [55, 96, 362] (see Fig. 2.7 in Chapter 2).

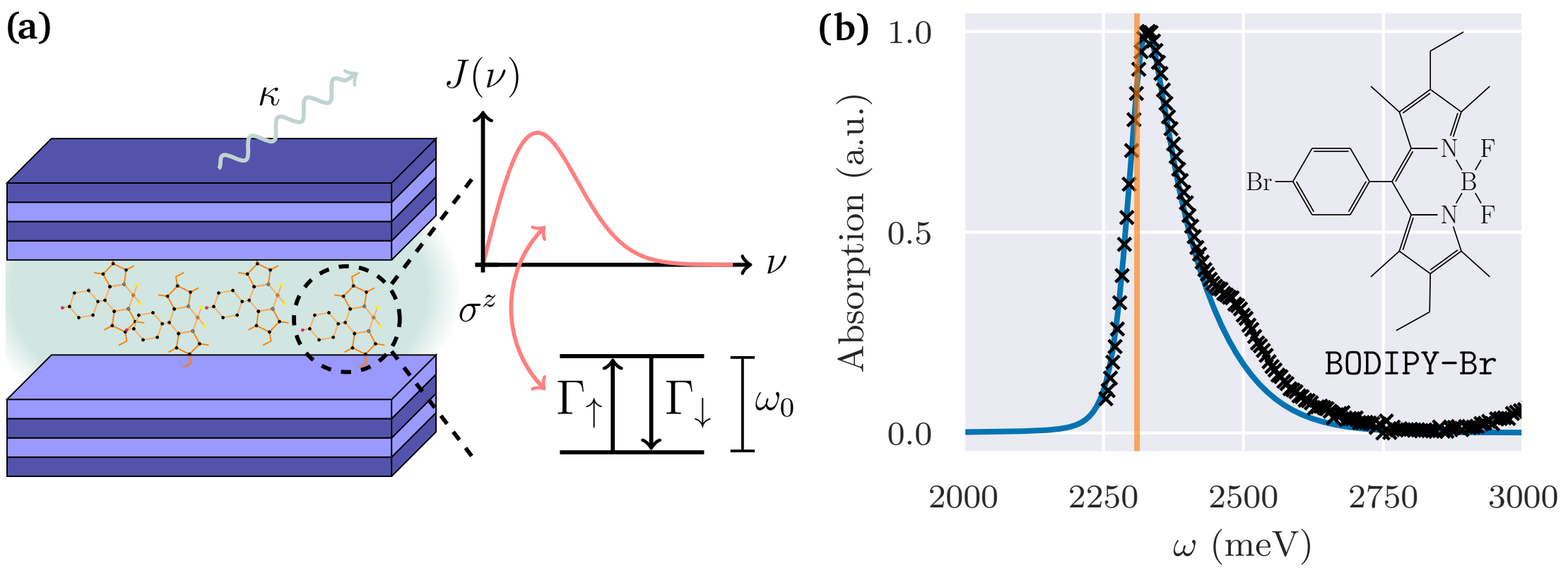

**Figure 4.1:** (a) Model of an organic lasing compressing a molecular ensemble in an optical microcavity. Each molecule is treated as a driven-dissipative two-level system with a diagonal coupling to a harmonic environment. The spectral density $J(\nu)$ of the environment is chosen to match (b) absorption data [55] for BODIPY-Br at 300K (black crosses: experimental data, blue curve: model spectrum, orange line: $\epsilon = 2310$ meV). For the Ohmic form Eq. (4.1.3) with dissipation $\Gamma_\downarrow = 10$ meV we obtained $\alpha = 0.25$ and $\nu_c = 150$ meV ($\hbar = 1$). Note calculating the absorption spectrum for our model used the analytical result for the spectral weight given later in Section 4.4.2. $\rightharpoonup_{\text{TOF}}$

Our model of a many-molecule–cavity system comprises two parts. First, $N$ two-level systems (Pauli matrices $\sigma_i^\alpha$) describe an electronic transition of the molecules near-resonant with a single cavity mode (bosonic operator $a$). These interact under the rotating-wave approximation according to the Tavis-Cummings Hamiltonian

$$H_S = \omega_c a^\dagger a + \sum_{i=1}^{N} \left[ \frac{\epsilon}{2}\sigma_i^z + \frac{\Omega}{2\sqrt{N}} \left( a\sigma_i^+ + a^\dagger \sigma_i^- \right) \right] \qquad (\hbar = 1)\,, \tag{4.1.1}$$

where $\epsilon$ and $\omega_c$ are the two-level system and cavity frequencies, and $\sigma_i^+$ ($\sigma_i^-$) the raising (lowering) operator for the $i$th spin. The collective coupling $\Omega$ controls the light-matter interaction such that the bright eigenstates of $H_S$, i.e., the polaritons, are split as $\pm\Omega/2$ at resonance.

The extension of the Tavis-Cummings model Eq. (4.1.1) to include a single vibrational mode, the Holstein-Tavis-Cummings model, has frequently been used to describe cavity bound organic emitters [54, 62–64, 66, 358, 359, 361]. We instead consider the interaction of each two-level system with a *continuum* of modes, represented by the harmonic environment

$$H_E^{(i)} = \sum_j \left[ \nu_j b_j^\dagger b_j + \frac{1}{2} \left( \xi_j b_j + \bar{\xi}_j b_j^\dagger \right) \sigma_i^z \right], \tag{4.1.2}$$

where $b_j$ is the annihilation operator for the $j$th mode of frequency $\nu_j$.

Each molecule then has a local vibrational environment independent from those of all other molecules. The system-environment coupling is characterized by a spectral density $J(\nu) = \sum_j |\xi_j/2|^2 \delta(\nu - \nu_j)$, taken to be Ohmic in the form

$$J(\nu) = 2\alpha\nu e^{-(\nu/\nu_c)^2}, \quad \nu > 0, \tag{4.1.3}$$

where $\alpha$ and $\nu_c$ are chosen to reproduce the leading structure of the absorption spectrum of BODIPY-Br at $T = 300$ K shown in Fig. 4.1b. This effectively captures the low frequency modes arising from the host matrix of the molecule. The realistic picture of vibrational dephasing it affords is the most significant advancement of our work.

The environment Eq. (4.1.2) is a collection of harmonic baths of the form introduced in the previous chapter (cf. Eq. (3.1.54)) with diagonal system-bath coupling $\hat{S} \to \sigma_i^z/2$. As explained there, the TEMPO and PT-TEMPO methods are designed to handle this form of environment, albeit for systems of small Hilbert space dimension ($N \sim 1$). Further, they may be expected to perform well where the environment has a broad, low-frequency structure such as described by $J(\nu)$ in Eq. (4.1.3). We note that in the limit that the system-environment coupling is weak one might look to make a Born-Markov approximation to derive a Redfield theory (Section 3.1.8). However, as we discuss in Section 4.3.3 below, this is difficult in the presence of strong light-matter coupling.

Finally we consider incoherent pump $\Gamma_\uparrow$ and dissipation $\Gamma_\downarrow$ of the two-level systems as well as photon losses $\kappa$. Since these are associated with baths at optical frequencies (e.g. $10^{15}$ Hz) they may be well approximated [4] by Markovian terms in the master equation for the total density operator $\rho$,

$$\partial_t \rho = -i\left[H_S + \sum_{i=1}^{N} H_E^{(i)}, \rho\right] + \kappa L[a] + \sum_{i=1}^{N} (\Gamma_\uparrow L[\sigma_i^+] + \Gamma_\downarrow L[\sigma_i^-]), \tag{4.1.4}$$

with $L[x] = x\rho x^\dagger - \{x^\dagger x, \rho\}/2$. If $H_E^{(i)}$ is absent one recovers the Tavis-Cummings model with pumping and decay which, as we discuss below, requires inversion $\Gamma_\uparrow > \Gamma_\downarrow$ to show lasing. In the following sections we fix $\Gamma_\downarrow$ and $\kappa$ and observe the transition of the system from a normal state, where the expectation $\langle a \rangle$ of the photon operator vanishes, to a lasing state, where $\langle a \rangle$ is non-zero and time dependent, as $\Gamma_\uparrow$ is increased from zero.

In summary, $N$ identical molecules with independent harmonic environments couple to a single cavity mode. A schematic for this many-to-one network in provided in Fig. 4.2a.

## 4.2 Mean-field TEMPO

In Section 3.2.7, we introduced the process tensor matrix product operator (PT-MPO) method PT-TEMPO [303] as a formulation of the TEMPO algorithm using the language of process tensors [313]. The process tensor (PT), a contraction of bath influence functionals, captures all possible effects of the environment on a system. The system Hamiltonian propagator, or any system operator, then forms a finite set of interventions that may be contracted with the PT to obtain the dynamics of any system observable or multi-time correlation function. The PT is represented efficiently as a matrix product operator and only needs to be calculated once for a given system-bath interaction and set of bath conditions. While this provides an efficient means to evolve a system including non-Markovian effects with long memory times, as with the original TEMPO method it is limited to systems of small Hilbert space dimension. Hence it, or any other exact numerical method, cannot be directly applied to the problem with a large number of molecules $N \gg 1$. Our strategy is to use mean-field theory to reduce the $N$-molecule–cavity system to a single molecule interacting with a coherent field which may be handled by the PT-TEMPO method without further approximation.

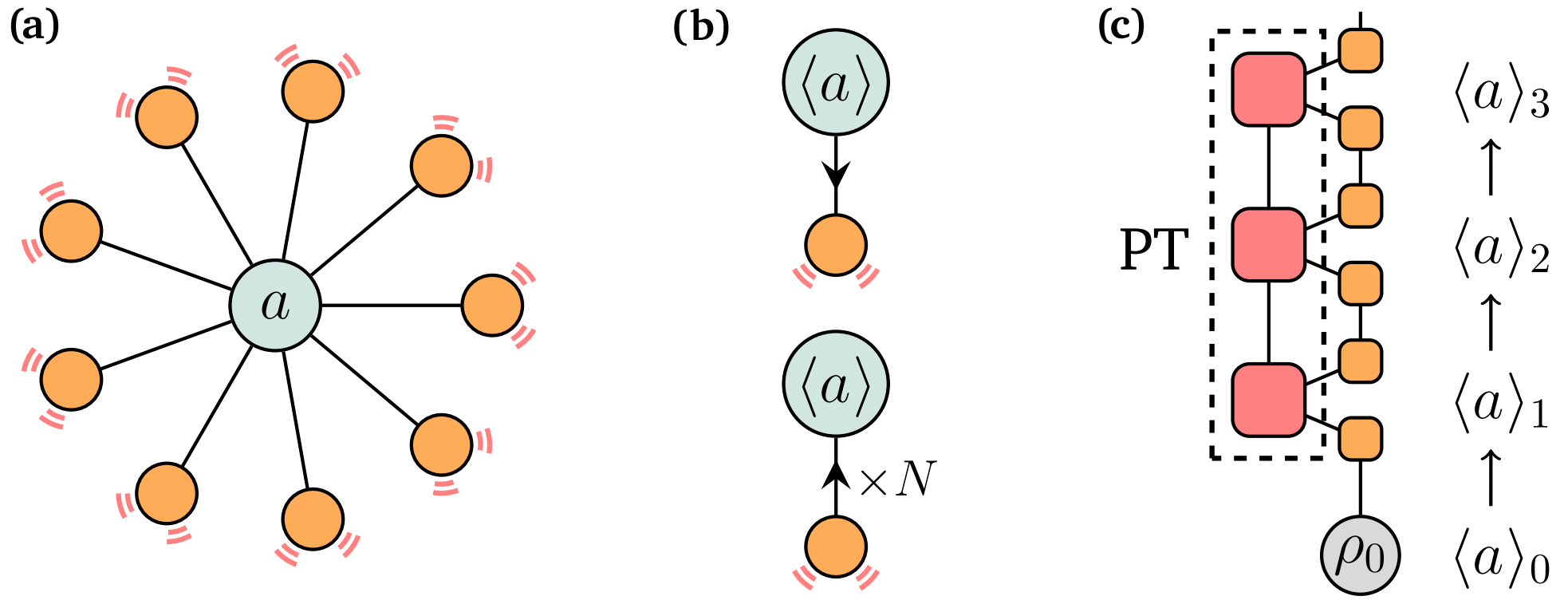

**Figure 4.2:** (a) Schematic for the many-molecule–cavity system. Each molecule (orange) couples to the central cavity mode and has its own vibrational environment (red dashes). (b) Mean-field theory reduction to an effective model where a single emitter evolves according to the mean-value $\langle a \rangle$ of the field which in turn evolves according to $N$ copies of molecular expectations. (c) The mean-field PT-TEMPO method. The field expectation is evolved concurrently with the molecular dynamics provided by the PT-MPO method. ⇀TOF

### 4.2.1 Mean-field equations

According to mean-field theory, we assume a product state for the many-body density operator $\rho$, i.e., a factorisation between the photon and individual molecules:

$$\rho = \rho_a \otimes \bigotimes_{i=1}^{N} \rho_i. \tag{4.2.1}$$

The reduced density matrix $\rho_a = \mathrm{Tr}_{\otimes i}\, \rho$ is obtained from the partial trace taken over the Hilbert space of all two-level systems labelled $i = 1, \dots, N$, and $\rho_i = \mathrm{Tr}_{a, \otimes j \neq i}\, \rho$ from the partial trace over the photonic degree of freedom and all but the $i$th two-level system. As commented in Section 3.3.1, this ansatz is known to be exact for the central boson model considered in the limit $N \to \infty$ [332, 333].

While in our calculations we consider the simple case where all of the molecules are identical, so that only a single $\rho_i$ needs to be calculated, the mean-field ansatz does not require this. The approach we describe below can be applied to models where each molecular site has different parameters, at the cost of requiring separate PT-TEMPO simulations for each $\rho_i$. We also note that even when all sites are equivalent, the assumption of identical $\rho_i$ is not the same as restriction to the totally symmetric Hilbert space, particularly when incoherent processes are present. The consequences of this are reflected in the role of dark exciton states within mean-field theory, which we discuss in Section 4.5.1.

In our approach both the non-Markovian environment and Markovian pumping and loss for each molecule are handled by the PT-TEMPO method. Therefore, we only need to consider the mean-field decoupling for the simpler model,

$$\partial_t \rho = -i[H_S, \rho] + \kappa L[a], \tag{4.2.2}$$

and later reintroduce the molecular dissipation and environments in the PT-TEMPO simulation.

Here $H_S$ is the Tavis-Cummings Hamiltonian from above,

$$H_S = \omega_c a^\dagger a + \sum_{i=1}^{N} \left[ \frac{\epsilon}{2}\sigma_i^z + \frac{\Omega}{2\sqrt{N}} \left( a\sigma_i^+ + a^\dagger \sigma_i^- \right) \right]. \tag{4.2.3}$$

The equations of motion for the reduced density matrices follow as

$$\partial_t \rho_a = -i \operatorname{Tr}_{\otimes i}[H_S, \rho] + \kappa \operatorname{Tr}_{\otimes i} L[a], \tag{4.2.4}$$

$$\partial_t \rho_i = -i \operatorname{Tr}_{a,\otimes j\neq i}[H_S, \rho] + \kappa \operatorname{Tr}_{a,\otimes j\neq i} L[a]. \tag{4.2.5}$$

The partial traces can be performed given the mean-field ansatz Eq. (4.2.1), recalling that the reduced density matrices are individually normalised to one. Noting also that the partial trace over a subsystem ($\operatorname{Tr}_\alpha$, $\alpha \in \{a, i\}$) of a commutator involving operators acting only on that subsystem will vanish,

$$-i \operatorname{Tr}_{\otimes i}[\omega_c a^\dagger a, \rho] = -i\omega_c [a^\dagger a, \rho_a], \tag{4.2.6}$$

$$-i \operatorname{Tr}_{\otimes j} \left[ \sum_{i=1}^{N} \frac{\epsilon}{2}\sigma_i^z, \rho \right] = 0, \tag{4.2.7}$$

$$\kappa \operatorname{Tr}_{\otimes i} L[a] = \kappa L_a[a], \tag{4.2.8}$$

where $L_a[x] = x\rho_a x^\dagger - \{x^\dagger x, \rho_a\}/2$ is the Lindblad operator for the photon density matrix, and

$$-i \operatorname{Tr}_{a,\otimes j\neq i}[\omega_c a^\dagger a, \rho] = 0, \tag{4.2.9}$$

$$-i \operatorname{Tr}_{a,\otimes j\neq i} \left[ \sum_{k=1}^{N} \frac{\epsilon}{2}\sigma_k^z, \rho \right] = -i \left[ \frac{\epsilon}{2}\sigma_i^z, \rho_i \right], \tag{4.2.10}$$

$$\kappa \operatorname{Tr}_{a,\otimes j\neq i} L[a] = 0. \tag{4.2.11}$$

It remains to determine the terms arising from the light-matter interaction in $H_S$. For the contribution to the evolution of the photon degree of freedom Eq. (4.2.4), one has

$$-i \sum_{i=1}^{N} \frac{\Omega}{2\sqrt{N}} \operatorname{Tr}_{\otimes j}[a\sigma_i^+ + \text{H.c.}, \rho] = -i\frac{\Omega\sqrt{N}}{2} \left( \langle \sigma^+ \rangle [a, \rho_a] + \langle \sigma^- \rangle [a^\dagger, \rho_a] \right). \tag{4.2.12}$$

For the evolution of the matter degree of freedom Eq. (4.2.5), the contribution is instead

$$-i \sum_{k=1}^{N} \frac{\Omega}{2\sqrt{N}} \operatorname{Tr}_{a,\otimes j\neq i}[a\sigma_k^+ + \text{H.c.}, \rho] = -i\frac{\Omega}{2\sqrt{N}} \left( \langle a \rangle [\sigma_i^+, \rho_i] + \langle a^\dagger \rangle [\sigma_i^-, \rho_i] \right). \tag{4.2.13}$$

From the above we find that the equation of motion for each molecule $\rho_i$ is

$$\partial_t \rho_i = -i[H_i, \rho_i], \tag{4.2.14}$$

where

$$H_i = H_{\text{MF}} = \frac{\epsilon}{2}\sigma_i^z + \frac{\Omega}{2\sqrt{N}} (\langle a \rangle \sigma_i^+ + \overline{\langle a \rangle} \sigma_i^-). \tag{4.2.15}$$

defines the mean-field Hamiltonian $H_{\text{MF}}$ for any one of emitters (note $\langle a^\dagger \rangle = \overline{\langle a \rangle}$ is the complex conjugate of $\langle a \rangle$). For the full dissipative model, one adds Lindblad terms $\Gamma_\uparrow L[\sigma_i^+]$, $\Gamma_\downarrow L[\sigma_i^-]$ to the system Liouvillian $\mathcal{L}_S$ to construct the system propagators for the PT-TEMPO method.

In Eq. (4.2.15) the only required property of the photon state $\rho_a$ is the expectation $\langle a \rangle$. The evolution of this expectation can be determined from the equation of motion for $\rho_a$,

$$\partial_t \rho_a = -i[H_a, \rho_a] + \kappa L_a[a], \tag{4.2.16}$$

where the Hamiltonian

$$H_a = \omega_c a^\dagger a + \frac{\Omega\sqrt{N}}{2}\left(a\langle\sigma^+\rangle + a^\dagger\langle\sigma^-\rangle\right), \tag{4.2.17}$$

follows from Eq. (4.2.4), given Eqs. (4.2.6), (4.2.8) and (4.2.12). The coupling term here picks up a factor of $N$ compared to that in $H_{\mathrm{MF}}$ due to their being $N$ molecules but only one photon. The Heisenberg equation for $\langle a \rangle$ is then[1]

$$\begin{aligned}
\partial_t\langle a\rangle &= \mathrm{Tr}_a\left(a\partial_t\rho_a\right) &(4.2.18)\\
&= -i\omega_c \,\mathrm{Tr}_a\left(a[a^\dagger a, \rho_a]\right) - i\frac{\Omega\sqrt{N}}{2}\langle\sigma^-\rangle\,\mathrm{Tr}_a\left(a[a^\dagger, \rho_a]\right)\\
&\quad + \kappa\,\mathrm{Tr}_a\left(a a\rho_a a^\dagger - a a^\dagger a\rho_a/2 - a\rho_a a^\dagger a/2\right)\\
&= -\left(i\omega_c + \frac{\kappa}{2}\right)\langle a\rangle - i\frac{\Omega\sqrt{N}}{2}\langle\sigma^-\rangle. &(4.2.19)
\end{aligned}$$

Thus, by propagating a *single* two-level system (spin) with $H_{\mathrm{MF}}$ and subject to the vibrational environment and individual losses in Eq. (4.1.4), we can effectively simulate the $N$-molecule system using the PT-TEMPO method provided that at each timestep we also evolve $\langle a \rangle$ according to Eq. (4.2.19). This method is depicted in Fig. 4.2c.

### 4.2.2 Implementation

The spin Hamiltonian $H_{\mathrm{MF}}$ depends on the mean field $\langle a \rangle$, which in turn depends on the state $\langle \sigma^- \rangle$ of the spin. Therefore the two must be integrated in a self-consistent fashion.

For the $n$th timestep we firstly evolve the spin $\rho_i(t_{n-1})$ using the PT-TEMPO method with $H_{\mathrm{MF}}(t) = H_{\mathrm{MF}}(\langle a(t)\rangle)$ set using the value of the field $\langle a\rangle_{n-1}$ at the beginning of the timestep. Having applied the total $(\mathcal{L}_S + \mathcal{L}_E)$ propagator to $\rho_i(t_{n-1})$ the remainder of the PT, describing evolution under $H_E$ for $t > t_n$, may be traced over to yield the state $\rho_i(t_n)$ (the calculation branches here; a copy of the evolution with the PT for $t > t_n$ is available for the next timestep).

The state $\rho_i(t_n)$ provides the spin expectation $\langle\sigma^-\rangle_n$. This is used in conjunction with both the field $\langle a\rangle_{n-1}$ and spin $\langle\sigma^-\rangle_{n-1}$ expectations at $t_{n-1}$ to integrate $\langle a\rangle_{n-1}$ using a second-order Runge-Kutta method [363]:

$$\langle a\rangle_n = \langle a\rangle_{n-1} + \frac{\delta t}{2}\left(k_{n_1} + k_{n_1}\right), \tag{4.2.20}$$

where

$$\begin{aligned}
k_{n_1} &= f\left(\langle a\rangle_{n-1}, \langle\sigma^-\rangle_{n-1}\right), &(4.2.21)\\
k_{n_2} &= f\left(\langle a\rangle_{n-1} + \delta t\cdot k_{n_1}, \langle\sigma^-\rangle_n\right), &(4.2.22)
\end{aligned}$$

with $f(\langle a\rangle, \langle\sigma^-\rangle) \equiv \partial_t\langle a\rangle$ from Eq. (4.2.19) and $\delta t$ the fixed timestep length. The value $\langle a\rangle_n$ can then be used to construct the system (spin) propagators for the next timestep. In fact, since

[1] You can check the shortcut Eq. (3.1.94) described in the previous chapter gives the same result. That result will be used to derive the numerous mean-field and cumulant equations in later chapters.

PT-TEMPO can handle explicit time-dependence in the system Hamiltonian, to achieve faster convergence under $\delta t \to 0$ we can use for $H_{\text{MF}}(\langle a \rangle)$ a linearisation of the field

$$\langle a(t) \rangle = \langle a \rangle_n + f\left(\langle a \rangle_{n-1}, \langle \sigma^- \rangle_{n-1}\right) \cdot (t - t_{n-1}), \qquad t \in [t_{n-1}, t_n]. \tag{4.2.23}$$

This form can be integrated exactly to construct the system propagators in the PT-TEMPO network for the next timestep.

#### Field rescaling

In the lasing phase of the Tavis-Cummings model $\langle a \rangle$ scales with $\sqrt{N}$ [105]. It is hence convenient to work with the rescaled quantity $\langle \tilde{a} \rangle = \langle a \rangle / \sqrt{N}$ such that Eqs. (4.2.15) and (4.2.19) become

$$\partial_t \langle \tilde{a} \rangle = -\left(i\omega_c + \frac{\kappa}{2}\right) \langle \tilde{a} \rangle - i\frac{\Omega}{2} \langle \sigma^- \rangle \tag{4.2.24}$$

and

$$H_{\text{MF}} = \frac{\epsilon}{2}\sigma^z + \frac{\Omega}{2}\left(\langle \tilde{a} \rangle \sigma^+ + \overline{\langle \tilde{a} \rangle} \sigma^-\right). \tag{4.2.25}$$

Then only a single parameter $\Omega$ is used to specify the light-matter interaction. The number of molecules $N$ does not appear anywhere in our computation, although $N \gg 1$ is implicit for the mean-field approximation. In Section 4.3 we calculate the dynamics of the rescaled photon number $|\langle \tilde{a} \rangle|^2 \equiv n/N$ using the mean-field plus PT-TEMPO method.

### 4.2.3 Original TEMPO method and code availability

While in this chapter we primarily use the PT-MPO form of the TEMPO method, that is PT-TEMPO, the mean-field integration been implemented for both PT-TEMPO [303] and the original (non PT-MPO) TEMPO method [162]. These are available as part of the OQuPy Python 3 package (see Section 3.2.8) with documentation on readthedocs. Together these implementations they may be referred to as 'mean-field TEMPO.'

The original TEMPO method is used in two instances our of work here: when checking convergence of computational parameters (see Appendix D.1) and for one instance in Section 4.4.3 to save time computing an additional process tensor for a single calculation.

## 4.3 Threshold for organic lasing

### 4.3.1 Effect of detuning and light-matter coupling

We now investigate the threshold for polariton lasing using the mean-field TEMPO approach. Importantly, the construction of the PT capturing the influence of the bath, which is the costly part of the calculation, only needs to performed once for a given spectral density Eq. (4.1.3) and bath temperature $T$. It can then be reused with many different system Hamiltonians or parameters. This is particularly advantageous in allowing us to determine how the threshold for lasing varies with cavity detuning $\Delta = \omega_c - \epsilon$ and collective light-matter coupling strength $\Omega$.

First, to illustrate the dynamics, Fig. 4.3a shows time evolution simulations at $\Omega = 200$ meV and a small negative detuning $\Delta = \omega_c - \epsilon = -20$ meV. For each run the bath was prepared in a thermal state at $T = 300$ K and the spin pointing down, with a small initial field to avoid the trivial fixed point of Eqs. (4.2.15) and (4.2.19). The dynamics were generated up to a time $t_f = 1.3$ ps

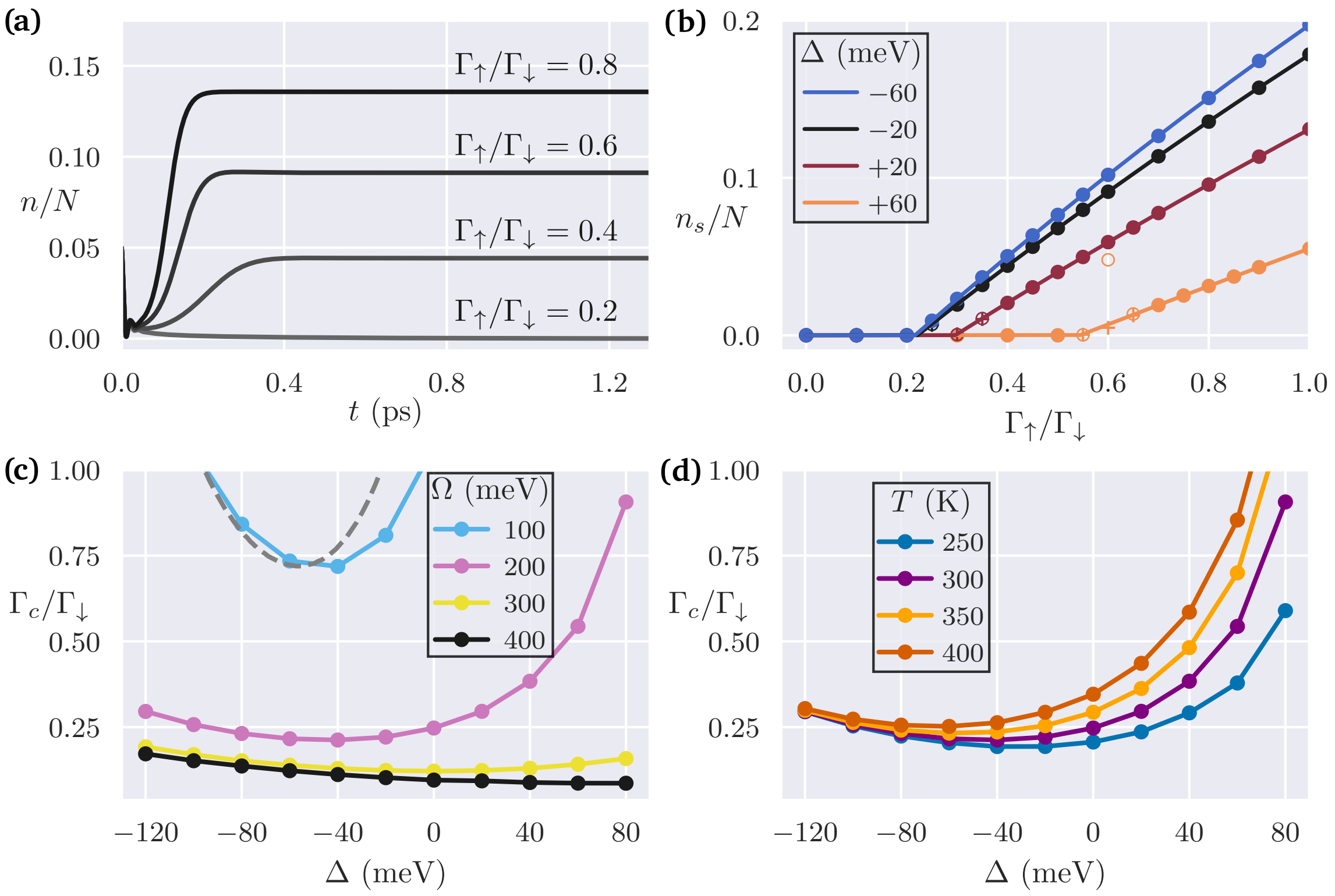

**Figure 4.3:** Determining the threshold of an organic laser. (a) Example dynamics of the scaled photon number $n/N = |\langle a\rangle|^2/N$ below ($\Gamma_\uparrow = 0.2\Gamma_\downarrow$) and above ($\Gamma_\uparrow \geq 0.4\Gamma_\downarrow$) the lasing transition for $\Omega = 200$ meV, $T = 300$ K and $\Delta = \omega_c - \epsilon = -20$ meV. The losses were fixed at $\kappa/2 = \Gamma_\downarrow = 10$ meV. Initial conditions: the system was prepared in a $\sigma^z$-down state with $n_0/N = 0.05$ and the bath in a thermal state. Computational parameters and convergence information are provided in Appendix D.1. (b) Steady-state photon number with pump strength at $\Omega = 200$ meV, $T = 300$ K for several different detunings (filled circle: steady-state value obtained from a valid fit of late time behaviour, open circle: invalid fit, cross: final value). Data from (a) contributes points on the black $\Delta = -20$ meV curve here. Fitting a curve to the data at each detuning provided an estimate of the threshold $\Gamma_c$ (when $n_s/N \to 0^+$). This was repeated for different $\Omega$ and $T$ to produce (c) and (d), respectively. The result of a weak-coupling calculation, performed in Section 4.3.4, is included for $\Omega = 100$ meV in (c) as a dashed grey line. ↩TOF

and the final value $\langle a\rangle_f$ recorded. This gave the steady-state field except near the phase boundary where, due to the critical slowing down associated with a second-order transition, $\langle a\rangle$ was still changing at $t_f$. To accommodate this, an exponential fit was made to the late time dynamics yielding an estimate of the steady-state value indicated by filled circles in Fig. 4.3b. Where this was not possible (i.e., the fitting failed), the final value of the field is marked with a cross and the attempted fit with an open circle. An automated procedure was used to assess fit validity and any point with an invalid fit was not used in subsequent calculations. This procedure as well as values of the computational parameters used are detailed in Appendix D.

Having obtained the steady-state field for a number of pump strengths encompassing the transition (Fig. 4.3b), a second fit was performed to extract the threshold pump $\Gamma_c$ at each detuning. This was repeated for different light-matter coupling strengths and temperatures to produce the phase diagrams Figs. 4.3c and 4.3d.

In Fig. 4.3c, we study the evolution of the threshold $\Gamma_c$ as the coupling $\Omega$ increases. At the smallest coupling considered, $\Omega = 100$ meV, the threshold is high and for $\Gamma_\uparrow \leq \Gamma_\downarrow$ there is only a small window of detunings for which lasing is supported—i.e., the photon frequency coincides with a region of net gain in the spectrum [364]. The grey dashed line shows the prediction of weak light-matter coupling theory explained later in Section 4.3.4. Disagreement of the exact mean-field result with this curve, most apparent nearer zero detuning, reflects the fact that $\Omega = 100$ meV is already beyond weak light-matter coupling.

We note the observed behaviour cannot be described by a weak system-bath coupling model in which the coupling to the bath is replaced by Markovian (temperature dependent) dephasing. As explained in Section 4.3.3, such a model requires $\Gamma_\uparrow > \Gamma_\downarrow$ for lasing and predicts a phase diagram that is symmetric about $\Delta = 0$. The same is true for models that completely neglect the effect of vibrational modes [105]. The existence of lasing for $\Gamma_\uparrow < \Gamma_\downarrow$ within our model is a consequence of the vibrational bath. The detuning for minimum threshold evolves with $\Omega$ and is not simply set by the peak of the molecular emission spectrum; this is due to reabsorption of cavity light playing a role for the parameters we consider [103].

As the light-matter coupling increases, faster emission into the cavity mode sees the threshold reduce before eventually saturating. The threshold becomes less dependent on detuning as lasing is now dictated by whether the frequency of the lower polariton formed coincides with a region of gain in the spectrum, and this occurs for a larger range of cavity frequencies. Similar observations were made in models with sharp vibrational resonances [62]. In that work re-entrance under $\Gamma_\uparrow$ was seen—behaviour absent here because of the broader molecular spectrum we consider.

### 4.3.2 Effect of bath temperature

A key question in the study of organic polaritons is to what extent thermalisation occurs, and thus how temperature affects the threshold [5, 103]. Motivated by this and the range of temperatures accessible in organic polariton experiments, we examine the dependence of threshold on environmental temperature $T$ at fixed $\Omega = 200$ meV. Changing $T$ shifts, and increases the width of, the molecular spectrum. The result for the phase diagram, shown in Fig. 4.3d, is a suppression of lasing with increasing $T$, most significantly for positive detunings where the lower polariton is more excitonic. This temperature dependence is one aspect of the phase diagram that cannot generally be captured by simplified models with a few vibrational modes, as we demonstrate in Section 4.3.5 below. Note that, in contrast to the curves in Fig. 4.3c, each curve here required a separate PT to be computed since a property of the bath ($T$) was changed rather than a system parameter.

We next consider approximate treatments of the model that may be taken in limiting cases: weak vibrational coupling and weak light-matter coupling, as well as a simplified model with a single vibrational mode. We discuss the applicability of these approaches and their ability to capture the features of the full, non-Markovian treatment of organic polariton lasing observed in Figs. 4.3 and 4.6 above.

### 4.3.3 Weak system-environment coupling

In the limit where the electronic coupling to the vibrational environment is weak, one might expect it would be possible to derive and use an accurate time-local (Markovian) description. In this section we discuss the challenges in doing this and explain why, even in this weak system-bath coupling limit, the PT-MPO approach (PT-TEMPO) may still be valuable.

When the system-environment coupling is weak, one can apply standard methods to derive a Redfield theory describing the low frequency vibrational environment, such as presented in the microscopic derivation in Section 3.1.8. As discussed there, in appropriate cases a secular approximation may also be made to give a density matrix equation of motion of Lindblad form. Considering the coupling to harmonic baths in Eq. (4.1.2), the incoherent contribution to the density matrix equation from the vibrational environments takes the form (cf. Eq. (3.1.85)):

$$\partial_t \rho|_{\text{vib.}} = \sum_{i,\omega} 2\,\mathrm{Re}\,\Gamma(\omega)(1/4)\left(\varsigma_{i,\omega}\rho\varsigma_{i,\omega}^\dagger - \frac{1}{2}\{\varsigma_{i,\omega}^\dagger\varsigma_{i,\omega}, \rho\}\right). \tag{4.3.1}$$

Here $\Gamma(\omega) = \int_0^\infty ds\, e^{i\omega s}C(s)$, where $C(s)$ describes correlations of the bath operators $x_j = \xi_j b_j + \bar{\xi}_j b_j^\dagger$ that couple to the system,

$$C(s) \equiv \sum_j |\xi_j|^2 \langle x_j(t)x_j(t-s)\rangle \tag{4.3.2}$$

$$= \int_0^\infty d\nu J(\nu)\left[\coth\left(\frac{\nu}{2T}\right)\cos(\nu s) - i\sin(\nu s)\right]. \tag{4.3.3}$$

The operators $\varsigma_{i,\omega}^z$ are the eigen-operator decomposition of $\sigma_i^z$. Note the system operator in $H_E^{(i)}$ is $\sigma_i^z/2$, hence the factor of $(1/4)$ in Eq. (4.3.1). They obey $[H_S, \varsigma_{i,\omega}^z] = -\omega\varsigma_{i,\omega}^z$ where $H_S$ is as in Eq. (4.1.1), and satisfy $\sum_\omega \varsigma_{i,\omega}^z = \sigma_i^z$. Formally, they can be found using the eigenstates of $H_S\,|n\rangle = \epsilon_n\,|n\rangle$, by writing a restricted sum over transitions with energy difference $\omega$

$$\varsigma_{i,\omega}^z = \sum_{\substack{m,p \\ \epsilon_p - \epsilon_m = \omega}} |m\rangle\langle m|\sigma_i^z|p\rangle\langle p|\,. \tag{4.3.4}$$

Evaluating this however presents a severe problem for the Tavis-Cummings model with strong light-matter coupling, as it requires expressions for the complete spectrum of eigenstates and energies. In general, for many-body problems, this is not available.

There do exist some special cases where one can give explicit forms of the dissipation. The simplest case—which recovers the phenomenological picture of vibrations causing dephasing—is to neglect light-matter coupling in deriving $\varsigma_{i,\omega}^z, \omega$. In this case the eigenstates are simply the excited $|1\rangle$ and ground $|0\rangle$ state of each molecule (energies $\pm\epsilon/2$) and, since $\sigma_i^z$ is diagonal in this basis, there is only one eigen-operator, when $\omega = 0$:

$$\sum_{\substack{m,p\in\{0,1\} \\ \epsilon_p - \epsilon_m = 0}} |m\rangle\langle m|\sigma_i^z|p\rangle\langle p| = |0\rangle\langle 0|\,\sigma_i^z\,|0\rangle\langle 0| + |1\rangle\langle 1|\,\sigma_i^z\,|1\rangle\langle 1| \tag{4.3.5}$$

$$= |1\rangle\langle 1| - |0\rangle\langle 0| \tag{4.3.6}$$

$$= \sigma_i^z\,. \tag{4.3.7}$$

Therefore

$$\partial_t \rho|_{\text{vib.}} = \sum_i (\mathrm{Re}\,\Gamma(0)/2)\left(\sigma_i^z\rho\sigma_i^z - \{\sigma_i^z\sigma_i^z, \rho\}/2\right) \tag{4.3.8}$$

$$= (\mathrm{Re}\,\Gamma(0)/2)\sum_i L[\sigma_i^z]. \tag{4.3.9}$$

The result is a pure dephasing process acting on each molecule with rates $\mathrm{Re}\,\Gamma(0)/2$. Note that the Lamb-shift Hamiltonian, which we omitted from Eq. (4.3.1), is proportional to the identity in this case ($\sigma^z\sigma^z = I$) and so does not affect the dynamics. Determining the coefficient $\mathrm{Re}\,\Gamma(0)$ is subtle due to the pole of coth. In Appendix D.4 we show that

$$\frac{1}{2}\mathrm{Re}\,\Gamma(0) = \frac{1}{4}\pi T \lim_{\omega\to 0}\left(\frac{2J(\omega)}{\omega}\right) \qquad (k_B = 1). \tag{4.3.10}$$

For the Ohmic spectrum $J(\nu)$ defined in Eq. (4.1.3),

$$\frac{1}{2}\,\mathrm{Re}\,\Gamma(0) = \pi\alpha T. \tag{4.3.11}$$

The behaviour of the driven-dissipative Tavis-Cummings model with dephasing has been extensively studied elsewhere (see e.g. Ref. [41]). In such a model lasing only occurs for $\Gamma_\uparrow > \Gamma_\downarrow$, and the threshold ratio $\Gamma_\uparrow/\Gamma_\downarrow$ is symmetric around cavity-molecule detuning $\Delta = 0$. Both of these features are notably different to the results seen in Fig. 4.3. We may also observe that the same statements apply when there is no effect of the vibrational bath at all. In that case our model becomes the Tavis-Cummings model with only pumping $\Gamma_\uparrow$, and decay $\Gamma_\downarrow, \kappa$ processes. As discussed extensively in previous works [102, 105], this model also requires $\Gamma_\uparrow > \Gamma_\downarrow$ for lasing to occur. However, polariton splitting is suppressed at large pumps strengths, so such models cannot provide a description of experiments demonstrating polariton lasing in the strong light-matter coupling regime [55, 81, 91, 94, 96, 97, 362].

Another case where explicit results can be derived is at weak excitation, when the saturable two-level operators $\sigma_i^\pm$ can be replaced by bosonic operators $b_i^\dagger, b_i$. This yields a system Hamiltonian that is quadratic in bosonic operators, and can be solved exactly as we show in Chapter 6. However, neglecting saturation of the two-level system is not valid when considering strong driving and lasing.

The fact that microscopic derivation of dissipation requires knowledge of the eigenspectrum of the system Hamiltonian in fact provides further motivation for methods such as the mean-field PT-MPO approach. That is, even when a weak coupling approach might be valid, it may not be practical to evaluate the eigen-operators and values. Approaches based on the PT-MPO remove this requirement, enabling one to study the dynamics of many-body systems coupled to structured environments.

### 4.3.4 Weak light-matter coupling theory

In Fig. 4.3c we included a weak light-matter coupling prediction for the phase boundary at $\Omega = 100$ meV. We now provide the supporting calculation and explain its failure to reproduce the observed boundary. The mismatch is a consequence of the conditions for lasing being outwith the weak light-matter coupling regime. Throughout this section 'weak-coupling' should be interpreted as meaning weak light-matter coupling.

The weak-coupling limit of the model has been considered in Ref. [103]. In that paper the authors worked to second order in the light-matter coupling to derive a weak-coupling master equation of the form

$$\begin{aligned}\partial_t\rho = -i\,[H_0,\rho] + \kappa L[a] + \sum_{i=1}^{N}\big(&\Gamma_\uparrow L[\sigma_i^+] + \Gamma_\downarrow L[\sigma_i^-]\\ &+\Gamma_A(\Delta)L[a\sigma_i^+] + \Gamma_E(\Delta)L[a^\dagger\sigma_i^-]\big)\,,\end{aligned} \tag{4.3.12}$$

where the free Hamiltonian $H_0 = \Delta a^\dagger a$ ($\Delta = \omega_c - \epsilon$), and $\Gamma_{A,E}$ define rates of absorption and emission processes, given by

$$\Gamma_{A,E}(\Delta) = \frac{\Omega^2}{4N}\int_{-\infty}^{\infty} dt e^{\pm i\Delta t}\langle\sigma^-(t)\sigma^+(0)\rangle_0. \tag{4.3.13}$$

Here $\langle\sigma^-(t)\sigma^+(0)\rangle_0$ is the correlator for a free molecule, measured in the absence of light-matter coupling. In Ref. [103], to calculate these quantities, it was assumed that the vibrational environment relaxes fast. This means that Eq. (4.3.13) can be calculated starting from an equilibrium state of the molecules, an approximation known as Kasha's rule [365]. For our parameters, this approximation does not necessarily hold (except for the special case of $\Gamma_\uparrow = 0$), so we use the PT-MPO method applied to an individual molecule to calculate $\Gamma_{A,E}$.

By making the mean-field factorisation approximation, as discussed above, one can assume $\langle a^\dagger a \sigma^+ \sigma^- \rangle \approx \langle a^\dagger a \rangle \langle \sigma^+ \sigma^- \rangle$ between the photon number and spin operators (implicit also is an assumption of Gaussianity; see Section 3.3). The resulting equation of motion for $n = \langle a^\dagger a \rangle$ is

$$\partial_t n = -\kappa n + N \left[ \Gamma_E(\Delta)(1+n)\langle \sigma^+ \sigma^- \rangle - \Gamma_A(\Delta) n (1 - \langle \sigma^+ \sigma^- \rangle) \right]. \tag{4.3.14}$$

At threshold ($\Gamma_\uparrow = \Gamma_c$), the coefficient of $n$ on the right-hand side of this equation changes from negative to positive. Combining this with the steady-state population of excited molecules, $\langle \sigma^+ \sigma^- \rangle = \Gamma_\uparrow / (\Gamma_\uparrow + \Gamma_\downarrow)$, we have the critical condition

$$-\kappa + N \left[ \Gamma_E(\Delta) \frac{\Gamma_c}{\Gamma_\downarrow + \Gamma_c} - \Gamma_A(\Delta) \frac{\Gamma_\downarrow}{\Gamma_\downarrow + \Gamma_c} \right] = 0, \tag{4.3.15}$$

from which

$$\frac{\Gamma_c}{\Gamma_\downarrow} = \frac{\kappa + N\Gamma_A(\Delta)}{N\Gamma_E(\Delta) - \kappa}. \tag{4.3.16}$$

Since the rates $\Gamma_{A,E}$ themselves depend on $\Gamma_\uparrow$ through $\langle \sigma^-(t) \sigma^+(0) \rangle_0$, we solved Eq. (4.3.16) iteratively for $\Gamma_\uparrow = \Gamma_c$, taking advantage of the efficiency with which many sets of system dynamics can be computed using a single PT. Setting $\Omega = 100$ meV, at each step $\Gamma_\uparrow$ was incremented and $\Gamma_{A,E}(\Delta)$ evaluated on the range $\Delta \in [-100, -20]$ meV. The first time equality resulted between the two sides of Eq. (4.3.16) for a particular $\Delta$ provided $\Gamma_c(\Delta)$ and hence a single point on the weak-coupling phase boundary in Fig. 4.3c.

As is visible in Fig. 4.3c, even at the smallest $\Omega$ used, the weak-coupling theory does not match the predictions of the full model. Reducing $\Omega$ much further leads to a regime where lasing never occurs—the collective cooperativity becomes too small [366]. As such, to verify that the full model does match the weak-coupling predictions, we must consider a different method of comparison. We choose to do this by comparing the photon absorption rates of unexcited molecules. This can be done by setting $\Gamma_\uparrow = 0$ and preparing an initial state with unexcited molecules and a small photon field. We then compare the rates at which this field decays.

Equation (4.3.14) provides an effective decay rate $\gamma_w$ for the photon number. When $\Gamma_\uparrow = 0$ this is simply

$$\gamma_w = \kappa + N\Gamma_A(\Delta), \tag{4.3.17}$$

and, since an analytical expression for $\Gamma_A(\Delta)$ is known for $\Gamma_\uparrow = 0$ (see Eq. (4.4.10) below), we can calculate $\gamma_w$ exactly for any $\Omega$ and $\Delta$, and compare to the rate $\gamma$ measured by recording the early time decay ($t \in [0, 400]$ fs) of $n/N$ in a PT-MPO simulation with the same parameters. This was done for several different detunings up to $\Omega = 25$ meV to produce Fig. 4.4a. We see the observed rate (cyan) deviates from the weak-coupling prediction (gray, dashed) from $\Omega = 10$ meV onwards. The breakdown of the weak-coupling approximation is made clear in Fig. 4.4b where we perform a fourth order polynomial fit to the difference $\gamma - \gamma_w$: the dominant $\Omega^4$ part, which we note increases with $\Delta$, cannot be captured by the second-order weak-coupling theory.

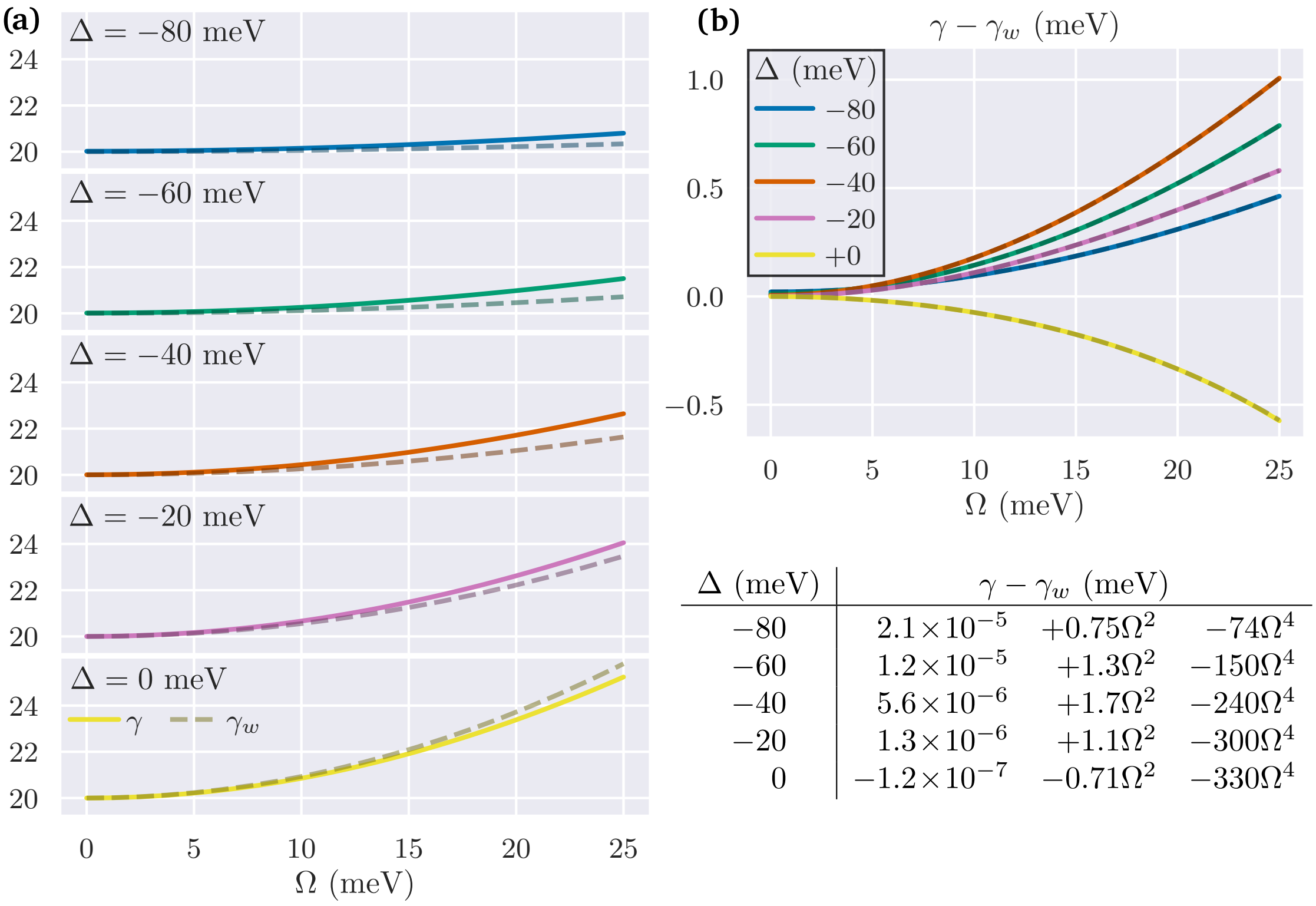

| Δ (meV) | $\gamma - \gamma_w$ (meV) | | |
|---|---|---|---|
| −80 | $2.1\times10^{-5}$ | $+0.75\Omega^2$ | $-74\Omega^4$ |
| −60 | $1.2\times10^{-5}$ | $+1.3\Omega^2$ | $-150\Omega^4$ |
| −40 | $5.6\times10^{-6}$ | $+1.7\Omega^2$ | $-240\Omega^4$ |
| −20 | $1.3\times10^{-6}$ | $+1.1\Omega^2$ | $-300\Omega^4$ |
| 0 | $-1.2\times10^{-7}$ | $-0.71\Omega^2$ | $-330\Omega^4$ |

**Figure 4.4:** (a) Dependence of effective decay rate $\gamma$ (solid) on light-matter coupling $\Omega$ for five different detunings when $\Gamma_\uparrow = 0$. The initial conditions and all other parameters were the same as used to produce Fig. 4.3c (in particular $\kappa = 20$ meV$= \gamma(\Omega = 20)$). The weak-coupling prediction $\gamma_w$ for the rate, Eq. (4.3.17), is indicated with a dashed line. (b) The difference $\gamma - \gamma_w$ at each detuning with a quartic fit (dashed) recorded in the table shown. Numerical error contributes a small constant and a small $\Omega^2$ term; it is the fourth-order term that describes behaviour beyond the weak-coupling theory. Note the dependence on $\Omega$ is weaker for more negative detunings, providing an explanation for the varying error of the weak-coupling prediction for the phase boundary in Fig. 4.3c. ↑TOF

### 4.3.5 Holstein-Tavis-Cummings model

We next make comparisons to a simplified model [62, 64] with a single vibrational mode and find that the simplified model cannot account for the temperature dependence of the phase boundary shown in Fig. 4.3d.

We consider the Holstein–Tavis-Cummings (HTC) Hamiltonian,

$$H = \omega_c a^\dagger a + \sum_{i=1}^{N} \left[ \frac{\epsilon}{2}\sigma_i^z + \frac{\Omega}{2\sqrt{N}}(a\sigma_i^+ + a^\dagger \sigma_i^-) \right] + \sum_{i=1}^{N} \omega_\nu \left[ b_i^\dagger b_i + \sqrt{S}(b_i^\dagger + b_i)\sigma_i^z \right], \qquad (4.3.18)$$

where $b_i^\dagger$ creates vibrational excitations of frequency $\omega_\nu$ on the $i$th molecule. These excitations couple to the electronic state of the molecule with strength $\omega_\nu\sqrt{S}$. Note that in contrast to Ref. [62] we make the rotating wave approximation and so do not include a diamagnetic $A^2$ term.

Incoherent processes are then included as Markovian terms in the master equation

$$\begin{aligned} \partial_t \rho = -i[H,\rho] + \kappa L[a] + \sum_{i=1}^{N} (\Gamma_\uparrow L[\sigma_i^+] + \Gamma_\downarrow L[\sigma_i^-] \\ + \Gamma_z L[\sigma_i^z] + \gamma_\uparrow L[b_i^\dagger + \sqrt{S}\sigma_i^z] + \gamma_\downarrow L[b_i + \sqrt{S}\sigma_i^z]). \end{aligned} \qquad (4.3.19)$$

In addition to the pump $\Gamma_\uparrow$, dissipation $\Gamma_\downarrow$ and cavity field decay $\kappa/2$ already considered we introduced dephasing of the electronic transition at rate $\Gamma_z$ and vibrational damping. The latter is due to relaxation of the vibrational mode to thermal equilibrium, accounting for the electronic-state-dependent vibrational displacement [62]. This occurs at temperature $T$ with rates $\gamma_\uparrow = \gamma_\nu n_B(T)$, $\gamma_\downarrow = \gamma_\nu(n_B(T)+1)$ where $n_B(T) = [\exp(\omega_\nu/T) - 1]^{-1}$, and approximately captures the effects of the additional vibrational degrees of freedom not strongly coupled to the electronic transition.

Compared to the Tavis-Cummings model considered above there are four extra parameters: the vibrational frequency $\omega_\nu$, the coupling $S$, and the rates $\Gamma_z$ and $\gamma_\nu$. There are several different approaches one might take to decide these parameters. We choose to set $\omega_\nu = 140$ meV according to the shoulder of the absorption spectrum of BODIPY-Br (Fig. 4.5a) and proceed to choose $S$, $\Gamma_z$, $\gamma_\nu$ so as to minimize the sum of squared deviations of the model's spectrum from the experimental data [55]. This is consistent with the use of the molecular absorption data to determine values of the parameters $\alpha$ and $\nu_c$ for the spectral density Eq. (4.1.3).

In Fig. 4.5c we show the phase boundaries (overlapping dashed lines) for the HTC model at $T = 300$ K and $T = 400$ K, calculated using code publicly available with Ref. [62]. Alongside we repeat the curves from Fig. 4.3d for the phase boundary of the full model at these temperatures. While the HTC model does allow for lasing without inversion, the boundary occurs at a noticeably higher pump strength over the majority of the region, and has a minimum controlled largely by the mode frequency $\omega_\nu = 140$ meV [62]. Most notably, the HTC model shows no dependence on temperature over the range we consider; this is in marked contrast to the results of the model with a continuum of low frequency vibrational modes. This occurs because the relaxation rates $\gamma_\uparrow$, $\gamma_\downarrow$ depend on temperature via the occupation $n_B = [\exp(\omega_\nu/T) - 1]^{-1}$ of the vibrational mode, but $\omega_\nu = 140$ meV far exceeds $T = 300\,\text{K} \sim 26$ meV and $T = 400\,\text{K} \sim 35$ meV hence $n_B(T) \sim 0$ for these and indeed all experimentally relevant temperatures. In contrast, the mean-field TEMPO approach used a continuum of low-frequency vibrational modes; the population of those modes can vary significantly over the relevant temperature range.

To complete this section, we note coupling to additional discrete vibrational modes could be handled by the mean-field TEMPO method, by including those modes in the system Hamiltonian $H_S$ directly. One would normally expect the computational cost of the TEMPO method to increase rapidly with the dimension of the system, limiting this approach to describe a single, high-frequency mode restricted to low excitation numbers in a Fock space truncated to a small number $N_\nu$ of levels. However, as explained in Section 3.2.6, degeneracies in the sums and differences of the bath coupling operator can be used to reduce the dimension of the bath tensors. For the case of a non-Markovian environment coupled to the electronic system, i.e., a coupling of the form $\sigma^\alpha \otimes I_{N_\nu}$, the resulting dimensions of the bath tensors are the same as that of the problem for the electronic system alone. Therefore we in fact expect additional discrete modes that do not couple directly to the non-Markovian environment can be easily included in this way, without restriction to small numbers of vibrational excitations.

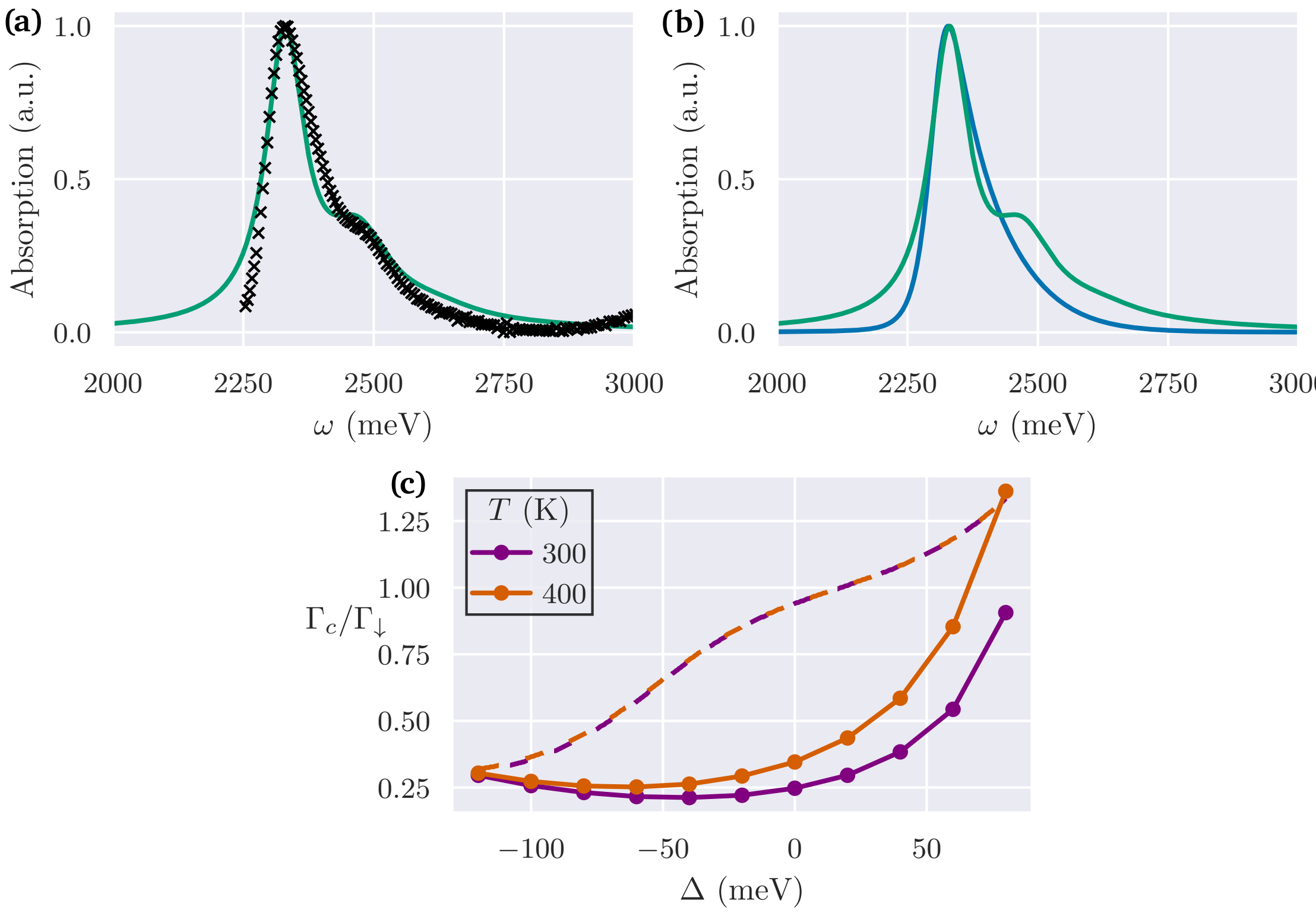

**Figure 4.5:** Molecular absorption spectrum of the HTC model (green curve), Eqs. (4.3.18) and (4.3.19), compared to (a) absorption data [55] for BODIPY-Br at $T = 300$ K (black crosses) and (b) the spectrum of our model (blue curve, Fig. 4.1b). Note as the HTC model has a smaller Stokes shift, a slightly higher two-level system frequency $\epsilon = 2330$ meV was required to match the absorption data ($\epsilon = 2310$ meV in our model). (c) Lasing threshold $\Gamma_c/\Gamma_\downarrow$ against detuning at $T = 300$ K and $T = 400$ K. Dashed lines indicate the phase boundary predicted by the HTC model for each temperature, and solid lines those of our model. Besides $\epsilon$, all parameters matched those used for Fig. 4.3d ($\Omega = 200$ meV and $\kappa/2 = \Gamma_\downarrow = 10$ meV). ↑TOF

## 4.4 Photoluminescence spectra ↑

Having demonstrated the utility of mean-field TEMPO for extracting steady-state information we now show how it can be used to determine absorption and emission dynamics. Specifically we calculate the spectral weight and the photoluminescence (PL) spectrum, the latter of which is the actual measured observable in polariton lasing experiments (see Section 2.2). This requires studying quadratic fluctuations about the mean field, as described by two-time correlations and their Fourier transforms. Multi-time correlations are naturally accessible within the PT-MPO framework, allowing us to calculate spectra spectra without recourse to the quantum regression theorem.

While so far we have considered a model with a single photon mode for which mean-field theory is exact as $N \to \infty$, it is straightforward to extend our analysis to include multiple photon modes. This allows us to calculate momentum-dependent spectra. As discussed below, mean-field theory can still provide a good approximation in this case.

### 4.4.1 Multimode model and momentum-dependent spectra

Including multiple photon modes, the system Hamiltonian Eq. (4.1.1) becomes

$$H_S = \sum_{\boldsymbol{k}} \omega_{\boldsymbol{k}} a_{\boldsymbol{k}}^\dagger a_{\boldsymbol{k}} + \sum_{i=1}^{N} \left[ \frac{\epsilon}{2}\sigma_i^z + \frac{\Omega}{2\sqrt{N}} \sum_{\boldsymbol{k}} \left( a_{\boldsymbol{k}} e^{-i\boldsymbol{k}\cdot\boldsymbol{r}_i} \sigma_i^+ + a_{\boldsymbol{k}}^\dagger e^{i\boldsymbol{k}\cdot\boldsymbol{r}_i} \sigma_i^- \right) \right]. \tag{4.4.1}$$

Here $\omega_{\boldsymbol{k}}$ is the cavity dispersion which we take to be quadratic, $\omega_{\boldsymbol{k}} = \omega_c + k^2/(2m_{\text{ph}})$ ($k = |\boldsymbol{k}|$) with an effective photon mass $m_{\text{ph}} = \omega_c/c^2$.

The form of the mean-field equations in the multimode case remains similar to those derived above. Indeed, if one assumes that only the $k = 0$ photon mode acquires a non-zero occupation, the mean-field equations are unchanged from those previously considered. The validity of this assumption is discussed shortly.

The spectrum of the nonequilibrium system is fully characterised by two independent photon Green's functions (see Appendix A.2). We use the retarded $D_{\boldsymbol{k}}^R(\omega)$ and Keldysh $D_{\boldsymbol{k}}^K(\omega)$ Green's functions, which may be written in terms of the photon dispersion $\omega_{\boldsymbol{k}}$ and exciton self-energies $\Sigma_{\boldsymbol{k}}^{-+}$, $\Sigma_{\boldsymbol{k}}^{--}(\omega)$ as [87]

$$D_{\boldsymbol{k}}^R(\omega) = \frac{1}{\omega - \omega_{\boldsymbol{k}} + i\kappa/2 - \Sigma_{\boldsymbol{k}}^{-+}(\omega)}, \tag{4.4.2}$$

$$D_{\boldsymbol{k}}^K(\omega) = -\frac{\Sigma_{\boldsymbol{k}}^{--}(\omega) + i\kappa}{\left|\omega - \omega_{\boldsymbol{k}} + i\kappa/2 - \Sigma_{\boldsymbol{k}}^{--}(\omega)\right|^2}. \tag{4.4.3}$$

For a translation-invariant system, the self-energies are diagonal in momentum [65, 87, 155]:

$$\Sigma_{\boldsymbol{k}}^{-+}(\omega) = -\frac{i\Omega^2}{4N} \sum_{i,j=1}^{N} \int_0^\infty dt e^{i\omega t} \langle [\sigma_i^-(t), \sigma_j^+(0)] \rangle e^{i(\boldsymbol{r}_i - \boldsymbol{r}_j)\cdot\boldsymbol{k}}, \tag{4.4.4}$$

$$\Sigma_{\boldsymbol{k}}^{--}(\omega) = -\frac{i\Omega^2}{4N} \sum_{i,j=1}^{N} \int_{-\infty}^\infty dt e^{i\omega t} \langle \{\sigma_i^-(t), \sigma_j^+(0)\} \rangle e^{i(\boldsymbol{r}_i - \boldsymbol{r}_j)\cdot\boldsymbol{k}}. \tag{4.4.5}$$

Below threshold, where the expectations $\langle \sigma_i^-(t) \rangle$, $\langle \sigma_j^+(0) \rangle$ vanish, only terms with $i = j$ survive within our mean-field approximation. In this case the self-energies are independent of $\boldsymbol{k}$,

$$\Sigma^{-+}(\omega) = -\frac{i\Omega^2}{4} \int_0^\infty dt e^{i\omega t} \langle [\sigma^-(t), \sigma^+(0)] \rangle, \tag{4.4.6}$$

$$\Sigma^{--}(\omega) = -\frac{i\Omega^2}{4} \int_{-\infty}^\infty dt e^{i\omega t} \langle \{\sigma^-(t), \sigma^+(0)\} \rangle. \tag{4.4.7}$$

In other words, the self-energies reduce to those of the model with a single ($k = 0$) photon mode. Hence, by calculating the correlators $\langle\sigma^-(t)\sigma^+(0)\rangle$ and $\langle\sigma^+(t)\sigma^-(0)\rangle$ of the single-mode model using the PT-MPO approach (Section 3.2.7), we can find the Green's functions $D^R_{\boldsymbol{k}}$ and $D^K_{\boldsymbol{k}}$ characterising the spectrum. For simplicity, and to avoid complications that may arise above threshold, for the purpose of calculating $k$-dependent spectra we will only consider the normal state.

Above threshold, it is still true that the commutator in Eq. (4.4.4) vanishes for $i \neq j$ within mean-field theory, giving a $k$-independent expression. For the anti-commutator in Eq. (4.4.5), we must now take into account that the expectation $\langle\sigma_i^-(t)\rangle$ is non-zero. For the lasing state this term in fact oscillates at the lasing frequency, which we will denote $\mu$, i.e., $\langle\sigma_i^-(t)\rangle = \langle\sigma_i^-(0)\rangle e^{-i\mu t}$. When lasing occurs at $k = 0$, this expectation is identical on all sites, so the anti-commutator expectation takes the form:

$$\langle\{\sigma_i^-(t), \sigma_j^+(0)\}\rangle = 2|\langle\sigma^-\rangle|^2 e^{-i\mu t} + \mathcal{A}_c(t)\delta_{ij}, \tag{4.4.8}$$

where $\mathcal{A}_c(t) = \langle\{\sigma_i^-(t), \sigma_i^+(0)\}\rangle - 2|\langle\sigma^-\rangle|^2$ is the connected part of the expectation. Here we have used the fact that within mean-field theory, the connected part exists part only for $i = j$. From Eq. (4.4.8) in Eq. (4.4.5) we find:

$$\Sigma^{--}_{\boldsymbol{k}}(\omega) = -\frac{i\Omega^2}{4}\left[2\pi N\delta_{\boldsymbol{k},0}\delta(\omega-\mu)2|\langle\sigma^-\rangle|^2 + \int_{-\infty}^{\infty} dt e^{i\omega t}\mathcal{A}_c(t)\right]. \tag{4.4.9}$$

The first term here is the source of a delta-singularity which we observe in the photoluminescence spectra below. This singularity exists only at the lasing wavevector, here taken to be $k = 0$.

Finally, we address the validity of a mean-field plus fluctuation treatment for the multimode model. As has been discussed extensively (see e.g. Refs. [5, 63]), such a treatment is valid provided the number of molecules is large compared to the number of relevant photon modes—those with energies sufficiently close the molecular transition energy.

To make this concrete, consider a finite system of area $A$. Denoting the areal density of molecules by $\rho$, the number of molecules is $\rho A$. To count photon modes, we use the mode spacing $k = 2\pi/\sqrt{A}$ , and count the number of modes with energy less than $E$: $N_{\text{ph}} = m_{\text{ph}}AE/(2\pi)$ (recall $\hbar = 1$). Hence the number of molecules per relevant photon mode is $N/N_{\text{ph}} = E_\rho/E$ where $E_\rho = 2\pi\rho/m_{\text{ph}}$. For typical molecular densities [11] we find $E_\rho \sim 10^7$ eV. This is many orders of magnitude greater than any relevant energy scale in the problem, notably the Rabi frequency $\Omega \sim 100$ meV. Therefore there are indeed many more molecules than relevant photon modes, so the mean-field plus fluctuation treatment is expected to be accurate.

A separate question for a multimode model is whether it is indeed the $k = 0$ mode which condenses. This question, which is beyond the scope of this work, is discussed in Refs. [62, 63] for the simpler Holstein-Tavis-Cummings model. It is found there that for $\Delta$ small or positive, condensation in $k = 0$ near threshold is typical.

### 4.4.2 Spectral weight and photoluminescence

We first consider the system without pumping ($\Gamma_\uparrow = 0$), since in this case exact expressions for the correlators may be found, allowing us to benchmark the mean-field PT-TEMPO method. When $\Gamma_\uparrow = 0$ the system is in the normal state, so $\langle\sigma^+(t)\sigma^-(0)\rangle \equiv 0$, while an exact expression for the other correlator was determined by Ref. [103] to be $\langle\sigma^-(t)\sigma^+(0)\rangle = e^{-i\epsilon t-\phi(t)-(\Gamma_\downarrow/2)t}$ where

$$\phi(t) = \int_{-\infty}^{\infty} d\omega \frac{J(\omega)}{\omega^2}\left[2\coth\left(\frac{\omega}{2T}\right)\sin^2\left(\frac{\omega t}{2}\right) + i\sin(\omega t)\right]. \tag{4.4.10}$$

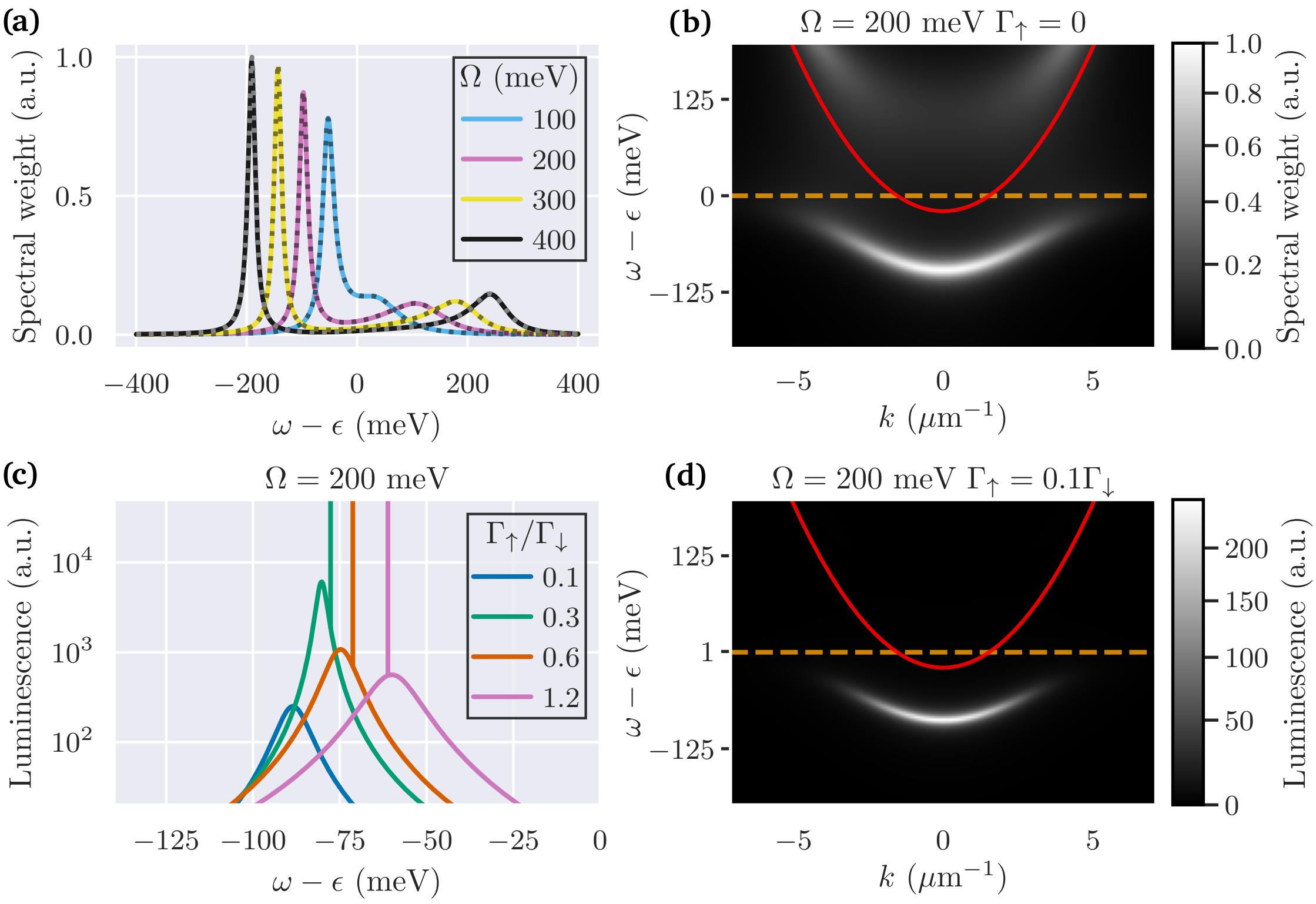

**Figure 4.6:** (a) Spectral weight, Eq. (4.4.11), at $k = 0$ when $\Gamma_\uparrow = 0$. At each light-matter coupling, results from the analytic self-energy are shown as a dotted line, and results from PT-TEMPO as a solid line. (b) $k$-dependent spectral weight for $\Omega = 200$ meV. The bare molecular energy $\epsilon$ is shown in orange and the photon dispersion $\omega_k$ in red (the photon mass $m_{\text{ph}} = \omega_c/c^2$). (c) Photoluminescence, Eq. (4.4.12), at $k = 0$ on a logarithmic scale for four different pump strengths at $\Omega = 200$ meV. Above threshold the spin-spin correlators have a non-zero long time value giving a delta singularity, i.e., a lasing peak, in the spectrum, indicated here as a vertical line. Cross sections at smaller and larger $\Omega$ are given in Fig. 4.7. (d) $k$-dependent photoluminescence below threshold at $\Omega = 200$ meV and $\Gamma_\uparrow = 0.1\Gamma_\downarrow$m with red and orange lines as in (b). All panels were produced at $\Delta = -20$ meV and $T = 300$ K, with losses $\kappa/2 = \Gamma_\downarrow = 10$ meV. ⤴TOF

To make a comparison to the PT-TEMPO numerics, we consider the spectral function

$$\varrho_{\boldsymbol{k}}(\omega) = -2\text{Im}D^R_{\boldsymbol{k}}(\omega). \tag{4.4.11}$$

In the limit of vanishing cavity losses $\varrho_{\boldsymbol{k}}(\omega)$ directly corresponds to the absorption spectrum. We refer to this function as the spectral weight (or density of states), however, as the absorption spectrum of a general lossy cavity is a more complicated expression of $D^R_{\boldsymbol{k}}$ (see Appendix A.2).

Figure 4.6a shows excellent agreement between the spectral weight derived from the analytical result Eq. (4.4.10) and that from measurement of the correlator using the PT-MPO method at $k = 0$ across the range of light-matter coupling strengths $100 \leq \Omega \leq 400$ meV considered in the previous section. The $k$ dependence of $\varrho_{\boldsymbol{k}}(\omega)$ at $\Omega = 200$ meV is illustrated in Fig. 4.6b. Weight is seen clearly at both the lower polariton, below the molecular energy $\omega = \epsilon$, and (less strongly) at the upper polariton above $\epsilon$ and the cavity dispersion. In Section 4.4.3 we examine more closely how $\varrho_{\boldsymbol{k}}$ at $k = 0$ changes as $\Gamma_\uparrow$ increases.

When the system is pumped, i.e., $\Gamma_\uparrow \neq 0$, no analytical results are available and it is nec-essary to determine both the spectrum and its occupation numerically. Here we calculate the photoluminescence [87],

$$\mathcal{L}_{\boldsymbol{k}}(\omega) = \frac{i}{2}\left(D^K_{\boldsymbol{k}}(\omega) - D^R_{\boldsymbol{k}}(\omega) + \overline{D^R_{\boldsymbol{k}}(\omega)}\right). \tag{4.4.12}$$

Figure 4.6c shows $\mathcal{L}_{\boldsymbol{k}=0}(\omega)$ at fixed detuning $\Delta = -20$ meV and $\Omega = 200$ meV for four different pump strengths. At the weakest pump strength, $\Gamma_\uparrow = 0.1\Gamma_\downarrow$, the system is below threshold yet $\mathcal{L}_{\boldsymbol{k}}(\omega)$ does not vanish since, in contrast to the mean-field calculation of the steady-state photon number, the photoluminescence contains an incoherent part. Plotting the $k$ dependence of the spectrum in this case (Fig. 4.6d) makes clear this arises from the lower polariton.

At higher pump strengths, $\Gamma_\uparrow = 0.3, 0.6, 1.2\Gamma_\downarrow$ in Fig. 4.6c, the system is above threshold, with the coherent lasing contribution indicated by a delta peak superimposed on the spectrum. As expected, the polariton spectrum is blueshifted as $\Gamma_\uparrow$ increases throughout this range. We observe in particular that, at $\Gamma_\uparrow = 0.3\Gamma_\downarrow$ and $0.6\Gamma_\downarrow$, the lasing frequency occurs noticeably to the right of the peak luminescence: the conditions to maximize $\mathcal{L}_{\boldsymbol{k}}$, which depends on both the density of states and their populations, do not, in general, coincide with the point at which the lasing instability develops. To explore this further, we now examine the real and imaginary parts of the inverse Green's functions as the transition is approached.

### 4.4.3 Inverse Green's functions in the normal state

Below threshold, the inverse retarded and Keldysh Green's functions provide insight into the normal state excitation spectra and distributions. For reference we show in Fig. 4.7 the photoluminescence $\mathcal{L}_{\boldsymbol{k}=0}(\omega)$, Eq. (4.4.12), at different pump strengths for light-matter couplings $\Omega = 100$ to $\Omega = 300$ meV, including that at $\Omega = 200$ meV given previously in Fig. 4.6. To simplify the discussion, we work at $k = 0$ throughout this section.

Spectra and occupation functions in the normal state may be expressed in terms of the components of the inverse Green's functions. We define the components $A(\omega)$, $B(\omega)$, $C(\omega)$ via

$$\left[D^R(\omega)\right]^{-1} = A(\omega) + iB(\omega), \tag{4.4.13}$$

$$\left[D^{-1}(\omega)\right]^K = iC(\omega), \tag{4.4.14}$$

where $\left[D^{-1}\right]^K$ is such that $D^K = -D^R\left[D^{-1}\right]^K D^A$. The spectral weight (density of states) $\varrho(\omega) = -2\,\mathrm{Im}\,D^R(\omega)$ and mode occupation function $2n(\omega) + 1 = iD^K(\omega)/\varrho(\omega)$ may then be written [367]

$$\varrho(\omega) = \frac{2B(\omega)}{A^2(\omega) + B^2(\omega)}, \tag{4.4.15}$$

$$n(\omega) = \frac{1}{2}\left[\frac{C(\omega)}{2B(\omega)} - 1\right], \tag{4.4.16}$$

and the photoluminescence

$$\mathcal{L}(\omega) = \frac{C(\omega) - 2B(\omega)}{2\left[A(\omega)^2 + B(\omega)^2\right]} \equiv \varrho(\omega)n(\omega). \tag{4.4.17}$$

The function $B(\omega)$ has the role of an effective linewidth for the normal modes whose position is determined by the zeros of $A(\omega)$. In the absence of light-matter coupling ($\Sigma^{-+} \equiv \Sigma^{--} \equiv 0$), $B(\omega) = \kappa/2$ is a constant and $A(\omega) = \omega - \omega_c$. In general it is possible for the distribution to diverge

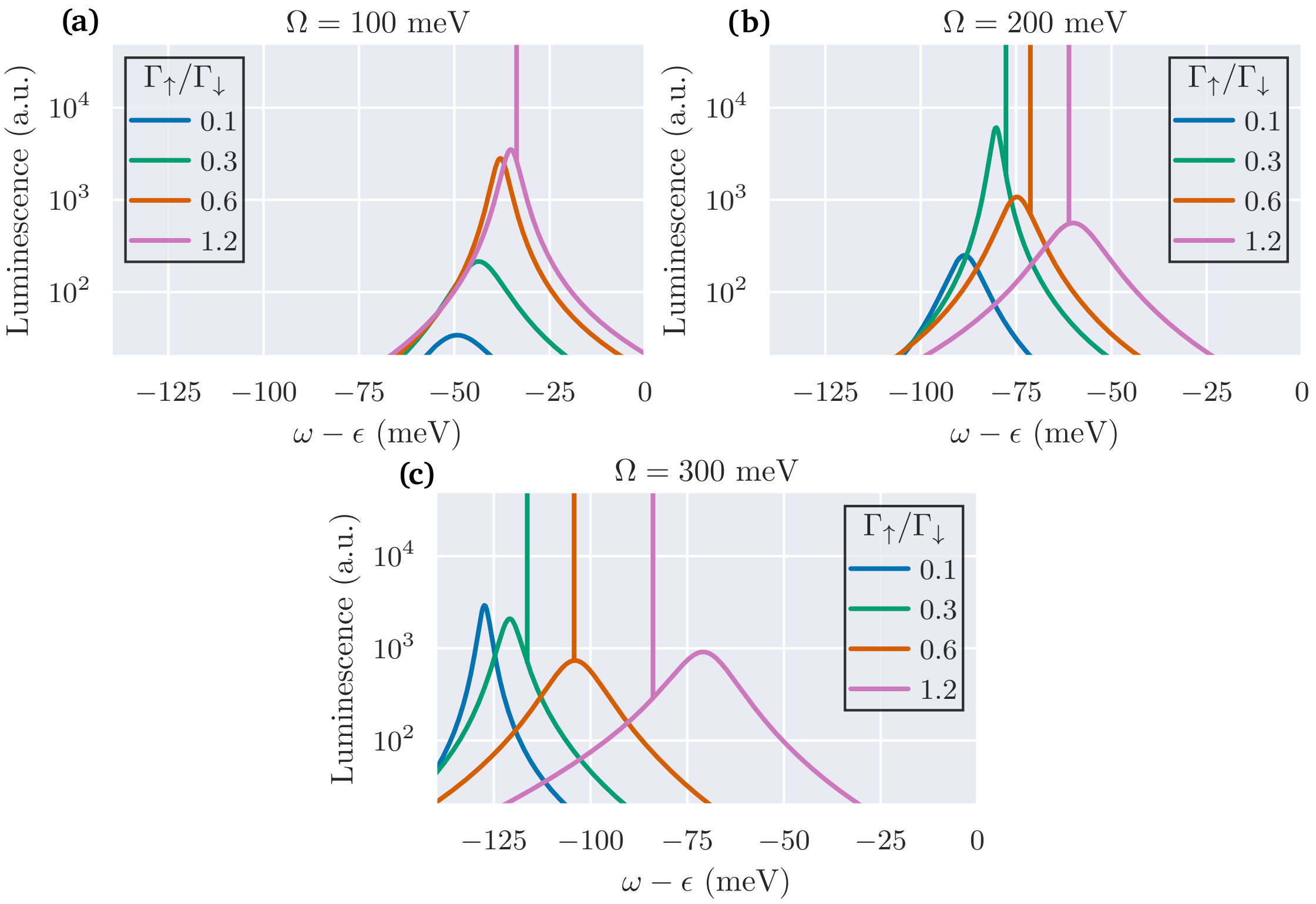

**Figure 4.7:** Photoluminescence Eq. (4.4.12) at $k = 0$ for four different pump strengths when (a) $\Omega = 100$ meV, (b) $\Omega = 200$ meV (repeat of Fig. 4.6c) and (c) $\Omega = 300$ meV. All other parameters match those used in Fig. 4.6c. A vertical line indicates a lasing peak in the spectrum. Note that, at $\Omega = 100$ meV, only the highest pump strength considered, $\Gamma_\uparrow = 1.2\Gamma_\downarrow$, is sufficient to induce lasing. Increasing the light-matter coupling both reduces the threshold and redshifts the spectrum. ↑TOF

as $n(\omega) \sim 1/(\omega - \omega^*)$, where $\omega^*$: $B(\omega^*) = 0$ defines an effective boson chemical potential, while the luminescence remains finite. Instead a condition for a divergence of $\mathcal{L}(\omega)$, i.e., a transition from the normal state to the lasing state, is a simultaneous zero of $A(\omega)$ and $B(\omega)$.

In the top row of Fig. 4.8 we show the components $A$, $B$ and $C$, as well as the derived $\varrho$, $n$ and $\mathcal{L}$ as a function of $\omega$ at $\Omega = 100$ meV for three pump strengths $\Gamma_\uparrow/\Gamma_\downarrow = 0.1$, 0.6 and 0.75 below threshold at $\Delta = -20$ meV ($\Gamma_c = 0.81\Gamma_\downarrow$ from Fig. 4.6c). As $\Gamma_\uparrow$ is increased we see the onset of a divergence in $n(\omega)$, which is established *before* the transition, as the graph of $B(\omega)$ (blue dotted line) moves downwards to develop two zeros (blue arrows), one of which is just left of the zero of $A(\omega)$ (red arrow).

At higher light-matter coupling strengths $\Omega = 200$ meV and 300 meV (bottom row of Fig. 4.8), the approach to the transition follows the same narrative albeit with more spectral weight—including additional zeros of $A(\omega)$ at $\Omega = 300$ meV—at the upper polariton $\sim (\omega - \epsilon)/\Omega = 0.5$.

### Use in determining the phase boundary

In retrospect, we note the condition $A(\omega) = B(\omega)$ at the transition offers a more efficient means to map out the phase boundaries presented in the previous section (Fig. 4.3). First, the PT-MPO method is used to calculate the molecular correlator $\langle \sigma^-(t)\sigma^+(0)\rangle$ in the normal state for e.g. a fixed grid of different pump strengths $\Gamma_\uparrow$ at a particular detuning $\Delta$. As $\langle a \rangle \equiv 0$, a model of a single molecule with time-independent $H_S = (\epsilon/2)\sigma^z$ can be used, which does not require the mean-field approach. This model has a steady state with $\langle \sigma^z \rangle = (\Gamma_\uparrow - \Gamma_\downarrow)/(\Gamma_\uparrow + \Gamma_\downarrow)$ [105]. The Green's function components can then be evaluated on the grid to determine the condition for lasing, e.g. using binary search. Noting the light-matter coupling $\Omega$ and cavity frequency $\omega_c$ only appear in the calculation at the level of the Green's function, the same set of correlator data could be used to determine the threshold for lasing at many different coupling strengths and detunings at effectively no cost, since no further PT-MPO computations would be required.

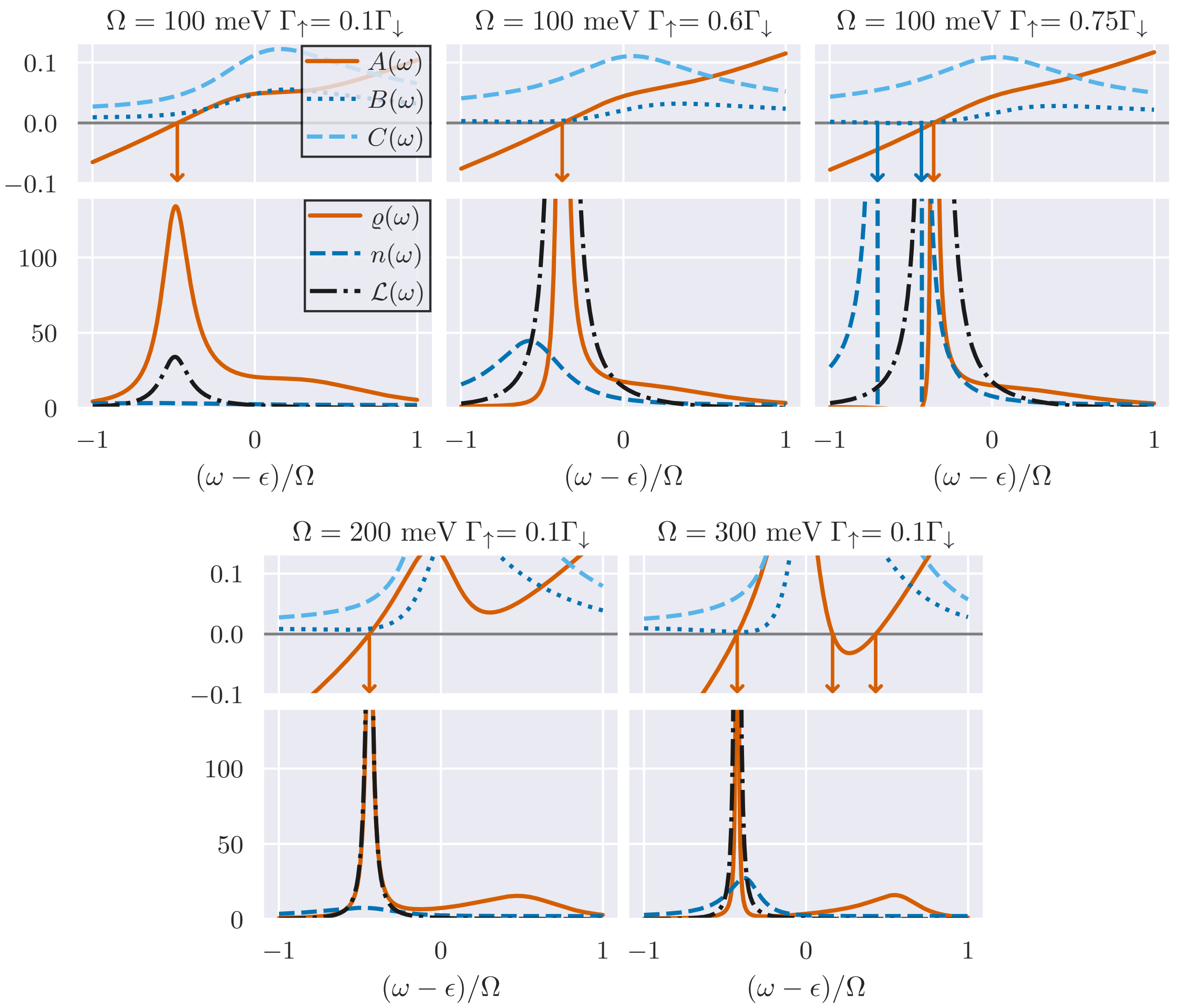

**Figure 4.8:** Real and imaginary parts of the inverse retarded and Keldysh Green's functions (top axis in each panel) as defined in Eqs. (4.4.13) and (4.4.14) and the corresponding spectral weight, occupation and photoluminescence (bottom axis). *Top row:* $\Gamma_\uparrow/\Gamma_\downarrow = 0.1, 0.6, 0.75$ at $\Omega = 100$ meV ($\Delta = -20$ meV, $T = 300$ K). The first two pump strengths correspond to the blue and red curves in Fig. 4.7a. The third, $\Gamma_\uparrow = 0.75\Gamma_\downarrow$, consists of separate data obtained using the non-PT TEMPO method (a longer time $t_f \sim 16$ ps was required to reach the steady state at this $\Gamma_\uparrow$ and it was more efficient to perform a one-off calculation than compute an additional, longer PT). Red and blue arrows indicate, respectively, zeros of the real and imaginary parts $A(\omega)$ and $B(\omega)$ of $\left[D^R\right]^{-1}$. As the threshold $\Gamma_c = 0.81\Gamma_\downarrow$ (see Fig. 4.3c) is approached, the imaginary part $B(\omega)$ decreases and develops two zeros (blue arrows). Of these, the rightmost is bound to reach the zero of $A(\omega)$ at $\Gamma_c$, at which point there is a real value $\omega^*$ such that $A(\omega^*) = B(\omega^*) = 0$, signaling instability of the normal state [87, 367]. *Bottom row:* $\Gamma_\uparrow/\Gamma_\downarrow = 0.1$ at $\Omega = 200$ meV and $\Omega = 300$ meV. Note $A(\omega)$ has two additional zeros at $\Omega = 300$ meV, a feature often taken to signal the strong coupling regime. Although the occupation function for this coupling is peaked on the right side of the first zero of $A(\omega)$, one expects this will move to the other side before the threshold (now at $\Gamma_c = 0.12\Gamma_\downarrow$) is reached. ↑TOF

## 4.5 Bright and dark exciton states in mean-field theory

For our final analysis of the model we examine the role that bright and dark excitonic states have within the mean-field approach. As discussed in Chapter 2, for a model of $N$ molecules coupled to a single photon mode, one can divide excitons into a single optically 'bright' mode—the spatially uniform superposition which couples to the cavity mode—and $N-1$ 'dark' modes which are orthogonal to the bright mode. The bright modes hybridize with the cavity mode to form polaritons, while the dark modes remain at the bare exciton energy. Extensions of this concept can also be made for models including a continuum of in-plane cavity modes; a similar division survives as long as the number of low energy photon modes is much smaller than the number of molecules [5, 368–370].

When the molecules are disordered (e.g. different on-site energies), this mixes the bright and dark states [371], leading to a non-vanishing spectral weight from the dark modes. Since our model has no disorder, one might expect the dark modes are absent. However, as we discuss here, one can directly show that within a mean-field treatment, both bright and dark states are occupied. Furthermore, despite the absence of static disorder, the vibrational environment provides a form of dynamical disorder which makes the dark modes optically active [43, 54, 65, 360].

### 4.5.1 Exciton populations

We first show how one can extract exciton populations from the mean-field theory, and show that both the $k=0$ bright states, as well as the $k \neq 0$ dark states are populated.

Firstly, the total exciton population is:

$$P_{\text{tot.}} = \sum_{i=1}^{N} \langle \sigma_i^+ \sigma_i^- \rangle = \frac{N}{2}\left(1 + \langle \sigma^z \rangle\right), \tag{4.5.1}$$

where we write $\langle \sigma^z \rangle$ for the expectation at any one of the $N$ identical sites. To find the bright and dark state populations, we can consider exciton modes with defined momenta corresponding to creation operators $\sum_i \sigma_i^+ e^{-i\boldsymbol{k}\cdot\boldsymbol{r}_i}/\sqrt{N}$. Following this, the $k=0$ exciton population is defined as

$$P_{\boldsymbol{k}=0} = \frac{1}{N} \sum_{i,j=1}^{N} \langle \sigma_i^+ \sigma_j^- \rangle. \tag{4.5.2}$$

Using the mean-field decoupling $\langle \sigma_i^+ \sigma_j^- \rangle = \langle \sigma_i^+ \rangle \langle \sigma_j^- \rangle$ for distinct sites $i \neq j$ and the properties of Pauli operators for $i = j$, the $k=0$ (bright) population is readily calculated as

$$\begin{aligned} P_{\boldsymbol{k}=0} &= \frac{1}{N} \sum_{i=1}^{N} \frac{1}{2}\left(1 + \langle \sigma^z \rangle\right) + \frac{1}{N} \sum_{j \neq i} \langle \sigma_i^+ \rangle \langle \sigma_j^- \rangle \\ &= \frac{1}{2}\left(1 + \langle \sigma^z \rangle\right) + (N-1)\left|\langle \sigma^+ \rangle\right|^2. \end{aligned}$$

By completeness of any $k$-space representation, the total population of dark states can then be found as $P_{\boldsymbol{k}\neq 0} = P_{\text{tot.}} - P_{\boldsymbol{k}=0}$. Since $P_{\boldsymbol{k}=0} \neq P_{\text{tot.}}$ one may clearly see that the mean-field approximation does not neglect the dark state population. The expressions for bright and dark mode populations simplify when we consider the limit of large $N$. In this case we may write:

$$P_{\boldsymbol{k}=0} \simeq N\left|\langle \sigma^+ \rangle\right|^2, \tag{4.5.3}$$

$$P_{\boldsymbol{k}\neq 0} \simeq \frac{N}{2}\left(1 + \langle \sigma^z \rangle - 2\left|\langle \sigma^+ \rangle\right|^2\right). \tag{4.5.4}$$

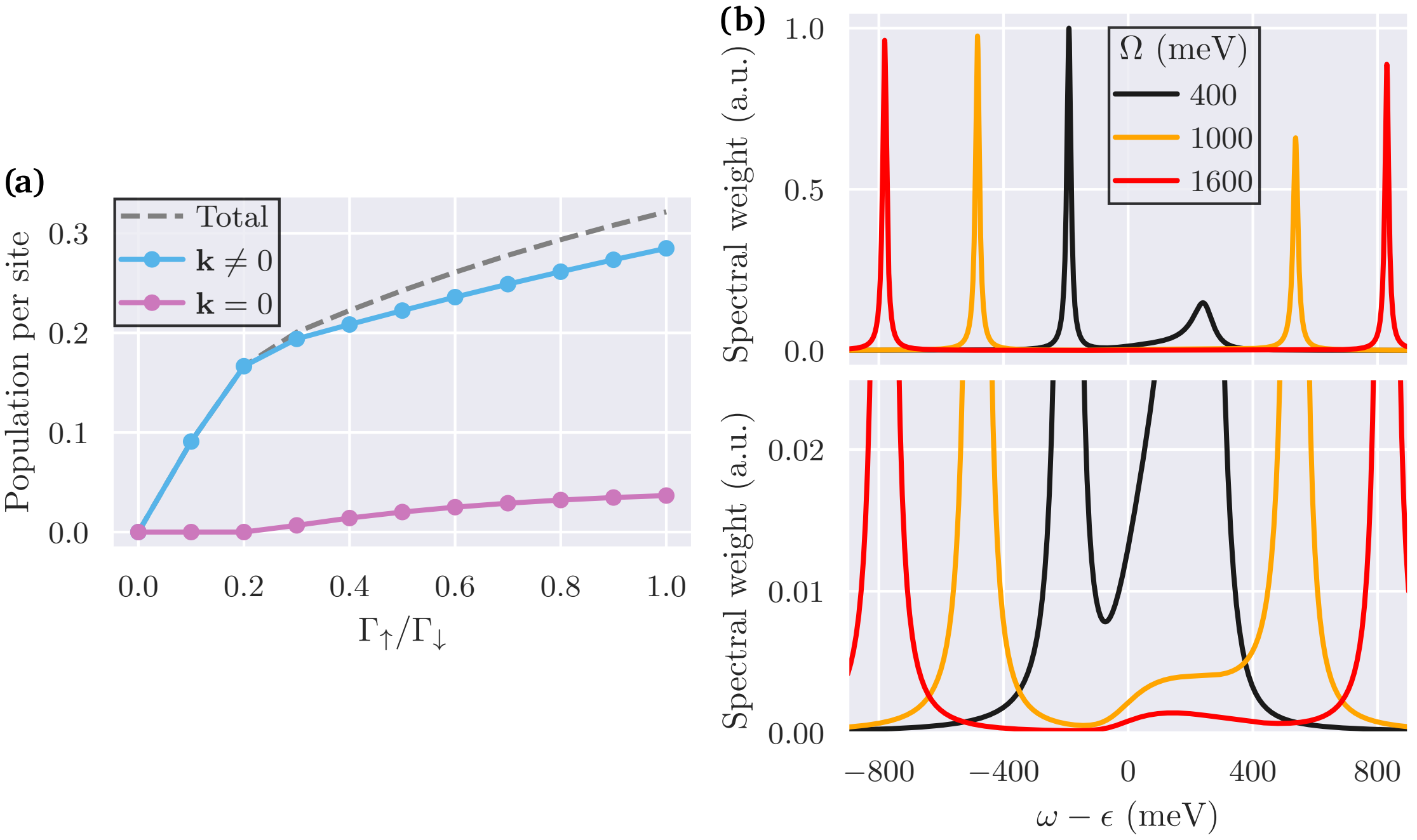

**Figure 4.9:** (a) Exciton populations per site in the steady state obtained using the PT-MPO method at $\Omega = 200$ meV. Below threshold the population per site of the $k = 0$ mode (or any single mode) vanishes as $1/N$. The $k = 0$ population becomes macroscopic above threshold. (b) Spectral weight, showing the existence of a residual excitonic peak at $\Omega = 1000, 1600$ meV. Both panels show the same data on different vertical scales. No residual peak is seen in the curve at $\Omega = 400$ meV, which was the largest light-matter coupling strength considered previously. This is due to the proximity of the upper polariton whose tail swamps the residual peak. Note the frequency structure of the vibrational environment means that this feature occurs at frequencies just above the zero-phonon line $\omega = \epsilon$. In this figure, the values of other parameters match those used in Fig. 4.3a ($\Delta = -20$ meV, $T = 300$ K, $\kappa/2 = \Gamma_\downarrow = 10$ meV).

In Fig. 4.9a we plot these steady-state populations as a function of pump strength, across the transition. When rescaled by $1/N$, the $k = 0$ mode has vanishing population in the normal state that becomes non-zero when macroscopic coherence arises in the lasing state.

### 4.5.2 Dark exciton spectral weight

An established signature of excitonic dark states in coupled light-matter systems is a residual peak in the absorption spectrum at the exciton energy [43, 54, 65, 360, 371, 372]. This occurs when either static [371, 372] or dynamic [43, 54, 65, 360] disorder can mix the bright and dark states. Mathematically, this arises due to the structure of the imaginary part of the molecular self-energy $\Sigma^{-+}$, Eq. (4.4.6). One finds that the weight of any residual peak decreases as the light-matter coupling $\Omega$ increases. This may be understood by considering the imaginary part of Eq. (4.4.2) for which $|\Sigma^{-+}|^2 \propto \Omega^4$ appears in the denominator. On the other hand, at small values of $\Omega$ the residual peak cannot be separated from the upper and lower polariton. The values of $\Omega$ in Fig. 4.6a were too small to separate the residual peak from the upper polariton. In Fig. 4.9b we show that by further increasing $\Omega$ this residual dark exciton peak may be clearly observed.

## 4.6 Summary and outlook

In this chapter we have developed an approach for calculating the non-Markovian dynamics of a many-body open system using mean-field theory and PT-MPO methods. We applied this method to model the polariton lasing of an organic dye in a microcavity including many molecules with realistic vibrational spectra. This provided the steady state of the driven-dissipative system and, via the measurement of two-time correlations, its spectrum. We first determined the dependence of the threshold for lasing on cavity detuning under different light-matter coupling strengths and environmental temperatures. Second, we observed how the photoluminescence and lasing frequency of the model evolved with pump strength. Further physical insight was obtained from examining the inverse retarded Green's function in the normal state as well as the exciton populations and their spectral contribution.

A main outcome of this work is a new exact numerical method, available in OQuPy Python 3 package [298], that may be used by others studying open many-body systems not limited to the systems of organic polaritons considered here. With this in mind, we conclude by commenting on the broader applicability of the mean-field PT-TEMPO method. In addition, in Chapter 7 we discuss combining higher-order cumulant expansions with PT-TEMPO as an extension to the mean-field approach.

### 4.6.1 Applicability of mean-field TEMPO

Within the classes of many-to-one and all-to-all models there are many instances in which mean-field theory is known to be exact as $N \to \infty$ (see Chapter 5) so that the mean-field PT-MPO method gives exact results for sufficiently large systems. As discussed in Chapter 2, systems with many-to-one coupling arise frequently in the context of cavity-QED, including molecules or cold atoms in single-mode optical cavities, and systems with all-to-all coupling in the same contexts when adiabatic elimination of the cavity mode is possible. All-to-all connectivity can also become a good approximation in systems with long-range interactions where the components couple to many, but not necessary all, others. In the next chapter we discuss exact results for mean-field behaviour within the many-to-one and all-to-all classes. We also consider another type of many-to-one model, the central spin model, which is important for many applications including spectroscopy, sensing, and quantum information, although the particular instance we consider is treated within the Markovian approximation hence we do not use the mean-field TEMPO method.

More generally, there are many physically relevant situations for which the mean-field theory is not exact but offers a good approximation, and so our method may be applied. The validity of mean-field approximations has been widely considered in equilibrium condensed matter physics [373]. In the equilibrium case it is known that for high enough dimensions mean-field theory can be a good approximation to the problem (i.e., the upper critical dimension; see Section 3.3). In particular, the effect of fluctuations beyond mean-field theory is controlled by the density of states for low energy modes. Similar questions have been explored in some open quantum systems. These include models of polariton condensation with multiple modes [63], non-equilibrium spin models (e.g. Refs. [374–377]), or cold atoms in multimode cavities [378, 379].

While we applied the mean-field PT-MPO method to a system of identical emitters, it can also be applied in the case of non-identical system Hamiltonians $H_S^{(i)}$. As mentioned in Section 4.2, the cost is a separate PT-MPO computation for each distinct type of system. Thus, the method would be suited to describe, for example, several different molecular species in a cavity but not a molecular population with a large distribution of energetic disorder. Code to handle multiple types of system was recently added to mean-field PT-TEMPO implementation in the OQuPy package [298].

As a broad principle, the approach described in this chapter can be applied in any context where: i. one can consider many systems, each of which has its own non-Markovian environment and ii. these systems couple to each other in a way that can be reasonably approximated by mean-field theory, i.e., systems couple via collective modes, or couple to many of their neighbours, such that a mean-field approximation may become good.

### 4.6.2 Collective dynamics using truncated equations

Finally, we note a recent method [137] for determining the collective behaviour of molecular polaritons as complementary to our approach. In collective dynamics using truncated equations (CUT-E) [137], a multi-configurational ansatz for the many-body wavefunction in the first-excitation manifold is applied to the single cavity mode–many molecule model. Working in a permutation symmetric basis, one finds a hierarchy of equations of motion where coupling between states conserving the number of molecules with vibrational excitations is collective, $\sim \Omega$, whereas processes that change the number of such molecules are proportional to the single-molecule coupling $\Omega/\sqrt{N}$. This affords a perturbative treatment in $\Omega/\sqrt{N}$ which mixes states with different number of ground state molecules with vibrational excitations. The result, at lowest order, is a model of the photon mode interacting with a single effective molecule, with higher-orders bringing corrections in powers of $1/\sqrt{N}$ via additional effective molecules. This dimension reduction is similar to that in our method, except we have effectively a single molecule interacting with a classical (i.e., coherent) mean field. For problems involving molecular polaritons it might be insightful to compare our mean-field ($N \to \infty$) approach to CUT-E, which provides explicit $1/\sqrt{N}$ corrections to the thermodynamic limit.

# Chapter 5

# Validity of cumulant expansions for central spin models ↑TOC

Everybody believes in the exponential law of errors: the experimenters, because they think it can be proved by mathematics; and the mathematicians, because they believe it has been established by observation.

Gabriel Lippmann

## Contents

In this chapter we investigate the validity of cumulant expansions for central spin models. We find, contrary to common expectations for models with many-to-one connectivity, behaviour that is not captured by mean-field theory in the thermodynamic limit. We also determine non-uniform convergence behaviour of the cumulant expansions for central spin models as well as central boson models. We explain how these behaviours arise in relation to the scaling of parameters and correlations with system size.

In Section 5.1 we motivate the need to establish the validity of cumulant expansions, and explain why mean-field behaviour may well be expected for a many-to-one model where $N$ satellites couple to a common central site. We discuss existing results for the exactness of mean-field theory within this class, and the limits of their applicability. The central spin model studied in this chapter is then described in Section 5.2. We explain how the permutation symmetry of the model may be used to obtain an exact solution at relatively large $N \sim 150$, and apply cumulant expansions up to third order. Results of these expansions at first and second order are firstly compared to exact data in Section 5.3. These reveal how the ability of mean-field theory to capture the $N \to \infty$ steady state of the full quantum model depends on the scaling of parameters in the model. Following this, higher-order expansions are investigated in Section 5.4, including results up to fifth order using the `QuantumCumulants.jl` package [352]. Here we observe non-uniform convergence between odd and even expansions orders. We discuss how this convergence behaviour arises in light of correlations present in the system and show that similar behaviour may be observed in models of light-matter interaction. Comparison with exact results across both Section 5.3 and 5.4 demonstrates how the error in cumulant expansion approximations does not generally decrease monotonically with $N$, nor with the order of expansion.

The results included in this chapter were presented in the publication P. Fowler-Wright et al., Determining the validity of cumulant expansions for central spin models, Phys. Rev. Research **5**, 033148 (2023) [13].

## 5.1 Cumulant expansions for many-to-one models

Cumulant expansions were introduced in Chapter 3 as a systematic way to go beyond mean-field theory by including successive orders of correlations[1]. This is a general approach capable of handling large many-body problems that provides, at finite $N$, corrections to mean-field results. Despite their widespread use, rigorous results for the validity of cumulant expansions are lacking. Even those that exist for mean-field theory are limited when considering the vast range of problems to which these methods are applied. Hence there is great need to establish the validity of cumulant expansions for different models of quantum many-body systems.

In previous applications of cumulant expansions, no general way of predicting whether correlations beyond a certain order will be significant to the dynamics has been established. As a result, it is often unclear whether a cumulant expansion of a given order will produce a satisfactory approximation. Whether this is the case will depend on factors such as the system size, parameter regime and timescale examined.

Often, one of two comparisons are made to verify the results of an order $M$ cumulant expansion: i., to exact numerics, at small $N$; ii., to the results of cumulant expansions at higher order $M$. The first comparison is predicated on the approximation provided by cumulant expansions improving with increasing $N$, and the second on this approximation improving with increasing $M$. In this chapter we provide a simple model where both of these assumptions may fail. To make clear the significance of these counter examples, we now explain why mean-field ($M = 1$) behaviour as $N \to \infty$ may be expected for the type of model considered.

[1] In the previous chapter we calculated fluctuations about the mean-field solution via linear response functions (the photon Green's functions). With cumulant expansions, beyond mean-field correlations are included from the outset.

### 5.1.1 Expectation for mean-field behaviour

When introducing mean-field theory (Section 3.4.5) we provided intuition for an 'average field' description being accurate for large systems with high connectivity. Here we make more concrete arguments for the effectiveness of mean-field theory for many-to-one models, and comment why they may fail. To be specific, we question the expectation for mean-field theory to capture exact behaviour as the number of satellite sites $N \to \infty$.

First, given $N$ identical satellites, monogamy of entanglement [380] restricts the entanglement between any two sites such that quantum correlations in the system vanish as $N \to \infty$. There is no similar restriction on *classical* correlations however. Second, in models with weak couplings to satellite sites, these may be treated as a harmonic bath for the central site with a linear response that becomes exact as $N \to \infty$ [187]. Third, for models with interactions between a large number of emitters and a *bosonic* mode, the mean-field equations can be well justified using saddle-point analysis [372]. Additional rigorous results regarding the exactness of mean-field theory as $N \to \infty$ for many-to-one and many-to-many models are discussed below.

In spite of these results, we present in this chapter simple examples where mean-field and higher-order cumulant expansions fail to capture $N \to \infty$ behaviour of a many-to-one model. The crux of why the above arguments breakdown is in the scaling of model parameters with system size $N$: our results show clearly how, in the presence of non-trivial scaling (i.e., parameters not constant functions of $N$), it is generally *not* sufficient to know the behaviour of correlations as $N \to \infty$, e.g., that they vanish, to determine whether they are relevant in this limit. Further, proofs of mean-field behaviour such as the linear response under weak coupling and saddle point analysis require individual satellite-central couplings dimmish sufficiently quickly as $N \to \infty$, usually as $1/\sqrt{N}$ or $1/N$. While this may be true for certain models (e.g. of electron-phonon interactions in solids [187]), for the central spin model we consider it is not.

### 5.1.2 Available exact results

A set of models for which mean-field theory *is* known to capture exact behaviour in the thermodynamic limit are Dicke-like central boson models [41]. Much work has been done proving [332, 333] as well as numerically investigating [105, 381–386] the accuracy of mean-field theory for this type of model. Rigorous results regarding the exactness of mean-field theory extend beyond Dicke models however, including other many-to-one models [387–389] and models with all-to-all coupling [327, 390–392].

We highlight the proof for open Dicke models in Ref. [333] as particularly clear and relevant for our application of mean-field approaches in Chapters 4 and 6. The proof establishes, for a dissipative multimode Dicke model, that the expectation of the average magnetisation or photon field follows dynamics governed by the mean-field Heisenberg equations of motion as $N \to \infty$ in the following sense: for an initial state with short-range correlations, at any finite time $t$ the error in the mean-field values vanishes under $N \to \infty$. As concisely summarised in Ref. [392], the essence of this and many other results is that the structure of the Liouvillian generator is such that, if the initial density matrix $\rho$ is a product state, it will remain a product for all times, in the thermodynamic limit. We now make further comments on on the applicability of these results.

After the obvious fact that many of the exact results apply only to central boson (not spin) models, there are several common ways in which they are restricted. Foremost, as mentioned above, a specific scaling with $N$ of couplings between sites is always required. This is normally taken to reflect physically an interaction spread uniformly over the parts of the system such that individual couplings become irrelevant in the thermodynamic limit. We note that the decay of individual couplings as $1/\sqrt{N}$, relevant to cavity mediated light-matter interactions, is not always sufficient.

Second, the proofs require at least an approximate mean-field initial state [333, 392]. This places a practical limitation on the type of dynamics that can be described: the mean-field equations require symmetry breaking and so cannot, without some form of trajectory averaging, describe behaviour of real systems where the dynamics is symmetry preserving. A related issue when determining the steady state is that the mean-field solution may not be uniquely defined, e.g., due to bistability [393–395]. It is worth pointing out that even when mean-field theory does correctly describe the average properties of the system at long-times, that does not mean quantum fluctuations do not persist or are physically irrelevant [396, 397].

Finally, there may be an issue of the ordering of limits. For example, in Ref. [333] and similar proofs for all-to-all models [392], a bound on the mean-field error of the form $B(N)e^{Ct}$ is determined. Here $C > 0$ is a constant and $B(N)$ is set by the correlations in the initial state. As the magnitude of $C$ is not specified one can imagine pathological cases where even small correlations grow rapidly over physically relevant timescales for very large (but finite) systems. Moreover, such a bound means the exactness of mean-field dynamics as $N \to \infty$ for finite $t$ may not guarantee results for the steady state $t \to \infty$ obtained for the $N \to \infty$ model[2].

Beyond all of these points, the above results regard mean-field theory specifically. No statements are made regarding the validity of higher-order cumulant expansions that we examine in this chapter. Surprisingly, we find that even when mean-field theory *does* capture exact $N \to \infty$ behaviour, higher-order cumulant expansions may fail to converge to the same result.

## 5.2 Central spin model

We consider a single spin-1/2 (Pauli matrices $\sigma_0^\alpha$) interacting with $N$ spin-1/2 satellites ($\sigma_n^\alpha$) according to

$$H = \frac{\omega}{2}\sigma_0^z + \sum_{n=1}^{N}\left[\frac{\epsilon}{2}\sigma_n^z + g\left(\sigma_0^+\sigma_n^- + \sigma_0^-\sigma_n^+\right)\right]. \tag{5.2.1}$$

Here $\omega$ and $\epsilon$ are on-site energies for the central and a satellite spin, and $g$ the interaction strength. This may be referred to as the XX Hamiltonian which is known to be integrable [398–400]. It can be obtained from the Tavis-Cummings Hamiltonian considered in the previous chapter with the replacement of the central boson by a spin. Note well however our choice to write the coupling strength as $g$ not $\Omega/(2\sqrt{N})$: we do not necessarily assume individual interactions diminish with $1/\sqrt{N}$.

In addition we consider dissipation with rate $\kappa$ from the central site as well as incoherent pump $\Gamma_\uparrow$ and loss $\Gamma_\downarrow$ for each satellite. These are included as Markovian terms in the master equation for the total density matrix $\rho$,

$$\partial_t\rho = -i\,[H,\rho] + \kappa L[\sigma_0^-] + \sum_{n=1}^{N}\left(\Gamma_\uparrow L[\sigma_n^+] + \Gamma_\downarrow L[\sigma_n^-]\right), \tag{5.2.2}$$

with $L[x] = x\rho x^\dagger - \{x^\dagger x, \rho\}/2$. Schematics for the system and these processes are given in Figs. 5.1a and 5.1b.

[2] One might expect this to be an issue near a phase transition, and so effect a small number of parameter choices only.

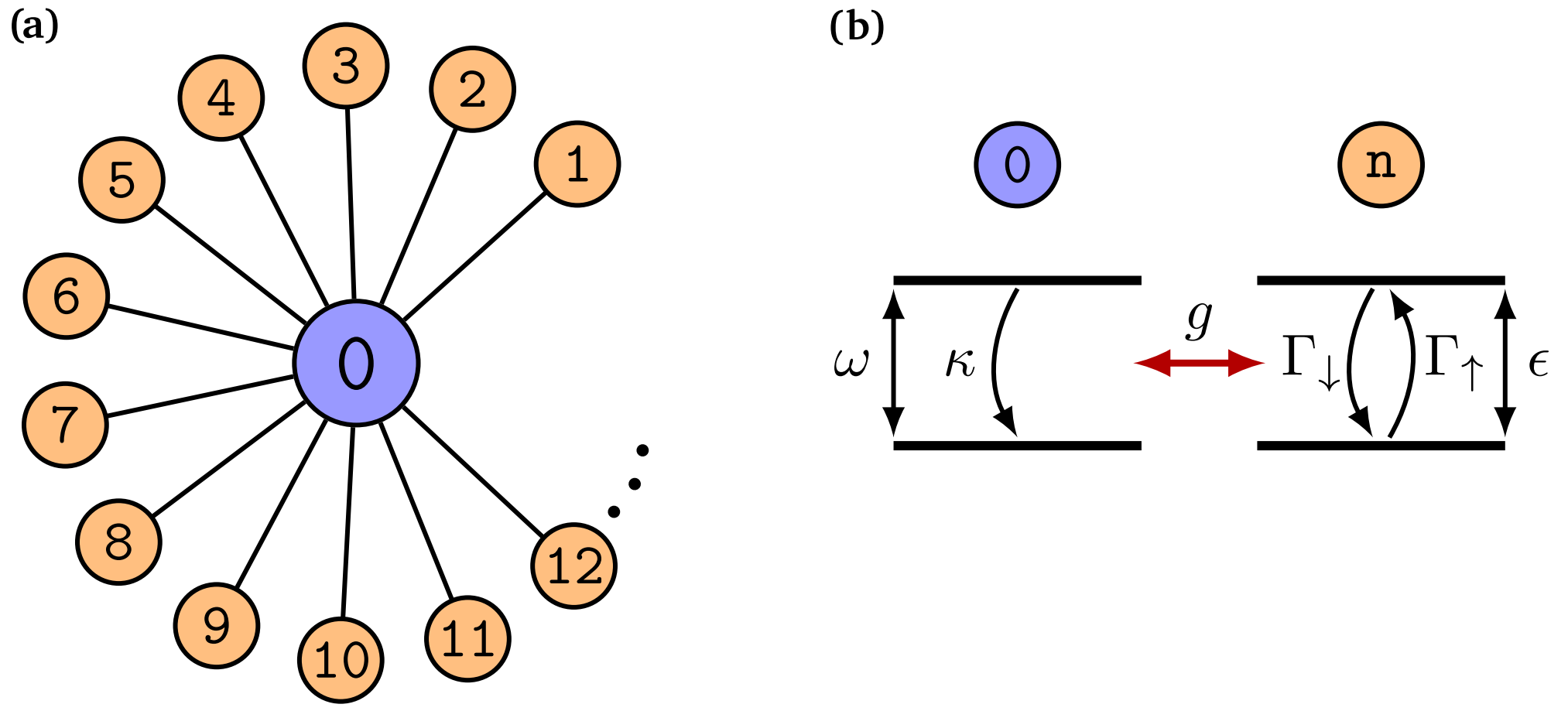

**Figure 5.1:** (a) Network of the model: a central site (index 0) couples to $N$ identical satellites ($n = 1, \ldots, N$). (b) Each site is a two-level system (spin-1/2) subject to decay ($\kappa$ or $\Gamma_\downarrow$) and, in the case of the satellites, pump $\Gamma_\uparrow$. Compared to the central boson model in the previous chapter, the photon mode $a$ has been replaced by a spin. ⇀TOF

The anisotropic interactions in Eq. (5.2.1) arise in resonant dipolar spin systems, for example between the nitrogen-vacancy center and the $^{13}$C nuclear spins in diamond [401]. This system has been extensively studied for its potential role in emerging quantum technologies including spectroscopy [27–29], quantum sensing [30–32], and computing [37, 38]. For our purpose the model serves as a minimal formulation of the open many-to-one problem to investigate mean-field theory and cumulant expansions. In certain cases, such as the absence of dissipation, or when the satellite dissipation is collective, there exist analytical or other efficient numerical methods capable of accessing large-$N$ behaviour of central spin models [398, 402–406]. However, for the case we consider with individual dephasing these methods do not apply.

The model Eq. (5.2.2) has cumulant equations that are analytically tractable up to third order whilst also allowing exact calculations for relatively large system sizes. Below, to compare approximations, we focus on properties of the central site which is the relevant degree of freedom in many of the applications. In particular we calculate the central-site population $p_0^\uparrow$ in the steady state. This relates to the polarization $\langle \sigma_0^z \rangle$ via $p_0^\uparrow \equiv (1 + \langle \sigma_0^z \rangle)/2$ and increases from zero as the satellite pump ratio $\Gamma_\uparrow/\Gamma_T$ ($\Gamma_T = \Gamma_\uparrow + \Gamma_\downarrow$) is increased.

### 5.2.1 Exact calculation in the permutation symmetric basis ⇀

The invariance of the model under the interchange of satellite spins allows one to work in a permutation symmetric basis when performing exact calculations [384, 407–411]. In this approach, instead of solving for all elements of $\rho$ one only needs to solve for a single representative of each set of elements related by the interchange of satellite spins, since those elements must be equivalent (Fig. 5.2a). This provides a *combinatoric* reduction in the size of the Liouvillian $\mathcal{L}$. In our case it allows finding the eigenvector of $\mathcal{L}$ with eigenvalue $0$, i.e., the steady state $\rho_0$, up to $N = 150$. No information is lost by working in this basis. In particular, all correlations can be computed exactly and compared to the prediction of the cumulant expansions. To generate the exact data presented in this chapter, we used the implementation [412] made publicly available with Ref. [384].

We now explain the application of the cumulant expansion method to the model at first (mean-field) and second order. We also give expressions for the third-order equations that will be used in conjunction with results at fourth and fifth order from the `QuantumCumulants.jl` [352] package in Section 5.4. A complete derivation of the equations is provided in Appendix E.

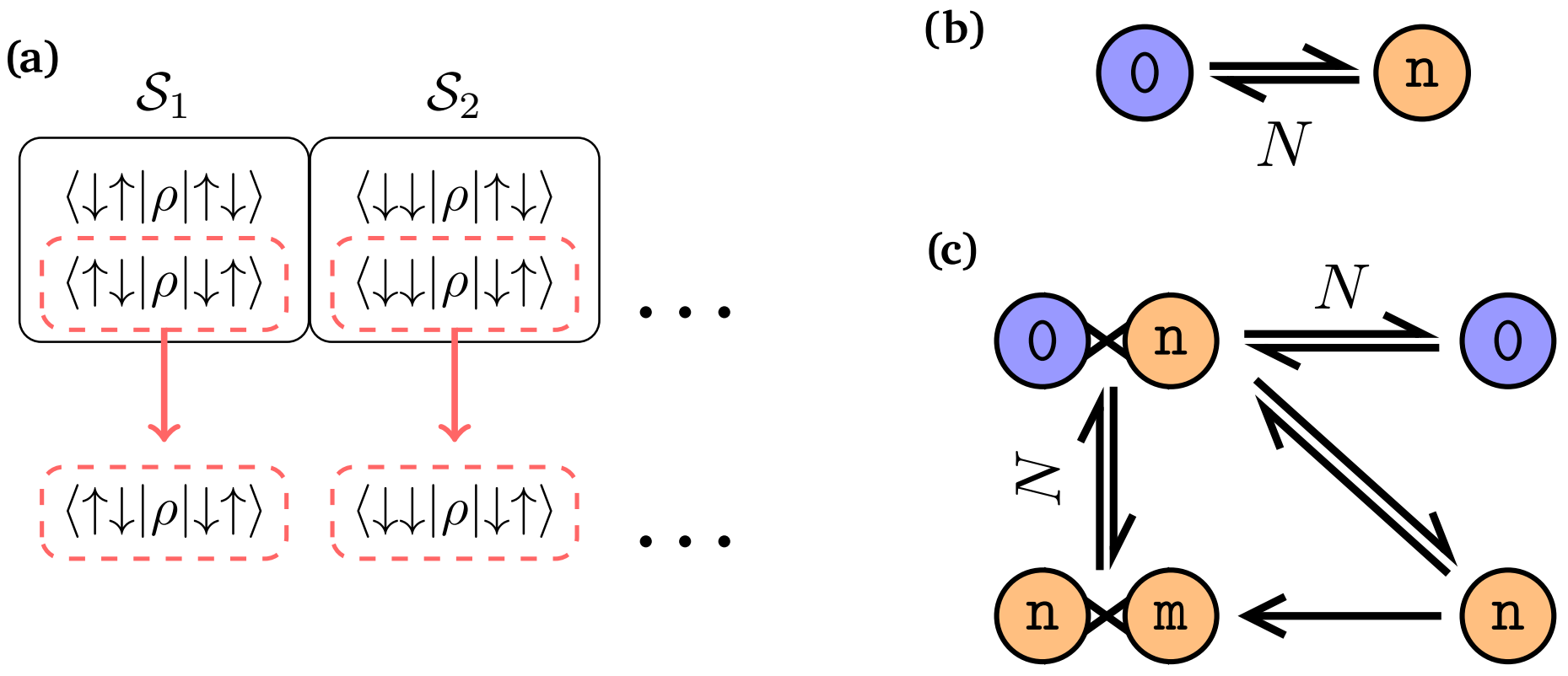

**Figure 5.2:** (a) Permutation symmetric subspaces $\mathcal{S}_1, \mathcal{S}_2, \ldots$ with matrix elements related by the interchange of a pair of satellite spins (here $N = 2$ and the central spin has been omitted). Only one element from each subspace needs to be considered, providing a combinatoric reduction of the dimension of the problem to be solved exactly: for $N = 2$ this is from $2^4 \cdot 2^2 = 64$ to $10 \cdot 2^2 = 40$ elements, where the last $2^2$ corresponds to the central spin in each case. (b) In mean-field theory, the reduction is to a two-body problem where expectations of a satellite satellite evolve according to expectations of the central site (⇀) which in turn evolve according to $N$ copies of the satellite expectations (↽). (c) In the second-order cumulant expansion central-satellite and satellite-satellite expectations couple into the system. ↵TOF

### 5.2.2 Mean-field theory ↵

From the master equation, Eq. (5.2.2), equations of motion for single-site expectations can be derived (Appendix E.1):

$$\partial_t\langle\sigma_0^z\rangle = -\kappa\left(\langle\sigma_0^z\rangle + 1\right) + 4gN\,\mathrm{Im}\left[\langle\sigma_0^+\sigma_n^-\rangle\right], \tag{5.2.3}$$

$$\partial_t\langle\sigma_n^z\rangle = -\Gamma_T\langle\sigma_n^z\rangle + \Gamma_\Delta - 4g\,\mathrm{Im}\left[\langle\sigma_0^+\sigma_n^-\rangle\right], \tag{5.2.4}$$

$$\partial_t\langle\sigma_0^+\rangle = \left(i\omega - \frac{\kappa}{2}\right)\langle\sigma_0^+\rangle - igN\langle\sigma_0^z\sigma_n^+\rangle, \tag{5.2.5}$$

$$\partial_t\langle\sigma_n^+\rangle = \left(i\epsilon - \frac{\Gamma_T}{2}\right)\langle\sigma_n^+\rangle - ig\langle\sigma_0^+\sigma_n^z\rangle, \tag{5.2.6}$$

with $\Gamma_\Delta = \Gamma_\uparrow - \Gamma_\downarrow$ and $\Gamma_T = \Gamma_\uparrow + \Gamma_\downarrow$. This set of equations is not closed since, for example, $\partial_t\langle\sigma_0^z\rangle$ depends on $\langle\sigma_0^+\sigma_n^-\rangle$. The equation for $\langle\sigma_0^+\sigma_n^-\rangle$ will in turn depend on expectations of operators from three different sites, and so on, resulting in an exponential (in $N$) number of equations involving operators on all sites.

To close these equations using a first-order cumulant expansion, that is mean-field theory, second-order moments are factorised into products, $\langle\sigma_0^\alpha\sigma_n^\beta\rangle \approx \langle\sigma_0^\alpha\rangle\langle\sigma_n^\beta\rangle$ (see Chapter 3). The resulting equations describe an effective two-body problem depicted in Fig. 5.2b:

$$\partial_t\langle\sigma_0^z\rangle = -\kappa\left(\langle\sigma_0^z\rangle + 1\right) + 4gN\,\mathrm{Im}\left[\langle\sigma_0^+\rangle\langle\sigma_n^-\rangle\right], \tag{5.2.7}$$

$$\partial_t\langle\sigma_n^z\rangle = -\Gamma_T\langle\sigma_n^z\rangle + \Gamma_\Delta - 4g\,\mathrm{Im}\left[\langle\sigma_0^+\rangle\langle\sigma_n^-\rangle\right], \tag{5.2.8}$$

$$\partial_t\langle\sigma_0^+\rangle = \left(i\omega - \frac{\kappa}{2}\right)\langle\sigma_0^+\rangle - igN\langle\sigma_0^z\rangle\langle\sigma_n^+\rangle, \tag{5.2.9}$$

$$\partial_t\langle\sigma_n^+\rangle = \left(i\epsilon - \frac{\Gamma_T}{2}\right)\langle\sigma_n^+\rangle - ig\langle\sigma_0^+\rangle\langle\sigma_n^z\rangle. \tag{5.2.10}$$

Solving these equations for the steady state one finds $\langle\sigma_0^z\rangle = -1$ for $\Gamma_\uparrow/\Gamma_T$ below a critical pump ratio $R_c \equiv (1 + \Gamma_T\kappa/4g^2N)/2$, while for $\Gamma_\uparrow/\Gamma_T > R_c$:

$$\langle\sigma_0^z\rangle = -\frac{1}{2}\left(1 - \frac{\Gamma_\Delta N}{\kappa}\right) - \frac{1}{2}\sqrt{\left(1 - \frac{\Gamma_\Delta N}{\kappa}\right)^2 + \frac{\Gamma_T^2}{g^2}}, \tag{5.2.11}$$

$$\langle\sigma_n^z\rangle = -\frac{\kappa\Gamma_T}{4g^2N\langle\sigma_0^z\rangle}, \quad \langle\sigma_n^+\rangle = \frac{i\kappa}{2gN\langle\sigma_0^z\rangle}\langle\sigma_0^+\rangle, \tag{5.2.12}$$

where the magnitude of $\langle\sigma_0^+\rangle$ satisfies

$$\left|\langle\sigma_0^+\rangle\right|^2 = -\langle\sigma_0^z\rangle\left(1 + \langle\sigma_0^z\rangle\right)/2. \tag{5.2.13}$$

Here and in the following for simplicity we set $\omega = \epsilon$, but we have checked our conclusions do not change off resonance. Detailed working for the solution (5.2.11)–(5.2.13) including the case $\omega \neq \epsilon$ is provided in Appendix E.2. As noted there the phase of $\sigma_0^+$ (or $\sigma_n^+$) is not fixed by Eq. (5.2.13). In fact, in general a non-stationary solution with a time-dependent phase is expected.

Although the model has U(1) symmetry, i.e., Eq. (5.2.2) is invariant under $\sigma^\pm \to \sigma^\pm e^{\pm i\theta}$, it was necessary to retain the symmetry-breaking terms $\langle\sigma_0^+\rangle$ and $\langle\sigma_n^+\rangle$ when performing the mean-field approximation in order to obtain a non-trivial solution (cf. Section 3.4.4): the state $\langle\sigma_0^z\rangle = -1$ is always a solution to the mean-field equations that only becomes unstable when $\Gamma_\uparrow/\Gamma_T > R_c$.

### 5.2.3 Second-order cumulant expansion

Breaking symmetry is not necessary at second order where $\langle\sigma_0^+\sigma_n^-\rangle$ can be non-zero whilst respecting the symmetry. The required equations for second-order moments are (Fig. 5.2c)

$$\begin{aligned}\partial_t\langle\sigma_0^+\sigma_n^-\rangle = &\left(i(\omega-\epsilon) - \frac{\kappa+\Gamma_T}{2}\right)\langle\sigma_0^+\sigma_n^-\rangle + \frac{ig}{2}\langle\sigma_n^z\rangle \\ &- \frac{ig}{2}\langle\sigma_0^z\rangle - ig(N-1)\langle\sigma_0^z\rangle\langle\sigma_n^+\sigma_m^-\rangle,\end{aligned} \tag{5.2.14}$$

$$\partial_t\langle\sigma_n^+\sigma_m^-\rangle = -\Gamma_T\langle\sigma_n^+\sigma_m^-\rangle + 2g\,\mathrm{Im}\left[\langle\sigma_0^+\sigma_n^-\rangle\right]\langle\sigma_n^z\rangle, \tag{5.2.15}$$

where $n \neq m$. Here we set third cumulants to zero and use the U(1) symmetry to write $\langle\sigma_0^z\sigma_n^+\sigma_m^-\rangle \approx \langle\sigma_0^z\rangle\langle\sigma_n^+\sigma_m^-\rangle$, $\langle\sigma_0^+\sigma_n^-\sigma_m^z\rangle \approx \langle\sigma_0^+\sigma_n^-\rangle\langle\sigma_n^z\rangle$. Equations (5.2.3), (5.2.4), (5.2.14) and (5.2.15) can also be solved exactly, albeit not explicitly, to find $p_0^\uparrow = (1 + \langle\sigma_0^z\rangle)/2$.

### 5.2.4 Higher-order cumulant expansions

#### Third-order equations

At third order, three equations are needed for the non-trivial moments $\langle\sigma_0^z\sigma_n^+\sigma_m^-\rangle$, $\langle\sigma_0^+\sigma_n^-\sigma_m^z\rangle$, and $\langle\sigma_n^z\sigma_m^+\sigma_k^-\rangle$. In addition, two further second-order moments $\langle\sigma_n^z\sigma_m^z\rangle$ and $\langle\sigma_0^z\sigma_n^z\rangle$ now couple into the dynamics (Appendix E.4).

In the following, $n$, $m$, and $k$ label distinct satellite sites.

$$\partial_t\langle\sigma_0^z\rangle = -\kappa\left(\langle\sigma_0^z\rangle + 1\right) + 4gN\,\mathrm{Im}\big[\langle\sigma_0^+\sigma_n^-\rangle\big] \tag{5.2.16}$$

$$\partial_t\langle\sigma_n^z\rangle = -\Gamma_T\langle\sigma_n^z\rangle + \Gamma_\Delta - 4g\,\mathrm{Im}\big[\langle\sigma_0^+\sigma_n^-\rangle\big] \tag{5.2.17}$$

$$\partial_t\langle\sigma_0^+\sigma_n^-\rangle = \left(i(\omega-\epsilon) - \frac{\kappa+\Gamma_T}{2}\right)\langle\sigma_0^+\sigma_n^-\rangle + \frac{ig}{2}\langle\sigma_n^z\rangle - \frac{ig}{2}\langle\sigma_0^z\rangle - ig(N-1)\langle\sigma_0^z\sigma_n^+\sigma_m^-\rangle \tag{5.2.18}$$

$$\partial_t\langle\sigma_n^+\sigma_m^-\rangle = -\Gamma_T\langle\sigma_n^+\sigma_m^-\rangle + 2g\,\mathrm{Im}\big[\langle\sigma_0^+\sigma_n^-\sigma_m^z\rangle\big] \tag{5.2.19}$$

$$\partial_t\langle\sigma_0^z\sigma_n^+\sigma_m^-\rangle = -(\kappa+\Gamma_T)\langle\sigma_0^z\sigma_n^+\sigma_m^-\rangle - \kappa\langle\sigma_n^+\sigma_m^-\rangle + 2g\,\mathrm{Im}\big[\langle\sigma_0^+\sigma_n^-\rangle\big] + 8g(N-2)\,\mathrm{Im}\big[\langle\sigma_0^+\sigma_n^-\rangle\big]\,\langle\sigma_n^+\sigma_m^-\rangle \tag{5.2.20}$$

$$\begin{aligned}\partial_t\langle\sigma_0^+\sigma_n^-\sigma_m^z\rangle = &\left(i(\omega-\epsilon) - \frac{\kappa+3\Gamma_T}{2}\right)\langle\sigma_0^+\sigma_n^-\sigma_m^z\rangle + \Gamma_\Delta\langle\sigma_0^+\sigma_n^-\rangle - ig\langle\sigma_n^+\sigma_m^-\rangle + \frac{ig}{2}\langle\sigma_n^z\sigma_m^z\rangle - \frac{ig}{2}\langle\sigma_0^z\sigma_n^z\rangle \\ &- ig(N-2)\Big(\langle\sigma_0^z\sigma_n^+\sigma_m^-\rangle\langle\sigma_n^z\rangle + \langle\sigma_0^z\rangle\langle\sigma_n^z\sigma_m^+\sigma_k^-\rangle + \langle\sigma_0^z\sigma_n^z\rangle\langle\sigma_n^+\sigma_m^-\rangle - 2\langle\sigma_0^z\rangle\langle\sigma_n^z\rangle\langle\sigma_n^+\sigma_m^-\rangle\Big)\end{aligned} \tag{5.2.21}$$

$$\begin{aligned}\partial_t\langle\sigma_n^z\sigma_m^+\sigma_k^-\rangle = &-2\Gamma_T\langle\sigma_n^z\sigma_m^+\sigma_k^-\rangle + \Gamma_\Delta\langle\sigma_n^+\sigma_m^-\rangle - 8g\,\mathrm{Im}\big[\langle\sigma_0^+\sigma_n^-\rangle\big]\,\langle\sigma_n^+\sigma_m^-\rangle \\ &+ 2g\Big(\mathrm{Im}\big[\langle\sigma_0^+\sigma_n^-\rangle\big]\,\langle\sigma_n^z\sigma_m^z\rangle - 2\,\mathrm{Im}\big[\langle\sigma_0^+\sigma_n^-\rangle\big]\,\langle\sigma_n^z\rangle^2 + 2\,\mathrm{Im}\big[\langle\sigma_0^+\sigma_n^-\sigma_m^z\rangle\big]\,\langle\sigma_n^z\rangle\Big)\end{aligned} \tag{5.2.22}$$

$$\partial_t\langle\sigma_n^z\sigma_m^z\rangle = -2\Gamma_T\langle\sigma_n^z\sigma_m^z\rangle + 2\Gamma_\Delta\langle\sigma_n^z\rangle - 8g\,\mathrm{Im}\big[\langle\sigma_0^+\sigma_n^-\sigma_m^z\rangle\big] \tag{5.2.23}$$

$$\partial_t\langle\sigma_0^z\sigma_n^z\rangle = -(\kappa+\Gamma_T)\langle\sigma_0^z\sigma_n^z\rangle - \kappa\langle\sigma_n^z\rangle + \Gamma_\Delta\langle\sigma_0^z\rangle + 4g(N-1)\,\mathrm{Im}\big[\langle\sigma_0^+\sigma_n^-\sigma_m^z\rangle\big]\,. \tag{5.2.24}$$

In writing Eqs. (5.2.20) to (5.2.22), fourth-order moments were approximated by setting the fourth-order cumulants to zero (Appendix C.3):

$$\langle\langle\sigma_0^+\sigma_n^-\sigma_m^+\sigma_k^-\rangle\rangle = 0, \quad \langle\langle\sigma_0^z\sigma_n^z\sigma_m^+\sigma_k^-\rangle\rangle = 0, \quad \langle\langle\sigma_0^+\sigma_n^-\sigma_m^z\sigma_k^z\rangle\rangle = 0, \tag{5.2.25}$$

where

$$\begin{aligned}\langle\langle\sigma_a^\alpha\sigma_b^\beta\sigma_c^\gamma\sigma_d^\delta\rangle\rangle := &\langle\sigma_a^\alpha\sigma_b^\beta\sigma_c^\gamma\sigma_d^\delta\rangle - \langle\sigma_a^\alpha\sigma_b^\beta\rangle\langle\sigma_c^\gamma\sigma_d^\delta\rangle - \langle\sigma_a^\alpha\sigma_c^\gamma\rangle\langle\sigma_b^\beta\sigma_d^\delta\rangle - \langle\sigma_a^\alpha\sigma_d^\delta\rangle\langle\sigma_b^\beta\sigma_c^\gamma\rangle \\ &-\langle\sigma_a^\alpha\rangle\langle\sigma_b^\beta\sigma_c^\gamma\sigma_d^\delta\rangle - \langle\sigma_b^\beta\rangle\langle\sigma_a^\alpha\sigma_c^\gamma\sigma_d^\delta\rangle - \langle\sigma_a^\alpha\sigma_b^\beta\sigma_d^\delta\rangle\langle\sigma_c^\gamma\rangle - \langle\sigma_a^\alpha\sigma_b^\beta\sigma_c^\gamma\rangle\langle\sigma_d^\delta\rangle \\ &+2\langle\sigma_a^\alpha\rangle\langle\sigma_b^\beta\rangle\langle\sigma_c^\gamma\sigma_d^\delta\rangle + 2\langle\sigma_a^\alpha\rangle\langle\sigma_b^\beta\sigma_c^\gamma\rangle\langle\sigma_d^\delta\rangle + 2\langle\sigma_a^\alpha\rangle\langle\sigma_b^\beta\sigma_d^\delta\rangle\langle\sigma_c^\gamma\rangle \\ &+2\langle\sigma_a^\alpha\sigma_b^\beta\rangle\langle\sigma_c^\gamma\rangle\langle\sigma_d^\delta\rangle + 2\langle\sigma_a^\alpha\sigma_c^\gamma\rangle\langle\sigma_b^\beta\rangle\langle\sigma_d^\delta\rangle + 2\langle\sigma_a^\alpha\sigma_d^\delta\rangle\langle\sigma_b^\beta\rangle\langle\sigma_c^\gamma\rangle \\ &-6\langle\sigma_a^\alpha\rangle\langle\sigma_b^\beta\rangle\langle\sigma_c^\gamma\rangle\langle\sigma_d^\delta\rangle.\end{aligned} \tag{5.2.26}$$

Note that many of these terms vanish for the model with U(1) symmetry.

#### Fourth and fifth-order numerical results

To calculate results at fourth and fifth orders, we firstly used the `QuantumCumulants.jl` [352] to generate the cumulant equations for the system. As this package does not provide a way to impose symmetries during the derivation, these included equations of motion for all moments, with symmetry-breaking terms. The number of derived equations was hence large: 54 at fourth order and 90 at fifth order[3]. The modelling toolkit `ModelingToolkit.jl` [413] and differential equation solver suite `DifferentialEquations.jl` [414] was then used to numerically integrate the system using the 5th order Tsitouras (`Tsit5()`) method to late times. This provided an approximate steady state for each set of system parameters (namely system size $N$) set out below.

As noted in the following results, the highest order solutions were susceptible to noise or instability at large $N$. These numerical issues have been observed in other recent works using high order cumulant expansions [351]. They would likely be mitigated by an implementation in which the vanishing of non-symmetry respecting moments was imposed, ideally from the outset, i.e., in the derivation of the equations themselves. On the other hand, having access to the more complicated symmetry-breaking equations allowed us to test the convergence of the cumulant expansions with symmetry breaking, including at lower orders, as is done in Section 5.4.3. For this a small amount of symmetry breaking was introduced in the initial conditions. Results from `QuantumCumulant.jl` were also used to check our symmetry-preserving equations (up to third order) that were derived by hand.

## 5.3 Results for mean-field and second-order cumulant expansions

In this section we compare the mean-field result Eq. (5.2.11) and the solution to the second-order equations Eqs. (5.2.3) to (5.2.15) to the exact steady state. We do this this under two possible choices for scaling parameters in the model as $N \to \infty$.

### 5.3.1 Fixed $g\sqrt{N}$

Figure 5.3a shows $p_0^\uparrow$ vs $1/N$ when fixing $g\sqrt{N}$. As discussed in Chapter 2, this scaling is often relevant in the context of light-matter coupling, where coupling strength $g$ is inversely proportional to the square root of mode volume: as the system becomes larger, both $N$ and mode volume grow, but $g\sqrt{N}$ remains fixed. This was the scaling used for the Tavis-Cummings model in the previous chapter, for which mean-field theory is known to be exact as $N \to \infty$ [332, 333]. Here however we see there is no agreement between exact and approximate results, each taking different $N \to \infty$ limits. This is in marked contrast not only the Tavis-Cummings but more general Dicke models, where both mean-field and second-order cumulant approximations converge to the exact steady state as $N \to \infty$ for this scaling [105]. Below we explain how the convergence of second-order cumulants to mean-field theory is precluded when $g \propto 1/\sqrt{N}$ for the central spin model.

### 5.3.2 Fixed $\kappa/N$

If instead the ratio $\kappa/N$ is kept fixed, Fig. 5.3b, mean-field and second-order cumulants have a common limit that captures the exact behaviour. Note Fig. 5.3b is plotted for parameters where non-zero $p_0^\uparrow$ is expected (a phase diagram is provided in Fig. 5.3c which we comment on shortly). This scaling may be understood to realize the limit of strong continuous measurement of the central site [415]. It has the feature[4], seen in Eqs. (5.2.11) to (5.2.13), that expectations of satellite and central-site quantities are of the same order, $O(1)$, as $N \to \infty$. In Section 5.3.3 we show that correlations are also non-vanishing and $O(1)$ as $N \to \infty$. Then the asymptotic form of Eq. (5.2.14) ($\kappa \sim N$),

$$\partial_t \langle \sigma_0^+ \sigma_n^- \rangle = N \left( -\frac{\kappa}{2N} \langle \sigma_0^+ \sigma_n^- \rangle - ig \langle \sigma_0^z \rangle \langle \sigma_n^+ \sigma_m^- \rangle \right) + O(1), \tag{5.3.1}$$

may be observed to match that predicted by mean-field theory,

$$\partial_t \left( \langle \sigma_0^+ \rangle \langle \sigma_n^- \rangle \right) = N \left( -\frac{\kappa}{2N} \langle \sigma_0^+ \rangle \langle \sigma_n^- \rangle - ig \langle \sigma_0^z \rangle \langle \sigma_n^+ \rangle \langle \sigma_m^- \rangle \right) + O(1). \tag{5.3.2}$$

[3]Whilst the fifth-order equations took only a minute to derive on a modern Desktop CPU, at sixth order (139 equations) this had increased to the best part of an hour. We did not pursue a numerical solution to the sixth-order equations.

[4]In fact, Eqs. (5.2.11) to (5.2.13) are *independent* of $N$ for this scaling, but this changes off-resonance (Appendix E.2).

Indeed:

$$\partial_t \left( \langle \sigma_0^+ \rangle \langle \sigma_n^- \rangle \right) = \left( \partial_t \langle \sigma_0^+ \rangle \right) \langle \sigma_n^- \rangle + \langle \sigma_0^+ \rangle \left( \partial_t \langle \sigma_n^- \rangle \right) \tag{5.3.3}$$

$$= \left( i\omega - \frac{\kappa}{2} \right) \langle \sigma_0^+ \rangle \langle \sigma_n^- \rangle - igN \langle \sigma_0^z \rangle \langle \sigma_n^+ \rangle \langle \sigma_n^- \rangle + \left( -i\epsilon - \frac{\Gamma_T}{2} \right) \langle \sigma_0^+ \rangle \langle \sigma_n^- \rangle + ig \langle \sigma_0^+ \rangle \langle \sigma_n^- \rangle \langle \sigma_n^z \rangle \tag{5.3.4}$$

$$= N \left( -\frac{\kappa}{2N} \langle \sigma_0^+ \rangle \langle \sigma_n^- \rangle - ig \langle \sigma_0^z \rangle \langle \sigma_n^+ \rangle \langle \sigma_m^- \rangle \right) + O(1), \tag{5.3.5}$$

having used $\kappa \sim N$ and that all expectations are order 1. The same is true for Eq. (5.2.15) and its mean-field analog, hence the second-order and mean-field equations have identical structures as $N \to \infty$ at fixed $\kappa/N$.

In contrast at fixed $g\sqrt{N}$ the correlations $\langle \sigma_n^+ \sigma_m^- \rangle$ do not remain finite as $N \to \infty$ but decay faster than $1/\sqrt{N}$. Consequently the terms $\sim g\langle \sigma_0^z \rangle, g\langle \sigma_n^z \rangle$ in Eq. (5.2.14), which are not present in mean-field theory, cannot be discounted as $N \to \infty$. This difference leads to distinct limits in Fig. 5.3a. Note equations for higher-order moments involving the central site will contain additional terms inconsistent with mean-field theory. Thus, while higher-order expansions *may* provide an improved approximation of the exact results, they will generally have distinct limits. This result also illustrates how knowledge that certain correlations vanish at large $N$ is not sufficient to determine if they become irrelevant as $N \to \infty$. Instead, the scaling with $N$ of parameters multiplying these correlations must also be taken into account.

Returning to Fig. 5.3, panel (d) shows the error in the mean-field and second-order results in the case $\kappa/N$ is fixed. This reveals that, even when cumulant expansions do capture exact behaviour as $N \to \infty$, the error in these approximations is not generally a simple function of $1/N$. Indeed, the error at second order is not monotonic with $N$ and even exceeds that of mean-field theory for $N \gtrsim 80$. The non-monotonicity is inevitable given this approach captures the exact $N \to \infty$ limit and must also be exact at $N = 2$, when all correlations are fully captured. As such, the second-order expansion provides an approximation that is only asymptotically[5] matched to the exact result at the two limits, and care must be taken in between.

Finally the phase diagram in Fig. 5.3c illustrates how the accuracy of both approximations vary with satellite pump $\Gamma_\uparrow/\Gamma_T$ at fixed $N = 50$. The discontinuous switch-on of the mean-field result as $\Gamma_\uparrow/\Gamma_T = R_c \approx 0.53$ is characteristic of a mean-field transition in the driven-dissipative Tavis-Cummings model as we studied (Chapter 4) as well as other models of lasing [348, 416]. The second-order cumulant result provides a more accurate approximation below the mean-field transition, at weak pumping, and steadily deviates from the exact result as $\Gamma_\uparrow/\Gamma_T$ is increased. Referring to Fig. 5.3b, if $N$ is increased from 50 the second-order cumulant curve will tend downwards towards the mean-field value, with the exact result clamped in between.

[5] A notion that must unfortunately be dispelled is that second-order cumulants provide a $1/N$ correction to mean-field theory: the asymptotic expansion of the second-order result about $N \to \infty$ is not well captured by a single $1/N$ term, but $1/N^2$, $1/N^3$,... have significant coefficients too. In other words, the red curve near 0 in Fig. 5.3b is not close to linear.

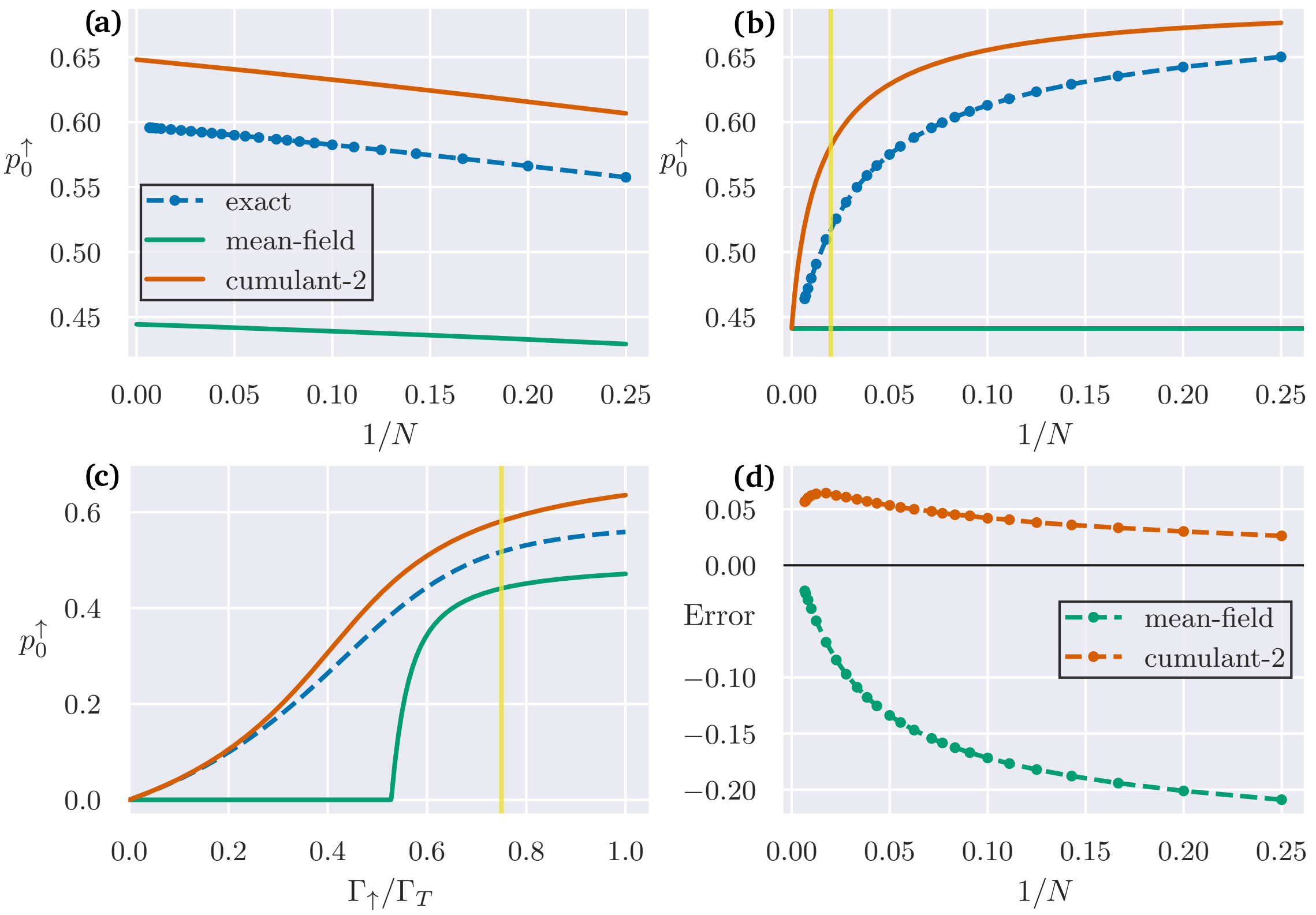

**Figure 5.3:** (a) Central-site population, $p_0^{\uparrow} = (1 + \langle \sigma_0^z \rangle)/2$, in mean-field and second-order cumulant (cumulant-2) approximations when $g\sqrt{N} = 3$ is fixed and $\kappa = 1$ in units of $\omega$. Exact data (blue dots) is included up to $N = 150$. The horizontal scale $1/N$ is such that $N \to \infty$ to the left. (b) Mean-field, second-order cumulant and exact results when $\kappa/N = 1/16$ is fixed and $g = 3/4$. The yellow vertical line at $N = 50$ indicates points equivalent to those along the corresponding line in (c). Note the mean-field limits in (a) and (b) are close but not identical: $p_0^{\uparrow} \to 4/9 \approx 0.444$ and $p_0^{\uparrow} \to 17/4 - \sqrt{2089}/12 \approx 0.441$, respectively. (c) $p_0^{\uparrow}$ vs $\Gamma_{\uparrow}/\Gamma_T$ at fixed $\kappa/N = 1/16$, $g = 3/4$, and $N = 50$. The low cost of the exact calculation at this $N$ allowed a continuous line to be plotted. (d) Error in mean-field and second-order results from (b). Other parameters used in these panels were $\epsilon = \omega = 1$, $\Gamma_T = 2$, and (except (c)) $\Gamma_{\uparrow} = 3/2$. ↑TOF

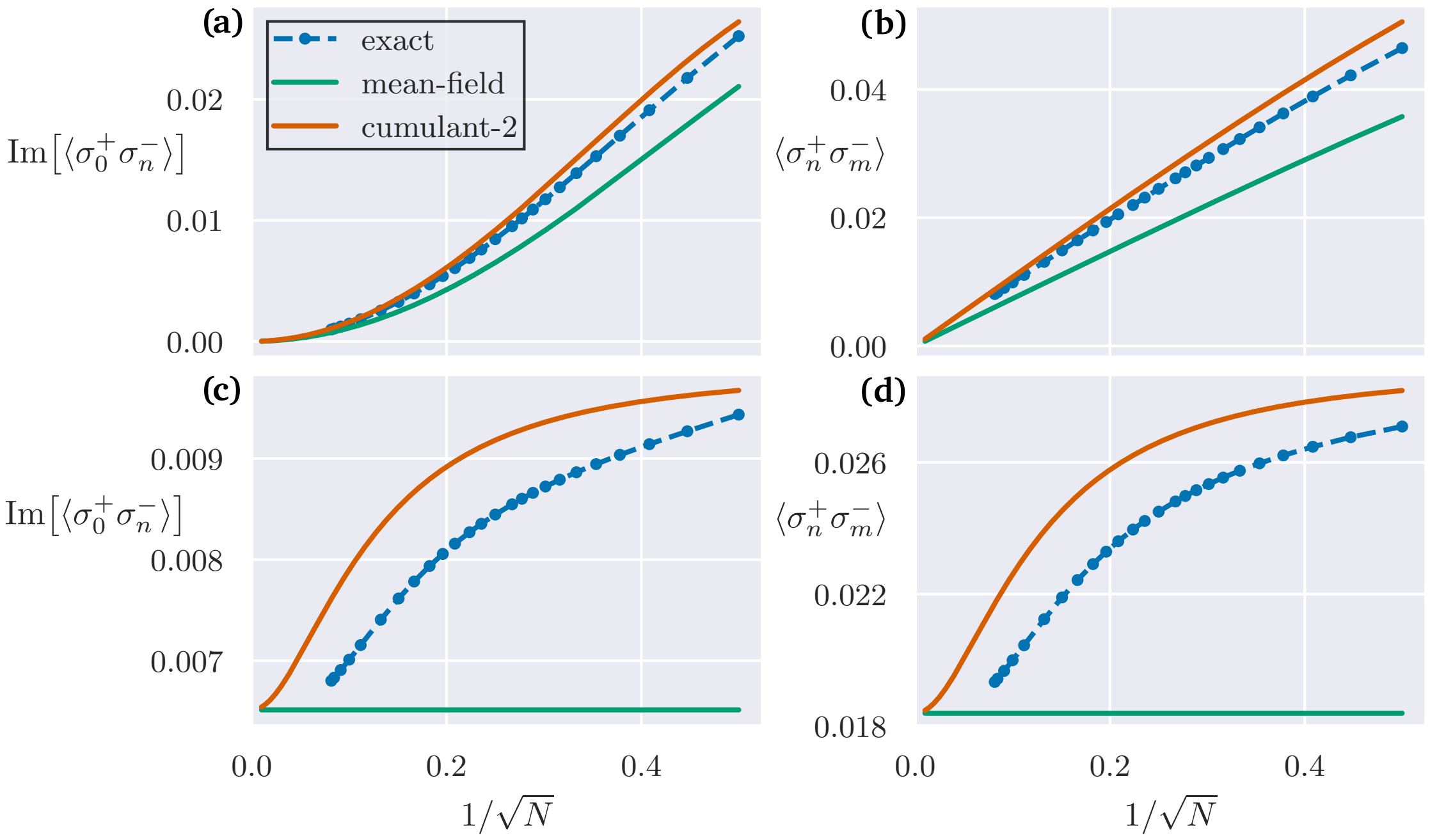

**Figure 5.4:** (a) Satellite-satellite and (b) central-satellite correlations in the steady state with $g\sqrt{N}$ fixed and parameters as in Fig. 5.3a ($g\sqrt{N} = 3$, $\kappa = 1$). Exact data (blue dots) up to $N = 150$ and second-order cumulant results for the correlations are included, as well as the mean-field approximations $\langle\sigma_n^+\sigma_m^-\rangle \approx |\langle\sigma_n^+\rangle|^2$ and $\langle\sigma_0^+\sigma_n^-\rangle \approx \langle\sigma_0^+\rangle\langle\sigma_n^-\rangle$. Note the use of $1/\sqrt{N}$ on the horizontal axis: for this scaling $\langle\sigma_n^+\sigma_m^-\rangle = o(1/\sqrt{N})$ and $\mathrm{Im}\left[\langle\sigma_0^+\sigma_n^-\rangle\right] = O(1/\sqrt{N})$ as $N \to \infty$ (the real part of $\langle\sigma_0^+\sigma_n^-\rangle$ vanishes at resonance). (c), (d) Correlations when instead $\kappa/N$ is fixed with parameters as in Fig. 5.3b ($\kappa/N = 1/16$, $g = 3/4$). In this case both pairs of correlations remain finite for all $N$. ↑TOF

### 5.3.3 Behaviour of correlations as $N \to \infty$ ↑

To support the above arguments for convergence, in Fig. 5.4 we show the behaviour of pairwise correlations as $N \to \infty$ for the central spin model.

Figures 5.4a and 5.4b include satellite-satellite $\langle\sigma_n^+\sigma_m^-\rangle$ and central-satellite $\langle\sigma_0^+\sigma_n^-\rangle$ correlations against $1/\sqrt{N}$ for the model at fixed $g\sqrt{N}$. We plot exact results up to $N = 150$ as well as the prediction of second-order cumulants and mean-field theory. Notice in particular that $\langle\sigma_n^+\sigma_m^-\rangle$ decays faster than $1/\sqrt{N}$ as $N \to \infty$ (vanishing gradient at $1/\sqrt{N} \to 0$ in Fig. 5.4a). As such, at large $N$, the terms $\sim g\langle\sigma_0^z\rangle, \sim g\langle\sigma_n^z\rangle$ present in the second-order equation Eq. (5.2.14) (but not mean-field theory) are dominant compared to the final term $\sim g\langle\sigma_0^z\rangle\langle\sigma_n^+\sigma_m^-\rangle$ occurring there.

When instead considering the correlations at fixed $\kappa/N$, shown in Figs. 5.4c and 5.4d, we see that both tend to finite limits, allowing for the reduction of the second-order cumulant equations to mean-field theory when $N \to \infty$ as argued above.

Thus we have both a case where correlations vanish as $N \to \infty$ but mean-field and second-order cumulants do not have a well-defined limit (Fig. 5.3a), and a case where they remain finite yet the two approaches have a common limit capturing the exact behaviour (Fig. 5.3b). This makes clear the fact that knowledge of the behaviour of correlations as $N \to \infty$ is not sufficient to conclude the correctness of mean-field theory or the convergence of higher-order cumulant expansions in this limit.

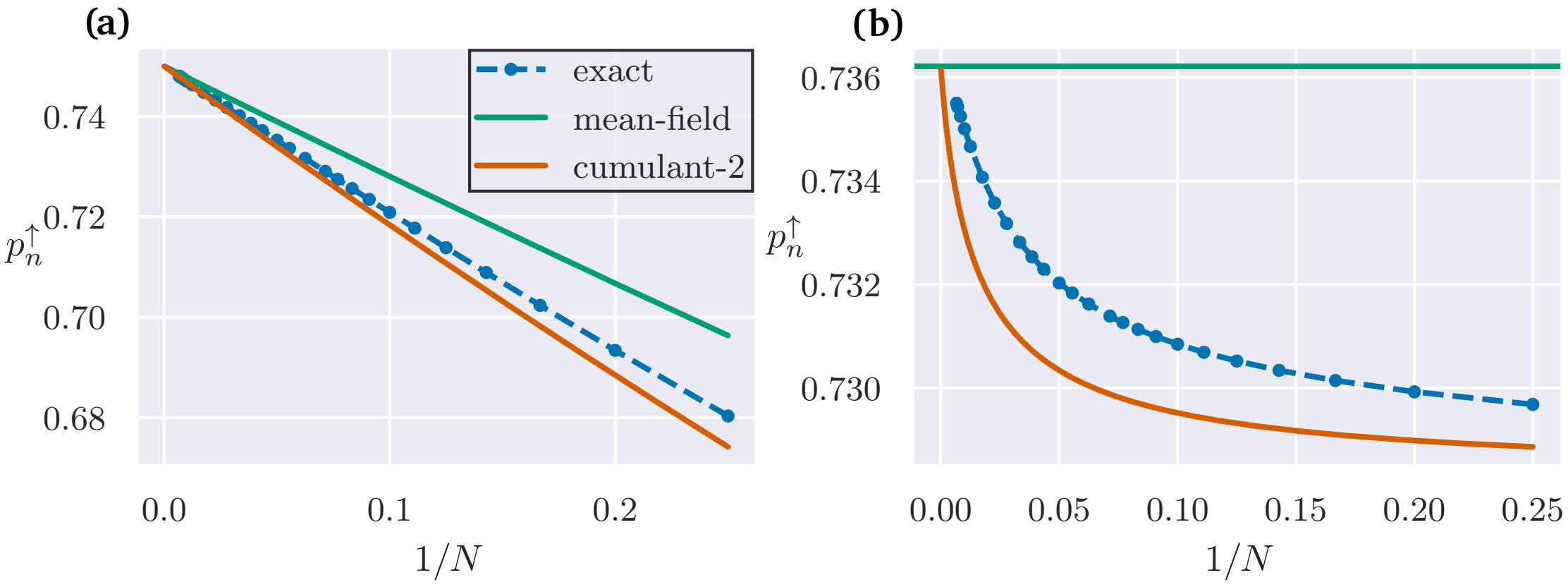

**Figure 5.5:** Satellite population $p_n^\uparrow = (1 + \langle\sigma_n^z\rangle)/2$ when (a) $g\sqrt{N} = 3$ is fixed ($\kappa = 1$) and (b) $\kappa/N = 1/16$ is fixed ($g = 3/4$), with exact data up to $N = 150$ (blue dots). Other parameters as in Fig. 5.3 ($\epsilon = \omega = 1$, $\Gamma_T = 2$, $\Gamma_\uparrow = 3/2$). As the central population $p_0^\uparrow$ is constant in the mean-field solution at fixed $\kappa/N$ (Fig. 5.3b), the satellite population is too (cf. Eq. (5.2.11)). N.B. (a), (b) have different vertical scales. ↑TOF

With knowledge of the scaling of correlations, we can re-examine the mean-field equations and notice that, when $g\sqrt{N}$ is fixed, Eqs. (5.2.4) and (5.2.6) for the satellite quantities decouple from the system as $N \to \infty$, since in these equations the coupling terms $g \to 0$ do not involve a factor of $N$. In particular, $\langle\sigma_n^z\rangle \to \Gamma_\Delta/\Gamma_T = 1/2$ in the steady state from Eq. (5.2.4). This is true for both mean-field theory and any higher-order cumulant expansion. Therefore, these approaches should predict the same behaviour for the expected satellite population $p_n^\uparrow = (1 + \langle\sigma_n^z\rangle)/2$ in the steady state, and this is exactly what is found (Fig. 5.5a). It is further seen in Fig. 5.5a that this common limit also captures the exact $N \to \infty$ behaviour of the satellite observable. We comment that in all cases where mean-field and second-order cumulant results agree at $N \to \infty$, we found the exact result to also approach the same limit (the same is certainly not true for higher-order cumulants, as we show below). Convergence to the exact result also occurs for the model with $\kappa/N$ fixed (Fig. 5.5b) where, as explained above, all of the cumulant and mean-field equations are commensurate as $N \to \infty$.

## 5.4 Results for higher-order expansions ↑

### 5.4.1 Central spin model ↑

Having established a well defined limit up to second order at fixed $\kappa/N$, we now investigate higher-order cumulant expansions for this scaling. As explained in Section 5.2.4, we use the `QuantumCumulants.jl` Julia package [352] to obtain fourth and fifth-order results in addition to the numerical solution to the third-order equations presented in that section. Surprisingly, we see in Fig. 5.6a that, whilst the fourth-order expansion provides an improved approximation on the entire range of $N$, the third-order expansion does not. Instead it converges to a limit far separated from the true result. Hence there is some $N$ beyond which the second-order (or mean-field) result provides a better approximation. Similarly the fifth-order result, despite being exact up to $N = 5$ and the best approximation at very small $N$, fails to capture the exact $N \to \infty$ limit.

To understand the dependence of convergence on order parity, the previous argument for the asymptotic reduction of the second-order equations to mean field as $N \to \infty$ can be extended to all even orders. First, note that before any factorisation is made the equations for moments involving

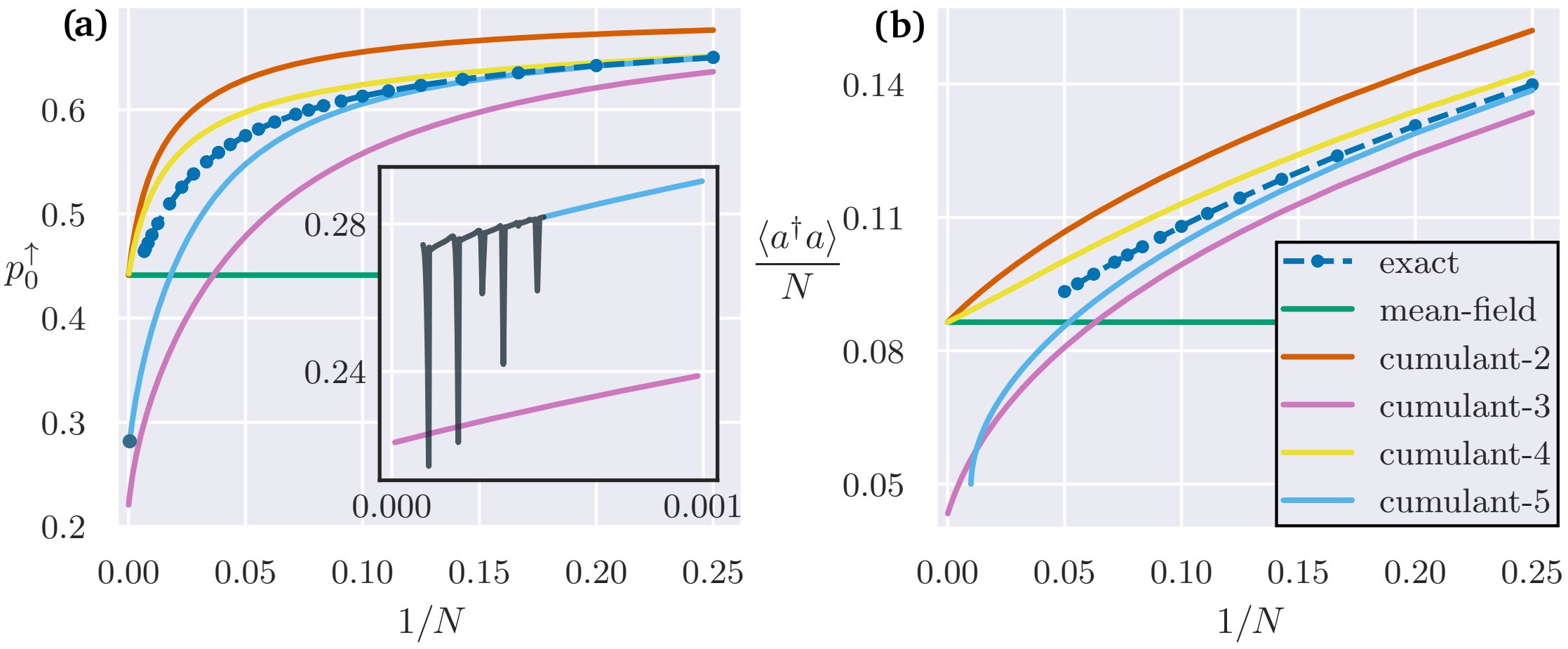

**Figure 5.6:** (a) Central-site population $p_0^\uparrow$ in mean-field and cumulant approximations up to fifth order at fixed $\kappa/N$ (parameters and exact data as in Fig. 5.3b). Results at fourth and fifth order were derived using QuantumCumulants.jl [352]. Inset: the fifth-order solution has numerical noise beyond $N \gtrsim 2,500$, but is approaching a value distinct from the third-order limit. (b) Mean-field and cumulant results for the scaled photon number $\langle a^\dagger a \rangle / N$ in the driven-dissipative Tavis-Cummings model using QuantumCumulants.jl. Exact results following a Fock-space truncation are included up to $N = 20$ ($N_{\text{phot}} = 20$ levels were sufficient to achieve convergence). Here the parameters were $g\sqrt{N} = 9/10$, $\epsilon = \omega = \kappa = 1$, $\Gamma_T = 1/2$, and $\Gamma_\uparrow = 3\Gamma_T/4$. The fifth-order solution became unstable for $N \gtrsim 100$. ↑TOF

satellite sites only match mean-field theory in structure since $H$ (Eq. (5.2.1)) is linear in these sites. When the central site is involved, this is no longer the case. However, the terms that survive as $N \to \infty$ at fixed $\kappa/N$ are those that arise from the commutator of a central operator with $\sigma_0^+\sigma_n^-$ or $\sigma_0^-\sigma_n^+$ followed by a sum $\sim N$ over the satellites. These terms have the same structure for both the cumulant equations and mean-field theory. Second, there is a key point about the coefficients associated with the cumulant expansion of a given term. As shown in Appendix C.3, by definition, the coefficients of any cumulant expansion should sum to 1. However, when some terms are eliminated because they do not respect the symmetries of the model, this statement may or may not remain true. When moments are factorised at even orders of expansion, the number of non-vanishing terms under U(1) symmetry does sum to 1[6]. As this matches the mean-field prediction for the number of terms, the asymptotic structure of even-order equations are compatible with mean-field theory.

On the other hand, closing the equations at odd orders requires factorising moments such as $\langle \sigma_0^+\sigma_n^-\sigma_m^+\sigma_k^- \ldots \rangle$ involving raising and lowering operators only. These produce a set of non-vanishing terms with coefficients that do *not* sum to 1. For example, when constructing the third-order equations setting the cumulant $\langle\langle \sigma_0^+\sigma_n^-\sigma_m^+\sigma_k^- \rangle\rangle$ to zero gives

$$\langle \sigma_0^+\sigma_n^-\sigma_m^+\sigma_k^- \rangle \approx 2\langle \sigma_0^+\sigma_n^- \rangle \langle \sigma_m^+\sigma_k^- \rangle. \tag{5.4.1}$$

It is the factor of 2 occurring here that is incongruent with mean-field theory. The number of terms produced by these type of factorisations varies with successive odd orders $(2, -3, 34, -455\ldots)$, so each can be expected to converge on its own limiting value at $N \to \infty$, as observed in Fig. 5.6a for third and fifth orders.

[6] This was checked explicitly up to 14th order.

A consequence of these observations is that symmetry-broken versions of the odd-order equations, for which no terms of the approximation for $\langle\sigma_0^+\sigma_n^-\sigma_m^+\sigma_k^-\ldots\rangle$ vanish, can produce the correct limit. In Section 5.4.3 we show this is indeed the case for our model. However, we note that at finite $N$ the exact solution never shows symmetry breaking, and that the symmetry-broken approximation is not necessarily a reliable improvement.

### 5.4.2 Tavis-Cummings model

Similar convergence behaviour between even and odd orders is observed in central boson models. Figure 5.6b includes results for the driven-dissipative Tavis-Cummings model up to fifth order of the cumulant expansion. As already noted, the Tavis-Cummings Hamiltonian is obtained from Eq. (5.2.1) by replacing the central spin with a bosonic operator. Note in the derivation of the cumulant equations the `QuantumCumulants.jl` software factorises multiple instances of the bosonic operator $a$ between moments when performing the cumulant expansion. This relies on an additional assumption of Gaussianity (see Section 3.3). An alternative approach is to consider a truncation of the photon space to $N_{\text{phot}}$ levels and work in the basis e.g. of generalised Gell-Mann matrices [417, 418]. This increases the number of equations at any order of expansion but removes the reliance on additional assumptions. We discuss this approach further in Chapter 7.

For these results $g\sqrt{N}$ is fixed since, as explained above, this scaling *does* provide matching exact and mean-field $N \to \infty$ limits for the steady state of the Tavis-Cummings model [384]. Despite rigorous proofs regarding the exactness of mean-field theory as $N \to \infty$ for this type of model (Section 5.1.2), there is still a complete failure of the odd ordered expansions at large $N$. In the following we show how this behaviour can also be corrected for by the introduction of symmetry breaking terms.

### 5.4.3 Third-order results with symmetry breaking

Finally, we provide results for third-order cumulant expansions with symmetry-breaking terms for the central spin and Tavis-Cummings models calculated using `QuantumCumulants.jl` [352]. Retaining moments such as $\langle\sigma_0^+\sigma_n^+\rangle$, in the equations of motion that would otherwise vanish under $\mathrm{U}(1)$ symmetry significantly increases the number of equations required, from $9$ (Eqs. (5.2.16) to (5.2.24)) to $29$.

As explained above, when one sets a cumulant to zero to obtain an approximation for a high-order moment, the number of terms in the approximation for that moment, accounting for their signs, is $1$. That is, provided no terms in the cumulant vanish due to symmetry considerations. Consequently, in the presence of symmetry breaking there is no longer disparity between the asymptotic form of odd-order cumulant equations and mean-field theory as $N \to \infty$. In line with this argument, Fig. 5.7a shows a common $N \to \infty$ limit for the third-order equations with symmetry breaking (dotted line) and mean-field theory. At finite $N$, there is a range ($N \leq 26$ in Fig. 5.7a) for which symmetry breaking is not present in the obtained steady state (Fig. 5.7b). In this case, the symmetry-preserving and symmetry-breaking third-order results match.

Even with symmetry breaking the third-order results cannot be relied upon to provide a better approximation than a second-order expansion. This is clearly seen in Fig. 5.7c, which shows $p_0^\uparrow$ against $\Gamma_\uparrow/\Gamma_\downarrow$ at $N = 50$. We point out the agreement of all cumulant expansions at pump strengths well below the mean-field threshold, where $p_0^\uparrow$ must vanish as $N \to \infty$. Note also the crossing of the third-order (symmetry-preserving) and mean-field curves which marks the transition to the symmetry-broken steady state. This is inevitable at large $N$, where the third-order result is below the mean-field prediction.

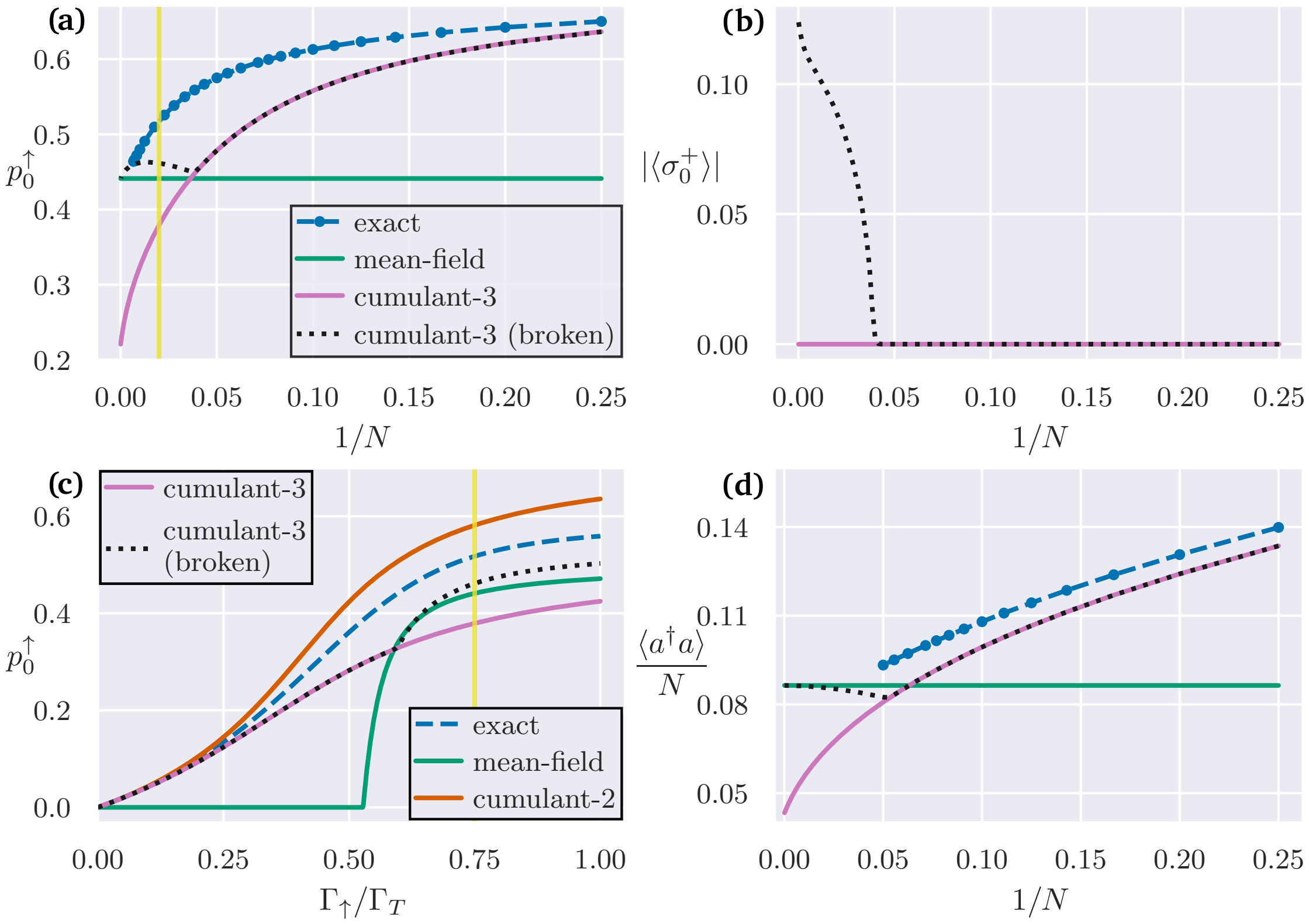

**Figure 5.7:** (a) Exact, mean-field, and third-order results for the steady state central-site population $p_0^\uparrow = (1 + \langle\sigma_0^z\rangle)/2$ of the central spin model at fixed $\kappa/N$ (parameters as in Fig. 5.3b). Third-order results retaining symmetry-breaking terms in the equations are indicated with a dotted black line. (b) At $N = 26$ symmetry breaking $\langle\sigma_0^+\rangle \neq 0$ occurs in the steady state of these equations. (c) Central population versus $\Gamma_\uparrow/\Gamma_T$ at $N = 50$ as in Fig. 5.3c, now including third-order results with and without symmetry-breaking terms. The yellow vertical line indicates data from (a). (d) Exact, mean-field, and third-order results for the scaled photon number in the Tavis-Cummings model with parameters as in Fig. 5.6b. ↑TOF

Last, in Fig. 5.7d we observe similar behaviour with the third-order equations with symmetry-breaking terms for the Tavis-Cummings model, although in this case the mean-field limit is approached from below. While here and for the central spin model we considered only the third-order expansion[7], for both models we expect similar modified convergence at higher-order odd expansions following the introduction of symmetry-breaking terms.

## 5.5 Summary and outlook

In this chapter we examined the convergence of mean-field and cumulant expansions at $N \to \infty$ as well as their accuracy at intermediate $N$. We considered the class of many-to-one models for which mean-field theory may be expected to be robust. Yet for our central spin model we demonstrated that whether mean-field theory captures the exact steady state as $N \to \infty$ depends on the scaling of parameters in the model. Further, even when mean-field theory does capture exact $N \to \infty$ behaviour, higher-order cumulant expansions may not converge to the same result. Comparison to exact results up to $N = 150$ allowed us to verify the large-$N$ behaviour and show the error of cumulant expansions is not monotonic with $N$.

[7] Implementing a consistent set of initial conditions would be laborious for the 139 symmetry-breaking equations at fifth-order. An alternative approach would be to introduce symmetry breaking in the Hamiltonian by adding, for example, a weak drive $\sim \lambda(a + a^\dagger)$.

The model considered here has been directly applied to study defect centres in diamond [27, 28, 419] and quantum dot systems [419, 420], but our reasoning may be applied quite generally to central spin models including, for example, other anisotropic or isotropic couplings or coherent drive [416, 421–426]. We have also seen that our results are relevant to models of collective light-matter coupling where cumulant expansions are an increasingly popular choice for analysing both small and large systems [40, 63, 105, 236, 237, 345, 346, 348, 349, 351, 352].

We now discuss two directions for future work.

### 5.5.1 Non-Gaussian dynamics of quantum fluctuations

In response to our work, and in conjunction with valuable discussions, F. Carollo performed a complementary study of the central spin model [389]. We provide a brief summary. First, if one considers the model with $g \propto 1/N$, then the central spin couples to the average behaviour of the satellite spins and mean-field theory applies to the dynamics. This statement may be related to the observation in our model at fixed $\kappa/N$ by a rescaling of time $t \to tN$ (also $\Gamma_\uparrow, \Gamma_\downarrow \to \Gamma_\uparrow/N, \Gamma_\downarrow/N$). Second, for the case $g \propto 1/\sqrt{N}$, that is the first scaling we considered, the central spin couples not to the average behaviour (mean-field) of the satellite spins but their *quantum fluctuations*. The result is an effective two-site model where the central spin couples to a bosonic degree of freedom associated with fluctuation operators for the satellite spin ensemble. Carollo provided exact results for the non-Gaussian dynamics that emerge in this regime.

Following this study, an immediate direction of research is to use the cumulant expansions to examine the *dynamics* of open central spin models. This would be done in light of other recent work [392] from Carollo and collaborators regarding the scope of mean-field theory to capture dynamics as $N \to \infty$. It would go beyond previous studies of central spin dynamics for closed systems [424, 427–432].

### 5.5.2 All-to-all models

A second direction is to look to apply our reasoning to models with all-to-all connectivity. These may describe, for example, lattice-spin models realised with cold atoms in optical lattices [328, 433, 434] or light interacting with atoms in two-dimensional arrays [236, 435]. The efficacy of cumulant expansions for the former class has been considered in Ref. [328]. For the latter, mean-field [436] and higher-order cumulant expansions [236, 237, 350, 437, 438] have been applied with varying levels of success: the approximations often become inaccurate in certain parameter regimes (for static quantities) or at late times (for dynamics). More serious breakdowns of the approximations also occur, with non-physical results, e.g., negative correlations functions [236]. This highlights that the cumulant equations do not necessarily obey physical constraints [236]. Generally, one returns to the fact that there is no way, a priori, of knowing whether correlations beyond a certain order will be important for a given set of parameters. There is hence great scope for work delimiting the validity of cumulant expansions for this class of model.

Ultimately, the results in this chapter results highlight the need to assess the validity of cumulant expansions in these and other applications, and prompt further exploration of how reliable higher-order expansions can be found.

# Chapter 6

# Organic polariton transport ↑TOC

See how they run.

Jonathan Keeling & Graham Turnbull

## Contents

In this chapter we develop a method to study polariton transport in organic materials. This involves a second-order cumulant expansion applied to a multimode Holstein-Tavis-Cummings (HTC) and the calculation of spatially resolved dynamics.

We start by reviewing recent observations of exciton-polariton transport in organic materials and the transport mechanism proposed to explain them. We also explain how our approach goes beyond previous descriptions and so may provide new insight into this open problem. The HTC model, which describes the interaction between many cavity models and a collection of vibrationally dressed emitters, is defined in Section 6.2. Solving this model to probe the transport mechanism requires firstly developing a new implementation of cumulant expansions that accommodates both vibrational dressing and spatial inhomogeneity of organic emitters, and secondly determining how relevant physical observables, as well as bright and dark exciton populations, can be calculated within this approach. In Section 6.3 we present results for these quantities following the expansion of a polariton cloud, and investigate how propagation velocity is affected by the vibronic coupling. Results consistent with the proposed transport mechanism are obtained, but these findings are preliminary: the scope of future work is set out in Section 6.4.

## 6.1 Anomalous transport in organic materials

### 6.1.1 Observed properties

In Chapter 2 we introduced the study of transport in organic materials with the motivation of devices realising long-range, polariton enhanced energy transport. As noted there, recent experimental studies [71, 84, 85, 116–118] directly imaging ultrafast polariton dynamics has revealed the varied nature of transport in organic materials. We focus on the 2023 Nature Materials publication by Balasubrahmaniyam et al. [116] presenting the culmination of several key experimental and theoretical observations. In this work, the propagation of Bloch surface wave polaritons (BSWPs), prepared by non-resonant excitation, was recorded in an organic semiconductor using ultrafast pump-probe microscopy.

The main experimental data from Ref. [116] are included in Fig. 6.1. First, there are regimes of both diffusive transport, where the polariton cloud expands slowly as $\sim Dt^{1/2}$, and ballistic transport, where expansion occurs $\sim v^{\text{obs}}t$ quickly at a constant velocity (Fig. 6.1a). The occurrence of these behaviours is dictated by the photonic weight $|\alpha_{\text{phot}}|^2$ of the polariton. Here $|\alpha_{\text{phot}}|^2$ is the Hopfield coefficient inferred from the measured BSWP dispersion at the emission energy and momentum; the corresponding excitonic weight is $|\alpha_x|^2 = 1 - |\alpha_{\text{phot}}|^2$. Transport was found to be purely diffusive at small $|\alpha_{\text{phot}}|^2$, ballistic at large $|\alpha_{\text{phot}}|^2$, and a cross-over between behaviours occurred around $|\alpha_{\text{phot}}|^2 \sim 60\%$.

Second (Fig. 6.1b), within each regime the rate of transport—either the coefficient of diffusion $D$ or ballistic propagation velocity $v^{\text{obs}}$—is dependent on $\left|\alpha_{\text{phot}}\right|^2$. In particular, the ballistic propagation occurs at speeds significantly below the calculated *group velocity* of the polaritons. That is, the exciton content of the polariton appeared to strongly renormalise the transport velocity.

While microscopic models with disorder scattering [71, 78, 121, 439, 440][1] can predict the emergence of diffusive behaviour of otherwise freely propagating polaritons, the fundamental mechanism for transport in organic materials which results in the strong dependence with photonic weight is still debated. A useful mechanism would account for the critical observation of sub-group-velocity propagation, but also any role the dark exciton states, which become weakly

[1]We point out the most recent work [121] showed a short-range exciton scattering potential could well explain transport dependence on $|\alpha_{\text{phot}}|^2$ in a bosonic exciton-polariton model.

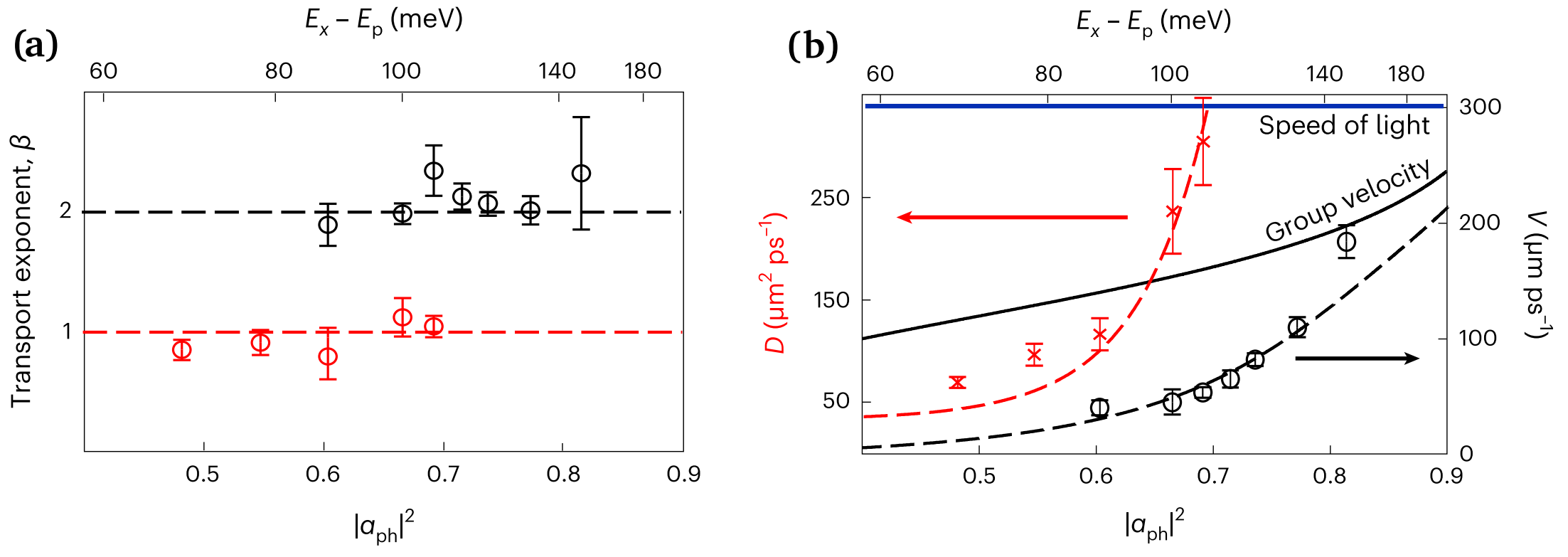

**Figure 6.1:** Organic polariton transport in a thin film organic semiconductor (TDBC) [116]. (a) Transport exponent $\beta$ as a function of photonic fraction. This exponent is defined by $\sigma_t^2 - \sigma_0^2 = Dt^\beta$ where $\sigma_t$ is the width (standard deviation) of the Gaussian-like polariton distribution at time $t$ with $\beta = 1$ and $\beta = 2$ describing diffusive and ballistic propagation, respectively. In the former case $D$ is the coefficient of diffusion. Fractions with two data points indicate the polariton cloud switched from ballistic to diffusive motion before reaching its stationary width, which occurred in $\sim 1$ ps in all cases. (b) Diffusion coefficient (red crosses) and expansion velocity (black circles) extracted from the data. The solid black line shows the theoretical group velocity of the lower polariton branch. Dashed lines indicate the prediction of the kinetic model Eqs. (6.1.3) and (6.1.4). The outlier at $|\alpha_{\text{ph}}|^2 = 0.82$ was suggested as due to non-thermal behaviour at such high photonic weight [116]. Figures reproduced with permission from Springer Nature. ↑TOF

coupled to light in real systems, have in the dynamics. Our aim in developing a simple quantum model of organic polariton transport is not only to account for the range of qualitative behaviours that have been observed but also provide insight into the underlying mechanism, hence a theoretical basis for understanding and predicting transport properties in organic materials.

### 6.1.2 Proposed mechanism ↑

The framework for understanding the discrepancy between the observed transport and group velocities $v_k^L$ of polaritons in organic materials is light-like ballistic propagation moderated by scattering and dark state interactions. For a clean system free of static or dynamic disorder, propagation occurs freely at $v_k^L$. However, in a real system motion is impeded by polariton scattering, be that induced by molecular disorder, exciton-phonon interactions or other means. Notably the large number of dark states do not remain perfectly dark but become optically active to a certain extent. Conversion between the propagating polaritons and these dark states, which are stationary (or slowly diffusive), contributes below $v_k^L$ transport [86, 119]. It is the relevance of this interconversion process in slowdown that we intend to examine with our methods.

The development of this picture followed the realisation by Groenhof and co-workers [122] that interactions with the dark states are not restricted to one way transfer to the dark populations. That is, appreciable transfer may also occur back to the bright states. Groenhof et al. investigated the role of reversible dark state population transfer in relaxation and wavepacket dynamics in subsequent studies using multi-scale molecular dynamics simulation methods [57, 86, 119, 120]. Reversible population transfer has also been a feature in other models of experimental data [71, 84, 123]. These works provide context for the kinetic model presented by Balasubrahmaniyam et al. [116], which we now summarise.

The authors [116] found the trends in diffusion coefficient and ballistic propagation velocity shown in Fig. 6.1b could be well described by a model with *thermally-activated* transfer to and from the dark states. One assumes the rate of transfer between the polariton ($\mathcal{P}$) and dark states ($\mathcal{D}$) follows an Arrhenius-type law (i.e., scattering is sufficient for equilibration to occur), so that ratio of *time* spent in dark versus polariton states at temperature $T$ satisfies

$$\frac{\tau_{\mathcal{D}}}{\tau_{\mathcal{P}}} = Ge^{-(\epsilon_{\mathcal{D}}-\epsilon_{\mathcal{P}})/T} \qquad (k_B = 1)\,, \tag{6.1.1}$$

where $\epsilon_{\mathcal{D}}$ and $\epsilon_{\mathcal{P}}$ are the exciton and polariton energies and $G$ a constant. If the dark states are stationary and the polariton state moves at the group velocity $v^{\mathrm{grp}}$, the reduction in propagation velocity is set simply by the fraction of time spent in the dark states:

$$\frac{v^{\mathrm{obs}}}{v^{\mathrm{grp}}} = \frac{\tau_{\mathcal{P}}}{\tau_{\mathcal{P}}+\tau_{\mathcal{D}}} = \frac{1}{1+\tau_{\mathcal{D}}/\tau_{\mathcal{P}}}, \tag{6.1.2}$$

Hence

$$v^{\mathrm{obs}} = \frac{v^{\mathrm{grp}}}{1+Ge^{-(\epsilon_{\mathcal{D}}-\epsilon_{\mathcal{P}})/T}}. \tag{6.1.3}$$

A temperature-dependent expression for the diffusion coefficient in terms of $v^{\mathrm{grp}}$ may be obtained from

$$D = \frac{1}{2\gamma^*}(v^{\mathrm{obs}})^2, \tag{6.1.4}$$

where the scattering rate $\gamma^*$ depends linearly on the excitonic weight $|\alpha_x|^2$ [116].

The predictions of Eqs. (6.1.3) and (6.1.4) are plotted as dashed lines in Fig. 6.1b. Despite these results, and those of other studies realising the dark state interconversion mechanism [71, 84, 86, 119, 120, 123], the necessity of this process in understanding polariton transport remains contested. On the contrary, in a study of exciton-polariton propagation in halide perovskite microcavities using momentum-resolved imaging, Xu et al. [85] have argued that renormalised polariton velocities can be explained on account of scattering by lattice phonons alone. To supplement their experimental work and support this conclusion, the authors used a trajectory based approach to simulate a model of the dynamics in which the electronic and photonic degrees of freedom were treated quantum mechanically, and the phonon environment classically. We chose to use parameters from this work as a basis for the fully quantum model we develop below.

Compared to this mixed quantum-classical model in Ref. [85], the molecular dynamics simulations of Groenhof et al. [86, 119, 120], and other models [71, 116] of polariton transport, our approach has three main features that may enable new understanding:

1. A fully quantum treatment of vibrational dynamics and hence dark state coupling
2. Bright and dark state populations may be directly accessed from second-order cumulants
3. No limitation to the first excitation subspace

The computational demands of our approach do not scale with the number of molecules, but the desired spatial resolution. The results presented in this chapter represent exploratory work. Yet they clearly show how the interplay between dark and bright states, as well as the velocity renormalisation, can be realised in a microscopic model.

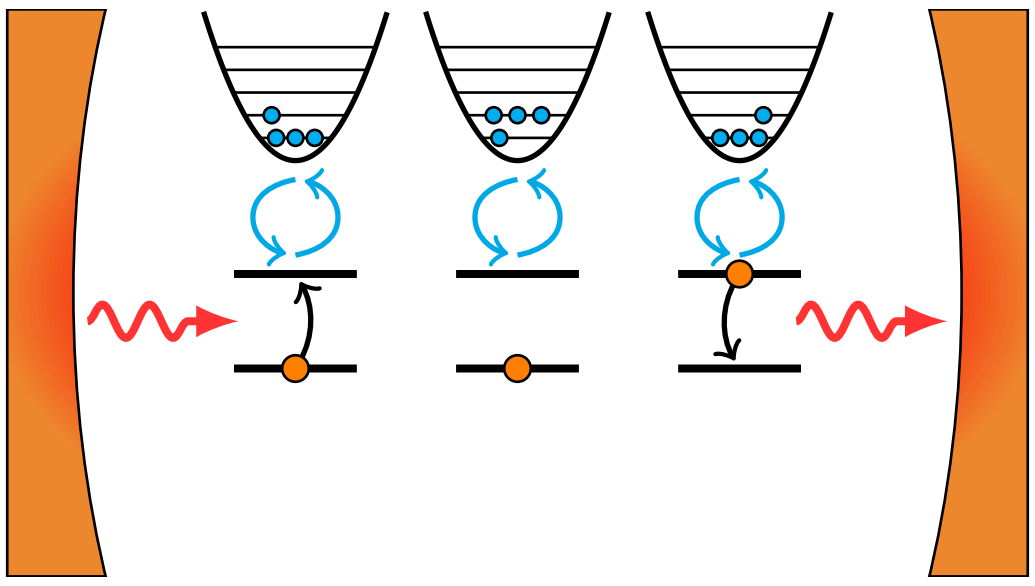

**Figure 6.2:** Holstein-Tavis-Cummings (HTC) model. A large number of emitters interact with a common cavity mode. Each emitter is also coupled to a single harmonic vibrational mode. We consider the extension of the model where there are $N_{\text{phot}}$ cavity modes. In Section 6.2.1 we introduce cavity and emitter loss as well as thermalisation of the vibrational modes. ↩TOF

## 6.2 Multimode Holstein-Tavis-Cummings model ↩

The Holstein-Tavis-Cummings (HTC) model is obtained from the Tavis-Cummings Hamiltonian by the addition of a single vibrational mode for each molecule. It hence describes many molecules with a vibrationally dressed electronic transition interacting with a common photon mode. As explained in Chapter 2, despite being the minimal inclusion of vibrational physics in the many-body Hamiltonian, it has proven effective in many problems involving organic polaritons [43]. These include the description of organic lasing and spectra discussed in Chapter 4 as well as the potential for cavity-mediated chemical reactions [9, 10, 66, 67]. It has also recently been used to model transport in 1-dimensional molecular aggregates [68, 69].

We consider an extension of the HTC model considered in previous chapters, which had a single photon mode, to the case where $N_{\text{phot}}$ photon modes are near resonant with the excitons ($N_{\text{m}}$ is now used to distinguish the number of molecules):

$$
\begin{aligned}
H = \sum_{k=1}^{N_{\text{phot}}} \omega_k a_k^\dagger a_k + \sum_{n=1}^{N_{\text{m}}} \left[ \frac{\epsilon}{2}\sigma_n^z + \sum_{k=1}^{N_{\text{phot}}} g(a_k \sigma_n^+ e^{-ikr_n} + a_k^\dagger \sigma_n^- e^{ikr_n}) \right] \\
+ \sum_{n=1}^{N_{\text{m}}} \omega_\nu \left[ b_n^\dagger b_n + \sqrt{S}(b_n^\dagger + b_n)\sigma_n^z \right].
\end{aligned}
\tag{6.2.1}
$$

Here $b_n^\dagger$ creates a vibrational excitation—a phonon—on molecule $n$, which is of frequency $\omega_\nu$ and couples to the electronic system with strength set by the Huang-Rhys parameter $S$.

The inclusion of multiple momentum states (modes $k$) is essential for modelling non-trivial spatial dynamics; note the light-matter coupling varies with molecular position $r_n$ according to the phase factors $e^{ikr_n}$. For simplicity, we consider a one-dimensional system, i.e., a line of molecules, but the generalisation to higher dimensions (vector $r_n$) is not difficult.

The model Eq. (6.2.1) was recently employed in studies of multimode organic polariton lasing [63, 441], where second-order cumulant expansions were used to capture beyond mean-field fluctuations and mode switching. These studies worked under the assumption of spatial homogeneity of the molecular population. This assumption simplifies the cumulant equations (in particular terms $\langle a_{k'}^\dagger a_k \rangle$ for $k' \neq k$ can be neglected) but cannot be made in our work: we are looking to capture time-dependent spatial *variations*.

While molecular disorder is not present in Eq. (6.2.1) ($\epsilon_n = \epsilon$ for all $n$), our method can readily accommodate this feature. We focus on discerning the effect of the vibrational coupling on the dynamics. Recall this is sufficient to make the dark excitonic states optically active (see Sections 2.2.3 and 4.5). We further note compared to the model of Ref. [85] we do not consider exciton hopping, given delocalisation due to strong coupling to light should dominate [64, 442].

### 6.2.1 Master equation

#### Vibrational thermalisation and incoherent processes

In conjunction with the unitary dynamics generated by Eq. (6.2.1) we include photon loss (rate $\kappa$) and exciton decay ($\Gamma_\downarrow$) as Markovian processes. To account for thermalisation of the vibrational modes due to coupling to the external environment we consider stimulated absorption $\gamma_\nu n_B$ and stimulated and spontaneous emission $\gamma_\nu(n_B+1)$ processes for each mode. Here $n_B = n_B(\omega_\nu/T)$ is the Bose-Einstein occupation function for the mode at temperature $T$ and $\gamma_\nu$ characterises the overall thermalisation rate. Following recent work [443] on exciton dynamics in an aggregate we use a model in which the vibrational mode is damped via its *momentum* coordinate, $p_n \sim (b_n - b_n^\dagger)$. This amounts to the thermalisation processes being brought in via the dissipators[2]

$$\gamma_\nu n_B L[b_n^\dagger] + \gamma_\nu(n_B+1)L[b_n], \tag{6.2.2}$$

and a compensating Lamb shift $H_{LS}^{(n)} = -(i\gamma_\nu/4)\left(b_n^\dagger b_n^\dagger - b_n b_n\right)$ of the system Hamiltonian. All together, the master equation is

$$\begin{aligned}\partial_t \rho = -i\left[H + H_{LS}, \rho\right] + \sum_{k=1}^{N_{\text{phot}}} \kappa L[a_k] + \sum_{n=1}^{N_{\text{m}}} \left[\Gamma_\downarrow L[\sigma_n^-] + \Gamma_z L[\sigma_n^z]\right] \\ + \sum_{n=1}^{N_{\text{m}}} \left[\gamma_\nu n_B L[b_n^\dagger] + \gamma_\nu(n_B+1)L[b_n]\right]\end{aligned} \tag{6.2.3}$$

where $H_{LS} = \sum_n H_{LS}^{(n)}$. In anticipation of a comparison to a model of pure dephasing we added a term $\Gamma_z L[\sigma_n^z]$: in this chapter we will discuss the model with vibrational coupling $S > 0$ and $\Gamma_z = 0$ as well as the model with $S = 0$ and $\Gamma_z > 0$.

In Section 6.3 we introduce values of the parameters relevant to a model in perovskite materials. An explicit form of the dispersion $\omega_k$ is also discussed. Before then we explain an efficient numerical implementation of a second-order cumulant expansion and the quantities whose dynamics we intend to study.

### 6.2.2 Second-order cumulant expansion—preliminaries

In order to obtain a set of equations that are numerically tractable two practical simplifications are made prior to the cumulant expansion. First, a coarse-graining of the molecular system to $N_k$ points, with $N_k \ll N_{\text{m}}$. Second, a truncation of the vibrational mode $b_n$ of each molecule to $N_\nu$ levels[3].

#### Coarse-graining and two-index notation

The system length $L$ is divided into a grid of $N_k = L/\Delta r$ points such that $r_n = n\Delta r$ ($n = 1, 2, \ldots, N_k$) is the position of an *ensemble* of $N_E = N_{\text{m}}/N_k$ molecules with the same on-site

[2] Note this differs from the damping of the HTC model in Section 4.3.5, which was used in previous works [62, 63].

[3] In contrast, the bosonic operators $a_k$ of the photon modes are retained in the description explicitly. As explained in Chapter 3 the accuracy of the cumulant expansion relies on Gaussian behaviour of these modes.

$$\textbf{(a)} \quad \lambda^{i_0} = \begin{pmatrix} * & \mathbf{0}_{N_\nu} \\ \mathbf{0}_{N_\nu} & * \end{pmatrix} \qquad \lambda^{i_+} = \begin{pmatrix} \mathbf{0}_{N_\nu} & * \\ \mathbf{0}_{N_\nu} & \mathbf{0}_{N_\nu} \end{pmatrix} \qquad \lambda^{i_-} = \begin{pmatrix} \mathbf{0}_{N_\nu} & \mathbf{0}_{N_\nu} \\ * & \mathbf{0}_{N_\nu} \end{pmatrix}$$

$$\textbf{(b)} \quad I_2 = \begin{pmatrix} 1 & 0 \\ 0 & 1 \end{pmatrix} \qquad \lambda^0 = \begin{pmatrix} 1 & 0 \\ 0 & -1 \end{pmatrix} \qquad \lambda^+ = \sqrt{2}\begin{pmatrix} 0 & 1 \\ 0 & 0 \end{pmatrix} \qquad \lambda^- = \sqrt{2}\begin{pmatrix} 0 & 0 \\ 1 & 0 \end{pmatrix}$$

**Figure 6.3:** (a) Block structure of the Hilbert space $\mathcal{H}_{e\nu} = \mathcal{H}_2 \otimes \mathcal{H}_{N_\nu}$ of a single molecule. Here an asterisk denotes possible non-zero values and $\boldsymbol{O}_{N_\nu}$ is the $N_\nu \times N_\nu$ zero matrix. (b) Basis when $N_\nu = 1$. ↑TOF

properties. $N_k$ defines the spatial resolution and hence number of modes in reciprocal space. For convenience we take this to equal the number of photon modes $N_{\text{phot}}$ in the model. Thus, $N_k$ should be large enough to capture both the spatial variations of the molecular populations and the number of relevant photon modes.

This coarse-graining reduces the number of cumulant equations from extensive in $N_{\text{m}}$ ($\sim N_{\text{m}}^2$ at second order) to extensive in $N_k$—essential given $N_{\text{m}}$ may be many thousands or millions in a real system. No pertinent information is lost since we are not concerned with granularity on the level of individual molecules. However, when deriving the equations care must be taken to account for coherences between molecules in the same ensemble correctly. For this it will useful to introduce a two-index notation $n\text{x}$ for molecular operators e.g. $\sigma^z_{n\text{x}}$ such that $n = 1, 2, \ldots, N_k$ indicates the ensemble (position $r_n$) and $\text{x} = 1, 2, \ldots, N_E$ the particular molecule from that ensemble.

### Finite representation of $\mathcal{H}_\nu$

A finite representation of the vibrational space $\mathcal{H}_\nu$ of each molecule is obtained by truncating $b$ to $N_\nu$ levels (note $b = b_{n\text{x}}$—we omit the subscript for operators that are the same for all molecules). Every operator in the exciton-phonon Hilbert space $\mathcal{H}_{e\nu} = \mathcal{H}_e \otimes \mathcal{H}_\nu$, for example $\sigma^z \otimes b$, is then a $2N_\nu \times 2N_\nu$ matrix. For $N_\nu$ large enough this provides an exact description of the exciton dynamics dressed by a harmonic mode. For the parameters chosen below however convergence under $N_\nu$ may not be expected, given the number of cumulants equations scales with $N_\nu^4$. In that case it may be more appropriate to say we are solving a model with $N_\nu$-level vibrational modes rather than approximating a model with harmonic vibrational degrees of freedom.

### Exciton-phonon basis

A complete basis for the $\mathcal{H}_{e\nu}$ space is provided by the identity $I_{2N_\nu}$ together with $(2N_\nu)^2 - 1$ generalised Gell-Mann (GGM) matrices $\lambda^i$ [417, 418]. GGM matrices have the appealing features of being traceless, Hermitian and of definite parity (symmetric or asymmetric). However, for the problem with U(1) symmetry, i.e., under the rotating wave approximation, it is useful to distinguish operators that add (+) or remove (−) an electronic excitation, or leave the electronic state unchanged (0). A suitable basis $\{\lambda^{i_0}, \lambda^{i_+}, \lambda^{i_-}\}$ is defined as follows:

1. $\lambda^{i_0}$ the $2N_\nu^2 - 1$ GGM matrices that leave the electronic state unchanged
2. $N_\nu^2$ matrices $\lambda^{i_+}$ that add one electronic excitation, constructed as $\lambda^{i_+} = (1/\sqrt{2})\left(\lambda^S + i\lambda^A\right)$ where $\lambda^S$ is a symmetric ($\sigma^x$-like) GGM matrix with a single unit value in the upper right and lower left quadrants, and $\lambda^A$ the corresponding antisymmetric ($\sigma^y$-like) matrix

3. $N_\nu^2$ matrices $\lambda^{i_-} = (\lambda^{i_+})^\dagger$ which destroy an electronic excitation

In each case the sign ($+$, $-$,0) of the superscript indicates the type of matrix that indexed, so $i_0$ takes $2N_\nu^2-1$ values and $i_+$, $i_-$ $N_\nu^2$ values. Unlike the $\lambda^{i_0}$ matrices, $\lambda^{i_+}$ and $\lambda^{i_-}$ are not Hermitian, but they are still traceless. Figure 6.3 illustrates the location of non-zero elements of each type of matrix. A set of structure factors $d_{i_0j_0p_0}$, $f_{i_0j_0p_0}$, $d^+_{i_0i_+j_+}$ and $f^+_{i_0i_+j_+}$ allow any product of matrices to be expressed as a linear combination of basis elements: if

$$Z_{i_0j_0p_0} = d_{i_0j_0p_0} + if_{i_0j_0p_0}, \tag{6.2.4}$$

$$Z^+_{i_0i_+j_+} = d^+_{i_0i_+j_+} + if^+_{i_0i_+j_+}, \tag{6.2.5}$$

and

$$Z^-_{i_0i_+j_+} = \overline{d}^+_{i_0i_+j_+} - i\overline{f}^+_{i_0i_+j_+} \tag{6.2.6}$$

then

$$\lambda^{i_0}\lambda^{j_0} = Z_{i_0j_0p_0}\lambda^{p_0} + \frac{1}{N_\nu}\delta_{i_0j_0}I_{2N_\nu} \tag{6.2.7}$$

$$\lambda^{i_0}\lambda^{i_+} = Z^+_{i_0i_+j_+}\lambda^{j_+} \tag{6.2.8}$$

$$\lambda^{i_0}\lambda^{i_-} = Z^-_{i_0i_+j_+}\lambda^{j_-} \tag{6.2.9}$$

$$\lambda^{i_+}\lambda^{j_-} = Z^+_{i_0i_+j_+}\lambda^{i_0} + \frac{1}{N_\nu}\delta_{i_+j_-}I_{2N_\nu}. \tag{6.2.10}$$

Here $\delta_{ij} = 1$ if $i = j$ and is zero otherwise (the one-one correspondence between $+$ and $-$ indices means $\delta_{i_+j_-}$ is unambiguous). Expressions for the structure factors as well as further properties are derived in Appendix F.1. We note the orthonormalisation condition

$$\mathrm{Tr}\big(\lambda^{i_\alpha}(\lambda^{j_\beta})^\dagger\big) = 2\delta_{i_\alpha j_\beta}\delta_{\alpha\beta}, \tag{6.2.11}$$

i.e., both the index value and sign must match. For the Hermitian $\lambda^{i_0}$, Eq. (6.2.11) is the usual trace orthogonality of GGM matrices.

### 6.2.3 Second-order cumulant expansion—derivation

We now explain the key steps in deriving a set of cumulant equations for the HTC model. Complete working is provided in Appendix F.2.

As with the Tavis-Cummings and central spin models in the previous chapters, the HTC model Eq. (6.2.1) has U(1) symmetry—the interaction is under the rotating wave approximation. Since the initial state we will consider (Section 6.3.4) will also have this symmetry, we only need to derive a symmetry preserving set of cumulant equations.

#### Formulation in new basis

Given the molecular basis $\{\lambda^{i_0}_{n\text{x}}, \lambda^{i_+}_{n\text{x}}, \lambda^{i_-}_{n\text{x}}\}$ and double-index scheme $n\text{x}$ described above ($n = 1, \ldots, N_k$, $\text{x} = 1, \ldots, N_E$), the first step is to rewrite the Hamiltonian Eq. (6.2.1) and master equation Eq. (6.2.3) in terms of these matrices:

$$H = \sum_k \omega_k a_k^\dagger a_k + \sum_{n\text{x}} \left[ A_{i_0}\lambda^{i_0}_{n\text{x}} + \frac{1}{\sqrt{N_\text{m}}}\sum_k \Big( B_{i_+}e^{-ikr_n}a_k\lambda^{i_+}_{n\text{x}} + \text{H.c.}\Big)\right], \tag{6.2.12}$$

$$\partial_t\rho = -i\,[H,\rho] + \sum_k \kappa L[a_k] + \sum_{n\text{x},\mu_0} L[\gamma^{\mu_0}_{i_0}\lambda^{i_0}_{n\text{x}}] + \sum_{n\text{x}} L[\gamma^+_{i_+}\lambda^{i_+}_{n\text{x}}] + \sum_{n\text{x}} L[\gamma^-_{i_-}\lambda^{i_-}_{n\text{x}}], \tag{6.2.13}$$

where

$$
\begin{aligned}
A_{i_0} &= \frac{1}{2}\,\mathrm{Tr}\left[\mathcal{A}\lambda^{i_0}\right], \quad \mathcal{A} = \frac{\epsilon}{2}\sigma^z \otimes I_{N_\nu} + \omega_\nu \left(I_2 \otimes b^\dagger b + \sqrt{S}\sigma^z \otimes \left(b^\dagger + b\right)\right) \\
&\qquad\qquad\qquad\qquad - \frac{i\gamma_\nu}{4} I_2 \otimes \left(b^\dagger b^\dagger - bb\right), \\
B_{i_+} &= \frac{1}{2}\,\mathrm{Tr}\left[\mathcal{B}\lambda^{i_-}\right], \quad \mathcal{B} = g\sqrt{N_\mathrm{m}}\sigma^+ \otimes I_{N_\nu}, \\
\gamma^{\mu_0}_{i_0} &= \frac{1}{2}\,\mathrm{Tr}\left[\mathcal{C}^{\mu_0}\lambda^{i_0}\right], \quad \mathcal{C}^1 = \sqrt{\Gamma_z}\sigma^z \otimes I_{N_\nu},\ \mathcal{C}^2 = \sqrt{\gamma_\uparrow} I_2 \otimes b^\dagger,\ \mathcal{C}^3 = \sqrt{\gamma_\downarrow} I_2 \otimes b, \\
\gamma^{\pm}_{i_\pm} &= \frac{1}{2}\,\mathrm{Tr}\left[\mathcal{D}^{\pm}\lambda^{i_\mp}\right], \quad \mathcal{D}^+ = \sqrt{\Gamma_\uparrow}\sigma^+ \otimes I_{N_\nu},\ \mathcal{D}^- = \sqrt{\Gamma_\downarrow}\sigma^- \otimes I_{N_\nu}, \\
\gamma_\uparrow &= \gamma_\nu n_B(T), \quad \gamma_\downarrow = \gamma_\nu\left(n_B(T)+1\right), \quad n_B(T) = \frac{1}{e^{\omega_\nu/T}-1} \quad (\hbar = k_B = 1).
\end{aligned}
\tag{6.2.14}
$$

As these coefficients are the same for all molecules, the molecular indices were dropped. Except for $A_{i_0}$ and $\gamma^{\mu_0}_{i_0}$, they are real. A few further comments on notation: repeated exciton-phonon indices ($i_0, j_0, \ldots, i_\pm, j_\pm, \ldots$) imply summation, but repeated mode ($k, p, \ldots$) or molecular indices ($nx$, $my$, $\ldots$) do not. An overline e.g. $\overline{f}^{+}_{i_0 i_+ p_+}$ continues to denote a complex conjugate. Note a factor of $\sqrt{N_\mathrm{m}}$ was ascribed to the $B_{i_+}$ coefficient so that it remains constant if different numbers $N_\mathrm{m}$ of molecules are considered at fixed $g\sqrt{N_\mathrm{m}}$.

Next we write down the equations of motion for moments involving the photon and molecular operators that are non-vanishing for the model with U(1) symmetry. At second order these are $\langle a^\dagger_{k'} a_k\rangle$, $\langle\lambda^{i_0}_{nx}\rangle$, $\langle a_k \lambda^{i_+}_{nx}\rangle$ and $\langle\lambda^{i_+}_{nx}\lambda^{i_-}_{my}\rangle$; $\langle\lambda^{i_0}_{nx}\lambda^{j_0}_{my}\rangle$ may also be non-zero but neither provides an observable of interest nor is required for the equations of other four. In fact, anticipating the final form of the equations we rescale the off-diagonal molecular matrices by a factor of $\sqrt{N_\mathrm{m}}$, $\hat\lambda^{i_\pm}_{nx} = \sqrt{N_\mathrm{m}}\lambda^{i_\pm}_{nx}$, indicated by a caret. This rescaling is made for convenience of the numerical implementation only—to calculate observables we will rescale back. The first three equations are (Appendix F.2)

$$
\begin{aligned}
\partial_t\langle a^\dagger_{k'} a_k\rangle &= \left[i(\omega_{k'}-\omega_k) - \kappa\right]\langle a^\dagger_{k'} a_k\rangle \\
&\quad + \frac{1}{N_\mathrm{m}}\sum_{nx}\left(iB_{i_+}e^{-ik' r_n}\langle a_k \hat\lambda^{i_+}_{nx}\rangle - iB_{i_-}e^{ikr_n}\langle a^\dagger_{k'}\hat\lambda^{i_-}_{nx}\rangle\right),
\end{aligned}
\tag{6.2.15}
$$

$$
\partial_t\langle\lambda^{i_0}_{nx}\rangle = \xi_{i_0 j_0}\langle\lambda^{j_0}_{nx}\rangle + \phi_{i_0} + \frac{1}{N_\mathrm{m}}\sum_k\left(2B_{i_+}f^{+}_{i_0 i_+ j_+}e^{-ikr_n}\langle a_k\hat\lambda^{j_+}_{nx}\rangle + \text{c.c.}\right),
\tag{6.2.16}
$$

$$
\begin{aligned}
\partial_t\langle a_k\hat\lambda^{i_+}_{nx}\rangle &= \left[\xi^{+}_{i_+ j_+} - (i\omega_k + \kappa/2)\delta_{i_+ j_+}\right]\langle a_k\hat\lambda^{j_+}_{nx}\rangle - \frac{1}{N_\mathrm{m}}\sum_{my} iB_{j_-}e^{ikr_m}\langle\hat\lambda^{i_+}_{nx}\hat\lambda^{j_-}_{my}\rangle \\
&\quad + 2B_{j_+}f^{+}_{i_0 i_+ j_+}\langle\lambda^{i_0}_{nx}\rangle\sum_{k'} e^{ik' r_n}\langle a^\dagger_{k'} a_k\rangle.
\end{aligned}
\tag{6.2.17}
$$

For the molecular coherences, when $nx \neq my$,

$$
\begin{aligned}
\partial_t\langle\hat\lambda^{i_+}_{nx}\hat\lambda^{j_-}_{my}\rangle &= \xi^{+}_{i_+ p_+}\langle\hat\lambda^{p_+}_{nx}\hat\lambda^{j_-}_{my}\rangle + \xi^{-}_{j_- p_-}\langle\hat\lambda^{i_+}_{nx}\hat\lambda^{p_-}_{my}\rangle \\
&\quad + \sum_k\left(2B_{p_+}\overline{f}^{+}_{i_0 j_+ p_+}e^{-ikr_m}\langle a_k\hat\lambda^{i_+}_{nx}\rangle\langle\lambda^{i_0}_{my}\rangle + 2B_{p_+}f^{+}_{i_0 i_+ p_+}\overline{e^{-ikr_n}\langle a_k\hat\lambda^{j_+}_{my}\rangle}\langle\lambda^{i_0}_{nx}\rangle\right),
\end{aligned}
\tag{6.2.18}
$$

whilst using Eq. (6.2.10)

$$
\langle\hat\lambda^{i_+}_{nx}\hat\lambda^{j_-}_{nx}\rangle = N_\mathrm{m}\left(Z^{+}_{i_0 i_+ j_+}\langle\lambda^{i_0}_{nx}\rangle + \frac{1}{N_\nu}\delta_{i_+ j_-}\right).
\tag{6.2.19}
$$

To derive the above equations we split up third-order moments e.g. $\langle a_k \hat{\lambda}^{i_+}_{nx} \lambda^{i_0}_{my} \rangle \approx \langle a_k \hat{\lambda}^{i_+}_{nx} \rangle \langle \lambda^{i_0}_{my} \rangle$ ($my \neq nx$) by setting the corresponding third cumulants to zero, discarding terms that vanish on account of the U(1) symmetry, i.e., do not conserve the total number of electron-*photon* excitations. The constant tensors $\xi$ and $\phi$ are

$$\begin{aligned}\xi_{i_0 j_0} &= 2 f_{i_0 p_0 j_0} A_{p_0} + 2 f_{i_0 r_0 q_0} \sum_{\mu_0} \mathrm{Im}\left[\overline{\gamma}^{\mu_0}_{r_0} \gamma^{\mu_0}_{p_0} Z_{q_0 p_0 j_0}\right] \\ &+ 2\gamma^+_{i_+} \gamma^+_{j_+} \mathrm{Im}\left[\overline{f}^+_{i_0 i_+ p_+} Z^-_{j_0 p_+ j_+}\right] + 2\gamma^-_{i_+} \gamma^-_{j_+} \mathrm{Im}\left[f^+_{i_0 i_+ p_+} Z^+_{j_0 p_+ j_+}\right]\end{aligned} \tag{6.2.20}$$

$$\phi_{i_0} = \frac{2 f_{i_0 j_0 q_0}}{N_\nu} \sum_{\mu_0} \mathrm{Im}\left[\overline{\gamma}^{\mu_0}_{j_0} \gamma^{\mu_0}_{q_0}\right] + \frac{2}{N_\nu} \mathrm{Im}\left[\gamma^+_{i_+} \gamma^+_{j_+} \overline{f}^+_{i_0 i_+ j_+} + \gamma^-_{i_+} \gamma^-_{j_+} f^+_{i_0 i_+ j_+}\right] \tag{6.2.21}$$

$$\begin{aligned}\xi^+_{i_+ j_+} &= -2 f^+_{i_0 i_+ j_+} A_{i_0} + i f^+_{i_0 i_+ p_+} \sum_{\mu_0} \left(\overline{\gamma}^{\mu_0}_{i_0} \gamma^{\mu_0}_{j_0} \overline{Z}^-_{j_0 p_+ j_+} - \overline{\gamma}^{\mu_0}_{j_0} \gamma^{\mu_0}_{i_0} Z^+_{j_0 p_+ j_+}\right) \\ &+ i f^+_{i_0 i_+ p_+} \left(\gamma^-_{q_+} \gamma^-_{p_+} \overline{Z}^-_{i_0 q_+ j_+} - \gamma^+_{p_+} \gamma^+_{q_+} Z^+_{i_0 q_+ j_+}\right)\end{aligned} \tag{6.2.22}$$

$$\begin{aligned}\xi^-_{i_+ j_+} &= -2 \overline{f}^+_{i_0 i_+ j_+} A_{i_0} + i \overline{f}^+_{i_0 i_+ p_+} \sum_{\mu_0} \left(\overline{\gamma}^{\mu_0}_{i_0} \gamma^{\mu_0}_{j_0} \overline{Z}^+_{j_0 p_+ j_+} - \overline{\gamma}^{\mu_0}_{j_0} \gamma^{\mu_0}_{i_0} Z^-_{j_0 p_+ j_+}\right) \\ &+ i \overline{f}^+_{i_0 i_+ p_+} \left(\gamma^+_{q_+} \gamma^+_{p_+} \overline{Z}^+_{i_0 q_+ j_+} - \gamma^-_{p_+} \gamma^-_{q_+} Z^-_{i_0 q_+ j_+}\right).\end{aligned} \tag{6.2.23}$$

#### Performing intra-ensemble sums

So far we have considered x, y to vary over all $N_E$ molecules of a given ensemble meaning Eqs. (6.2.15) to (6.2.18) form a system of equations extensive in $N_\mathrm{m}$ squared. The entire point of the grouping of molecules into ensembles however was that only a smaller set, $\sim N_k^2$, is required when molecular variations are significant over $\Delta r = L/N_k$, i.e., between ensembles, only. In other words, x or y should denote a single molecule with on-site properties representative of any molecule from the same ensemble. Sums over these indices should then be replaced by single terms appropriately weighted by $N_E$.

For Eq. (6.2.15), realising this is straightforward since only single-site molecular expectations appear under the sum:

$$\begin{aligned}&\frac{1}{N_\mathrm{m}} \sum_{nx} \left(i B_{i_+} e^{-ik' r_n} \langle a_k \hat{\lambda}^{i_+}_{nx} \rangle - i B_{i_-} e^{ikr_n} \langle a^\dagger_{k'} \hat{\lambda}^{i_-}_{nx} \rangle\right) \\ \to &\frac{1}{N_k} \sum_{n} \left(i B_{i_+} e^{-ik' r_n} \langle a_k \hat{\lambda}^{i_+}_{nx} \rangle - i B_{i_-} e^{ikr_n} \langle a^\dagger_{k'} \hat{\lambda}^{i_-}_{nx} \rangle\right) \qquad \left(\frac{N_E}{N_\mathrm{m}} = \frac{1}{N_k}\right).\end{aligned} \tag{6.2.24}$$

On the other hand, the sum over $m$, y in Eq. (6.2.17) involves molecular coherences, so care should be taken to account for the possibility $nx = my$, which must be handled separately. Recalling there are $N_E$ molecules in an ensemble (and so one $\mathrm{y}$ such that $my = n\mathrm{x}$ when $m = n$),

$$\begin{aligned}
&\frac{1}{N_\mathrm{m}} \sum_{my} iB_{j_-} e^{ikr_m} \langle \hat{\lambda}^{i_+}_{n\mathrm{x}} \hat{\lambda}^{j_-}_{my} \rangle \\
&\to \frac{N_E}{N_\mathrm{m}} \sum_{m \neq n} iB_{j_-} e^{ikr_m} \langle \hat{\lambda}^{i_+}_{n\mathrm{x}} \hat{\lambda}^{j_-}_{my} \rangle + \frac{1}{N_\mathrm{m}} (N_E - 1)\, iB_{j_-} e^{ikr_n} \overbrace{\langle \hat{\lambda}^{i_+}_{n\mathrm{x}} \hat{\lambda}^{j_-}_{n\mathrm{y}} \rangle}^{\mathrm{x}\neq\mathrm{y}} \\
&\qquad + \frac{1}{N_\mathrm{m}} iB_{j_-} e^{ikr_n} \langle \hat{\lambda}^{i_+}_{n\mathrm{x}} \hat{\lambda}^{j_-}_{n\mathrm{x}} \rangle
\end{aligned} \tag{6.2.25}$$

$$\begin{aligned}
&= \frac{1}{N_k} \sum_{m} iB_{j_-} e^{ikr_m} \langle \hat{\lambda}^{i_+}_{n\mathrm{x}} \hat{\lambda}^{j_-}_{my} \rangle - \frac{1}{N_\mathrm{m}} iB_{j_-} e^{ikr_n} \langle \hat{\lambda}^{i_+}_{n\mathrm{x}} \hat{\lambda}^{j_-}_{n\mathrm{y}} \rangle \\
&\qquad + iB_{j_+} e^{ikr_n} \left( Z^{+}_{i_0 i_+ j_+} \langle \lambda^{i_0}_{n\mathrm{x}} \rangle + \frac{1}{N_\nu} \delta_{i_+ j_-} \right).
\end{aligned} \tag{6.2.26}$$

In Eq. (6.2.26) and all following equations it should be implicit in writing the expectation of a product of $\lambda$-matrices that the molecules are distinct (otherwise the structure factors are used to remove the product immediately). Further, we drop the $\mathrm{x}$, $\mathrm{y}$ indices from our notation with the caveat that one must remember $\langle \hat{\lambda}^{i_+}_{n} \hat{\lambda}^{j_-}_{n} \rangle$ corresponds to distinct molecules, albeit from the same ensemble, as captured by Eq. (6.2.18).

The system Eqs. (6.2.15) to (6.2.18) reduces to a set of $\sim N_k^2 N_\nu^4$ equations[4]:

$$\begin{aligned}
\partial_t \langle a^\dagger_{k'} a_k \rangle &= \Big[ i(\omega_{k'} - \omega_k) - \kappa \Big] \langle a^\dagger_{k'} a_k \rangle \\
&\quad + \frac{1}{N_k} \sum_n \left( iB_{i_+} e^{-ik' r_n} \langle a_k \hat{\lambda}^{i_+}_n \rangle - iB_{i_-} e^{ikr_n} \langle a^\dagger_{k'} \hat{\lambda}^{i_-}_n \rangle \right),
\end{aligned} \tag{6.2.27}$$

$$\partial_t \langle \lambda^{i_0}_n \rangle = \xi_{i_0 j_0} \langle \lambda^{j_0}_n \rangle + \phi_{i_0} + \frac{1}{N_\mathrm{m}} \sum_k \left( 2B_{i_+} f^{+}_{i_0 i_+ j_+} e^{-ikr_n} \langle a_k \hat{\lambda}^{j_+}_n \rangle + \text{c.c.} \right), \tag{6.2.28}$$

$$\begin{aligned}
\partial_t \langle a_k \hat{\lambda}^{i_+}_n \rangle &= \left[ \xi^{+}_{i_+ j_+} - (i\omega_k + \kappa/2) \delta_{i_+ j_+} \right] \langle a_k \hat{\lambda}^{j_+}_n \rangle - \frac{1}{N_k} \sum_m iB_{j_-} e^{ikr_m} \langle \hat{\lambda}^{i_+}_n \hat{\lambda}^{j_-}_m \rangle \\
&\quad + \frac{1}{N_\mathrm{m}} iB_{j_-} e^{ikr_n} \langle \hat{\lambda}^{i_+}_n \hat{\lambda}^{j_-}_n \rangle - iB_{j_+} e^{ikr_n} \left( Z^{+}_{i_0 i_+ j_+} \langle \lambda^{i_0}_n \rangle + \frac{1}{N_\nu} \delta_{i_+ j_+} \right) \\
&\quad + 2B_{j_+} f^{+}_{i_0 i_+ j_+} \langle \lambda^{i_0}_n \rangle \sum_{k'} e^{ik' r_n} \langle a^\dagger_{k'} a_k \rangle
\end{aligned} \tag{6.2.29}$$

$$\begin{aligned}
\partial_t \langle \hat{\lambda}^{i_+}_n \hat{\lambda}^{j_-}_m \rangle &= \xi^{+}_{i_+ p_+} \langle \hat{\lambda}^{p_+}_n \hat{\lambda}^{j_-}_m \rangle + \xi^{-}_{j_- p_-} \langle \hat{\lambda}^{i_+}_n \hat{\lambda}^{p_-}_m \rangle \\
&\quad + \sum_k \left( 2B_{p_+} \overline{f}^{+}_{i_0 j_+ p_+} e^{-ikr_m} \langle a_k \hat{\lambda}^{i_+}_n \rangle \langle \lambda^{i_0}_m \rangle + 2B_{p_+} f^{+}_{i_0 i_+ p_+} \overline{e^{-ikr_n} \langle a_k \hat{\lambda}^{j_+}_m \rangle} \langle \lambda^{i_0}_n \rangle \right).
\end{aligned} \tag{6.2.30}$$

[4] Eqs. (6.2.27) to (6.2.30) describe $N_k^2$, $N_k(2N_\nu^2 - 1)$, $N_k^2 N_\nu^2$ and $N_k^2 N_\nu^4$ equations, respectively ($k$ has $N_k$ values and $i_0$ and $i_+$ take $2N_\nu^2 - 1$ and $N_\nu^2$ values).

#### Transform approach

A main computational overhead when integrating Eqs. (6.2.27) to (6.2.30) is performing the summations. To bypass this, note each sum is just a discrete Fourier transform (DFT):

$$c_{kk'} = \frac{1}{N_k}\sum_n e^{-ik'r_n} iB_{i_+}\langle a_k \hat{\lambda}_n^{i_+}\rangle \qquad \overline{c}^T_{kk'} = \frac{1}{N_k}\sum_n e^{ikr_n}\left(-iB_{i_-}\langle a^\dagger_{k'}\hat{\lambda}_n^{i_-}\rangle\right) \tag{6.2.31}$$

$$d^{j_+}_{mn} = \sum_k e^{-ikr_m}\langle a_k \hat{\lambda}_n^{i_+}\rangle \qquad \text{(Eq. (6.2.28) has } d^{j_+}_{nn}, \overline{d}^{j_+}_{nn}) \tag{6.2.32}$$

$$\beta^{i_+}_{nk} = \frac{1}{N_k}\sum_m e^{ikr_m}\left(-iB_{j_-}\langle \hat{\lambda}_n^{i_+}\hat{\lambda}_m^{j_-}\rangle\right) \qquad \alpha_{nk} = \sum_{k'} e^{ik'r_n}\langle a^\dagger_{k'} a_k\rangle \tag{6.2.33}$$

$$d^{i_+}_{mn} = \sum_k e^{-ikr_m}\langle a_k \hat{\lambda}_n^{i_+}\rangle \qquad \overline{d}^{j_+}_{nm} = \sum_k \overline{e^{-ikr_n}\langle a_k \hat{\lambda}_m^{j_+}\rangle}. \tag{6.2.34}$$

Thanks to the Fast Fourier Transform algorithm, it is far quicker ($O(N_k \ln N_k)$ vs. $O\left(N_k^2\right)$) to perform several DFTs at the start and of each timestep then to calculate the sums explicitly. This is a key part in the implementation being an efficient one. Note we included in the definition of $c_{kk'}$ and $\beta^{i_+}_{nk}$ the $B_{i_+}$ tensor: performing the contraction over the $i_+$ (or $j_-$) index first reduces the dimension of the array to be transformed.

### 6.2.4 Second-order cumulant expansion—equations

The fast and final form of the equations are

$$\partial_t\langle a^\dagger_{k'} a_k\rangle = \Big[i(\omega_{k'}-\omega_k)-\kappa\Big]\langle a^\dagger_{k'} a_k\rangle + c_{kk'} + \overline{c}^T_{kk'}, \tag{6.2.35}$$

$$\partial_t\langle \lambda_n^{i_0}\rangle = \xi_{i_0 j_0}\langle \lambda_n^{j_0}\rangle + \phi_{i_0} + \frac{1}{N_\mathrm{m}} 4B_{i_+} f^+_{i_0 i_+ j_+}\,\mathrm{Re}[d^{j_+}_{nn}], \tag{6.2.36}$$

$$\begin{aligned}\partial_t\langle a_k \hat{\lambda}_n^{i_+}\rangle = {} & [\xi^+_{i_+ j_+} - (i\omega_k + \kappa/2)\delta_{i_+ j_+}]\langle a_k \hat{\lambda}_n^{j_+}\rangle + \beta^{i_+}_{nk} \\ & + \frac{1}{N_\mathrm{m}} iB_{j_-} e^{ikr_n}\langle \hat{\lambda}_n^{i_+}\hat{\lambda}_n^{j_-}\rangle - iB_{j_+} e^{ikr_n}\left(Z^+_{i_0 i_+ j_+}\langle \lambda_n^{i_0}\rangle + \frac{1}{N_\nu}\delta_{i_+ j_+}\right) \\ & + 2B_{j_+} f^+_{i_0 i_+ j_+}\langle \lambda_n^{i_0}\rangle \alpha_{nk},\end{aligned} \tag{6.2.37}$$

$$\begin{aligned}\partial_t\langle \hat{\lambda}_n^{i_+}\hat{\lambda}_m^{j_-}\rangle = {} & \xi^+_{i_+ p_+}\langle \hat{\lambda}_n^{p_+}\hat{\lambda}_m^{j_-}\rangle + \xi^-_{j_- p_-}\langle \hat{\lambda}_n^{i_+}\hat{\lambda}_m^{p_-}\rangle \\ & + 2B_{p_+}\overline{f}^+_{i_0 j_+ p_+} d^{i_+}_{mn}\langle \lambda_m^{i_0}\rangle + 2B_{p_+} f^+_{i_0 i_+ p_+}\overline{d}^{j_+}_{nm}\langle \lambda_n^{i_0}\rangle,\end{aligned} \tag{6.2.38}$$

with transforms $c$, $d$, $\alpha$, $\beta$ defined in Eqs. (6.2.31) to (6.2.34) and constants $\xi$, $\phi$ in Eqs. (6.2.20) to (6.2.23). Einstein summation applies to the $i_0$, $i_\pm$ indices only, and $\hat{\lambda}_n^{i_\pm} = \sqrt{N_\mathrm{m}}\lambda_n^{i_\pm}$. Before presenting the result of solving these equations for a model of the expansion of a polariton cloud, we decide relevant physical quantities to extract from the dynamics.

### 6.2.5 Observables, bright and dark states

#### Photon and exciton populations

For constructing a picture of excitation dynamics in real space two physical observables are immediately accessibly from the cumulant equations. These are the photonic (phot) excitation,

$$n_{\text{phot}}(t, r_n) = \langle a^\dagger(r_n) a(r_n) \rangle \tag{6.2.39}$$

$$= \frac{1}{N_k} \sum_{kk'} \langle a^\dagger_{k'} a_k \rangle e^{i(k'-k)r_n}, \tag{6.2.40}$$

and the excitonic or molecular (m) population,

$$n_{\text{m}}(t, r_n) = \langle (\sigma^+ \sigma^-)(r_n) \rangle \tag{6.2.41}$$

$$= N_E \left( C^0_{i_0} \langle \lambda^{i_0}_n \rangle + D^0 \right), \tag{6.2.42}$$

where $C^0_{i_0} = (1/2)\,\mathrm{Tr}\left[\sigma^+ \sigma^- \lambda^{i_0}\right]$ and $D^0 = (1/2N_\nu)\,\mathrm{Tr}[\sigma^+\sigma^-] = 1/2$. The factor $N_E = N_{\text{m}}/N_k$ is required since $\langle \lambda^{i_0}_n \rangle$ provides the on-site value for $N_E$ molecules in the ensemble at $r_n$. Note for a closed system the sum $\sum_n \left(n_{\text{phot}}(t, r_n) + n_{\text{m}}(t, r_n)\right)$ is a constant set by the initial excitation, but decreases with time in the presence of losses such as $\kappa$ and $\Gamma_\downarrow$ in our model.

If we start from a localised purely exciton population (Section 6.3.4), plotting $n_{\text{phot}}$ will show how the excitation spreads out in the system via light-matter interaction. As we discuss further below, it will also allow us to define a mean deviation and wavefront for the excitation from which a propagation velocity may be determined. The dynamics of the total molecular population $n_{\text{m}}$ will be less revealing from this initial state, since it will be overwhelmingly dark. For this reason it will be useful to distinguish the fraction that actually interacts with light, the bright state population.

#### Bright and dark exciton states

Let $\sigma^\pm_k = (1/\sqrt{N_{\text{m}}}) \sum_{nx} \sigma^\pm_{nx} e^{\mp i k r_n}$ be the Fourier transform of $\sigma^\pm_{nx}$. The bright ($\mathcal{B}$) exciton, that is, the molecular population that couples to the photon modes in Eqs. (6.2.35) and (6.2.37), is

$$n_{\mathcal{B}}(t, r_n) = \frac{N_E}{N_{\text{m}}} \sum_{kk'} \langle \sigma^+_{k'} \sigma^-_k \rangle e^{i(k'-k)r_n} \tag{6.2.43}$$

$$= \frac{1}{N_k} \sum_{kk'} \langle \sigma^+_{k'} \sigma^-_k \rangle e^{i(k'-k)r_n}. \tag{6.2.44}$$

The coherences $\langle \sigma^+_{k'} \sigma^-_k \rangle$ are

$$\langle \sigma^+_{k'} \sigma^-_k \rangle = \varsigma_{i_+} \varsigma_{j_+} \frac{1}{N_{\text{m}}} \sum_{nx,my} \langle \lambda^{i_+}_{nx} \lambda^{j_-}_{my} \rangle e^{-ik' r_n} e^{ik r_m} \tag{6.2.45}$$

where $\varsigma_{i_+} = \frac{1}{2}\,\mathrm{Tr}[\sigma^+ \lambda^{i_-}]$. Evaluating the double sum occurring here requires similar handling of the coherences as in Eq. (6.2.25):

$$\langle \sigma^+_{k'} \sigma^-_k \rangle = \frac{N_E}{N_k} \sum_{n,m} \langle \lambda^{i_+}_n \lambda^{j_-}_m \rangle e^{-ik' r_n} e^{ik r_m} - \frac{1}{N_k} \sum_n \langle \lambda^{i_+}_n \lambda^{j_-}_n \rangle e^{i(k-k')r_n}$$
$$+ \frac{1}{N_k} \sum_n \left( Z^+_{i_0 i_+ j_+} \langle \lambda^{i_0}_n \rangle + \frac{1}{N_\nu} \delta_{i_+ j_+} \right) e^{i(k-k')r_n}. \tag{6.2.46}$$

Hence

$$n_{\mathcal{B}}(t, r_n) = \varsigma_{i_+}\varsigma_{j_+}\Bigg[\frac{N_E}{N_k^2}\sum_{kk'}\sum_{m_1,m_2}\langle\lambda_{m_1}^{i_+}\lambda_{m_2}^{j_-}\rangle e^{-ik'r_{m_1}+ikr_{m_2}+ik'r_n-ikr_n}$$
$$-\frac{1}{N_k^2}\sum_{kk'}\sum_{m}\langle\lambda_m^{i_+}\lambda_m^{j_-}\rangle e^{i(k-k')r_m+ik'r_n-ikr_n} \tag{6.2.47}$$
$$+\frac{1}{N_k^2}\sum_{kk'}\sum_{m}\left(Z^+_{i_0 i_+ j_+}\langle\lambda_m^{i_0}\rangle+\frac{1}{N_\nu}\delta_{i_+j_+}\right)e^{i(k-k')r_m+ik'r_n-ikr_n}\Bigg]$$
$$= \varsigma_{i_+}\varsigma_{j_+}\left[(N_E-1)\langle\lambda_n^{i_+}\lambda_n^{j_-}\rangle+\left(Z^+_{i_0 i_+ j_+}\langle\lambda_n^{i_0}\rangle+\frac{1}{N_\nu}\delta_{i_+j_+}\right)\right]. \tag{6.2.48}$$

It is worth reiterating that in Eq. (6.2.48) the expectation $\langle\lambda_n^{i_+}\lambda_n^{j_-}\rangle$ involves operators on *different* molecules; at each $i_+$, $j_-$ it is the diagonal entries of the $N_k \times N_k$ array $\langle\lambda_n^{i_+}\lambda_m^{j_-}\rangle$.

The dark ($\mathcal{D}$) population is just the total population minus the part that is bright:

$$n_{\mathcal{D}}(t, r_n) = n_{\mathrm{m}}(t, r_n) - n_{\mathcal{B}}(t, r_n) \tag{6.2.49}$$
$$= N_E\left(C^0_{i_0}\langle\lambda_n^{i_0}\rangle + D^0\right) - \left(\varsigma_{i_+}\varsigma_{j_+}Z^+_{i_0 i_+ j_+}\langle\lambda_n^{i_0}\rangle+\frac{1}{N_\nu}\varsigma_{i_+}\varsigma_{j_+}\right)$$
$$-(N_E-1)\varsigma_{i_+}\varsigma_{j_+}\langle\lambda_n^{i_+}\lambda_n^{j_-}\rangle. \tag{6.2.50}$$

In Appendix F.3 we show $\varsigma_{i_+}\varsigma_{j_+}Z^+_{i_0 i_+ j_+} \equiv C^0_{i_0}$ and $\varsigma_{i_+}\varsigma_{i_+} \equiv D^0$. Then

$$n_{\mathcal{D}}(t, r_n) = n_{\mathrm{m}}(t, r_n) - \frac{1}{N_E}n_{\mathrm{m}}(t, r_n) - (N_E-1)\varsigma_{i_+}\varsigma_{j_+}\langle\lambda_n^{i_+}\lambda_m^{j_-}\rangle \tag{6.2.51}$$
$$= (N_E-1)\left[\frac{n_{\mathrm{m}}(t, r_n)}{N_E} - \varsigma_{i_+}\varsigma_{j_+}\langle\lambda_n^{i_+}\lambda_n^{j_-}\rangle\right]. \tag{6.2.52}$$

### 6.2.6 Velocity measurements

In order to make comparisons for the speed at which the light-matter excitation propagates in the simulated dynamics, three points must be established:

1. How to define the extent of the polariton cloud and so extract a velocity $v^{\mathrm{obs}}$
2. A group velocity $v_k$ to make comparisons against
3. An appropriate wavevector $k_0$ at which to make the comparison

We now establish 1. Points 2 and 3 are discussed in Section 6.3.2.

Even without access to the polariton population at this stage we can make a meaningful comparison using the photonic component, $n_{\mathrm{phot}}$, that propagates with the polariton cloud. There are many ways one could chose to track the expansion of the excitation density using this function. We consider two possibilities. The first is by calculating the root-mean-squared deviation (RMSD) of $n_{\mathrm{phot}}(t, r_n)$ from the centre of the system, $\mathrm{RMSD}[n_{\mathrm{phot}}](t)$. This measure was used by Groenhof et al. in their molecular dynamics simulations [86, 119, 120]. We found this to provide a simple, unambiguous way to follow the expansion dynamics. A second is to calculate, at each timestep, the position beyond which a certain fraction of $n_{\mathrm{phot}}(t, r_n)$ has passed. This approach was taken by Xu et al. in Ref. [85], with $P_{\mathrm{cut}} = 4\,\%$ taken as the threshold fraction. Since we base our model

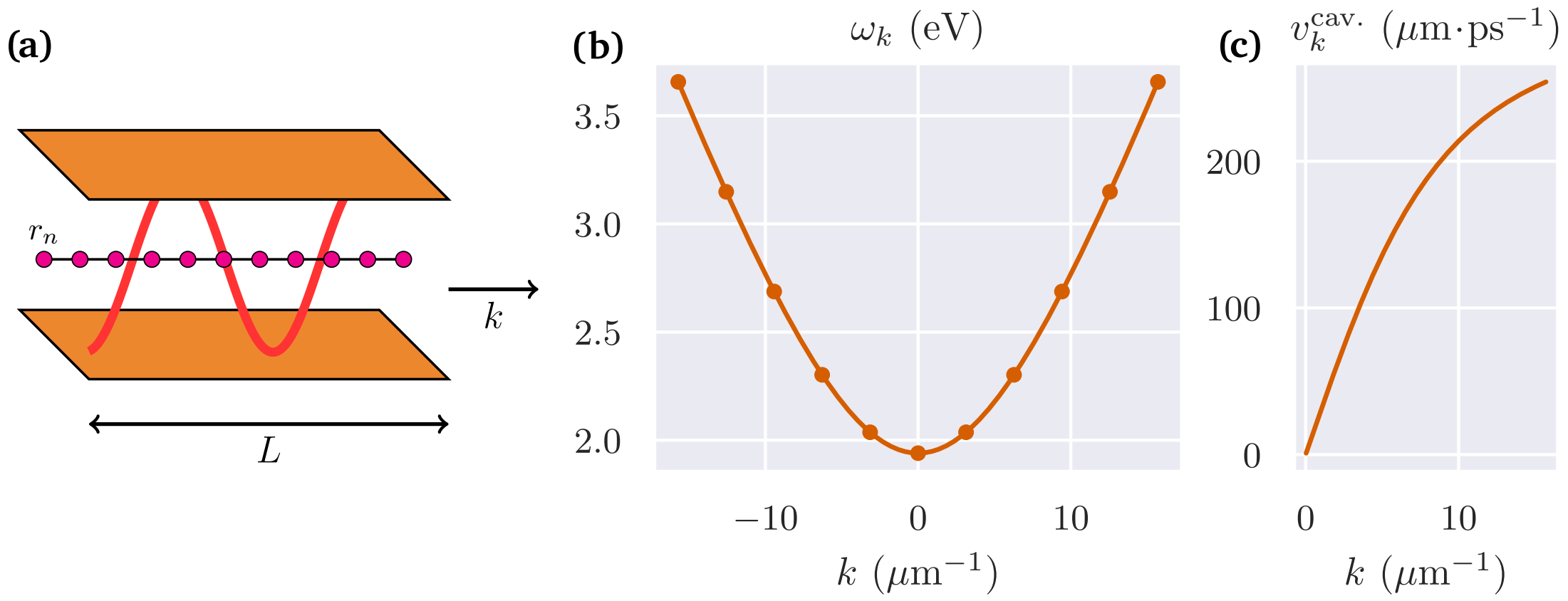

**Figure 6.4:** (a) Sketch of the system: a line of $N_{\text{m}}$ molecules (position $r_n$) in a multimode cavity with in-plane momentum $k$. (b) Cavity dispersion Eq. (6.3.1) when $\omega_c = 1.94$ eV. For illustration, dots mark $N_k = 11$ equally spaced modes ($Q_0 = 5$). (c) Corresponding group velocity for $k > 0$. ↩TOF

parameters on those in Ref. [85], we include this measure as an alternative calculation in our results. However, as we discuss further below, it was not as useful as the RMSD calculation for our work, yielding a velocity highly dependent on the choice of $P_{\text{cut}}$.

In either case one obtains as a function of time a position for the average deviation (RMSD) or wavefront ($P_{\text{cut}}$) of the excitation density from which a velocity can be extracted, for example, after making a polynomial fit of the position. Due to measuring $n_{\text{phot}}$, the overall motion will be modulated by oscillations at the Rabi frequency as excitation is transferred from $n_{\text{phot}}$ to the bright exciton and back. Hence before extracting a velocity we apply a lowpass filter to remove these oscillations and obtain a smooth signal. In all plots of the RMSD or $P_{\text{cut}}$ position below this filtering has been performed.

In line with the polariton experiments we are interested in early-time propagation after injecting an excitation into the system. Practically, we also need to ensure effects from the excitation reaching the edge of the simulated system do not influence the measured velocity. Thus, in the following we perform a power-law fit $Dt^{\beta}$ for $0 \leq t \leq 0.05$ ps of the recorded position and calculate the velocity at $t = 0.05$ ps. The total simulation time was $t = 0.3$ ps.

# 6.3 Propagation of a polariton cloud ↩

## 6.3.1 Cavity dispersion and physical parameters ↩

We use a model of a two-dimensional Fabry-Pérot cavity [57] with free propagation in a single direction such that the energy $\omega_k$ ($\hbar = 1$) is a function of a single momentum coordinate $k$,

$$\omega_k = \sqrt{\omega_c^2 + k^2c^2}, \tag{6.3.1}$$

where $\omega_c$ is the minimum cavity energy (Fig. 6.4). The cavity length $L$ in this direction dictates the quantisation of $k$ as

$$k = 0, \pm\frac{2\pi}{L}, \pm\frac{4\pi}{L}, \ldots, \pm\frac{2\pi}{L}Q_0. \tag{6.3.2}$$

| Description | Symbol | Value |
|---|---|---|
| Cavity energy | $\omega_c$ | 1.94 eV |
| Exciton energy | $\epsilon$ | 2.14 eV |
| Light-matter coupling | $g\sqrt{N_\mathrm{m}}$ | 0.15 eV |
| Cavity loss | $\kappa$ | 5 ps$^{-1}$ |
| Exciton decay | $\Gamma_\downarrow$ | $5 \times 10^{-4}$ ps$^{-1}$ |

**(a)** Photon and exciton parameters

| Description | Symbol | Value |
|---|---|---|
| Number of modes | $N_k$ | 301 |
| Number of molecules | $N_\mathrm{m}$ | 6001 |
| Cavity length | $L$ | 60 $\mu$m |

**(b)** System sizes

| Description | Symbol | Value |
|---|---|---|
| Energy | $\omega_\nu$ | $6 \times 10^{-3}$ eV |
| Temperature | $T$ | 300 K |
| Coupling (Huang-Rhys) | $S$ | 0, 7.11 |
| Thermalisation rate | $\gamma_\nu$ | 15 ps$^{-1}$ |

**(c)** Vibrational mode parameters ([125, 126])

**Table 6.1:** Parameter values. The values of $\omega_c$, $\epsilon$ and $g\sqrt{N_\mathrm{m}}$ match those used by Ref. [85] to model experimental data for transport in halide perovskite microcavities. The losses $\kappa$, $\Gamma_\downarrow$ were based on typical cavity and exciton lifetimes measured in those experiments at room temperature. The phonon energy $\omega_\nu$ and coupling $S$ were based on values for organic molecular crystals in Ref. [125, 126], whilst the thermalisation rate $\gamma_\nu$ was set at several times the photon decay rate[5].

Here the integer $Q_0$ imposes a cut-off such that the total number of modes is $N_k = 2Q_0+1$. Below we continue to set $\hbar = k_B = 1$ and refer to energies in electronvolts.

For the cavity and electronic parameters, we use those derived for halide perovskites at room temperature in the study by Xu et al. [85]. As discussed in Section 6.1, those authors used a mixed quantum-classical model of the dynamics to explain the velocity renormalisation observed in their experiment due to coupling to the phonon environment alone. For a comparable work, we set the number of modes $N_k = 301$ and molecules $N_\mathrm{m} = 6001$ equal those used in their theoretical model, but note our computation is not necessarily restricted to these values. In particular, increasing $N_\mathrm{m}$ has no computational cost in our approach, but increasing $N_k$ does—this is explained in Section 6.4.

For the phonon environment we adopt values for charge transport in organic molecular crystals [125, 126] to set the energy $\omega_\nu$ and coupling strength $S$ of the vibrational mode of each molecule. These values were also taken by Ref. [85], but to construct a *classical* description of the vibrational degrees of freedom.

The full set of parameter values are summarised in Table 6.1. Note the detuning $\Delta = \omega_c - \epsilon = -0.2$ eV is comparable in magnitude to the light-matter coupling strength $g\sqrt{N_\mathrm{m}} = 0.15$ eV. $\Delta$ and $g\sqrt{N_\mathrm{m}}$ set the scale for the range of relevant modes in the photon dispersion and, considering the bare energies $\omega_c$, $\epsilon$ and losses $\kappa$, $\Gamma_\downarrow$, place the system in the regime of strong, but not ultra strong, light-matter coupling. Note a finite value of $S = 7.11$ is stated but we intend to investigate the effect of turning this parameter on from zero. Later a pure dephasing term $\Gamma_z \neq 0$ will also be considered (cf. Eq. (6.2.3)).

### 6.3.2 Low density polariton dispersion

Although the exact forms of the lower $L_k^\dagger$ and upper $U_k^\dagger$ polariton creation operators of the open many-body system are not known, expressions for these operators in the low-excitation density limit in the case of no vibrational coupling are straightforward to derive. While only approximate, this will provide a polariton group velocity for comparisons.

[5] This is not necessarily representative of the halide perovskite system (Ref. [85] provides no value). On the contrary, $\gamma_\nu 0.1$ ps$^{-1}$ may have been a more realistic choice [43] (see Table 2.1).

The starting point of the calculation is the light-matter Hamiltonian

$$H = \sum_k \omega_k a_k^\dagger a_k + \sum_n \left[ \frac{\epsilon}{2}\sigma_n^z + \sum_k g\left( a_k \sigma_n^+ e^{-ikr_n} + a_k^\dagger \sigma_n^- e^{ikr_n} \right) \right], \tag{6.3.3}$$

where we returned to the original, single-index notation (the $\lambda$ reduce to Pauli matrices when $N_\nu = 1$).

At low excitation densities the the two-level system operators $\sigma_n^+, \sigma_n^-$ can be approximated by bosonic operators $s_n^\dagger$, $s_n$:

$$H = \sum_k \omega_k a_k^\dagger a_k + \sum_n \left[ \epsilon s_n^\dagger s_n + \sum_k g\left( a_k s_n^\dagger e^{-ikr_n} + a_k^\dagger s_n e^{ikr_n} + \right) \right] + \text{const.} \tag{6.3.4}$$

Fourier transforming $s_n^\dagger = \frac{1}{\sqrt{N_m}} \sum_p s_p^\dagger e^{ipr_n}$, $H$ can be written[6]

$$H = \sum_k \begin{pmatrix} s_k^\dagger & a_k^\dagger \end{pmatrix} \begin{pmatrix} \epsilon & g\sqrt{N_\text{m}} \\ g\sqrt{N_\text{m}} & \omega_k \end{pmatrix} \begin{pmatrix} s_k \\ a_k \end{pmatrix}. \tag{6.3.5}$$

From here the unitary transformation

$$\begin{pmatrix} U_k \\ L_k \end{pmatrix} = \begin{pmatrix} \cos\theta_k & \sin\theta_k \\ -\sin\theta_k & \cos\theta_k \end{pmatrix} \begin{pmatrix} s_k \\ a_k \end{pmatrix} \tag{6.3.6}$$

yields the diagonal form

$$H = \sum_k \begin{pmatrix} U_k^\dagger & L_k^\dagger \end{pmatrix} \begin{pmatrix} \epsilon_k^U & 0 \\ 0 & \epsilon_k^L \end{pmatrix} \begin{pmatrix} U_k \\ L_k \end{pmatrix} \tag{6.3.7}$$

provided

$$\cos\theta_k = \frac{1}{\sqrt{2}}\sqrt{1 + \frac{\epsilon - \omega_k}{2\zeta_k}}, \quad \sin\theta_k = \frac{1}{\sqrt{2}}\sqrt{1 - \frac{\epsilon - \omega_k}{2\zeta_k}}, \tag{6.3.8}$$

where

$$\zeta_k = \frac{1}{2}\sqrt{(\epsilon - \omega_k)^2 + 4g^2 N_\text{m}}. \tag{6.3.9}$$

The upper ($U$, $+$) and lower ($L$, $-$) polariton energies in Eq. (6.3.7) are

$$\epsilon_k^{U/L} = \frac{\epsilon + \omega_k}{2} \pm \zeta_k, \tag{6.3.10}$$

and corresponding group velocities

$$v_k^{U/L} = \frac{\partial \epsilon_k^{U/L}}{\partial k} = \frac{1}{2} v_k^\text{cav.} \left[ 1 \pm \frac{\omega_k - \epsilon}{2\zeta_k} \right]. \tag{6.3.11}$$

[6] See Section 2.2 for the equivalent quadratic form for the single-mode model.

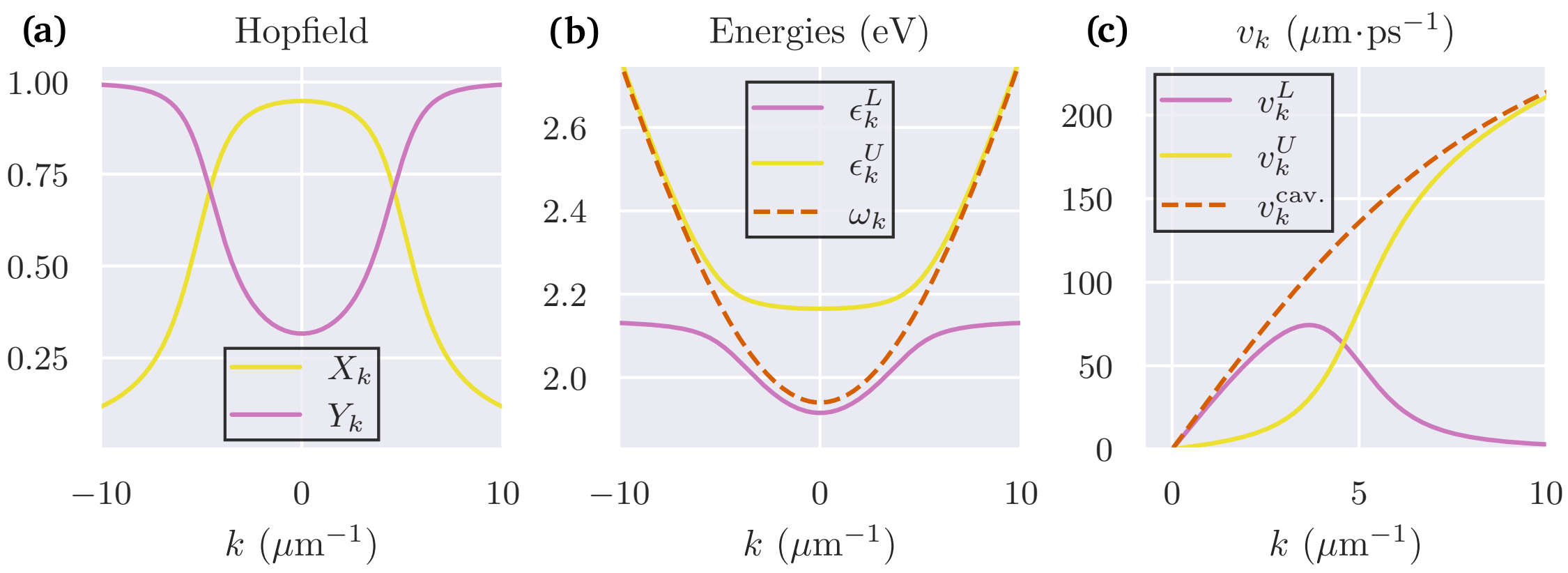

**Figure 6.5:** (a) Photon ($X_k$) and exciton ($Y_k$) Hopfield coefficients at low-excitation densities in the absence of vibrational coupling. (b) Polariton energies and (c) corresponding group velocities in this approximation. The cavity dispersion and velocity is included as a dashed line. Parameters $\omega_c = 1.94$ eV, $\epsilon = 2.14$ eV and $g\sqrt{N_\text{m}} = 0.15$ eV as in Table 6.1. ↗TOF

Here

$$v_k^{\text{cav.}} = \frac{\partial \omega_k}{\partial k} = \frac{kc^2}{\sqrt{\omega_c^2 + k^2c^2}}, \tag{6.3.12}$$

is the photon group velocity calculated from the cavity dispersion Eq. (6.3.1), and we used

$$\frac{\partial \zeta_k}{\partial k} = \frac{1}{8\zeta_k} \cdot 2(\epsilon - \omega_k) \cdot \left(-\frac{\partial \omega_k}{\partial k}\right) = v_k^{\text{cav.}} \frac{\omega_k - \epsilon}{4\zeta_k}. \tag{6.3.13}$$

### Hopfield coefficients and polariton operators

The Hopfield coefficients $X_k = \cos\theta_k$, $Y_k = \sin\theta_k$, which are real and normalised as $X_k^2 + Y_k^2 = 1$, dictate the ratio of optical to material character of the excitations. For the lower polariton, this ratio is $X_k^2/Y_k^2$, as evident from the associated creation operator

$$L_k^\dagger = -Y_k s_k^\dagger + X_k a_k^\dagger \tag{6.3.14}$$

$$\approx -\frac{Y_k}{\sqrt{N_m}} \sum_n \sigma_n^+ e^{-ikr_n} + X_k a_k^\dagger, \tag{6.3.15}$$

which is an antisymmetric superposition of the $k^{\text{th}}$ photon mode and bright state. $L_k^\dagger$ creates a polariton with a specific momentum $k$ and energy $\epsilon_k^L$ on the lower branch. In Fig. 6.5 the Hopfield coefficients, as well as the polariton energies and group velocities, are plotted for the system parameters in Table 6.1. Comparing to the notation of Ref. [116] in Fig. 6.1, we have $|\alpha_{\text{ph}}|^2 = X_k^2$ at the in-plane momentum $k$ of the polaritonic state recorded.

## 6.3.3 Characteristic wavevector ↗

For a group velocity to compare the expansion velocity $v^{\text{obs}}$ measured in the simulation we take the velocity $v_k^L$ of the lower polariton, since it is the lower polariton that should be predominately populated by a weak initial excitation.

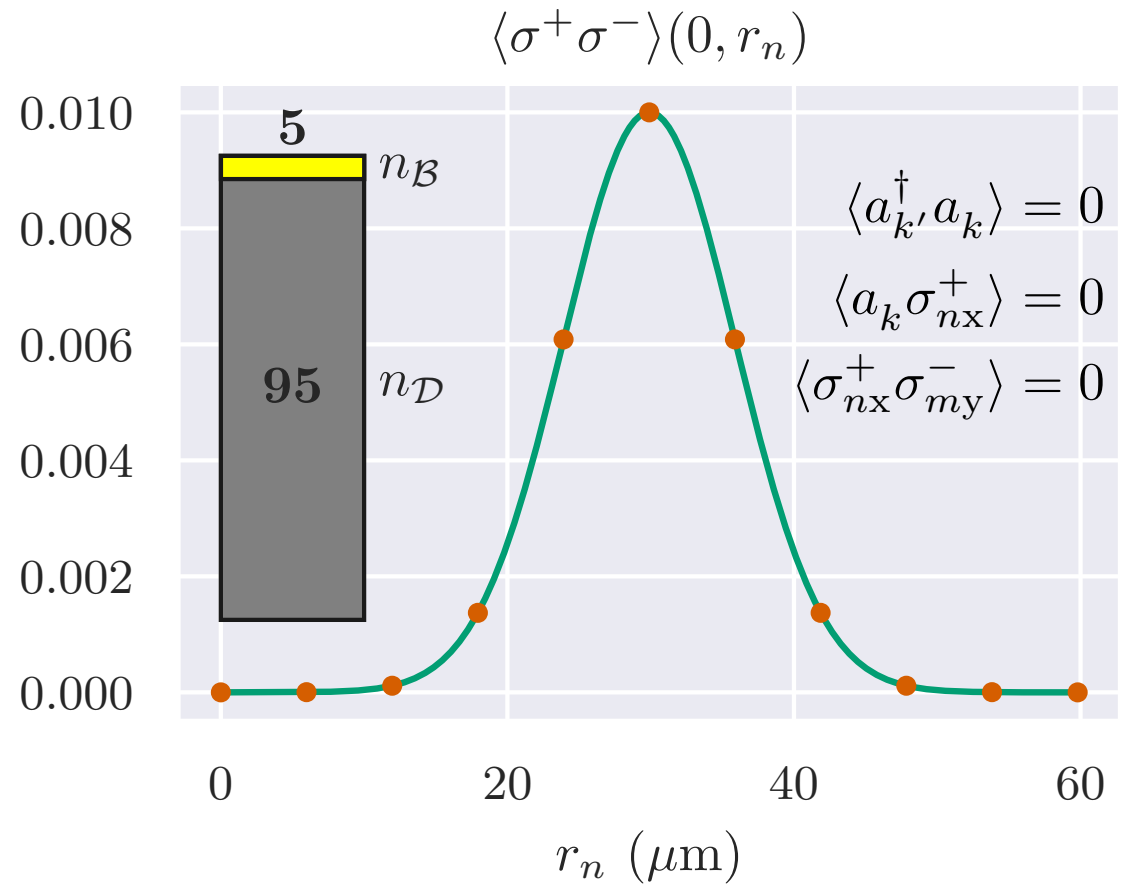

**Figure 6.6:** Initial excitation density profile $p_{\text{ex}}(r_n) = 0.01e^{-(r_n - L/2)^2/(2(L/10)^2)}$, with grid points for $N_k = 11$ ensembles. The incoherent state Eq. (6.3.16) leads to a molecular population roughly 95% dark and 5% bright, as $1/N_E = 301/6001 \approx 1/20$ (inset). All other variables in the initial state are zero. ↑TOF

As the state we propagate in the simulation will not be a wavepacket of well defined momentum (Section 6.3.4), a scheme must be devised to determine a wavevector $k_0$ that that defines a dominant or average momentum. For this we use the root-mean-square deviation of the photon number $n_k = \langle a^\dagger_k a_k\rangle$ at the time $t = 0.05$ ps at which $v^{\text{obs}}$ is to be calculated. We reason this as a sensible choice in that it should give a typical momentum $k_0$ both when the state is strongly peaked around a single momentum and when it is not, which is the situation we find below.

We comment further on the limitations of this characterisation as well as the choice of $v^L_k$ for the comparison in light of observed data in later sections.

### 6.3.4 Initial conditions ↑

We replicate the non-resonant excitation used by Balasubrahmaniyam et al. [116] and many other experiments [85, 112, 113, 117, 118, 123, 444] to initiate polariton propagation. This creates an *incoherent* population of excitons, which may be realised as the product state

$$\rho_{e\nu} = \bigotimes_{n=1}^{N_k} \rho^e_n \otimes \rho^\nu_n, \tag{6.3.16}$$

where both the electronic $\rho^e_n$ and vibrational $\rho^\nu_n$ parts are diagonal. This state respects U(1) symmetry and so is compatible with the set (6.2.35)–(6.2.38) of cumulant equations, which are symmetry preserving.

For the single-site electronic density matrix we write

$$\rho^e_n = \begin{pmatrix} p_{\text{ex}}(r_n) & 0 \\ 0 & 1 - p_{\text{ex}}(r_n) \end{pmatrix}, \tag{6.3.17}$$

where $p_{\text{ex}}(r_n)$ describes the initial excitation density. We take this to be a Gaussian of magnitude $p_{\text{ex}}(L/2) = 0.01$ at the centre of the system with a standard deviation of $L/10$ (Fig. 6.6). Note whilst the total initial molecular population is then $n_{\text{mtot.}}(0) = \sum_n p_{\text{ex}}(r_n)$, only a portion $n_{\mathcal{B}\text{tot.}}(0) = n_{\text{mtot.}}(0)/N_E$ is bright. Here the fraction controlling the ratio of bright to dark states, $1/N_E$, is set physically by the ratio of number of relevant photon modes to number of molecules [63] as $1/N_E = N_k/N_{\rm m}$: for a given $L$ we require $N_k$ such that all relevant photon modes $k \leq k_{\text{max.}} \sim N_k/L$ are included.

The vibrational density matrix on the other hand is taken to be a thermal state,

$$\rho_n^\nu = \rho^{\text{th.}}(T) = \frac{1}{\sum_{n_\nu=1}^{N_\nu-1} e^{-n(\omega_\nu/T)}} \sum_{n_\nu=0}^{N_\nu-1} e^{-n_\nu(\omega_\nu/T)} |n_\nu\rangle\langle n_\nu| \,, \tag{6.3.18}$$

where $|n_\nu\rangle$ is the Fock state with $n_\nu$ phonons.

No coherence means $\langle \sigma_n^+ \sigma_m^- \rangle = \langle a_k \sigma_n^+ \rangle = 0$, or equivalently $\langle \lambda_n^{i+} \lambda_m^{j-} \rangle = \langle a_k \lambda_n^{i+} \rangle = 0$, whilst $\langle a_{k'}^\dagger a_k \rangle = 0$ for an initially empty cavity. Then all that needs to be calculated are the initial expectations

$$\langle \lambda_n^{i_0} \rangle = \text{Tr} \left[ \rho_n^e \otimes \rho^{\text{th.}}(T) \lambda_n^{i_0} \right] . \tag{6.3.19}$$

Since the initial state is diagonal in both electronic *and* vibrational parts, this trace will be non-trivial only for those $\lambda_n^{i_0}$ which are diagonal in the combined $\mathcal{H}_{e\nu}$ space.

### 6.3.5 Results for early-time dynamics

In Fig. 6.7 we show the photon $n_{\text{phot}}$ and molecule $n_{\text{m}}$ populations in the system starting from an incoherent exciton population, Eq. (6.3.19), for two simulations. The data in the top row of this figure is for the model without vibronic coupling, $S = 0$, and the bottom row for the model with $N_\nu = 4$ vibrational levels ($S = 7.11$). As the molecular population is dominated by dark states, which are static, we plot the difference in $n_{\text{m}}$ above its initial value: dark regions of Figs. 6.7b and 6.7d correspond to *depletion* of the initial molecular population.

The physical picture is the following. The initial excitation Eq. (6.3.19) is purely excitonic, so at $t = 0$ there are no photons in the cavity (black horizontal line in Fig. 6.7a). Strong light-matter coupling results in transfer to, and then from, $n_{\text{phot}}$ on a time scale set by the Rabi frequency, $1/(2g\sqrt{N_{\text{m}}}) \sim 0.01$ ps. So the bright lines in Fig. 6.7a correspond to dark lines in Fig. 6.7b of excitonic depletion. The polariton in this process spreads outwards from the initial spot, which is roughly confined to the middle third of the cavity ($20 < r_n < 40$ $\mu$m). The dashed and dotted lines in Fig. 6.7a trace the root-mean-squared (RMSD) and wavefront ($P_{\text{cut}} = 4\%$) positions, respectively. The latter suggests beyond $t \sim 0.1$ ps the dynamics may be influenced by the edge of the system.

Switching on strong coupling to the vibrational modes, $S = 7.11$ in Figs. 6.7c and 6.7d, dampens the excitation dynamics: Rabi oscillations become less clearly defined at later times and the wavefront expansion reduces. In addition, molecular depletion in Fig. 6.7d is increased at late times compared to the case $S = 0$, corresponding to more photonic excitation in Fig. 6.7c. As we discuss below, this is due to dark states feeding the polariton population.

To examine the role of bright and dark states in both cases ($S = 0$, $S = 7.11$) we plot population dynamics for these states in Fig. 6.8.

Consider first $S = 0$. In Fig. 6.8a we sum over all ensemble positions $r_n$ to obtain the total change of populations from the initial conditions as a function of time. As expected without coupling to the vibrational modes, the dark states (purple line) are decoupled from the dynamics. These states decay over a timescale $1/\Gamma_\downarrow \sim 10^3$ ps far longer than that of the simulation. The change in bright state population (orange) mirrors the photon population, and also describes the change in overall molecular population (green) given the dark state inactivity. The fact that the total excitation $n_{\text{phot}} + n_{\text{m}}$ (black, dashed) decreases is a result of cavity leakage $1/\kappa \sim 0.2$ ps.

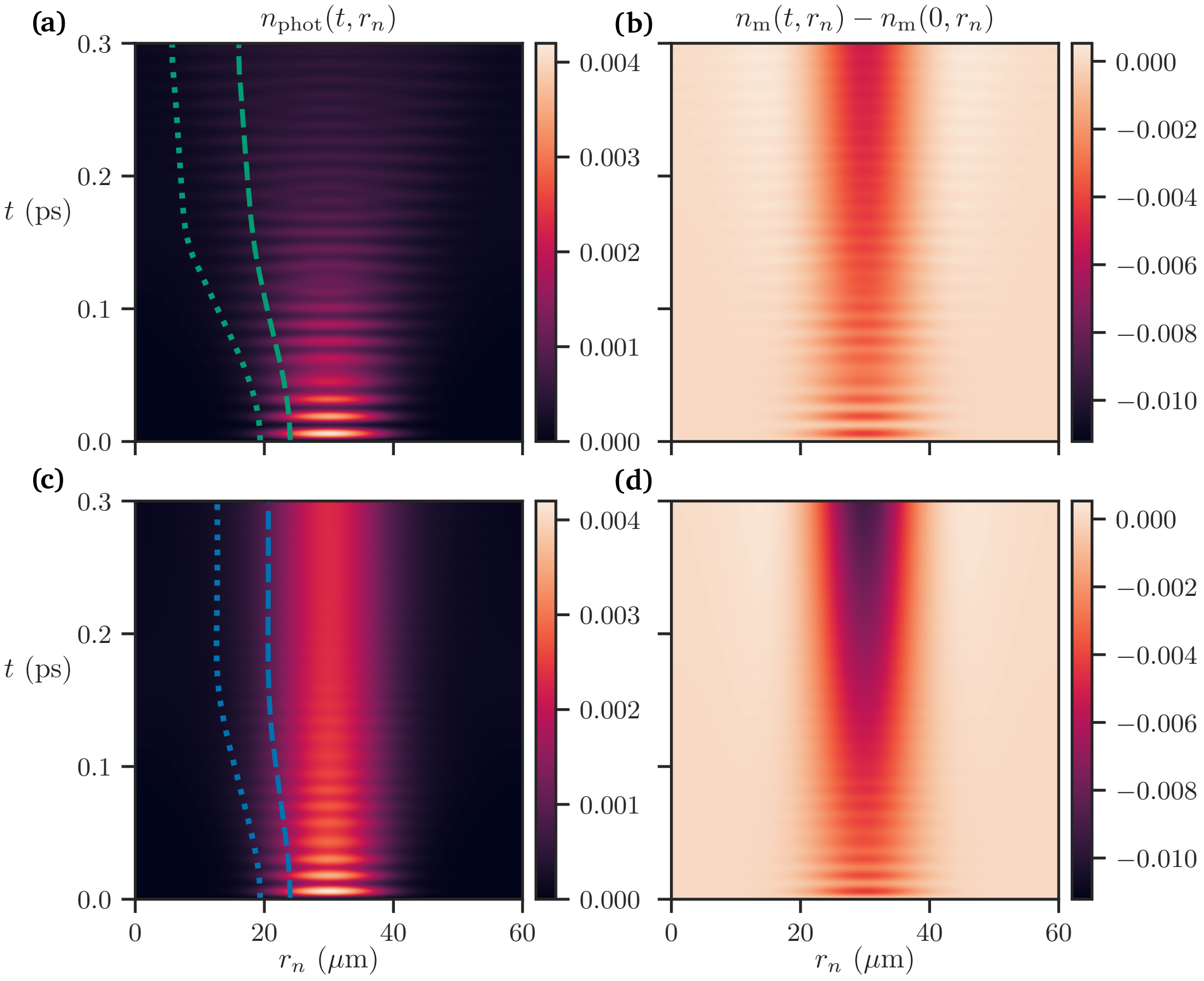

**Figure 6.7:** Photon and excess molecular populations starting from an incoherent exciton population (Eq. (6.3.19)) for the model with (a), (b) no vibronic coupling ($S = 0$) and (c), (d) coupling $S = 7.11$ of each emitter to a vibrational mode with $N_\nu = 4$ levels. Dashed and dotted lines indicate, respectively, root-mean-square (RMSD) and $P_{\text{cut}} = 4\%$ wavefront positions defined in Section 6.2.6. Note these curves were smoothed using a lowpass filter. The physical parameters used here and all the figures in this section are—except possibly $S$—as in Table 6.1. The computation requirements of these simulations are discussed at the end of the chapter (Section 6.4.4). ↑TOF

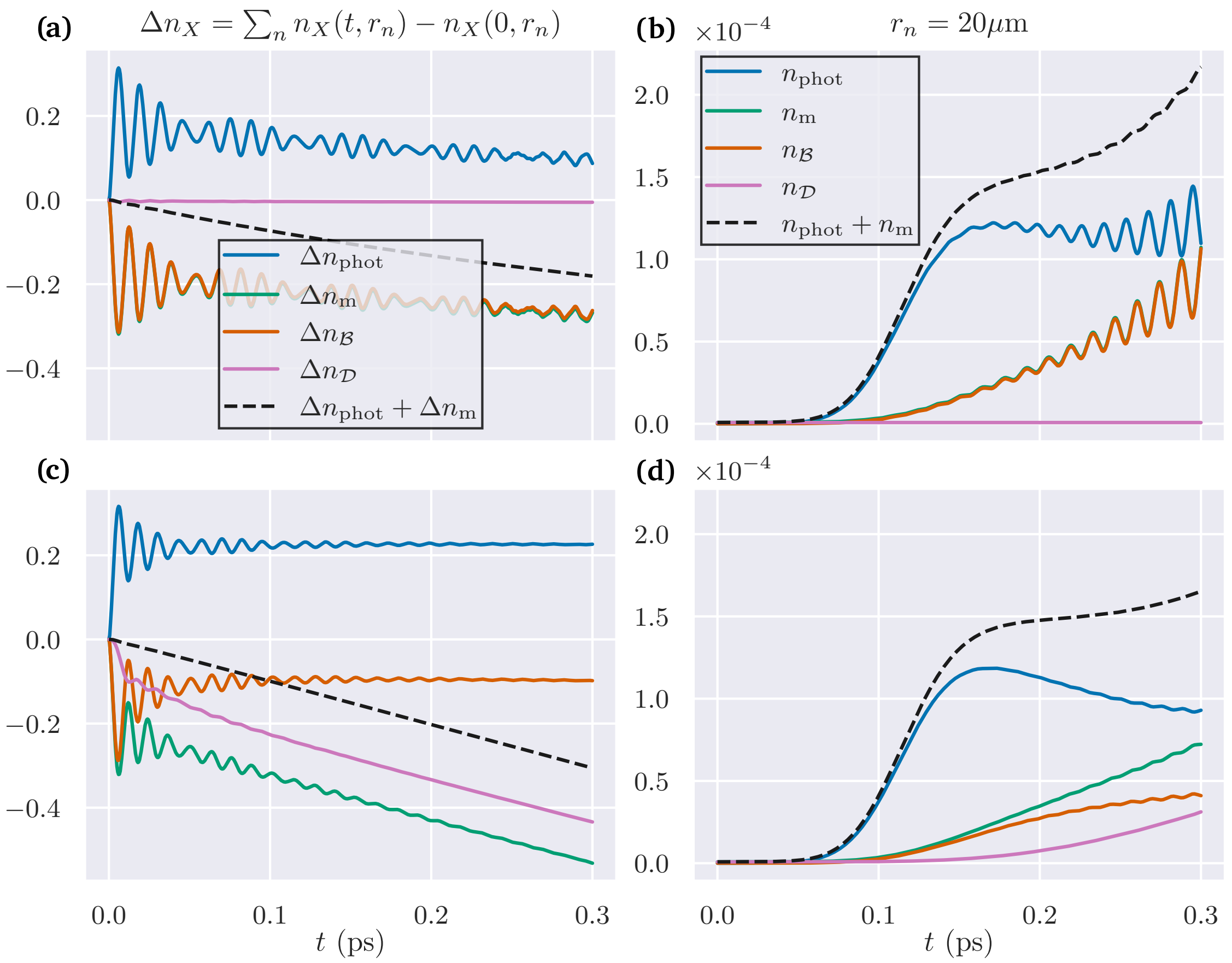

**Figure 6.8:** (a) Change in total photon $n_{\text{phot}}$, molecule $n_{\text{m}}$, bright $n_{\mathcal{B}}$ and dark $n_{\mathcal{D}}$ state populations for the dynamics shown in Fig. 6.7 when $S = 0$ (refer to Eqs. (6.2.40) to (6.2.50) for definitions of these quantities). The total initial excitation was split roughly $95\% : 5\%$ for dark:bright excitons (Fig. 6.6). (b) Populations at $r_n = 20\ \mu\text{m}$. (c), (d) Same when $S = 7.11$ ($N_\nu = 4$). ↑TOF

Fig. 6.8b shows the population at a single molecular ensemble at $r_n = 20\ \mu\text{m}$ when $S = 0$. From Fig. 6.7 we see this is ensemble is at the very edge of the initial excitation. Hence both bright $n_{\mathcal{B}}$ and photon $n_{\text{phot}}$ populations increase from zero as the initial excitation spreads out past this position. The growth in the two populations does not match each other however—the outward propagation is not uniform in this sense—but at later times we do see paired oscillations and a common magnitude being approached.

Moving now to $S = 7.11$, bottom row of Fig. 6.8, the picture changes significantly. As a result of the dynamical disorder induced by the vibrational modes, the dark states become weakly coupled into the dynamics. These states, which we recall are far outweigh (approx. 20 times) the bright populations, can then act as a reservoir. This is seen clearly in Fig. 6.8c as $n_{\mathcal{D}}$ continually depletes to the effect of stabilising the photon and bright state populations: provided there remains a sufficient dark state population, cavity losses are effectively negated. The damping effect of the vibronic coupling is seen in this panel, but also clearly in the $r_n = 20\ \mu\text{m}$ slice Fig. 6.8d. The dynamics of the dark states dynamics in this final panel is more subtle: at this position there is initially no excitation, so the dark state population *increases* at the cost of reduced growth of $n_{\mathcal{B}}$

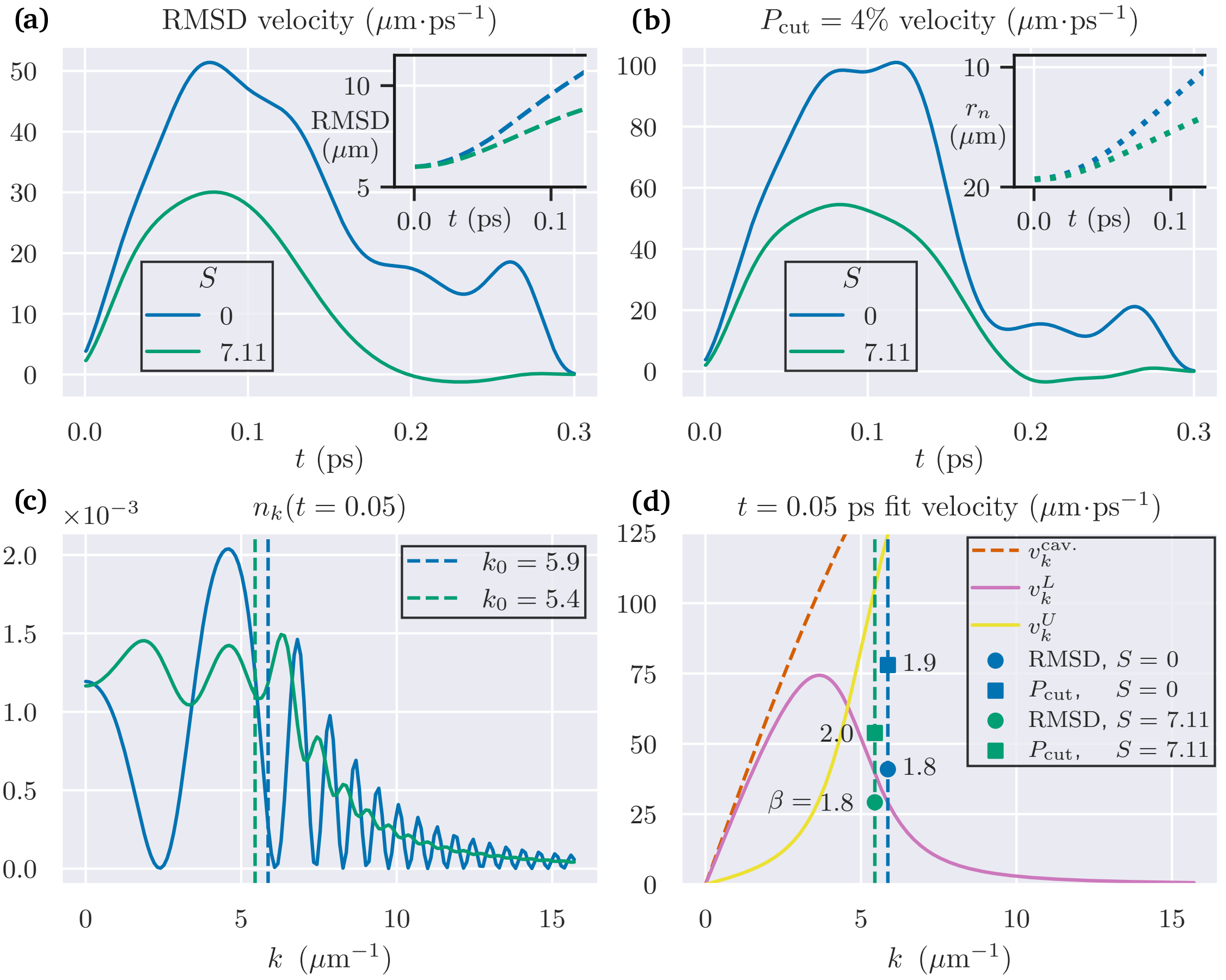

**Figure 6.9:** Expansion velocities. (a) Instantaneous velocities obtained as the gradient of the root-mean-square deviation (RMSD, inset) of $n_{\text{phot}}$ for $S = 0$ and $S = 7.11$ ($N_\nu = 4$). Note these curves were smoothed using a lowpass filter, and this smoothing was done prior to calculating the velocities. (b) Instantaneous velocities obtained from the wavefront position defined by the $P_{\text{cut}} = 4\%$ probability for $n_{\text{phot}}$. (c) Photon number $n_k = \langle a_k^\dagger a_k \rangle$ at $t = 0.05$ ps for $S = 0$ (blue) and $S = 7.11$ (green). The RMSD value is indicated by a vertical dashed line in each case. These lines are copied to panel (d) which includes scatter points for each velocity $v^{\text{obs}}$ calculated at $t = 0.05$ ps from a polynomial fit $\sim t^\beta$ to the RMSD or wavefront ($P_{\text{cut}}$) positions for $0 \leq t \leq 0.05$ ps. The transport exponent $\beta$ is shown next to each data point. The power-law fits were good and the $v^{\text{obs}}$ close to the instantaneous values read from Fig. 6.9a or 6.9b at $t = 0.05$ ps. ↑TOF

compared to Fig. 6.8b.

So far we have made qualitative observations of the bright and dark excitation profiles and found them to be consistent with our intuition of exciton-polariton dynamics with vibrational dressing. Next we extract propagation velocities from the dynamics of $n_{\text{phot}}$.

### 6.3.6 Expansion velocity calculation ↑

In Fig. 6.9a the outward velocity of the RMSD position calculated using finite differences is plotted for $S = 0$ (blue) and $S = 7.11$ (green). Portions of the corresponding dashed lines from Fig. 6.7a and 6.7c are repeated in the inset.

Coupling to vibrational modes slows the expansion considerably, reducing the maximum velocity from about 50 $\mu$m ps$^{-1}$ to 30 $\mu$m ps$^{-1}$, although the behaviour $t \lesssim 0.05$ ps is more similar. Using instead the wavefront measure ($P_{\text{cut}} = 4\%$), Fig. 6.9b, the comparison is similar. Note however the vertical scale on this panel is doubled compared to Fig. 6.9a. We found this measure of velocity to be highly dependent on the choice of $P_{\text{cut}}$, with smaller $P_{\text{cut}}$ selective of higher momentum states. Beyond $t \sim 0.08$ ps the expansion slows, with some contraction occurring at late times for $S = 7.11$. We are mainly interested in the initial, early-time expansion dynamics and so do not discuss the slowdown further, but note it is likely influenced by the fastest excitations reaching the edges of the simulated region—periodic boundary conditions are implicit in our modelling.

In order to make a comparison with a reference group velocity, in Fig. 6.9c we plot the photon number $n_k$ at $t = 0.05$ ps as a function of wavevector $k$ for $S = 0$ (blue) and $S = 7.11$ (green), and in each case indicate the root-mean-square value with a dashed vertical line. This provided a characteristic wavevector $k_0 = 5.9$ $\mu$m$^{-1}$, $k_0 = 5.4$ $\mu$m$^{-1}$ for the two models. The dashed vertical lines are copied in Fig. 6.9d where the recorded expansion velocities $v^{\text{obs}}$ are plotted for comparison with the polariton dispersions. In addition, the transport exponent $\beta$ of the fit $\sim t^{\beta}$ on $0 \leq t \leq 0.05$ ps for each RMSD (circle) or wavefront ($P_{\text{cut}}$, square) position is indicated.

First, the velocities derived from the wavefront with $P_{\text{cut}} = 4\%$ are significantly higher, above the lower polariton group velocity $v^L_{k_0}$ (they are still well below the upper polariton group velocity $v^U_{k_0}$, however). As mentioned above, we believe small $P_{\text{cut}}$ selected higher energy polaritons at the leading edge of the wavefront. We did not think this is compatible with our root-mean-square characterisation of $k_0$, or choice of lower polariton group velocity for the comparison. Henceforth we discuss the RMSD-derived velocities only.

The transport exponent $\beta \sim 1.8$ is close to 2 at both $S = 0$ and $S = 7.11$, indicating ballistic propagation. That we did not observe diffusive behaviour in the presence of strong vibrational coupling may be surprising given the Hopfield coefficient $|\alpha_{\text{ph}}|^2 = X^2_{k_0} \approx .3$ at $k_0 = 5.4$ $\mu$m$^{-1}$ corresponds to an exciton content $\gtrsim 70\%$, far in excess of the fraction observed in other models [85] and experiment [85, 116] to produce diffusive transport (e.g. $|\alpha_{\text{ph}}|^2 = 0.54$ in Fig. 6.1a). It must be considered that the Hopfield coefficients followed an approximate derivation for the system without vibronic coupling but, more critically, that they are changing rapidly around $k \sim 5$ $\mu$m$^{-1}$ (Fig. 6.5a). Hence the predicted composition will depend strongly on the $k_0$ chosen. Yet as seen in Fig. 6.9c the polariton is *not* peaked around a single wavevector, but involves a broad range of momentum states. Whilst we have chosen what we believe to be a sensible method (root-mean-square of $\langle a^{\dagger}_k a_k \rangle$) to determine $k_0$, one that should allow for meaningful comparisons *between* observed velocities, it does not necessarily provide an accurate indicator for the total excitonic weight.

### Further limitations

In addition to uncertainties associated with the characterisation of $k_0$, our ability to observe diffusive behaviour may be limited by the short duration of the measurement window $t \lesssim 0.05$ ps. For example, at $|\alpha_{\text{ph}}|^2 = 0.6$, Ref. [116] observed ballistic propagation that only became diffusive after $t \approx 0.3$ ps. These timescales are not directly applicable to our model, but this does motivate investigating dynamics over longer times than in our current analysis. We discuss the computational implications of doing so and other ways to look for diffusive behaviour in Section 6.4.

The broad momentum distribution of the excitation described above means a direct comparison of the $v^{\text{obs}}$ to the proposed polariton group velocity may not be warranted. Moreover, although one might physically expect rapid decay from the upper polariton branch [368], seeing Fig. 6.9d made us realise that the propagating excitation may well have a significant upper polariton pop-ulation. This would also explain propagation speeds exceeding $v^L_k$. A refined comparison would be to construct a combination of $v^L_{k_0}$ and $v^U_{k_0}$ weighted according to the lower $n^L_{k_0} = \langle L^\dagger_{k_0} L_{k_0} \rangle$ and upper $n^U_{k_0} = \langle U^\dagger_{k_0} U_{k_0} \rangle$ polariton populations at $k_0$, or, further, across all wavevectors:

$$v^{\text{com}} = \frac{1}{\sum_k (n^L_k + n^U_k)} \sum_k n^L_k v^L_k + n^U_k v^U_k \,. \tag{6.3.20}$$

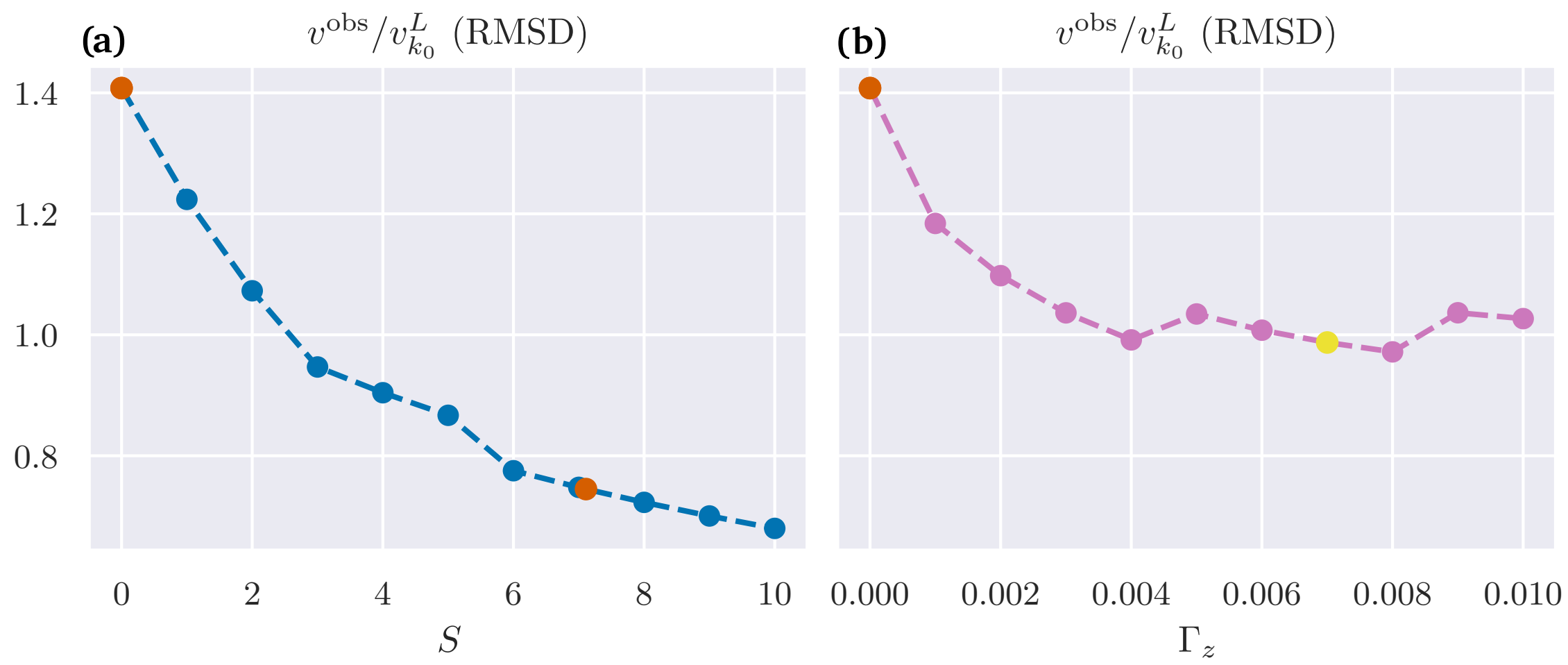

**Figure 6.10:** (a) Ratio of velocity $v^{\text{obs}}$ at $t = 0.05$ ps from the RMSD position of $n_{\text{phot}}$ to the lower polariton group velocity $v^L_{k_0}$ (Eq. (6.3.11)) for vibronic coupling strengths $S$ increasing from 0 ($N_\nu = 4$). All other parameters remain at the values specified in Table 6.1a, and $k_0$ is the characteristic wavevector at $t = 0.05$ ps described above. Even at $S = 10$ the early-time transport was still ballistic. Points in red correspond to data at $S = 0$, $S = 7.11$ featured in previous figures. (b) Ratios for a model where the vibrational mode of each molecule is replaced by a pure dephasing term $\Gamma_z L[\sigma^z_n]$ in the master equation. Uncertainties introduced in the selection of $k_0$ were presumably responsible for the uneven trend seen beyond $\Gamma_z = 0.004$. The yellow point at $\Gamma_z = 0.007$ corresponds to data shown in Fig. 6.11b. ↗TOF

We did not pursue this, given the added complexity and that it still only one possible choice of comparison, and not clearly the most accurate one (recall $L_k$, $U_k$ are only approximations for the model with $S = 0$). Instead, in the following section we compare $v^{\text{obs}}$ for a range of vibrational couplings $0 \leq S \leq 10$, which may still provide useful information even where a single absolute comparison to $v^L_k$ cannot. In Section 6.4 we also discuss a mean-field approach that could prepare wavepackets at a single momentum $k_0$ and so avoid these issues entirely.

While we cannot conclude the dark state interconversion, which is present at $S = 7.11$ but not $S = 0$, is necessarily connected to sub-group-velocity propagation, the lower $v^{\text{obs}}$ at $S = 7.11$ measurement does summarise what was already clear from the velocity times plots Figs. 6.9a to 6.9b: propagation is impeded by vibronic coupling. This is further illustrated by the trends observed with $S$ below. In Section 6.4 we discuss next steps that could be taken using this method to better determine the importance of dark states in the polariton transport.

### 6.3.7 Comparison to pure dephasing model

For the final part of our analysis we show results for a range of vibrational strengths and make comparisons to a model with pure dephasing.

In Fig. 6.10a, the ratio $v^{\text{obs}}/v^L_{k_0}$ is plotted for a range of vibrational couplings from $S = 0$ to $S = 10$, including the values $S = 0$, $S = 7.11$ that have already been considered (red points). Here $v^{\text{obs}}$ is velocity extracted from the RMSD position of $n_{\text{phot}}$ and $v^L_{k_0}$ the polariton group velocity at the wavevector $k_0$ determined using the procedure described above. The trend in lowering propagation velocities with increasing $S$ is clear and appears to continue beyond $S = 7.11$, albeit at a lessening rate. Note that the change in characteristic $k_0$ over the entire range of $S$ was small, decreasing from $k_0 = 5.9\ \mu\text{m}^{-1}$ $(S = 0)$ to $k_0 = 5.4\ \mu\text{m}^{-1}$ $(S = 7.11)$, so the observed trend holds for the absolute velocities too.

Figure 6.10b includes results for the model where the vibrational mode for each molecule is replaced $(S = 0)$ by a phenomelogical dephasing term $\Gamma_z L[\sigma^z_n]$ in the master equation. This term also induces dark state mixing, but in a completely incoherent way. Although initially $v^{\text{obs}}/v^L_{k_0}$ decreases with increasing $\Gamma_z$, the ratio quickly saturates.

The above results may be consistent with the kinetic model [116] of reversible scattering to the dark states. That model is realised when equilibration due to scattering occurs, which should be true for sufficient $S > 0$ or $\Gamma_z > 0$. Naturally, there will be a continuous cross-over from the case $S$, $\Gamma_z$ are zero, hence the smooth initial decreases in Figs. 6.10a and 6.10b. When equilibration is reached, further increasing $S$ or $\Gamma_z$ would not be expected further decrease in $v^{\text{obs}}$, as in Fig. 6.10b. However, this is only true so far as the system energies do not change: Eq. (6.1.3) involves $\epsilon_{\mathcal{P}} - \epsilon_{\mathcal{D}}$, and $v^L_{k_0}$ is calculated for the system with $S = 0$. That the curve in Fig. 6.10a does not level off may well be explained by the system eigenstates changing with $S$.

It is interesting that saturation for the pure dephasing model appears to occur around $v^{\text{obs}} = v^L_{k_0}$. Given the uncertainties regarding the comparison group velocity this is likely an artefact, i.e., does not lead to the conclusion that sub-group-velocity propagation is absent in this model. We comment further on this in the next section.

For completeness, we compare population dynamics of the two models in Fig. 6.11. This figure shows the total change in populations for the data points in Fig. 6.10a at $S = 7.11$ (repeat of Fig. 6.8a), and in Fig. 6.10b at a comparable dephasing strength $\Gamma_z = 0.007$ (yellow point in Fig. 6.10b). At this value of $\Gamma_z$ dark states are in fact depleted more effectively than in the model with $S = 7.11$. This conveys a larger photon population (blue line) in the pure dephasing model. Note this does not necessarily indicate a higher photonic weight (and so propagation speed)—that is dictated by the $k$-dependent Hopfield coefficients.

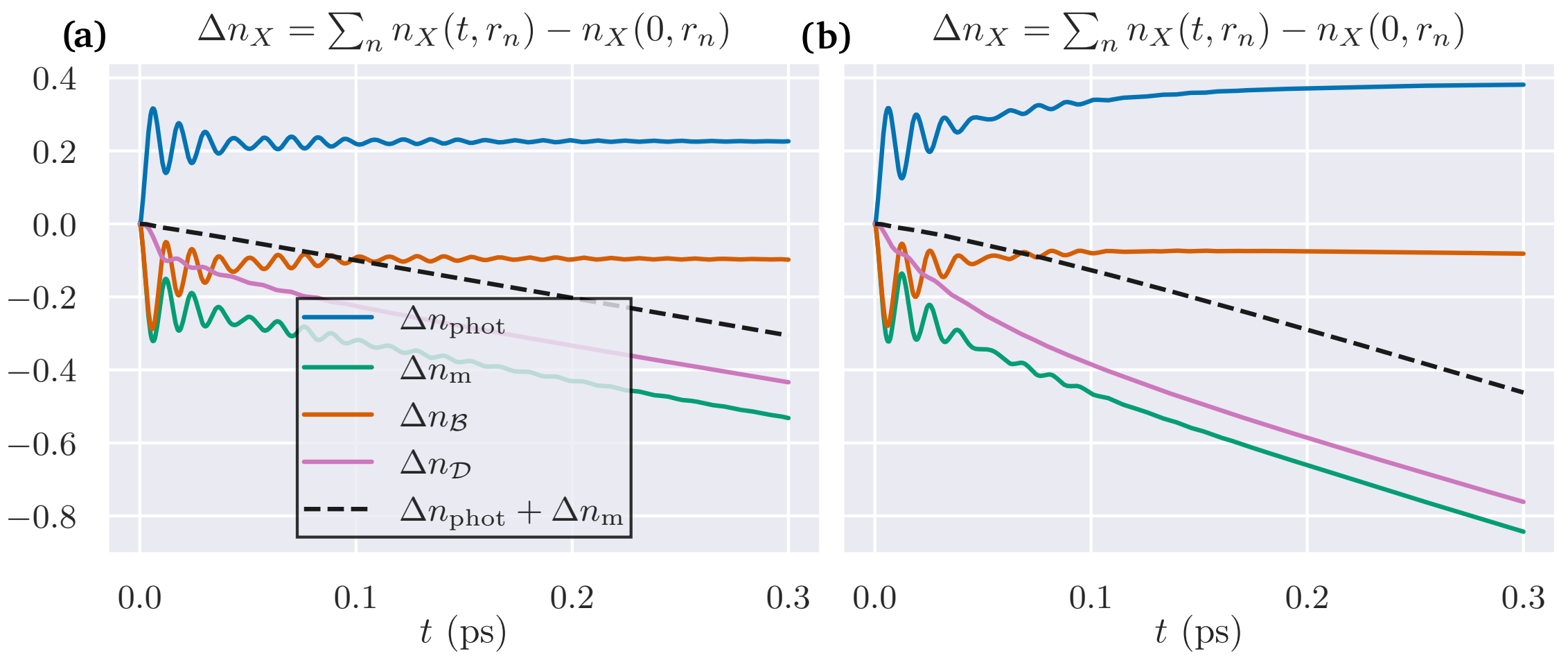

**Figure 6.11:** Change in total photon $n_{\text{phot}}$, molecule $n_{\text{m}}$, bright $n_{\mathcal{B}}$ and dark $n_{\mathcal{D}}$ state populations when (a) $S = 7.11$, $\Gamma_z = 0$ ($N_\nu = 4$) and (b) $S = 0$, $\Gamma_z = 0.007$. From Fig. 6.10, the corresponding propagation velocities are approximately $0.7v^L_{k_0}$ (second red point) and $1.0v^L_{k_0}$ (yellow point). ↩TOF

## 6.4 Summary and outlook ↩

In this chapter we developed a model of organic polariton transport including both many photon modes and a quantum vibrational degree of freedom for each emitter. An efficient implementation of a second-order cumulant expansion afforded spatially resolved dynamics for many molecules. The results presented in a first use of this method demonstrated the ability of the approach to simulate the expansion of a polariton cloud and resolve dark and bright exciton populations, but were limited in their power to make quantitative statements on sub-group-velocity propagation. We now discuss possible next steps to better discern the transport mechanism and other properties of polariton transport in organic materials.

### 6.4.1 Realising diffusive transport ↩

As discussed in the analysis of the results, it was surprising not to observe a cross-over to diffusive behaviour as the vibrational coupling $S$ was switched on. It should be checked whether this remains true for different initial exciton profiles, e.g., by varying the width and density of the Gaussian spot. Similarly, one may question whether the timescale $t \lesssim 0.05$ ps is too short to observe diffusive behaviour, although faithfully exploring late-time dynamics is challenging as we explain further below.

A case for which transport is ballistic at $S = 0$ but becomes diffusive at $S > 0$ would be consistent with the importance of dark state interconversion in the transport mechanism. If a diffusive regime were found, a fit of the RMSD position would yield the diffusion coefficient $D$. This could be compared to the prediction of the kinetic model Eq. (6.1.4). A further comparison could be made to the effect of introducing energetic disorder $\epsilon_n$, which may also produce diffusion.

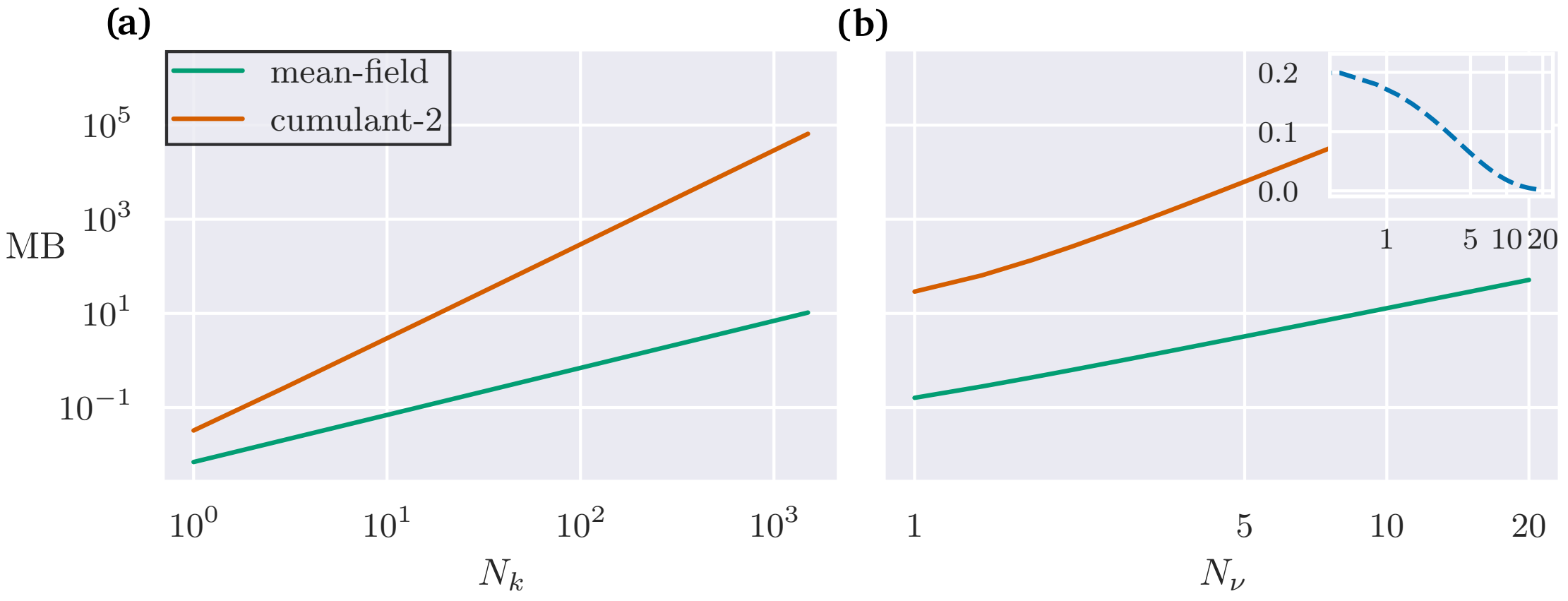

**Figure 6.12:** Memory required in MB ($1024^2$ bytes) to propagate the dynamics as a function of (a) number of modes $N_k$ at fixed $N_\nu = 4$ and (b) number of vibrational levels $N_\nu$ at fixed $N_k = 301$. The values $N_\nu = 4$, $N_k$ were those used for the results in this chapter. In each case we show requirements for both the second-order cumulant approach used in this chapter, and the alternative mean-field approach described in Section 6.4.5. The memory usage is an estimated lower bound based on the size of the system of differential equations at a given $N_k$ and $N_\nu$ and a solver [445] that retains 7 states of this system in memory when integrating the equations of motion. The insert in (b) shows additionally the probability of occupation of the $\omega_\nu = 6$ meV vibrational mode at $N_\nu$ according to the Bose-Einstein distribution at $T = 300$ K. TOF

### 6.4.2 Thermally activated scattering

At the start of this chapter we explained Ref. [116] found trends in transport velocity and diffusion coefficient that were well described by a model in which dark state transfer was thermally activated. An important line of investigation we did not yet explore is to what extent this behaviour can be captured in our model. The phonon temperature $T$ controls both the initial vibrational populations and thermalisation rates $\gamma_\uparrow(T)$, $\gamma_\downarrow(T)$. Hence the temperature dependence of the observed velocity or diffusion coefficient can be tested to follow an Arrhenius-type law (cf. Eqs. (6.1.3) and (6.1.4)). Considering the computational cost of simulation with large $N_\nu$ (see below), this would be most readily done for a range of temperatures well below 300 K, or a model with a higher vibrational frequency $\omega_\nu$, so that a small (e.g. $N_\nu = 4$) number of levels was sufficient to describe the entire vibrational population.

### 6.4.3 Initial photonic excitation

Recall the initial conditions Eq. (6.3.19) resulted in a large initial dark state populations. One might wonder how the picture obtained would change given an initial incoherent population of photons instead of excitons, or a combination of incoherent photons and excitons. Whilst not directly linked to experimental preparation technique, this is an easily accessible next step that may aid in understanding the role of the dark state reservoir when $S > 0$.

### 6.4.4 Computation cost and late-time dynamics

The resource requirements of the computation are determined by the number of modes $N_k$ and number of vibrational levels $N_\nu$. For the second-order cumulant approach, the number of cumulant equations—that is, state variables—scales as $N_k^2 N_\nu^4$ (from $\partial_t \langle \lambda_n^{i+} \lambda_m^{j-} \rangle$). In Fig. 6.12 we show an estimated lower bound for the memory (RAM) required to integrate the differential equations, given the solver used (Runge-Kutta fourth-order) stored multiple instances of the state in memory to calculate each timestep [445]. We now explain how $N_k$ and $N_\nu$ are determined and, in particular, how this makes exploring late-time dynamics challenging.

First, $N_k$ must satisfy the physical requirement of capturing all relevant photon modes and spatial variation of the molecules. Note the momentum cut-off $k_{\text{max}} = (\pi/L)(N_k - 1)$ scales with $N_k$, hence $N_k$ indeed sets the spatial resolution of the simulation. In practice, one checks for numerical convergence of the results under increases of $N_k$: we found $N_k = 301$ as used by Ref. [85] to be suitable for our system parameters and initial conditions. This can be easily seen, for example, in Fig. 6.9c, where momentum states near the cut-off have negligible occupation.

Faithfully simulating dynamics to longer times requires a larger system length $L$ to avoid boundary effects. As the momentum cut-off $k_{\text{max}} \sim N_k/L$, $N_k$ must increase proportionally with $L$ in order to capture the same range of momenta. In this way accessing dynamics at longer times rapidly increases the computational cost. This is unlikely to cause a problem at $S = 0$ (when $N_\nu = 1$), but coupling to even $N_\nu = 4$ levels for the current system size, $L = 60$ $\mu$m and $N_k = 301$, demanded in practice approximately $\sim 10$ GB in memory, making increases of $L$ (hence $N_k$) difficult.

Second, for a model of a harmonic vibrational mode, $N_\nu$ should be large enough so that the highest state included does not acquire a significant population during the dynamics. Due to the rapid scaling of resource requirement with $N_\nu$, it may be difficult to achieve convergence with respect to this parameter at room temperatures. For example, we found even $N_\nu = 6$ levels too demanding for a modern desktop computer with $64$ GB of RAM, yet at $T = 300$ K a thermal state for the vibrational mode $\omega_\nu$ has approximately $10\%$ occupation for the fourth level and $5\%$ for the sixth level (see inset in Fig. 6.12b).

There are several factors that may alleviate the limits of computation. One is access to high performance computing clusters that may have memory pools of many hundreds of GB. Another is that it may be possible to use a differential equation solver that requires few state evaluations in memory to propagate the dynamics. There is also always the option to look for easier sets of physical parameters, either to describe a different type of material or to simply explore the possible physics (i.e., not tied to a physical system at all). For example, decreasing the system energies should make a smaller number of photon modes ($N_k$) relevant to the dynamics, and raising the vibrational frequency $\omega_\nu$ would reduce the population of higher vibrational levels near $N_\nu$. It is quite likely that, given a combination of these factors, longer-time (larger $L$) dynamics or dynamics convergence with respect to $N_\nu$ may be realised. We also now discuss a mean-field approach for which the resource requirements are significantly lower.

### 6.4.5 Mean-field approach

A main difficulty in the analysis above was that the propagating excitation was not a well defined wavepacket but instead comprised a broad distribution of momentum states. While it may be possible to influence the distribution of momenta by varying the width and intensity of the initial exciton population, there is limited scope to change the initial conditions beyond that. This is because the initial state must respect $\mathrm{U}(1)$ symmetry to be compatible with the derived second-order cumulant equations, which are symmetry preserving. So whilst they can follow the off-resonant excitation in experiments, these equations cannot follow a on-resonant excitation, which creates coherences. The opposite is true for a first-order cumulant expansion, i.e., mean-field theory. There the equations must feature symmetry breaking for non-trivial dynamics.

When we first set out to develop a model of polariton transport, there was a choice of using mean-field theory or second-order cumulants. Both appeared to be viable with well motivated experimental initial conditions. We opted for second-order cumulants, given the non-resonant (incoherent) initial conditions used by the experiment [116] inspiring our investigation—well captured by symmetry preserving cumulant equations—as well as the possibility that correlations, for example between modes $\langle a^\dagger_{k'} a_k \rangle$, may be important in resolving ultrafast dynamics. The results have made clearer the potential value of the mean-field approach however, in being able to prepare a coherent wavepacket about a single momentum state $k_0$. The approximate form of the polariton creation operator $L^\dagger_k$, Eq. (6.3.15), could be used for this purpose, corresponding to an on-resonant excitation in experiment and also avoiding the potential complication of creating significant upper polariton populations under non-resonant excitation.

The first main advantage of this approach is that it avoids the ambiguities in having to choose an appropriate wavevector at which to make velocity comparisons. The second is that it provides a significant reduction of computational cost: the mean-field equations are far simpler and of a number that scales as $N_k N_\nu^2$ not $N_k^2 N_\nu^4$. This would make far larger system sizes and higher vibrational levels accessible, see Fig. 6.12. Hence, we believe it worthwhile to pursue in conjunction to further work with the second-order cumulant equations.

### 6.4.6 Pure dephasing model

Finally, further comparisons should be made with the model of pure dephasing $\Gamma_z$, be that with second-order cumulants or mean-field theory. The obvious question to address is whether the saturation observed in Fig. 6.10b does in fact occur at the polariton group velocity. A positive result here would be incompatible with the proposed transport mechanism, since for $\Gamma_z$ significantly non-zero scattering should be sufficient to realise the equilibrated transfer to and from the dark states. It would add to the deficiencies of pure dephasing models compared to exact treatments of vibrational environments—we saw in Chapter 4 how such models cannot account for basic features of organic lasing, for example. A subsequent question is what aspect exactly does the model lack that allows the HTC and other treatments of the vibrational environment [85] to describe sub-group-velocity—or at least slower—transport, and more generally in what way the vibrational environment is an essential component in any account of the transport properties of organic polaritons. These questions are surely not easy to answer.

# Chapter 7

# Conclusion ↑TOC

You've only got to look properly at
anything to see something in it.

John Piper

## Contents

## 7.1 Summary of results ↑

In this thesis, we developed and applied methods for many-body open quantum systems, focusing in particular on realistic models of organic polariton lasing and transport. Our methods were based on assumptions about the structure of, and so correlations in, the many-body state. Starting from the mean-field approximation, which assumes no correlations, higher-order effects were introduced either by calculating fluctuations about the mean-field or by incorporating inter-site correlations via cumulant expansions.

By reducing the dimensionality of the problem these expansions enabled the analysis of both static and dynamic properties of large many-body systems, including the possibility of strong coupling to multiple environments. We addressed the description of the vibrational degrees of freedom of organic molecules, which were included as discrete intramolecular modes or as a broad continuum of external modes. We also examined the validity and convergence behaviour of the mean-field and cumulant expansions for the class of many-to-one models. Understanding the convergence properties of these methods is crucial for assessing their reliability and applicability.

The results of the preceding chapters may be summarised:

- Chapter 4: a method was introduced for simulating open systems with strong coupling to multiple environments that combined a mean-field reduction with a process tensor matrix product operator method to calculate exact, non-Markovian dynamics. The method was demonstrated by determining the lasing threshold and absorption and emission spectra for a realistic model of an organic laser. The method has been implemented for the TEMPO tensor network [162] in the OQuPy Python 3 package [298].

- Chapter 5: the validity and convergence of the mean-field and cumulant expansion methods for a central spin model were investigated. This revealed the failure of mean-field theory to capture the $N \to \infty$ limit of the central spin model under a common $1/\sqrt{N}$ scaling of interaction strength, as well as non-uniform convergence of the cumulant expansions for many-to-one models across even and odd orders of expansion.
- Chapter 6: a second-order cumulant expansion method was developed to calculate spatially resolved dynamics of a multimode Holstein-Tavis-Cummings model. This was used to track the expansion of a polariton cloud, including the dark and bright state populations, in a study of transport in organic materials. Sub-group-velocity propagation was observed and comparisons made to a model of pure dephasing by the vibrational mode.

While we focused on the physics of organic polaritons and the class of many-to-one models, our methods are generally applicable to open many-body problems involving models of high connectivity where mean-field theory is expected to be accurate for sufficiently large systems.

## 7.2 Outlook

We conclude by outlining lines of research that may be pursued following the work of this thesis. These should be taken in addition to the outlook provided in each of Chapters 4 to 6. In particular, next steps to continue the investigation of organic polariton transport were detailed in Chapter 6, and we do not repeat these here.

### 7.2.1 Cumulant TEMPO

In Chapter 4 we showed how mean-field theory could be used in conjunction with the TEMPO tensor network method to efficiently calculate the dynamics of large systems with strong coupling to multiple environments. While the TEMPO method [162] provides an exact numerical calculation for the effective single molecule–mean-field system, the mean-field approximation includes no correlations and is only accurate for large systems $N \to \infty$.

One may naturally look to extend this approach by combining a second-order cumulant expansion with the TEMPO method. That is, starting from equations of motion for reduced density matrices for each type of site in the many-body system, apply the second-order (pairwise correlations) ansatz for the total density operator $\rho$. Then, derive effective evolution equations for the reduced density matrices—now including coherences—to be time evolved using the TEMPO method.

When considering the central boson model, a primary challenge is the infinite Hilbert space associated with the bosonic degree of freedom: TEMPO operates on systems (matrices) of finite dimension. To address this, the most straightforward approach is to truncate the bosonic space to a finite number of levels $N_a$. However, to avoid physical error, $N_a$ should be sufficiently large such that the bosonic mode does not accumulate significant population near the highest included level during the evolution. As the size of the state $\rho_a$ in memory grows with $N_a$, this approach may be limited to small numbers of emitters $N$[1]. Note the degeneracy simplification discussed in Section 3.2.6 could be applied, alleviating the computational cost.

[1] We note prior to the work on the central spin model (Chapter 5) we considered cumulant expansions for the truncated central boson model, and found unexpected (including non-physical) results in cases where $N \gtrsim N_a$.

Alternatively, in certain situations (for example, the dispersive regime of cavity dynamics [397, 446, 447]), one may adiabatically eliminate the bosonic mode to obtain an all-to-all model of coupled emitters. In this case, and for models that contain no bosonic degree of freedom such as all-to-all spin ensembles [328], the description would not be limited by system size $N$. This would allow one to explore corrections to the mean-field approach at arbitrary $N$.

### 7.2.2 Breakdown of Gaussianity and multi-time cumulants

Following the work in Chapter 5, there are further questions regarding the applicability of cumulant expansions that would be valuable to address.

The first arises when considering models with bosonic degrees of freedom, such as the central boson model. As discussed in Section 3.4, when applying a cumulant expansion to such a model one must split instances of the bosonic operator $a$ and its conjugate $a^\dagger$ between moments in order to obtain a closed set of equations. This is not consistent with the form of the cumulant ansatz discussed in this thesis, which prescribes factorisation based on distinct *sites* (Hilbert spaces), not operators. For example, $\langle a^\dagger a\rangle$ should be regarded as a second-order moment. The validity of this factorisation instead relies on the Gaussianity of the bosonic distribution [334]. The question therefore arises as to what implications an assumption of Gaussianity of the bosonic distribution has on the accuracy of the cumulant expansion method.

There is no simple way to avoid this assumption when working with the full representation of the bosonic degree of freedom, since products such as $a^\dagger a$, $a^\dagger a a^\dagger$, etc. cannot be written as linear combinations of single powers of $a$ and $a^\dagger$. However, as in the previous section, one can consider a finite representation of the bosonic site with $N_a$ levels. Then $a^\dagger$ and $a$ are described by $N_a \times N_a$ matrices and the equations close without further approximations. This is because with a finite basis, e.g., of generalised Gell-Mann matrices $\lambda^\alpha$, any product $\lambda^\alpha\lambda^\beta$ may be written as a linear combination of single operators $\lambda^\delta$. Again, $N_a$ must be large enough so as to not incur error, but small enough to allow a numerical implementation. We note that even at second order the number of cumulant equations to be solved scales with $N_a^4$.

To investigate the implications of the Gaussianity assumption, a comparison could be made between the result of a second-order cumulant expansion using the bosonic operator $a$ (and a Gaussianity assumption), and that using a finite representation at the same order. Extending this comparison to third-order would allow for non-Gaussian correlations in the cumulant dynamics, and so a more general exploration of the breakdown of Gaussianity, but would have a higher computational cost. The results of these comparisons would be important for the use of cumulant expansions in studies of light-matter interaction [40, 63, 105, 345–349]. Furthermore, they may provide insight into what physical phenomena can arise from the breakdown of Gaussianity in real systems.

The study of quantum fluctuations in the dynamics of driven-dissipative systems and the preservation of non-Gaussian states holds significant interest for applications in quantum metrology, computing, and information processing [351, 389, 397, 448–451].

A separate outstanding problem is how to correctly derive multi-time correlations within the cumulant expansion framework. Recall in Chapter 4 we determined the spectrum, i.e., the photon Green's functions, by calculating fluctuations around the mean-field solution. The question may then be posed: what would be the equivalent treatment of fluctuations for a higher-order cumulant expansion?

To understand the difficulties in addressing this question, consider an attempt to calculate the photon correlations $\langle a^\dagger(t)a(0)\rangle$ for the Tavis-Cummings model $H = g\sum_n \left(a\sigma_n^+ + a^\dagger\sigma_n^-\right)$ at resonance. Starting from the equations for the single-time expectations,

$$\partial_t \langle a^\dagger(t) \rangle = igN \langle \sigma^+(t) \rangle, \tag{7.2.1}$$
$$\partial_t \langle \sigma^+(t) \rangle = -ig \langle (a^\dagger \sigma^z)(t) \rangle, \tag{7.2.2}$$

one might look to apply the quantum regression theorem (Section 3.1.10) to obtain

$$\partial_t \langle a^\dagger(t) a(0) \rangle = igN \langle \sigma^+(t) a(0) \rangle, \tag{7.2.3}$$
$$\partial_t \langle \sigma^+(t) a(0) \rangle = -ig \langle a^\dagger(t) \sigma^z(t) a(0) \rangle. \tag{7.2.4}$$

This presents what appears to be a 'third-order' moment, $\langle a^\dagger(t)\sigma^z(t)a(0)\rangle$. Even putting aside the problem that $a^\dagger$ and $a$ act on the same physical Hilbert space[2], factorising this moment is not straightforward because $\sigma^z(t)$ and $a(0)$ do not necessarily commute, which should be true for a reliable application of a second-order cumulant expansion (this condition is only obviously satisfied for operators in the Schrödinger picture from different Hilbert spaces). If one ignores these issues and factorises this moment to $\langle a^\dagger(t)a(0)\rangle\langle\sigma^z(t)\rangle$, then the equations can be solved [63, 105], but it is not clear this gives the correct spectrum[3].

Determining a consistent way to apply cumulant expansions in this context would be valuable to investigate corrections to the mean-field spectrum [12, 88, 105, 361] and for the calculation of higher-order coherence functions in quantum optics [452–454].

### 7.2.3 Epilogue

The above questions, along with those concerning the nature of organic polariton transport, represent a selection of areas the author finds compelling to pursue in subsequent work. They are only a fraction of those to be explored within the fields of many-body open quantum systems and organic polaritons that have been the subject of this thesis. Collaboration and discussion with researchers interested in these or other questions is sincerely invited.

[2] Alas, this issue is also no longer simply resolved by a finite representation since e.g. $\lambda^\alpha(t)\lambda^\beta(t')$ cannot generally be simplified unless $t = t'$.

[3] We notice the matrix structure in Eq. (18) of Ref. [105] gives four poles for a generalised Dicke model, but considering the problem of resonance florescence [4] one might expect a five-pole structure. Hence by 'correct' we mean we believe the Green's function calculation (mean-field plus fluctuations) should capture the exact spectrum as $N \to \infty$.

# Appendices

# Appendix A

# Systems: mathematical results ↑TOC

> The domain of applied mathematics and theoretical physics, or mathematical physics as it is often called, may be viewed as a set of nested Chinese boxes in which it is not possible to get at the inner boxes without first opening the outer ones.
>
> Benjamin Moiseiwitsch

## A.1 Tavis-Cummings Eigenstates ↑

Starting from

$$H = \begin{pmatrix} b^\dagger & a^\dagger \end{pmatrix} \begin{pmatrix} \epsilon & g\sqrt{N} \\ g\sqrt{N} & \omega_c \end{pmatrix} \begin{pmatrix} b \\ a \end{pmatrix} \tag{A.1.1}$$

The unitary transformation

$$\begin{pmatrix} U \\ L \end{pmatrix} = \begin{pmatrix} \cos\theta & \sin\theta \\ -\sin\theta & \cos\theta \end{pmatrix} \begin{pmatrix} b \\ a \end{pmatrix} \tag{A.1.2}$$

yields the diagonal form

$$H = \begin{pmatrix} U^\dagger & L^\dagger \end{pmatrix} \begin{pmatrix} \epsilon^U & 0 \\ 0 & \epsilon^L \end{pmatrix} \begin{pmatrix} U \\ L \end{pmatrix} \tag{A.1.3}$$

provided

$$\cos\theta = \frac{1}{\sqrt{2}}\sqrt{1 + \frac{\epsilon - \omega_c}{2\zeta}}, \quad \sin\theta = \frac{1}{\sqrt{2}}\sqrt{1 - \frac{\epsilon - \omega_c}{2\zeta}} \tag{A.1.4}$$

where

$$\zeta = \frac{1}{2}\sqrt{(\epsilon - \omega_c)^2 + 4g^2N}. \tag{A.1.5}$$

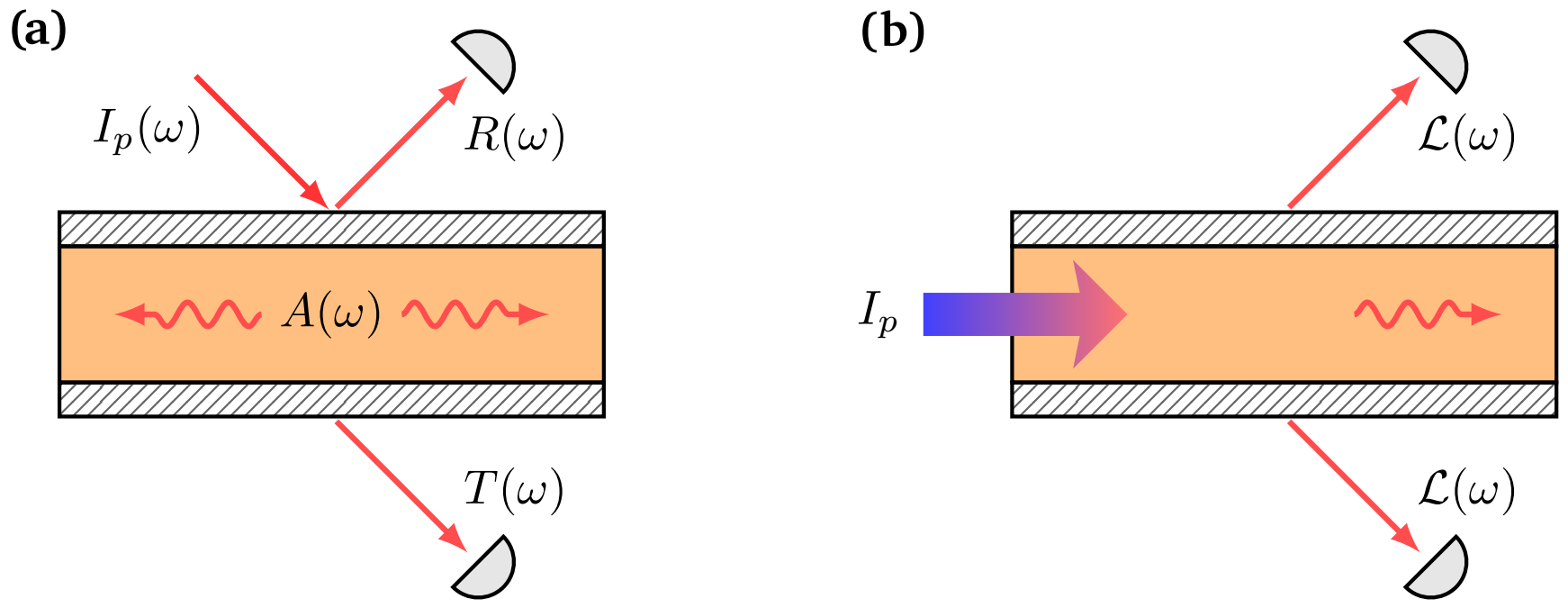

**Figure A.1:** Linear spectroscopy of organic polaritons [54]. (a) A monochromatic laser transfers photons through partially reflecting mirrors with subsequent measurements of reflectivity, transmissivity and absorption. (b) Photoluminescence (PL) experiment with non-resonant pumping. An incoherent exciton population is created which feeds the polariton states. The polaritons formed are not correlated to the pump light. ↑TOF

The upper $(U, +)$ and lower $(L, -)$ polariton energies in Eq. (A.1.3) are

$$\epsilon^{U/L} = \frac{\epsilon + \omega_c}{2} \pm \zeta = \frac{1}{2}\left[\epsilon + \omega_c \pm \sqrt{(\epsilon - \omega_c)^2 + 4g^2 N}.\right] \quad \text{(A.1.6)}$$

Defining $X = \cos\theta$ and $Y = \sin\theta$, the corresponding polariton creation operators are

$$U^\dagger = X \frac{1}{\sqrt{N}} \sum_{n=1}^{N} \sigma_n^+ + Y a^\dagger, \quad \text{(A.1.7)}$$

$$L^\dagger = -Y \frac{1}{\sqrt{N}} \sum_{n=1}^{N} \sigma_n^+ + X a^\dagger, \quad \text{(A.1.8)}$$

having used $b^\dagger = J^+/\sqrt{N}$ where $J^+ = \sum_n \sigma_n^+$ is the collective spin raising operator.

## A.2 Green's functions for optical spectra ↑

In this section we collate results [54, 65, 87, 88, 155] for how the spectroscopic observables of organic polaritons may be calculated via the photon Green's functions.

First, the optical spectra of an organic microcavity are defined as functions of the frequency $\omega$ of incident light in the experiment shown in Fig. A.1a: the reflection $R(\omega)$ and transmission $T(\omega)$ are collected on the near and far sides of the cavity, and that which is unaccounted for, i.e., is absorbed by the medium, is the absorption $A(\omega)$ such that $A(\omega) = 1 - R(\omega) - T(\omega)$ (the conservation of photon flux) [54]. In terms of intracavity dynamics, an incoming photon may be absorbed to create a polariton which subsequently decays producing either a propagating photon that may exit the cavity (contributing to $R$, $T$) or a bound mode excitation that does not $(A)$.

In a typical photoluminescence (PL) experiment on the other hand, Fig. A.1b, optical excitation occurs non-resonantly at high-energies. This produces a population of excitons that may scatter and relax to populate polariton states. The resulting emission from the cavity is the PL $\mathcal{L}(\omega)$. The PL is a most useful tool to study polariton condensation in providing access to polariton populations and (via the emission linewidth) their lifetimes [5].

The measurements are normally angle-resolved: during the excitation process polariton states with different in-plane momenta are populated. Then the PL $\mathcal{L}(\omega) = \mathcal{L}_{\boldsymbol{k}}(\omega)$, for example, provides energy *and* momentum resolved populations. Similarly $R$, $T$, $A$ are generally functions of the in-plane momentum $k$. For simplicity below we continue to write these as functions of energy (frequency) $\omega$ alone, in mind of, e.g., a single-mode Tavis-Cummings model, but similar relations apply to the general $k$-dependent spectra captured by a corresponding multimode model such as described in Chapter 4.

To summarise the polariton experiments discussed in Chapter 2, basic information of the polariton system—the dispersions—is obtained via measurements of the absorption, transmission or reflection spectra. Thereafter, polariton populations and condensation are studied in complementary photoluminescence experiments [56].

For the corresponding theoretical description, the spectra and its non-equilibrium occupation may be determined by two linearly independent photon Green's functions [87]. In the following we consider the 'normal' Green's functions suitable for Hamiltonians with number conserving terms, i.e., under the rotating wave approximation (RWA). For problems where the counter-rotating terms cannot be neglected, the Green's functions becomes matrices with normal (diagonal) and anomalous terms. Details may be found in the book chapter Ref. [87].

We consider the retarded ($R$) and Keldysh ($K$) Green's functions defined by [87]

$$D^R(t) = -i\Theta(t)\langle[a(t), a^\dagger(0)]\rangle, \quad D^K(t) = -i\langle\{a(t), a^\dagger(0)\}\rangle, \tag{A.2.1}$$

where $[\cdot]$ and $\{\cdot\}$ are commutators and anticommutators, and $\langle\cdot\rangle = \mathrm{Tr}[\cdot\rho]$ an expectation with respect to the non-equilibrium steady-state[1].

$D^R(t)$ and $D^K(t)$ are fundamentally linear response functions of the system, here capturing field fluctuations about the steady-state. One normally works with these functions in Fourier space,

$$D^R(\omega) = -\int_0^\infty dt\, e^{i\omega t}\langle[a(t), a^\dagger(0)]\rangle, \quad D^K(\omega) = -i\int_{-\infty}^\infty dt\, e^{i\omega t}\langle\{a(t), a^\dagger(0)\}\rangle. \tag{A.2.2}$$

We note in particular $\varrho(\omega) = -2\,\mathrm{Im}[D^R]$ has the interpretation of the spectral weight which is the density of states of excitations. Note for a system in equilibrium, there is only one independent Green's function and this is often taken as $D^R$.

For the non-interacting problem (subscript 0), meaning no light-matter coupling, the Green's functions take the simple form [87]

$$D_0^R(\omega) = \frac{1}{\omega - \omega_c + i\kappa/2}, \quad D_0^K(\omega) = -\frac{i\kappa}{(\omega-\omega_c)^2 + (\kappa/2)^2}, \tag{A.2.3}$$

where $\omega_c$ is the cavity frequency and $\kappa = \kappa_L + \kappa_R$ the cavity decay rate[2] due to losses from the two ('left' and 'right') cavity mirrors.

For the interacting problem, $g > 0$, the Green's functions may be determined by expanding the evolution operator $U$ in the interaction picture (cf. Section 3.1). The result [87] is a Dyson equation for the inverse retarded or Keldysh Green's functions of the form $D^{-1}(\omega) = D_0^{-1}(\omega) - \Sigma(\omega)$, where $\Sigma$ is a molecular self-energy. For the model of $N$ emitters interacting with a single

[1]Considering the response of the steady-state means the Green's functions are functions of a single time argument. Without stationarity, one has $D^R = D^R(t,t')$, $D^K = D^K(t,t')$ where, e.g., $D^R(t,t') = -i\Theta(t-t')\langle[a(t), a^\dagger(t')]\rangle$.

[2]To be specific, the field $\langle a\rangle$ decays at rate $\kappa/2$ and the photon number $\langle a^\dagger a\rangle$ at $\kappa$. This definition is in keeping with a Lindblad term $\kappa L[a]$ in the master equation used throughout the thesis. Note some authors [12] may chose to define $\kappa$ as the field decay rate corresponding to $+2\kappa L[a]$ in the master equation.

photon mode, in the large $N$ limit the relevant self-energies are determined by [65, 87]:

$$\Sigma^{-+}(\omega) = -ig^2 \int_0^\infty dt e^{i\omega t} \langle [\sigma^-(t), \sigma^+(0)] \rangle, \tag{A.2.4}$$

$$\Sigma^{--}(\omega) = -ig^2 \int_{-\infty}^\infty dt e^{i\omega t} \langle \{\sigma^-(t), \sigma^+(0)\} \rangle. \tag{A.2.5}$$

The Green's functions are then

$$D^R(\omega) = \frac{1}{\omega - \omega_c + i\kappa/2 - \Sigma^{-+}(\omega)}, \tag{A.2.6}$$

$$D^K(\omega) = \frac{\Sigma^{--}(\omega) - 2i\kappa}{|\omega - \omega_c + i\kappa/2 - \Sigma^{-+}(\omega)|^2}. \tag{A.2.7}$$

Corresponding expressions for the $k$-dependent spectra are provided in Chapter 4. Be aware there we use the Rabi splitting $\Omega = 2g\sqrt{N}$ to denote the light-matter coupling.

Expressions for the linear optical response spectra for organic polaritons have been determined using the input-output formalism [65, 88, 155, 455]. We have

$$T(\omega) = \kappa_L \kappa_R \left| D^R(\omega) \right|^2, \tag{A.2.8}$$

$$R(\omega) = 1 + 2\kappa_L \operatorname{Im} D^R(\omega) + \kappa_L^2 \left| D^R(\omega) \right|^2, \tag{A.2.9}$$

$$A(\omega) = -2\kappa_L \operatorname{Im} D^R(\omega) - \kappa_L \kappa \left| D^R(\omega) \right|^2, \tag{A.2.10}$$

If $\kappa_L = \kappa_R = \kappa/2$, these simplify to

$$T(\omega) = (\kappa/2)^2 \left| D^R(\omega) \right|^2, \tag{A.2.11}$$

$$R(\omega) = 1 + \kappa \operatorname{Im} D^R(\omega) + (\kappa/2)^2 \left| D^R(\omega) \right|^2, \tag{A.2.12}$$

$$A(\omega) = -\kappa \operatorname{Im} D^R(\omega) - (\kappa^2/2) \left| D^R(\omega) \right|^2. \tag{A.2.13}$$

This makes it clear how the spectral weight $\varrho = -2 \operatorname{Im} D^R$ gives the absorption of a good cavity ($\kappa^2 \ll \kappa$). In any case, $T + R + A \equiv 1$, as required by conservation of photon flux in the absence of other sources of pumping.

The photoluminescence on the other hand is defined by [87]

$$\mathcal{L}(\omega) = \frac{i}{2} \left( D^K(\omega) - D^R(\omega) + \overline{D^R}(\omega) \right). \tag{A.2.14}$$

In Chapter 4 we show that this may be written as

$$\mathcal{L}(\omega) = \varrho(\omega) n(\omega), \tag{A.2.15}$$

where $\varrho = -2 \operatorname{Im} D^R$ is the spectral weight as before and $n(\omega)$ defines the mode occupation, which satisfies

$$n(\omega) = \frac{1}{2} \left[ \frac{i D^K(\omega)}{\varrho(\omega)} - 1 \right]. \tag{A.2.16}$$

The occupation and spectral weight are inspected in the normal state of a driven-dissipative multimode Tavis-Cummings model in Chapter 4. Using the method developed in that chapter and the generalisation of the self-energies Eqs. (A.2.4) and (A.2.5) to the multimode case, $k$-dependent spectra are also calculated.

# Appendix B

# Open systems: mathematical results

TOC

> In this class, I hope you will learn not merely results, or formulae applicable to cases that may possibly occur in our practice afterwards, but the principles on which those formulae depend, and without which the formulae are mere mental rubbish.
>
> James Maxwell

## B.1 Propagator for time-dependent Hamiltonian

The standard derivation of the series solution for the evolution operator $U$ is via integration and iteration [2]. Without loss of generality we set $t_0 = 0$ and write $U(t,0) \equiv U(t)$. Starting from

$$\partial_t U(t) = -iH(t)U(t), \tag{B.1.1}$$

a single integration gives

$$U(t) = 1 - i\int_0^t dt_1 H(t_1)U(t_1), \tag{B.1.2}$$

having used $U(0) = U(0,0) = 1$ (the identity). Substituting the left-hand side of Eq. (B.1.2) into the right side repeatedly (the 'iteration' part) produces a series

$$U(t) = 1 - i\int_0^t dt_1 H(t) + (-i)^2\int_0^t dt_1 \int_0^{t_1} dt_2 H(t_1)H(t_2) + \dots \tag{B.1.3}$$

$$= 1 + \sum_{n=0}^{\infty}(-i)^n \int_0^t dt_1 \int_0^{t_1} dt_2 \dots \int_0^{t_{n-1}} dt_n H(t_1)H(t_2)\dots H(t_n), \tag{B.1.4}$$

where we note that under each integral we have $t_1 > t_2 > \ldots > t_n$. If we introduce the time ordering operator $\mathcal{T}$,

$$\int_0^t dt_1 \int_0^t dt_2 \mathcal{T} H(t_1)H(t_2) = \int_0^t dt_1 \int_0^{t_1} dt_2 H(t_1)H(t_2) + \int_0^t dt_2 \int_0^{t_2} dt_2 H(t_2)H(t_1), \quad \text{(B.1.5)}$$

where the limits of the integrals must be carefully noted. The terms on the right hand-side actually give the same result, since they are related by the change of integration variables $t_1 \to t_2$, $t_2 \to t_1$. Therefore we can write

$$\frac{1}{2!}\int_0^t dt_1 \int_0^t dt_2 \mathcal{T} H(t_1)H(t_2) = \int_0^t dt_1 \int_0^{t_1} dt_2 H(t_1)H(t_2). \quad \text{(B.1.6)}$$

It is possible to show for the $n^{\text{th}}$ term

$$\frac{1}{n!}\int_0^t dt_1 \int_0^t dt_2 \ldots \int_0^t dt_n \mathcal{T} H(t_1)H(t_2)\ldots H(t_n) = \int_0^t dt_1 \int_0^{t_1} dt_2 \ldots \int_0^{t_{n-1}} dt_n H(t_1)H(t_2)\ldots H(t_n). \quad \text{(B.1.7)}$$

In other words, the time-ordering operator allows us to make the upper limits of all the integrals equal to $t$, at the cost of a factorial $1/n!$ to avoid over-counting the result. But this this also allows the series solution Eq. (B.1.4) to be expressed as an exponential:

$$U(t) = 1 + \sum_{n=0}^{\infty}(-i)^n \int_0^t dt_1 \int_0^{t_1} dt_2 \ldots \int_0^{t_{n-1}} dt_n H(t_1)H(t_2)\ldots H(t_n) \quad \text{(B.1.8)}$$

$$= 1 + \sum_{n=0}^{\infty}\frac{(-i)^n}{n!} \int_0^t dt_1 \int_0^{t} dt_2 \ldots \int_0^{t} dt_n H(t_1)H(t_2)\ldots H(t_n) \quad \text{(B.1.9)}$$

$$= \mathcal{T}\exp\left[-i\int_0^t dt' H(t')\right]. \quad \text{(B.1.10)}$$

This is known as Dyson's formula after F. Dyson who first presented the series solution in 1949 [456]. While the time-ordered exponential appears rather formal, it is very useful for formulating the TEMPO method in the time-dependent case (Section 3.2). There the integral in the exponential may either by exactly performed, or otherwise well approximated by quadrature, so that the exponential can be computed by numerical methods [457].

## B.2 Commutator algebra

In the derivation of the Heisenberg equations it is important to avoid mistakes when evaluating commutators involving multiple operators. The rules to use here are

$$[AB, C] = A[B, C] + [A, C]B, \quad [A, CD] = C[A, D] + [A, C]D. \quad \text{(B.2.1)}$$

Or, in combination,

$$[AB, CD] = AC[B, D] + A[B, C]D + C[A, D]B + [A, C]DB \quad \text{(B.2.2)}$$

$$= CA[B, D] + A[B, C]D + C[A, D]B + [A, C]BD. \quad \text{(B.2.3)}$$

The last two being equivalent, that is, you can factor out the left $(AC)$ or right $(CD)$ parts of the commutator first, provided you are consistent with which side of the commutator you extract the operators to.

The previous formula simplifies when a subset of the operators commute. This happens in particular when $A$,$C$ act on one Hilbert space and $B$, $D$ on a second, i.e., $AB = A \otimes B$ and $CD = C \otimes D$ are tensor products. Then, since operators from different spaces commute (they act identically on each other's parts), $[B, C] = [A, D] = 0$ and

$$[AB, CD] = AC[B, D] + [A, C]DB = ACBD - CADB. \tag{B.2.4}$$

Note this is *not* equivalent to $[A, C] \otimes [B, D]$: you cannot factorise the commutator of tensor products as a tensor product of commutators.

# Appendix C

# Cumulant formulae

> This may look like a neatly arranged stack of numbers, but it's actually a mathematical treasure trove.
>
> Wajdi Ratemi

## C.1 Second-order cumulant ansatz

### C.1.1 Closed expression

The second-order ansatz Eq. (3.4.2) may be expressed as

$$\rho = \bigotimes_{i=1}^{N} \rho_i + \sum_{N_p=2}^{N//2} \left[ \left( \prod_{\mathrm{x}=0}^{N_p-1} \sum_{i_{2\mathrm{x}+1} > i_{2\mathrm{x}-1}}^{\prime} \sum_{i_{2\mathrm{x}+2} > i_{2\mathrm{x}+1}}^{\prime} \tau_{i_{2\mathrm{x}+1}, i_{2\mathrm{x}+2}} \right) \bigotimes_{k \neq i_{\mathrm{x}}} \rho_k \right], \tag{C.1.1}$$

where the prime over each sum indicates that all indices must be distinct and we define $i_{-1} = 0$ so that $i_1 > 0$, $i_2 > i_1$, $i_3 > i_1$, $i_4 > i_3$,.... The number of pairs $N_p$ runs from 2 to $N//2$ (integer floor division). Note carefully the placement of brackets in Eq. (C.1.1): at each $N_p$ there is a product of $2N_p$ sums, two for each pair of indices of a $\tau$ matrix, and then a tensor product of reduced density matrices for all those sites not included in the pairings.

For a term with $N_p$ pairs there are $\binom{N}{2}$ ways to chose the first pair of sites where ($\binom{n}{r}$) the binomial coefficient), $\binom{N-2}{2}$ to choose the second, and so on, such that the total number of *unique* terms at each $N_p$ is

$$\frac{1}{N_p!} \binom{N}{2} \binom{N-2}{2} \cdots \binom{N-2(N_p-1)}{2} = \frac{1}{N_p!} \frac{1}{2^{N_p}} \frac{N!}{(N-2N_p)!}. \tag{C.1.2}$$

### C.1.2 Compatibility with nested cumulant expansions

Consider calculating a fourth-order moment $\langle A_1A_2A_3A_4\rangle = \mathrm{Tr}\left(A_1A_2A_3A_4\rho\right)$. Since the trace of a $\tau_{ij}$ matrix without operators for both $i$ and $j$ vanishes, the only relevant part of the ansatz is

$$\begin{aligned}\rho \ni \bigotimes_{i=1}^{N}\rho_i + \sum_{i=1}^{4}\sum_{j>i}^{4}\tau_{ij}\bigotimes_{k\neq i,j}\rho_k \\ + \sum_{i=1}^{4}\sum_{j>i}^{4}\sum_{r>i}^{4}\sum_{s>r}^{4}\tau_{ij}\tau_{rs}\bigotimes_{k\neq i,j,r,s}\rho_k .\end{aligned} \tag{C.1.3}$$

Let's determine the contribution from the top line first. Remembering the reduced density matrices are normalised to 1, and $\mathrm{Tr}_{ij}\left(A_iA_j\tau_{ij}\right)$ is equal to the joint cumulant of $A_1$, $A_2$, we have

$$\begin{aligned}\langle A_1\rangle\langle A_2\rangle\langle A_3\rangle\langle A_4\rangle &+ \Big(\langle A_1A_2\rangle - \langle A_1\rangle\langle A_2\rangle\Big)\langle A_3\rangle\langle A_4\rangle \\ &+ \Big(\langle A_1A_3\rangle - \langle A_1\rangle\langle A_3\rangle\Big)\langle A_2\rangle\langle A_4\rangle \\ &+ \Big(\langle A_1A_4\rangle - \langle A_1\rangle\langle A_4\rangle\Big)\langle A_2\rangle\langle A_3\rangle \\ &+ \Big(\langle A_2A_3\rangle - \langle A_2\rangle\langle A_3\rangle\Big)\langle A_1\rangle\langle A_4\rangle \\ &+ \Big(\langle A_2A_4\rangle - \langle A_2\rangle\langle A_4\rangle\Big)\langle A_1\rangle\langle A_3\rangle \\ &+ \Big(\langle A_3A_4\rangle - \langle A_3\rangle\langle A_4\rangle\Big)\langle A_1\rangle\langle A_2\rangle\end{aligned} \tag{C.1.4}$$

$$\begin{aligned}= -5\langle A_1\rangle\langle A_2\rangle\langle A_3\rangle\langle A_4\rangle + \langle A_1A_2\rangle\langle A_3\rangle\langle A_4\rangle + \langle A_1A_3\rangle\langle A_2\rangle\langle A_4\rangle + \langle A_1A_4\rangle\langle A_2\rangle\langle A_3\rangle \\ + \langle A_2A_3\rangle\langle A_1\rangle\langle A_4\rangle + \langle A_2A_4\rangle\langle A_1\rangle\langle A_3\rangle + \langle A_3A_4\rangle\langle A_1\rangle\langle A_2\rangle .\end{aligned} \tag{C.1.5}$$

In the second line, there are actually only three terms:

$$\begin{aligned}&\mathrm{Tr}_{12}\left(A_1A_2\tau_{12}\right)\mathrm{Tr}_{34}\left(A_3A_4\tau_{34}\right) \\ &+ \mathrm{Tr}_{13}\left(A_1A_3\tau_{13}\right)\mathrm{Tr}_{24}\left(A_2A_4\tau_{24}\right) \\ &+ \mathrm{Tr}_{14}\left(A_1A_4\tau_{14}\right)\mathrm{Tr}_{23}\left(A_2A_3\tau_{23}\right)\end{aligned} \tag{C.1.6}$$

$$\begin{aligned}&= \Big(\langle A_1A_2\rangle - \langle A_1\rangle\langle A_2\rangle\Big)\Big(\langle A_3A_4\rangle - \langle A_3\rangle\langle A_4\rangle\Big) \\ &+ \Big(\langle A_1A_3\rangle - \langle A_1\rangle\langle A_3\rangle\Big)\Big(\langle A_2A_4\rangle - \langle A_2\rangle\langle A_4\rangle\Big) \\ &+ \Big(\langle A_1A_4\rangle - \langle A_1\rangle\langle A_4\rangle\Big)\Big(\langle A_2A_3\rangle - \langle A_2\rangle\langle A_3\rangle\Big).\end{aligned} \tag{C.1.7}$$

Combining Eqs. (C.1.5) and (C.1.7),

$$\begin{aligned}\langle A_1A_2A_3A_4\rangle_{\text{ans}} = &- 2\langle A_1\rangle\langle A_2\rangle\langle A_3\rangle\langle A_4\rangle + \langle A_1A_2\rangle\langle A_3A_4\rangle \\ &+ \langle A_1A_3\rangle\langle A_2A_4\rangle + \langle A_1A_4\rangle\langle A_2A_3\rangle ,\end{aligned} \tag{C.1.8}$$

where the label ans is to distinguish this as arising from the ansatz. This is to be compared to performing a nested cumulant expansion (nest). Using the expressions for the third and fourth order cumulants provided in Appendix C.2 below, $\langle\langle A_1A_2A_3A_4\rangle\rangle = 0$ implies

$$\begin{gathered}\langle A_1A_2A_3A_4\rangle_{\text{nest}} = \\ \langle A_1\rangle\langle A_2A_3A_4\rangle + \langle A_1A_3A_4\rangle\langle A_2\rangle + \langle A_1A_2A_3\rangle\langle A_4\rangle + \langle A_1A_2A_4\rangle\langle A_3\rangle \\ +\langle A_1A_2\rangle\langle A_3A_4\rangle + \langle A_1A_4\rangle\langle A_2A_3\rangle + \langle A_1A_3\rangle\langle A_2A_4\rangle - 3\langle A_1\rangle\langle A_2\rangle\langle A_3A_4\rangle \\ -2\langle A_1\rangle\langle A_2A_3\rangle\langle A_4\rangle - 2\langle A_1\rangle\langle A_2A_4\rangle\langle A_3\rangle - 2\langle A_1A_2\rangle\langle A_3\rangle\langle A_4\rangle \\ -2\langle A_1A_3\rangle\langle A_2\rangle\langle A_4\rangle - 2\langle A_1A_4\rangle\langle A_2\rangle\langle A_3\rangle + 6\langle A_1\rangle\langle A_2\rangle\langle A_3\rangle\langle A_4\rangle.\end{gathered} \tag{C.1.9}$$

Then

$$\langle\langle A_1A_2A_3\rangle\rangle = \langle\langle A_1A_2A_4\rangle\rangle = \langle\langle A_2A_3A_4\rangle\rangle = 0 \tag{C.1.10}$$

gives

$$\begin{gathered}\langle A_1A_2A_3A_4\rangle_{\text{nest}} = \\ \langle A_1\rangle\Big(\langle A_2\rangle\langle A_3A_4\rangle + \langle A_3\rangle\langle A_2A_4\rangle + \langle A_4\rangle\langle A_2A_3\rangle - 2\langle A_2\rangle\langle A_3\rangle\langle A_4\rangle\Big) \\ \langle A_2\rangle\Big(\langle A_3\rangle\langle A_1A_4\rangle + \langle A_1\rangle\langle A_3A_4\rangle + \langle A_4\rangle\langle A_1A_3\rangle - 2\langle A_1\rangle\langle A_3\rangle\langle A_4\rangle\Big) \\ \langle A_3\rangle\Big(\langle A_1\rangle\langle A_2A_4\rangle + \langle A_2\rangle\langle A_1A_4\rangle + \langle A_4\rangle\langle A_1A_2\rangle - 2\langle A_1\rangle\langle A_2\rangle\langle A_4\rangle\Big) \\ \langle A_4\rangle\Big(\langle A_1\rangle\langle A_2A_3\rangle + \langle A_2\rangle\langle A_1A_3\rangle + \langle A_3\rangle\langle A_1A_2\rangle - 2\langle A_1\rangle\langle A_2\rangle\langle A_3\rangle\Big) \\ +\langle A_1A_2\rangle\langle A_3A_4\rangle + \langle A_1A_4\rangle\langle A_2A_3\rangle + \langle A_1A_3\rangle\langle A_2A_4\rangle - 2\langle A_1\rangle\langle A_2\rangle\langle A_3A_4\rangle \\ -2\langle A_1\rangle\langle A_2A_3\rangle\langle A_4\rangle - 2\langle A_1\rangle\langle A_2A_4\rangle\langle A_3\rangle - 2\langle A_1A_2\rangle\langle A_3\rangle\langle A_4\rangle \\ -2\langle A_1A_3\rangle\langle A_2\rangle\langle A_4\rangle - 2\langle A_1A_4\rangle\langle A_2\rangle\langle A_3\rangle + 6\langle A_1\rangle\langle A_2\rangle\langle A_3\rangle\langle A_4\rangle\end{gathered} \tag{C.1.11}$$

$$\begin{aligned}&= -2\langle A_1\rangle\langle A_2\rangle\langle A_3\rangle\langle A_4\rangle + \langle A_1A_2\rangle\langle A_3A_4\rangle \\ &\quad + \langle A_1A_3\rangle\langle A_2A_4\rangle + \langle A_1A_4\rangle\langle A_2A_3\rangle\end{aligned} \tag{C.1.12}$$

$$= \langle A_1A_2A_3A_4\rangle_{\text{ans}} \tag{C.1.13}$$

So, the second-order ansatz permits a nested cumulant expansion can be made to factorise fourth-order moments. We leave it as an exercise to check the prediction for higher-order moments in the unlikely case those are required.

## C.2 Joint cumulants in terms of moments

### C.2.1 General expression

The joint cumulant for $n$ operators $A_1$, $A_2$,… $A_n$ on distinct sites can be expressed in terms of their mixed moments [344] (expectations) as

$$\langle\langle A_1, \ldots, A_n\rangle\rangle = \sum_{\pi\in\mathcal{P}_n} (-1)^{|\pi|-1}\,(|\pi|-1)! \prod_{B\in\pi}\left\langle \prod_{i\in B} A_i\right\rangle. \tag{C.2.1}$$

Here $\mathcal{P}_n$ is the collection of all possible partitions of $\{1,\ldots,n\}$, $B$ runs through all the blocks of the partition $\pi$, and $|\pi|$ is the number of parts in the partition (for example, for $\pi = \{\{1\},\{2,\ldots,n\}\} \in \mathcal{P}_n$, $B = \{1\}$ or $\{2,\ldots,n\}$ and $|\pi| = 2$). To get the approximation for the $n^{\text{th}}$ moment correspond to setting the $n^{\text{th}}$ cumulant to zero, simply exclude the partition with one part (the entire set) and add a minus sign:

$$\langle A_1,\ldots,A_n\rangle = \sum_{\substack{\pi\in\mathcal{P}_n \\ \pi\neq\{\{1,\ldots,n\}\}}} (-1)^{|\pi|}\,(|\pi|-1)!\prod_{B\in\pi}\left\langle\prod_{i\in B} A_i\right\rangle. \tag{C.2.2}$$

Formula Eq. (C.2.1) seems quite menacing, but is rather easy to follow in practice. Here are the first few orders:

### C.2.2 Second order

$$\langle\langle A_1A_2\rangle\rangle = \langle A_1A_2\rangle - \langle A_1\rangle\langle A_2\rangle$$

### C.2.3 Third order

$$\langle\langle A_1A_2A_3\rangle\rangle = \langle A_1A_2A_3\rangle - \langle A_1\rangle\langle A_2A_3\rangle - \langle A_1A_2\rangle\langle A_3\rangle - \langle A_1A_3\rangle\langle A_2\rangle + 2\langle A_1\rangle\langle A_2\rangle\langle A_3\rangle$$

### C.2.4 Fourth order

$$\begin{aligned}\langle\langle A_1A_2A_3A_4\rangle\rangle = \langle A_1A_2A_3A_4\rangle - \langle A_1\rangle\langle A_2A_3A_4\rangle - \langle A_1A_3A_4\rangle\langle A_2\rangle - \langle A_1A_2A_3\rangle\langle A_4\rangle \\ -\langle A_1A_2A_4\rangle\langle A_3\rangle - \langle A_1A_2\rangle\langle A_3A_4\rangle - \langle A_1A_4\rangle\langle A_2A_3\rangle - \langle A_1A_3\rangle\langle A_2A_4\rangle \\ +2\langle A_1\rangle\langle A_2\rangle\langle A_3A_4\rangle + 2\langle A_1\rangle\langle A_2A_3\rangle\langle A_4\rangle + 2\langle A_1\rangle\langle A_2A_4\rangle\langle A_3\rangle + 2\langle A_1A_2\rangle\langle A_3\rangle\langle A_4\rangle \\ +2\langle A_1A_3\rangle\langle A_2\rangle\langle A_4\rangle + 2\langle A_1A_4\rangle\langle A_2\rangle\langle A_3\rangle - 6\langle A_1\rangle\langle A_2\rangle\langle A_3\rangle\langle A_4\rangle\end{aligned}$$

To reiterate the main text, $A_1$, $A_2$,... should strictly be operators corresponding to different sites (Hilbert spaces). Additionally, many of the terms in the above expressions may vanish for models with known symmetries.

It seems unlikely higher-order cumulants would be needed if working by hand. The `QuantumCumulants.jl` package [352] is able to calculate cumulant equations of any desired order, but note this program does not automatically take into account of symmetries of the model, so even by third or fourth order the produced list of symbolic equations can be formidable to parse.

## C.3 Counting cumulant coefficients

To sum all the coefficients in a cumulant expansion we firstly need the number of ways to partition a set of $n$ elements in $k$ subsets. This is given by Stirling numbers of the second kind $S(n,k)$, which satisfy the boundary values [458]

$$S(n,n+1) = 0 \text{ and } S(n,0) = 0 \text{ for } n>1, \tag{C.3.1}$$

as well as the recursive formula

$$S(n,k) = kS(n-1,k) + S(n,k-1). \tag{C.3.2}$$

The sum over all coefficients in Eq. (C.2.1) may then be expressed as

$$\Sigma[n] = \sum_{k=1}^{n} (-1)^{k-1}(k-1)!S(n,k). \tag{C.3.3}$$

This may be proven [459] to vanish for $n > 1$ by induction ($\Sigma[1] = 1$ for the trivial first-order cumulant $\langle\langle A_1\rangle\rangle = \langle A_1\rangle$). Suppose $\Sigma[n] = 0$ for $n \in \mathbb{N}$. Using Eq. (C.3.2),

$$\Sigma[n+1] = \sum_{k=1}^{n+1} (-1)^{k-1}(k-1)!S(n,k) \tag{C.3.4}$$

$$= \sum_{k=1}^{n+1} (-1)^{k-1}(k-1)!\,(kS(n,k) + S(n,k-1)) \tag{C.3.5}$$

$$= \sum_{k=1}^{n+1} (-1)^{k-1}k!S(n,k) - \sum_{k=1}^{n+1} (-1)^{k}(k-1)!S(n,k-1) \tag{C.3.6}$$

$$= \sum_{k=1}^{n+1} (-1)^{k-1}k!S(n,k) - \sum_{k=0}^{n+1} (-1)^{k-1}k!S(n,k) \tag{C.3.7}$$

$$= (-1)^{n+1}(n+1)!S(n,n+1) - (-1)^0 0!S(n,0) \tag{C.3.8}$$

$$= 0 \text{ for } n > 1. \tag{C.3.9}$$

Therefore the coefficients in the definition of an $n^{\text{th}}$-order cumulant ($n > 1$) sums to $0$. Equivalently, the coefficients in the corresponding expansion for the $n^{\text{th}}$-order momentum sum to $1$.

There is the question of whether expressions for the sum of coefficients for the symmetry-preserving cumulants expansions can be derived, i.e., when moments of certain patterns of operators vanish. In particular, in Chapter 5 we found the cumulants $\langle\langle\sigma_0^+\sigma_n^-\sigma_m^+\sigma_k^- \ldots\rangle\rangle$ involving raising and lowering operators only had expansions with coefficients that did not sum to $0$, but (by direct evaluation), $-1$, $4$, $-33$, $456$... for the fourth, sixth,... cumulants (corresponding to the odd orders of expansion). Well here's a fun observation [460]:

Start writing out the values of Pascal's triangle, squared.

$$\begin{matrix} 1 \\ 1 & 1 \\ 1 & 2 & 1 \\ 1 & 3 & 3 & 1 \\ 1 & 4 & 6 & 4 & 1 \\ 1 & 5 & 10 & 10 & 5 & 1 \\ 1 & 6 & 15 & 20 & 15 & 6 & 1 \end{matrix} \quad \longrightarrow \quad \begin{matrix} 1 \\ 1 & 1 \\ 1 & 4 & 1 \\ 1 & 9 & 9 & 1 \\ 1 & 16 & 36 & 16 & 1 \\ 1 & 25 & 100 & 100 & 25 & 1 \\ 1 & 36 & 225 & 400 & 225 & 36 & 1 \end{matrix} \tag{C.3.10}$$

Arrange these in a matrix

$$A = \begin{pmatrix} 1 & 0 & 0 & 0 & 0 & 0 & 0 \\ 1 & 1 & 0 & 0 & 0 & 0 & 0 \\ 1 & 4 & 1 & 0 & 0 & 0 & 0 \\ 1 & 9 & 9 & 1 & 0 & 0 & 0 \\ 1 & 16 & 36 & 16 & 1 & 0 & 0 \\ 1 & 25 & 100 & 100 & 25 & 1 & 0 \\ 1 & 36 & 225 & 400 & 225 & 36 & 1 \end{pmatrix} \tag{C.3.11}$$

Now take a logarithm

$$\ln A = \begin{pmatrix} 0 & 0 & 0 & 0 & 0 & 0 & 0 \\ 1 & 0 & 0 & 0 & 0 & 0 & 0 \\ -1 & 4 & 0 & 0 & 0 & 0 & 0 \\ 4 & -9 & 9 & 0 & 0 & 0 & 0 \\ -33 & 64 & -36 & 16 & 0 & 0 & 0 \\ 456 & -825 & 400 & -100 & 25 & 0 & 0 \\ -9460 & 16416 & -7425 & 1600 & -225 & 36 & 0 \end{pmatrix} \tag{C.3.12}$$

and the first column appears to generate the sequence we observed! I was able to verify that -9460 was indeed the sum of coefficient for the $12^{\text{th}}$ cumulant of this type, but haven't made any progress in proving the connection. The sequence $a_n = 1, -1, 4, -33, \ldots$ is in fact related to the coefficients of the log-Bessel function [461] and has a nice recurrence formula [462],

$$a_n = n! b_n, \quad b_{n+1} = \sum_{k=1}^{n} \frac{n}{n+1} \binom{n-1}{k-1} b_k b_{n+1-k}, \quad b_1 = -1, \tag{C.3.13}$$

for $n > 0$.

# Appendix D

# Mean-field TEMPO ↑TOC

Mean-field theory's broken, and everyone uses it.

Martin Long

## D.1 Convergence of dynamics

In this section we provide values of the computational parameters relevant to the process tensor TEMPO algorithm (hereafter 'PT-TEMPO') used in Chapter 4, justified by convergence tests of the dynamics.

As discussed in Section 3.2, there are three computational parameters to consider in a PT-TEMPO calculation: timestep size $\delta t$, precision (relative singular value cutoff) $\epsilon_{\mathrm{rel}}$ and memory length $K$. The first requirement is for $K\delta t$ to be greater than physical correlation times of the system. In fact, we found that if the effective discontinuity introduced into the bath autocorrelation by truncating the PT after $K$ steps was significant on the scale set by $\epsilon_{\mathrm{rel}}$, then a large bond dimension resulted (Fig. D.1). That is, the cutoff effectively implies $C_{\mathrm{eff}}(t) = C(t)\Theta(K\delta t - t)$, and the sharp step function leads to the existence of many singular values of order $C(K\delta t)$ in the process tensor. When $C(K\delta t) \gtrsim \epsilon_{\mathrm{rel}}$, this significantly increases the bond dimension. At high precisions avoiding this issue required $K\delta t \gtrsim 80$ fs in excess of any correlation times in the system and hence the memory cutoff had no effect on the accuracy of our calculations (Fig. D.2a).

Figures D.2b and D.2c show, respectively, convergence tests under changes in $\delta t$ and $\epsilon_{\mathrm{rel}}$ where the value of the photon number $n/N$ was recorded (crosses) at $t = 0.66$ ps for one set of system parameters ($\Omega = 200$ meV, $\Delta = 20$ meV, $\Gamma_{\uparrow} = 0.4\Gamma_{\downarrow}$). In these panels the data corresponding to the computational parameters that were finally chosen, $\delta t = 0.4$ fs and $\epsilon_{\mathrm{rel}} = 5 \times 10^{-12}$, is indicated with a red circle. For comparison, we include results (filled circles) obtained using the original (non-PT) implementation [162] of the TEMPO method. Note that the accuracy of the two algorithms for a given set of computational parameters is not necessarily the same, because of the different ordering of tensor contractions in the two approaches. In particular, we noticed the error in the PT-TEMPO calculation become unstable below $\delta t = 0.4$ fs at $\epsilon_{\mathrm{rel}} = 5 \times 10^{-12}$ (Fig. D.2b) whilst the non-PT results remained stable down to $\delta t = 0.1$ fs at this precision. This issue could not be resolved by further increases in precision, likely due to operations required to calculate singular values reaching the limits of machine (floating point) precision. Similarly in

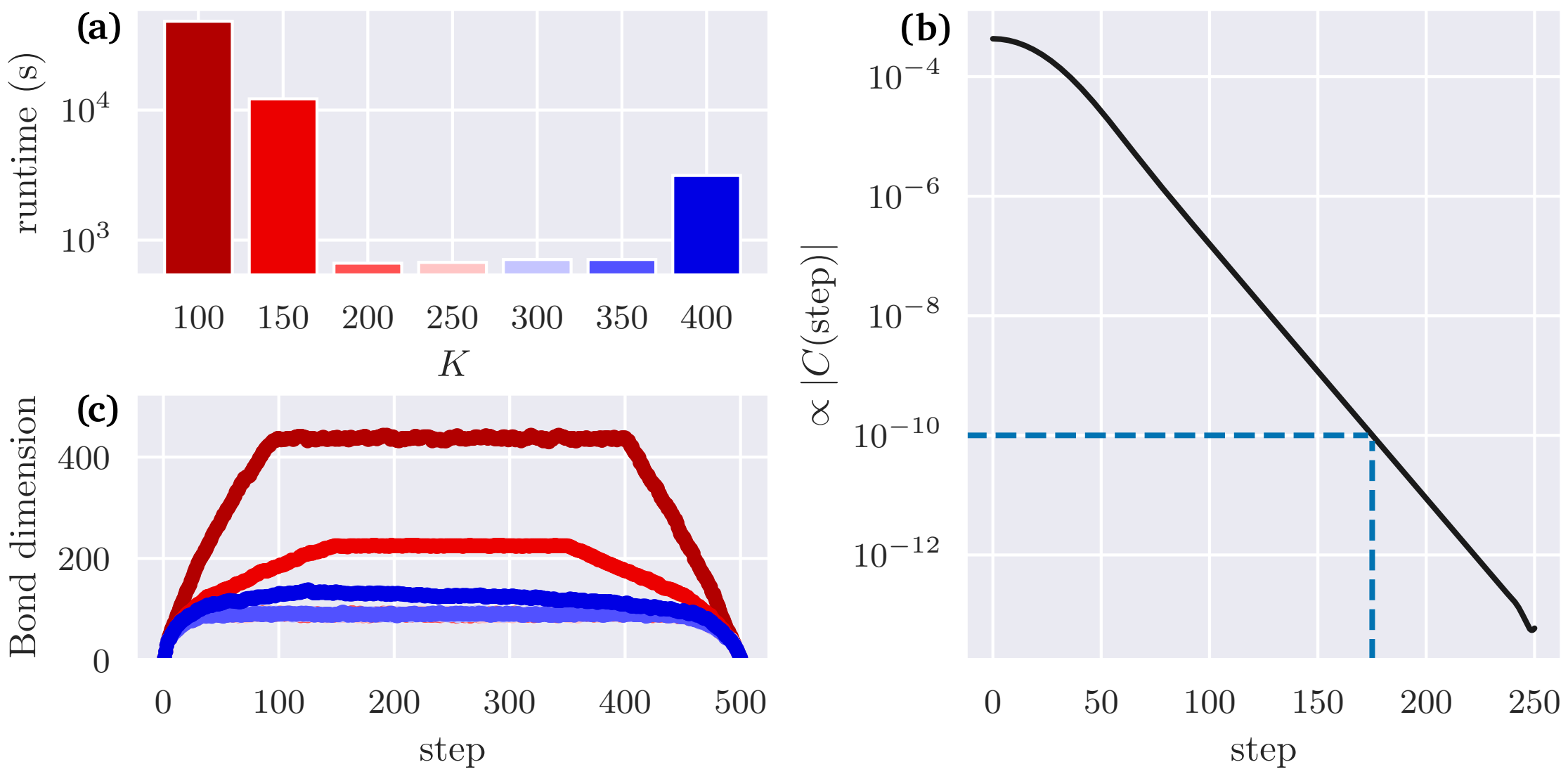

**Figure D.1:** To illustrate the effect of memory cutoff $K$ on PT computation in the PT-TEMPO method we show (a) the total computation time and (a) bond dimensions of a PT 500 timesteps in length for values of $K \in [100, 400]$ (in steps) and a precision of $\epsilon_{\mathrm{rel}} = 10^{-10}$ (recall PT-TEMPO works at a fixed precision rather than a fixed bond dimension). The timestep size $\delta t = 0.4$ fs and spectral density parameters matched those of the PT used in Chapter 4 at $T = 300$ K. Below $K = 200$ a sharp rise in computation time is observed corresponding to a growing bond dimension across the tensor. These effects grew with decreasing $K$ such that we were unable to construct a PT at $K < 100$ with available resources ($\sim$ 7 GB memory). (c) A hard cutoff on correlations after $K$ steps corresponds to a discontinuity in the bath autocorrelation function $C(\mathrm{step})$ at $K$, so we can use the absolute value of this function—here scaled such that $\epsilon_{\mathrm{rel}} = 10^{-10}$ coincides with the observed jump in computation time at around $K = 175$—to estimate the minimum $K$ required to avoid this issue at higher precisions. This suggests, for example, $K \sim 220$ should be sufficient for the precision $\epsilon_{\mathrm{rel}} = 5\times10^{-12}$ used in Chapter 4. ↑TOF

Fig. D.2c at $\delta t = 0.4$ fs we found no benefit in reducing $\epsilon_{\mathrm{rel}}$ below $5 \times 10^{-12}$, instead observing fluctuations in the PT-TEMPO results about the non-PT value. The discrepancy between the two implementations did allow us to quantify the error in the PT-TEMPO calculation at $\delta t = 0.4$ fs, $\epsilon_{\mathrm{rel}} = 5\times10^{-12}$, taking the $\delta t = 0.1$ fs non-PT result as an exact reference. This was done for three difference pump strengths at $\Omega = 200$ meV to produce Fig. D.2d. By $\delta t = 0.4$ fs, the estimated error is well below $0.5\%$ in each case.

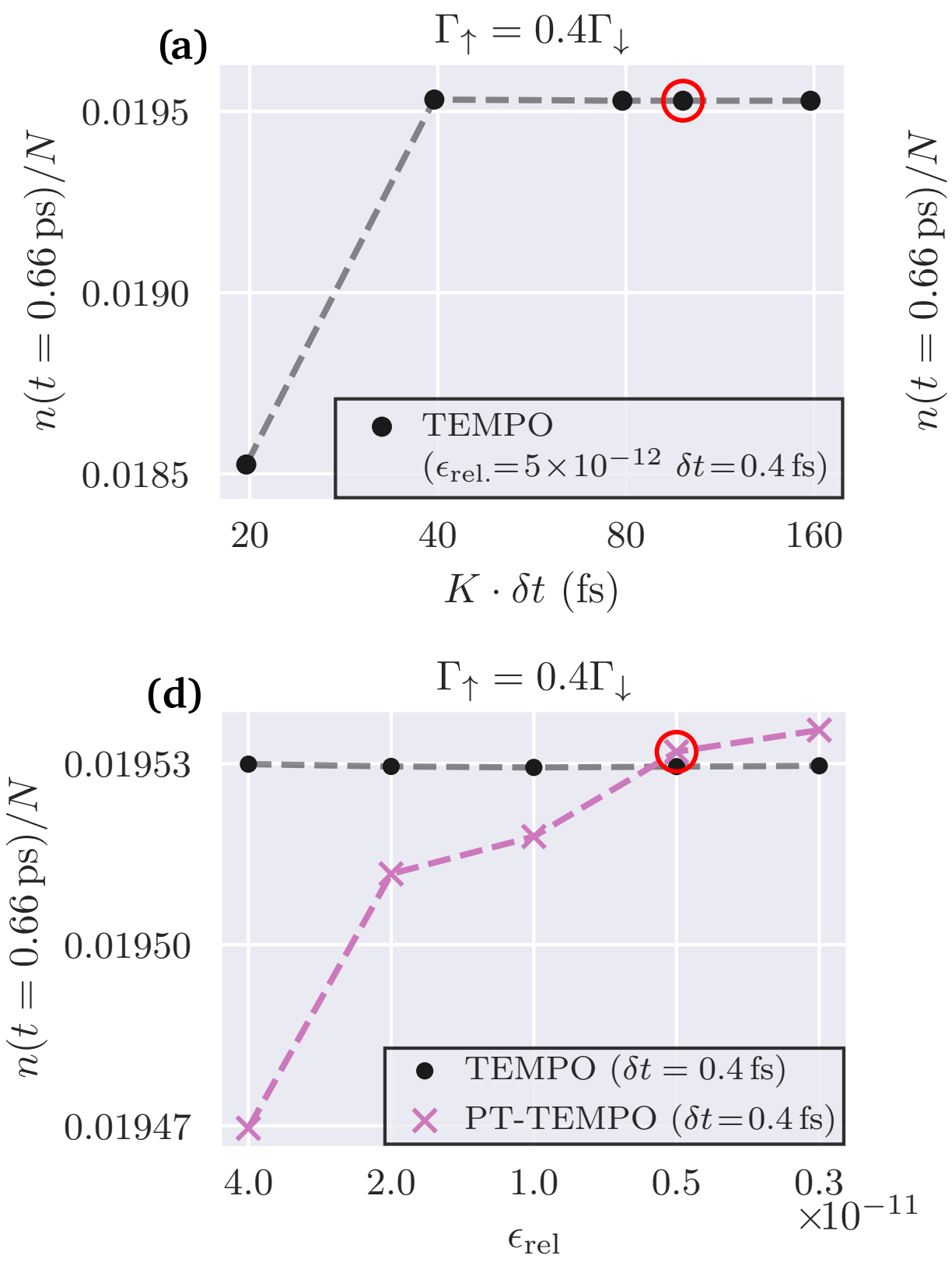

**Figure D.2:** Convergence tests for the computational parameters (a) $K$, (b) $\delta t$ and (b) $\epsilon_{\text{rel}}$. These panels show the $t = 0.66$ ps value of the scaled photon number $n/N$ in simulations using the PT-TEMPO (crosses) and non-PT TEMPO (filled circles) methods at $\Omega = 200$ meV, $\Delta = 20$ meV, $\Gamma_\uparrow = 0.4\Gamma_\downarrow$, and $T = 300$ K, with losses $\kappa/2 = \Gamma_\downarrow = 10$ meV as in Fig. 4.3. In each panel, the horizontal axis is ordered so that convergence occurs on moving to the right. In addition, a red circle indicates data corresponding to the computational parameters used in Chapter 4 ($K = 250$, $\delta t = 0.4$ fs, $\epsilon_{\text{rel}} = 5\times10^{-12}$). (a) The requirement on $K\delta t$ to attain a manageable bond dimension (see Fig. D.1) means our chosen memory length $K\delta t \sim 100$ fs is far beyond that at which any significant change in system dynamics is observed. (b) The PT-TEMPO result becomes unstable below $\delta t = 0.4$ fs whilst the change in the non-PT result continues to decrease linearly with timestep halvings. (c) The PT-TEMPO method appears to require a higher precision (smaller $\epsilon_{\text{rel}}$) for comparable accuracy. This is a trade-off of the gain of computational efficiency: the PT-TEMPO data point at $\epsilon_{\text{rel}} = 5\times10^{-12}$ here took less than 5 minutes to obtain compared to 3.5 hours using the non-PT method. (d) Error in the PT-TEMPO value at $\epsilon_{\text{rel}} = 5\times10^{-12}$ for $\Gamma_\uparrow/\Gamma_\downarrow = 0.2$, 0.4 and 0.6 relative to non-PT data with the smallest timestep $\delta t = 0.1$ fs at that precision. ↩TOF

## D.2 Computational cost

For the spectral density Eq. (4.1.3) ($\alpha = 0.25$, $\nu_c = 150$ meV) and temperature $T = 300$ K, and using the computational parameters described above, the PT took approximately 3-4 core hours to construct on a 2.1 GHz Intel® Xeon® processor. Calculations of similar length were required to construct PTs for the other three temperatures $T = 250$ K, $T = 350$ K and $T = 400$ K used in Fig. 4.3d. Having precomputed a PT, subsequent contraction with the chosen initial state, system propagators and control operators took only minutes to complete (we found 10 minutes typical).

## D.3 Fitting procedures for Figs. 4.3 and 4.6

In this section we detail the procedures used to extract the lasing threshold $\Gamma_c$ plotted in Figs. 4.3c and 4.3d. We also explain how we check that the steady state has been reached before applying the operators that allow us to calculate the two-time correlators used to determine the spectra in Fig. 4.6.

In order to obtain the steady-state scaled photon number $n_s/N$ for each set of system parameters ($\Omega$, $\Delta$, $\Gamma_\uparrow$) and environment temperature $T$, the dynamics were firstly calculated to a final time $t_f = 1.3$ ps using a pre-computed PT for that temperature. An exponential $a\exp(-bt) + c$ or constant ($a = b = 0$) fit was then made to the late time dynamics $t \geq 1$ ps. If the mean squared error of the fit, scaled by the magnitude of $a$ (or $c$ if $a = 0$), was less than $10^{-2}$, the fit was accepted and $c$ used as the value for $n_s/N$ (e.g. filled circles in Fig. 4.3b). On the contrary, if the error exceeded this cutoff the fit was deemed poor and the data not used in the subsequent threshold calculation (open circles in Fig. 4.3b). Note in the case $n(t_f)/N$ was less than $10^{-12}$ no fit was attempted and instead this final value was taken as the steady-state value.

Before extracting the threshold from the resulting plots of $n_s/N$ against $\Gamma_\uparrow$ such as those in Fig. 4.3b, it was ensured that there were sufficient ($> 5$) values of $\Gamma_\uparrow$ with valid fits in the lasing phase. A quadratic fit of the form $\Theta(x - \Gamma_c)\left[a_1(x - \Gamma_c) + a_2(x - \Gamma_c)^2\right]$ was then made to the steady-state values at each light-matter coupling, detuning and temperature, yielding the threshold $\Gamma_c$ at those parameters; a single point in Fig. 4.3c or Fig. 4.3d.

To produce Figs. 4.6a to 4.6d the dynamics were calculated to $t'_f = 1.6$ ps using the $T = 300$ K PT (only $4/5$ths of this tensor was used for Fig. 4.3). Firstly, to reach to steady state ($t_f = 1.3$ ps) and, secondly, to measure either the $\langle\sigma^+(t)\sigma^-(t_f)\rangle$ or $\langle\sigma^-(t)\sigma^+(t_f)\rangle$ correlator ($t_f \leq t \leq t'_f$). These measurements are performed by inserting a control operation $\sigma^-$ (or $\sigma^+$) in the tensor network at $t = t_f$ and subsequently recording the expectation of $\sigma^+$ (or $\sigma^-$). To ensure the system had reached the steady state by $t_f = 1.3$ ps, the exponential fitting described above was made up to $t_f$; then only if the fit was valid *and* close (within $1\%$ or $10^{-5}$ in absolute value) to the observed value $n(t_f)/N$ at this time was the state at $t_f$ deemed suitable for determining the two-time correlations.

## D.4 Weak system-environment coupling dephasing rates

In this section we evaluate the rate $\mathrm{Re}\,\Gamma(\omega)/2$ at $\omega = 0$, where

$$\Gamma(\omega) = \int_0^\infty ds\, e^{i\omega s} C(s), \quad C(s) = \int_0^\infty d\nu J(\nu) \left[\coth\left(\frac{\nu}{2T}\right)\cos(\nu s) - i\sin(\nu s)\right]. \tag{D.4.1}$$

As we only need the real part, it is useful to note

$$\frac{1}{2}\operatorname{Re}\Gamma(\omega) = \frac{1}{4}\left(\Gamma(\omega) + \overline{\Gamma(\omega)}\right) = \frac{1}{4}\int_{-\infty}^{\infty} ds\, e^{i\omega s} C(s). \tag{D.4.2}$$

Writing the trigonometric functions in terms of complex exponentials,

$$\Gamma(\omega) = \frac{1}{2}\int_0^{\infty} ds\, e^{i\omega s}\int_{-\infty}^{\infty} d\nu J(\nu)\left[\left(\coth\left(\frac{\nu}{2T}\right)+1\right)e^{-i\nu s} + \left(\coth\left(\frac{\nu}{2T}\right)-1\right)e^{i\nu s}\right], \tag{D.4.3}$$

where the lower limit of the integral was set to $-\infty$ on the condition[1] $J(\nu) = 0$ for $\nu < 0$. The integrals over $s$ then define Fourier transforms of $1$ with an exponential shift, giving [463]

$$\int_{-\infty}^{\infty} ds\, e^{i(\omega\pm\nu)s} = 2\pi\delta(\omega\pm\nu). \tag{D.4.4}$$

The delta functions evaluate the integrand of the remaining integral at $\nu = \mp\omega$,

$$\begin{aligned}
\frac{1}{2}\operatorname{Re}\Gamma(\omega) &= \frac{\pi}{4}\int_{-\infty}^{\infty} d\nu J(\nu)\left[\left(\coth\left(\frac{\nu}{2T}\right)+1\right)\delta(\omega-\nu) + \left(\coth\left(\frac{\nu}{2T}\right)-1\right)\delta(\omega+\nu)\right] && \text{(D.4.5)}\\
&= \frac{\pi}{4}\left[J(\omega)\left(\coth\left(\frac{\omega}{2T}\right)+1\right) + J(-\omega)\left(\coth\left(-\frac{\omega}{2T}\right)-1\right)\right] && \text{(D.4.6)}\\
&= \frac{\pi}{4}\left(J(\omega)-J(-\omega)\right)\left(\coth\left(\frac{\omega}{2T}\right)+1\right), && \text{(D.4.7)}
\end{aligned}$$

where we used that $\coth(\omega/2T)$ is odd. This function has a singularity at $0$, with series [458]

$$\coth\left(\frac{\omega}{2T}\right) = \frac{2T}{\omega} + O(\omega) \text{ as } \omega \to 0. \tag{D.4.8}$$

Then

$$\lim_{\omega\to 0}\left(J(\omega)-J(-\omega)\right)\left(\coth\left(\frac{\omega}{2T}\right)+1\right) = 2T\lim_{\omega\to 0^+}\left(\frac{J(\omega)}{\omega}\right) = 2T\lim_{\omega\to 0^-}\left(\frac{J(-\omega)}{-\omega}\right). \tag{D.4.9}$$

So the limit $\omega \to 0$ is well defined, and for $J(\nu) = 2\alpha\nu e^{-(\nu/\nu_c)^2}$,

$$\frac{1}{2}\operatorname{Re}\Gamma(0) = \frac{1}{2}\pi T\lim_{\omega\to 0}\left(\frac{J(\omega)}{\omega}\right) = \pi\alpha T. \tag{D.4.10}$$

[1] In other words, $J(\nu)$ is defined on $\mathbb{R}$ via multiplication by a Heaviside (step) function $\Theta(\nu)$. This sidesteps confusion on how to assign the weight of the delta function to the endpoint $\nu = 0$.

# Appendix E

# Central spin model cumulant equations ↑TOC

> The proof can be written on one line, provided you start sufficiently far to the left...
>
> Yulii Shikhmurzaev

> ...the University needs larger whiteboards.

In this appendix we provide complete working for the cumulant expansions for the central spin model in Chapter 5 up to third order. As all computations involving the central spin model boil down to the algebra of Pauli matrices, it will be useful to have the following results at hand

$$
\begin{gathered}
\sigma^+\sigma^- = \frac{1}{2}\left(1+\sigma^z\right), \quad \sigma^-\sigma^+ = \frac{1}{2}\left(1-\sigma^z\right), \\
\sigma^z\sigma^+ = \sigma^+, \quad \sigma^z\sigma^- = -\sigma^-, \quad \sigma^+\sigma^z = -\sigma^+, \quad \sigma^-\sigma^z = \sigma^-, \\
[\sigma^+,\sigma^-] = \sigma^z, \quad [\sigma^z,\sigma^\pm] = \pm 2\sigma^\pm.
\end{gathered} \tag{E.0.1}
$$

For reference we restate the Hamiltonian Eq. (5.2.1) and master equation Eq. (5.2.2):

$$
H = \frac{\omega}{2}\sigma_0^z + \sum_{n=1}^{N}\left[\frac{\epsilon}{2}\sigma_n^z + g\left(\sigma_0^+\sigma_n^- + \sigma_0^-\sigma_n^+\right)\right], \tag{E.0.2}
$$

$$
\partial_t\rho = -i\,[H,\rho] + \kappa L[\sigma_0^-] + \sum_{n=1}^{N}\left(\Gamma_\uparrow L[\sigma_n^+] + \Gamma_\downarrow L[\sigma_n^-]\right), \tag{E.0.3}
$$

with $L[x] = x\rho x^\dagger - \{x^\dagger x, \rho\}/2$.

## E.1 Mean-field equations

We start by deriving the equations

$$\partial_t\langle\sigma_0^z\rangle = -\kappa\left(\langle\sigma_0^z\rangle+1\right)+4gN\,\mathrm{Im}\left[\langle\sigma_0^+\sigma_n^-\rangle\right], \tag{E.1.1}$$
$$\partial_t\langle\sigma_n^z\rangle = -\Gamma_T\langle\sigma_n^z\rangle+\Gamma_\Delta-4g\,\mathrm{Im}\left[\langle\sigma_0^+\sigma_n^-\rangle\right], \tag{E.1.2}$$
$$\partial_t\langle\sigma_0^+\rangle = \left(i\omega-\frac{\kappa}{2}\right)\langle\sigma_0^+\rangle - igN\langle\sigma_0^z\sigma_n^+\rangle, \tag{E.1.3}$$
$$\partial_t\langle\sigma_n^+\rangle = \left(i\epsilon-\frac{\Gamma_T}{2}\right)\langle\sigma_n^+\rangle - ig\langle\sigma_0^+\sigma_n^z\rangle, \tag{E.1.4}$$

which will form a closed set after applying the mean-field approximation. We will make use of the result (see Section 3.1.9)

$$\partial_t\langle X\rangle = -i\big\langle[X,H]\big\rangle-(1/2)\sum_i\big\langle[X,Y_i^\dagger]\,Y_i+Y_i^\dagger\,[Y_i,X]\big\rangle \tag{E.1.5}$$
$$= -i\,\langle[X,H]\rangle-\sum_i\mathrm{Re}\left\langle[X,Y_i^\dagger]\,Y_i\right\rangle \quad \text{for } X \text{ Hermitian.} \tag{E.1.6}$$

### $\langle\sigma_0^z\rangle$

$$\partial_t\langle\sigma_0^z\rangle = -ig\sum_{n=1}^{N}\left\langle[\sigma_0^z,\sigma_0^+]\sigma_n^- - \text{H.c.}\right\rangle-\kappa\,\mathrm{Re}\langle[\sigma_0^z,\sigma_0^+]\sigma_0^-\rangle \tag{E.1.7}$$
$$= -2ig\sum_{n=1}^{N}\left(\langle\sigma_0^+\sigma_n^-\rangle-\text{H.c.}\right)-2\kappa\,\mathrm{Re}\langle\sigma_0^+\sigma_0^-\rangle \tag{E.1.8}$$
$$= -\kappa\left(\langle\sigma_0^z\rangle+1\right)+4gN\,\mathrm{Im}\left[\langle\sigma_0^+\sigma_n^-\rangle\right]. \tag{E.1.9}$$

### $\langle\sigma_n^z\rangle$

$$\partial_t\langle\sigma_n^z\rangle = -ig\left(\sigma_0^+[\sigma_n^z,\sigma_n^-]-\text{H.c.}\right)-\Gamma_\uparrow\,\mathrm{Re}\left\langle[\sigma_n^z,\sigma_n^-]\sigma_n^+\right\rangle-\Gamma_\downarrow\,\mathrm{Re}\left\langle[\sigma_n^z,\sigma_n^+]\sigma_n^-\right\rangle \tag{E.1.10}$$
$$= 2ig\left(\langle\sigma_0^+\sigma_n^-\rangle-\text{H.c.}\right)-(\Gamma_\uparrow+\Gamma_\downarrow)+2\Gamma_\uparrow\langle\sigma_n^-\sigma_n^+\rangle-2\Gamma_\downarrow\langle\sigma_n^+\sigma_n^-\rangle \tag{E.1.11}$$
$$= -4g\,\mathrm{Im}\left[\langle\sigma_0^+\sigma_n^-\rangle\right]+\Gamma_\uparrow\left(1-\langle\sigma_n^z\rangle\right)-\Gamma_\downarrow\left(1+\langle\sigma_n^z\rangle\right) \tag{E.1.12}$$
$$= -\Gamma_T\langle\sigma_n^z\rangle+\Gamma_\Delta-4g\,\mathrm{Im}\left[\langle\sigma_0^+\sigma_n^-\rangle\right] \tag{E.1.13}$$

with $\Gamma_T=\Gamma_\uparrow+\Gamma_\downarrow$, $\Gamma_\Delta=\Gamma_\uparrow-\Gamma_\downarrow$.

### $\langle\sigma_0^+\rangle$

$$\partial_t\langle\sigma_0^+\rangle = -i\frac{\omega}{2}\langle[\sigma_0^+,\sigma_0^z]\rangle-ig\sum_{n=1}^{N}\langle[\sigma_0^+,\sigma_0^-]\sigma_n^+\rangle-\frac{\kappa}{2}\langle\sigma_0^+[\sigma_0^-,\sigma_0^+]\rangle \tag{E.1.14}$$
$$= +i\omega\langle\sigma_0^+\rangle-igN\langle\sigma_0^z\sigma_n^+\rangle+\frac{\kappa}{2}\langle\sigma_0^+\sigma_0^z\rangle \tag{E.1.15}$$
$$= \left(i\omega-\frac{\kappa}{2}\right)\langle\sigma_0^+\rangle-igN\langle\sigma_0^z\sigma_n^+\rangle. \tag{E.1.16}$$

$\langle \sigma_n^+ \rangle$

$$\partial_t \langle \sigma_n^+ \rangle = -i\frac{\epsilon}{2}\langle [\sigma_n^+, \sigma_n^z] \rangle - ig\langle \sigma_0^+ [\sigma_n^+, \sigma_n^-] \rangle - \frac{\Gamma_\uparrow}{2}\langle [\sigma_n^+, \sigma_n^-]\sigma_n^+ \rangle - \frac{\Gamma_\downarrow}{2}\langle \sigma_n^+ [\sigma_n^-, \sigma_n^+] \rangle \tag{E.1.17}$$

$$= +i\epsilon\langle \sigma_0^+ \rangle - ig\langle \sigma_0^+ \sigma_n^z \rangle - \frac{\Gamma_\uparrow}{2}\langle \sigma_n^z \sigma_n^+ \rangle + \frac{\Gamma_\downarrow}{2}\langle \sigma_n^+ \sigma_n^z \rangle \tag{E.1.18}$$

$$= \left( i\epsilon - \frac{\Gamma_T}{2} \right) \langle \sigma_n^+ \rangle - ig\langle \sigma_0^+ \sigma_n^z \rangle. \tag{E.1.19}$$

Applying the mean-field approximation, $\langle \sigma_0^+ \sigma_n^- \rangle = \langle \sigma_0^+ \rangle \langle \sigma_n^- \rangle$ and $\langle \sigma_0^z \sigma_n^+ \rangle = \langle \sigma_0^z \rangle \langle \sigma_n^+ \rangle$, to Eqs. (E.1.1) to (E.1.4), gives

$$\partial_t \langle \sigma_0^z \rangle = -\kappa \left( \langle \sigma_0^z \rangle + 1 \right) + 4gN\,\mathrm{Im}\left[ \langle \sigma_0^+ \rangle \langle \sigma_n^- \rangle \right], \tag{E.1.20}$$

$$\partial_t \langle \sigma_n^z \rangle = -\Gamma_T \langle \sigma_n^z \rangle + \Gamma_\Delta - 4g\,\mathrm{Im}\left[ \langle \sigma_0^+ \rangle \langle \sigma_n^- \rangle \right], \tag{E.1.21}$$

$$\partial_t \langle \sigma_0^+ \rangle = \left( i\omega - \frac{\kappa}{2} \right) \langle \sigma_0^+ \rangle - igN\langle \sigma_0^z \rangle \langle \sigma_n^+ \rangle, \tag{E.1.22}$$

$$\partial_t \langle \sigma_n^+ \rangle = \left( i\epsilon - \frac{\Gamma_T}{2} \right) \langle \sigma_n^+ \rangle - ig\langle \sigma_0^+ \rangle \langle \sigma_n^z \rangle. \tag{E.1.23}$$

## E.2 Mean-field steady-state solution

The system of equations to be solved are

$$0 = -\kappa \left( \langle \sigma_0^z \rangle + 1 \right) + 4gN\,\mathrm{Im}\left[ \langle \sigma_0^+ \rangle \langle \sigma_n^- \rangle \right], \tag{E.2.1}$$

$$0 = -\Gamma_T \langle \sigma_n^z \rangle + \Gamma_\Delta - 4g\,\mathrm{Im}\left[ \langle \sigma_0^+ \rangle \langle \sigma_n^- \rangle \right], \tag{E.2.2}$$

$$0 = \left( i\omega - \frac{\kappa}{2} \right) \langle \sigma_0^+ \rangle - igN\langle \sigma_0^z \rangle \langle \sigma_n^+ \rangle, \tag{E.2.3}$$

$$0 = \left( i\epsilon - \frac{\Gamma_T}{2} \right) \langle \sigma_n^+ \rangle - ig\langle \sigma_0^+ \rangle \langle \sigma_n^z \rangle. \tag{E.2.4}$$

First note there is always a stationary solution with

$$\langle \sigma_0^+ \rangle = \langle \sigma_n^+ \rangle = 0$$
$$\langle \sigma_0^z \rangle = -1, \quad \langle \sigma_n^z \rangle = \frac{\Gamma_\Delta}{\Gamma_T}. \tag{E.2.5}$$

An additional solution may exist, but it is not necessarily stationary (time independent). For generality we look for a solution where all spins rotate about the spin-$z$ axis at rate $\mu$, $\langle \sigma_\alpha^\pm \rangle \to \langle \sigma_\alpha^\pm \rangle e^{\mp \mu t}$, with $\mu = 0$ indicating a stationary solution. From the equations of motion Eqs. (E.1.20) to (E.1.23) this amounts to a shift $\omega \to \tilde{\omega} = \omega - \mu$, $\epsilon \to \tilde{\epsilon} = \epsilon - \mu$ of energies. Hence we solve Eqs. (E.2.1) to (E.2.4) with $\tilde{\omega}$, $\tilde{\epsilon}$ in place of $\omega$, $\epsilon$. Our strategy is to use three equations to determine all expectations in terms of $\langle \sigma_0^z \rangle$, and then the final equation to determine a value for $\langle \sigma_0^z \rangle$.

From the last pair of equations Eq. (E.2.3) & (E.2.4),

$$\langle \sigma_0^+ \rangle = \frac{igN\langle \sigma_0^z \rangle \langle \sigma_n^+ \rangle}{(i\tilde{\omega} - \kappa/2)} = -\frac{g^2 N \langle \sigma_0^z \rangle \langle \sigma_n^z \rangle}{(i\tilde{\omega} - \kappa/2)\,(i\tilde{\epsilon} - \Gamma_T/2)} \langle \sigma_0^+ \rangle. \tag{E.2.6}$$

$\langle\sigma_0^+\rangle = 0$ gives the fully polarised solution above, otherwise $\langle\sigma_0^+\rangle$, $\langle\sigma_0^z\rangle$ and $\langle\sigma_n^z\rangle$ must all be non-zero and

$$\langle\sigma_n^z\rangle = -\frac{(i\tilde{\omega} - \kappa/2)\,(i\tilde{\epsilon} - \Gamma_T/2)}{g^2 N}\frac{1}{\langle\sigma_0^z\rangle}. \tag{E.2.7}$$

Next, return to Eq. (E.2.4) to determine

$$\langle\sigma_n^+\rangle = \frac{ig\langle\sigma_0^+\rangle}{(i\tilde{\epsilon} - \Gamma_T/2)}\langle\sigma_n^z\rangle \tag{E.2.8}$$

$$= \frac{(\tilde{\omega} + i\kappa/2)}{gN}\frac{\langle\sigma_0^+\rangle}{\langle\sigma_0^z\rangle}. \tag{E.2.9}$$

Using the conjugate of this expression,

$$\langle\sigma_0^+\rangle\langle\sigma_n^-\rangle = \frac{(\tilde{\omega} - i\kappa/2)}{gN}\frac{\left|\langle\sigma_0^+\rangle\right|^2}{\langle\sigma_0^z\rangle} \tag{E.2.10}$$

$$\Rightarrow \mathrm{Im}\left[\langle\sigma_0^+\rangle\langle\sigma_n^-\rangle\right] = -\frac{\kappa}{2gN}\frac{\left|\langle\sigma_0^+\rangle\right|^2}{\langle\sigma_0^z\rangle}. \tag{E.2.11}$$

Substituting into Eq. (E.2.1),

$$-\kappa\left(\langle\sigma_0^z\rangle + 1\right) - 2\kappa\frac{\left|\langle\sigma_0^+\rangle\right|^2}{\langle\sigma_0^z\rangle} = 0. \tag{E.2.12}$$

This provides condition (5.2.13) in the main text,

$$\left|\langle\sigma_0^+\rangle\right|^2 = -\langle\sigma_0^z\rangle\left(1 + \langle\sigma_0^z\rangle\right)/2, \tag{E.2.13}$$

but also $\langle\sigma_0^z\rangle < 0$, since $\left|\langle\sigma_0^+\rangle\right|^2$ must be non-negative (or positive, if solution (E.2.5) is discounted). Substituting Eqs. (E.2.7) and (E.2.11) into Eq. (E.2.2) gives a quadratic for $\langle\sigma_0^z\rangle$:

$$\Gamma_\Delta + \frac{(i\tilde{\omega} - \kappa/2)(i\tilde{\epsilon} - \Gamma_T/2)}{g^2 N}\frac{\Gamma_T}{\langle\sigma_0^z\rangle} - \frac{\kappa}{N}\left(1 + \langle\sigma_0^z\rangle\right) = 0 \tag{E.2.14}$$

$$\Rightarrow \langle\sigma_0^z\rangle^2 + \left(1 - \frac{\Gamma_\Delta N}{\kappa}\right)\langle\sigma_0^z\rangle - \frac{\Gamma_T(i\tilde{\omega} - \kappa/2)(i\tilde{\epsilon} - \Gamma_T/2)}{g^2\kappa} = 0. \tag{E.2.15}$$

The solutions are

$$\langle\sigma_0^z\rangle = -\frac{1}{2}\left(1 - \frac{\Gamma_\Delta N}{\kappa}\right) + \frac{1}{2}\left[\left(1 - \frac{\Gamma_\Delta N}{\kappa}\right)^2 + \frac{4\Gamma_T}{g^2\kappa}\left(i\tilde{\omega} - \frac{\kappa}{2}\right)\left(i\tilde{\epsilon} - \frac{\Gamma_T}{2}\right)\right]^{1/2}, \tag{E.2.16}$$

where $[\cdot]^{1/2}$ denotes the complex square root function . Remembering $\tilde{\omega} = \omega - \mu$, $\tilde{\epsilon} = \epsilon - \mu$, for the square root to yield real values we must at least have

$$\mathrm{Im}\left[\left(i(\omega - \mu) - \frac{\kappa}{2}\right)\left(i(\epsilon - \mu) - \frac{\Gamma_T}{2}\right)\right] = -\frac{1}{2}\left(\Gamma_T(\omega - \mu) + \kappa(\epsilon - \mu)\right) = 0 \tag{E.2.17}$$

or

$$\mu = \frac{\epsilon\kappa + \Gamma_T\omega}{\kappa + \Gamma_T}.$$

With this value,

$$\langle\sigma_0^z\rangle = -\frac{1}{2}\left(1-\frac{\Gamma_\Delta N}{\kappa}\right) - \frac{1}{2}\sqrt{\left(1-\frac{\Gamma_\Delta N}{\kappa}\right)^2 + \frac{\Gamma_T^2}{g^2}\left(1+4\frac{(\omega-\epsilon)^2}{(\Gamma_T+\kappa)^2}\right)}, \tag{E.2.18}$$

where $\sqrt{\cdot}$ is now the positive square root and we discarded the other solution on account of the condition $\langle\sigma_0^z\rangle < 0$ established above.

A couple of points should be made regarding this solution. As expected for the problem under the rotating wave approximation, only the detuning $(\omega-\epsilon)$ matters not the absolute energies. Although non-zero detuning adds a small correction resonant result[1], the qualitative behaviour of the solution does not change. Second, we need to impose $\langle\sigma_0^z\rangle \geq -1$ (i.e., $p_0^\uparrow \geq 0$) for a physical solution. If we consider switching on the pumping $\Gamma_\uparrow/\Gamma_T$ from $0$, solving $\langle\sigma_0^z\rangle \geq -1$ (now using Mathematica's [464] symbolic solver `Solve`) gives

$$\frac{\Gamma_\uparrow}{\Gamma_T} \geq \frac{1}{2}\left[1+\frac{\Gamma_T\kappa}{4g^2N}\left(1+\frac{(\omega-\epsilon)^2}{(\Gamma_T+\kappa)}\right)\right]. \tag{E.2.19}$$

When $\omega=\epsilon$ this provides the threshold pump ratio $R_c = (1+\Gamma_T\kappa/4g^2N)/2$[2].

## E.3 Second-order cumulant equations

We now derive the additional equations

$$\begin{aligned}\partial_t\langle\sigma_0^+\sigma_n^-\rangle &= \left(i(\omega-\epsilon) - \frac{\kappa+\Gamma_T}{2}\right)\langle\sigma_0^+\sigma_n^-\rangle + \frac{ig}{2}\langle\sigma_n^z\rangle \\ &\quad - \frac{ig}{2}\langle\sigma_0^z\rangle - ig(N-1)\langle\sigma_0^z\sigma_n^+\sigma_m^-\rangle, \qquad \text{(E.3.1)}\\ \partial_t\langle\sigma_n^+\sigma_m^-\rangle &= -\Gamma_T\langle\sigma_n^+\sigma_m^-\rangle + 2g\,\mathrm{Im}\left[\langle\sigma_0^+\sigma_n^-\sigma_n^z\rangle\right], \qquad \text{(E.3.2)}\end{aligned}$$

which following a second-order cumulant expansion will form a closed set with Eqs. (E.1.1) to (E.1.4).

### $\langle\sigma_0^+\sigma_n^-\rangle$

Firstly note that the contributions to the equation of motion for $\langle\sigma_0^+\sigma_n^-\rangle$ from terms in the master equation involving only the central site or a satellites site must have the same form as the corresponding contributions to the single-site equations $\partial_t\langle\sigma_0^+\rangle$, $\partial_t\langle\sigma_n^-\rangle$. That is,

$$\partial_t\langle\sigma_0^+\sigma_n^-\rangle \ni \left(i(\omega-\epsilon)-\frac{\kappa+\Gamma_T}{2}\right)\langle\sigma_0^+\sigma_n^-\rangle, \tag{E.3.3}$$

where $\ni$ denotes 'contains'.

[1]In particular we note $\langle\sigma_0^z\rangle$ is no longer exactly a constant at fixed $\kappa/N$. This dispels an hypothesis that in order to make mean-field theory correct you need a scaling for which it does not vary with $N$.

[2]Proving Eq. (E.2.18) provides the stable solution (not $\langle\sigma_0^z\rangle = -1$) above $R_c$ would require more work; we take the physical motivation of a non-zero population as justification.

From the interaction,

$$\partial_t\langle\sigma_0^+\sigma_n^-\rangle \ni -ig\sum_{m=1}^{N}\big\langle[\sigma_0^+\sigma_n^-,\sigma_0^+\sigma_m^-]+[\sigma_0^+\sigma_n^-,\sigma_0^-\sigma_m^+]\big\rangle \tag{E.3.4}$$

$$= -ig\sum_{m=1}^{N}\big\langle\sigma_0^+[\sigma_n^-,\sigma_0^-\sigma_m^+]+[\sigma_0^+,\sigma_0^-\sigma_m^+]\sigma_n^-\big\rangle \tag{E.3.5}$$

$$= -ig\langle\sigma_0^+\sigma_0^-[\sigma_n^-,\sigma_m^+]\rangle - ig\sum_{m=1}^{N}\langle[\sigma_0^+,\sigma_0^-]\sigma_m^+\sigma_n^-\rangle \tag{E.3.6}$$

$$= -i\frac{g}{2}\langle(1+\sigma_0^z)(-\sigma_n^z)\rangle - ig\sum_{m=1}^{N}\langle\sigma_0^z\sigma_m^+\sigma_n^-\rangle \tag{E.3.7}$$

$$= i\frac{g}{2}\langle\sigma_n^z\rangle + i\frac{g}{2}\langle\sigma_0^z\sigma_n^z\rangle - i\frac{g}{2}\langle\sigma_0^z\rangle - i\frac{g}{2}\langle\sigma_0^z\sigma_n^z\rangle - ig\sum_{m\neq n}\langle\sigma_0^z\sigma_m^+\sigma_n^-\rangle, \tag{E.3.8}$$

where in the final line we separated out the term $n = m$ from the sum. The terms involving $\langle\sigma_0^z\sigma_n^z\rangle$ cancel and then, including the contribution Eq. (E.3.3),

$$\partial_t\langle\sigma_0^+\sigma_n^-\rangle = \left(i(\omega-\epsilon)-\frac{\kappa+\Gamma_T}{2}\right)\langle\sigma_0^+\sigma_n^-\rangle + \frac{ig}{2}\langle\sigma_n^z\rangle$$
$$-\frac{ig}{2}\langle\sigma_0^z\rangle - ig(N-1)\langle\sigma_0^z\sigma_n^+\sigma_m^-\rangle, \tag{E.3.9}$$

as required. Setting the third-order cumulant $\langle\langle\sigma_0^z\sigma_n^+\sigma_m^-\rangle\rangle$ to zero and using the U(1) symmetry implies

$$\langle\sigma_0^z\sigma_n^+\sigma_m^-\rangle \approx \langle\sigma_0^z\rangle\langle\sigma_n^+\sigma_m^-\rangle \tag{E.3.10}$$

which when substituted in Eq. (E.3.9) gives the cumulant equation stated in the main text.

### $\langle\sigma_n^+\sigma_m^-\rangle$

For the terms in the master equation involving only a single satellite ($\sim\epsilon,\Gamma_\uparrow,\Gamma_\downarrow$), we can again infer them from the equation for $\langle\sigma_n^+\rangle$ and its conjugate, Eq. (E.1.4) (this relies on $n\neq m$):

$$\partial_t\langle\sigma_n^+\sigma_m^-\rangle \ni \left(i\epsilon-\frac{\Gamma_T}{2}\right)\langle\sigma_n^+\sigma_m^-\rangle + \left(-i\epsilon-\frac{\Gamma_T}{2}\right)\langle\sigma_n^+\sigma_m^-\rangle \tag{E.3.11}$$

$$= -\Gamma_T\langle\sigma_n^+\sigma_m^-\rangle. \tag{E.3.12}$$

The interaction clearly involves a sum over two terms which are conjugates,

$$\partial_t\langle\sigma_n^+\sigma_m^-\rangle \ni -ig\sum_{k=1}^{N}\big(\langle\sigma_0^+[\sigma_n^+,\sigma_k^-]\sigma_m^-\rangle + \langle\sigma_0^-\sigma_n^+[\sigma_m^-,\sigma_k^+]\rangle\big) \tag{E.3.13}$$

$$= -ig\langle\sigma_0^+\sigma_n^z\sigma_m^-\rangle + ig\langle\sigma_0^-\sigma_n^+\sigma_m^z\rangle \tag{E.3.14}$$

$$= 2g\,\mathrm{Im}\big[\langle\sigma_0^+\sigma_n^-\sigma_m^z\rangle\big]\,. \tag{E.3.15}$$

Adding contributions (E.3.12) and (E.3.15),

$$\partial_t\langle\sigma_n^+\sigma_m^-\rangle = -\Gamma_T\langle\sigma_n^+\sigma_m^-\rangle + 2g\,\mathrm{Im}\big[\langle\sigma_0^+\sigma_n^-\sigma_n^z\rangle\big]\,. \tag{E.3.16}$$

The cumulant expansion for the third-order moment here is simply $\langle\sigma_0^+\sigma_n^-\sigma_n^z\rangle \approx \langle\sigma_0^+\sigma_n^-\rangle\langle\sigma_n^z\rangle$. The full set of equations is then

$$\partial_t\langle\sigma_0^z\rangle = -\kappa\left(\langle\sigma_0^z\rangle+1\right) + 4gN\,\mathrm{Im}\big[\langle\sigma_0^+\sigma_n^-\rangle\big]\,, \tag{E.3.17}$$
$$\partial_t\langle\sigma_n^z\rangle = -\Gamma_T\langle\sigma_n^z\rangle + \Gamma_\Delta - 4g\,\mathrm{Im}\big[\langle\sigma_0^+\sigma_n^-\rangle\big]\,, \tag{E.3.18}$$
$$\partial_t\langle\sigma_0^+\rangle = \left(i\omega - \frac{\kappa}{2}\right)\langle\sigma_0^+\rangle - igN\langle\sigma_0^z\sigma_n^+\rangle, \tag{E.3.19}$$
$$\partial_t\langle\sigma_n^+\rangle = \left(i\epsilon - \frac{\Gamma_T}{2}\right)\langle\sigma_n^+\rangle - ig\langle\sigma_0^+\sigma_n^z\rangle, \tag{E.3.20}$$
$$\begin{aligned}\partial_t\langle\sigma_0^+\sigma_n^-\rangle = {}&\left(i(\omega-\epsilon) - \frac{\kappa+\Gamma_T}{2}\right)\langle\sigma_0^+\sigma_n^-\rangle + \frac{ig}{2}\langle\sigma_n^z\rangle\\ &- \frac{ig}{2}\langle\sigma_0^z\rangle - ig(N-1)\langle\sigma_0^z\rangle\langle\sigma_n^+\sigma_m^-\rangle,\end{aligned} \tag{E.3.21}$$
$$\partial_t\langle\sigma_n^+\sigma_m^-\rangle = -\Gamma_T\langle\sigma_n^+\sigma_m^-\rangle + 2g\,\mathrm{Im}\big[\langle\sigma_0^+\sigma_n^-\rangle\big]\langle\sigma_n^z\rangle. \tag{E.3.22}$$

Whilst possible to obtain a closed expression for the steady-state solution on resonance $\omega = \epsilon$, for convenience (and consistency with checking the result off-resonance) we used Mathematica's [464] `NSolve` function to solve for the steady state of Eqs. (E.3.17) to (E.3.22) numerically. This produced the second-order cumulant results in Figs. 5.3 to 5.6.

## E.4 Third-order equations

The equations derived so far are

$$\partial_t\langle\sigma_0^z\rangle = -\kappa\left(\langle\sigma_0^z\rangle+1\right) + 4gN\,\mathrm{Im}\big[\langle\sigma_0^+\sigma_n^-\rangle\big]\,, \tag{E.4.1}$$
$$\partial_t\langle\sigma_n^z\rangle = -\Gamma_T\langle\sigma_n^z\rangle + \Gamma_\Delta - 4g\,\mathrm{Im}\big[\langle\sigma_0^+\sigma_n^-\rangle\big]\,, \tag{E.4.2}$$
$$\partial_t\langle\sigma_0^+\rangle = \left(i\omega - \frac{\kappa}{2}\right)\langle\sigma_0^+\rangle - igN\langle\sigma_0^z\sigma_n^+\rangle, \tag{E.4.3}$$
$$\partial_t\langle\sigma_n^+\rangle = \left(i\epsilon - \frac{\Gamma_T}{2}\right)\langle\sigma_n^+\rangle - ig\langle\sigma_0^+\sigma_n^z\rangle, \tag{E.4.4}$$
$$\begin{aligned}\partial_t\langle\sigma_0^+\sigma_n^-\rangle = {}&\left(i(\omega-\epsilon) - \frac{\kappa+\Gamma_T}{2}\right)\langle\sigma_0^+\sigma_n^-\rangle + \frac{ig}{2}\langle\sigma_n^z\rangle\\ &- \frac{ig}{2}\langle\sigma_0^z\rangle - ig(N-1)\langle\sigma_0^z\sigma_n^+\sigma_m^-\rangle,\end{aligned} \tag{E.4.5}$$
$$\partial_t\langle\sigma_n^+\sigma_m^-\rangle = -\Gamma_T\langle\sigma_n^+\sigma_m^-\rangle + 2g\,\mathrm{Im}\big[\langle\sigma_0^+\sigma_n^-\sigma_n^z\rangle\big]\,. \tag{E.4.6}$$

To obtain a complete set of third-order equations, we start by deriving equations of motion for the two third-order moments in Eqs. (E.4.5) and (E.4.6). After a cumulant expansion, these equations involve other moments not included in Eqs. (E.4.1) to (E.4.6)L in total one must determine equations for $\langle\sigma_0^z\sigma_n^+\sigma_m^-\rangle$, $\langle\sigma_0^+\sigma_n^-\sigma_m^z\rangle$, $\langle\sigma_n^z\sigma_n^+\sigma_k^-\rangle$, $\langle\sigma_n^z\sigma_m^z\rangle$ and $\langle\sigma_0^z\sigma_n^z\rangle$ where $n$, $m$, $k$ are distinct satellite indices.

Given the equations to derive are for moments with operators from distinct sites, in each case the contribution from terms in the master equation involving only a single operator ($\omega$, $\epsilon$, $\kappa$, $\Gamma_\uparrow$ and $\Gamma_\downarrow$) follows directly from the equation of motion for the single-site observables Eqs. (E.1.1) to (E.1.4). In other words, we only really need to address the contribution from the interaction Hamiltonian.

$\langle \sigma_0^z \sigma_n^+ \sigma_m^- \rangle$

Considering the interaction Hamiltonian (note $0, n, m$ are distinct so the operators in this moment commute):

$$\partial_t \langle \sigma_0^z \sigma_n^+ \sigma_m^- \rangle \ni -ig \sum_{k=1}^{N} \Big( \langle [\sigma_0^z, \sigma_0^+] \sigma_k^- \sigma_n^+ \sigma_m^- \rangle + \langle [\sigma_0^z, \sigma_0^-] \sigma_k^+ \sigma_n^+ \sigma_m^- \rangle + \langle \sigma_0^z \sigma_0^+ [\sigma_n^+, \sigma_k^-] \sigma_m^- \rangle + \langle \sigma_0^z \sigma_0^- \sigma_n^+ [\sigma_m^-, \sigma_k^+] \rangle \Big) \tag{E.4.7}$$

$$= -2ig \sum_{k=1}^{N} \left( \langle \sigma_0^+ \sigma_k^- \sigma_n^+ \sigma_m^- \rangle - \langle \sigma_0^- \sigma_k^+ \sigma_n^+ \sigma_m^- \rangle \right) - ig \langle \sigma_0^+ \sigma_n^z \sigma_m^- \rangle - ig \langle \sigma_0^- \sigma_n^+ \sigma_m^z \rangle. \tag{E.4.8}$$

In the remaining sums we need to distinguish the cases $k = n$ and $k = m$ from $k \neq n, m$. If $k = n$ then $\langle \sigma_0^- \sigma_k^+ \sigma_n^+ \sigma_m^- \rangle = 0$ and

$$\langle \sigma_0^+ \sigma_k^- \sigma_n^+ \sigma_m^- \rangle = \frac{1}{2} \left( \langle \sigma_0^+ \sigma_m^- \rangle - \langle \sigma_0^+ \sigma_m^- \sigma_n^z \rangle \right), \tag{E.4.9}$$

whereas if $k = m$ then $\langle \sigma_0^+ \sigma_k^- \sigma_n^+ \sigma_m^- \rangle = 0$ and

$$\langle \sigma_0^- \sigma_k^+ \sigma_n^+ \sigma_m^- \rangle = \frac{1}{2} \left( \langle \sigma_0^- \sigma_n^+ \rangle + \langle \sigma_0^- \sigma_n^+ \sigma_m^z \rangle \right). \tag{E.4.10}$$

If $k \neq n, m$ then we have a fourth-order moments and make the cumulant expansions

$$\langle \sigma_0^+ \sigma_k^- \sigma_n^+ \sigma_m^- \rangle \approx \langle \sigma_0^+ \sigma_k^- \rangle \langle \sigma_n^+ \sigma_m^- \rangle + \langle \sigma_0^+ \sigma_m^- \rangle \langle \sigma_n^+ \sigma_k^- \rangle \tag{E.4.11}$$

$$= 2 \langle \sigma_0^+ \sigma_n^- \rangle \langle \sigma_n^+ \sigma_m^- \rangle, \tag{E.4.12}$$

where we relabelled satellite indices (recall the satellites are identical). Similarly when $k \neq n, m$,

$$\langle \sigma_0^- \sigma_k^+ \sigma_n^+ \sigma_m^- \rangle \approx 2 \langle \sigma_0^- \sigma_n^+ \rangle \langle \sigma_n^+ \sigma_m^- \rangle. \tag{E.4.13}$$

Noting this is the conjugate of Eq. (E.4.12),

$$\partial_t \langle \sigma_0^z \sigma_n^+ \sigma_m^- \rangle \ni 8g(N-2) \operatorname{Im}\left[ \langle \sigma_0^+ \sigma_n^- \rangle \langle \sigma_n^+ \sigma_m^- \rangle \right] - ig \left( \langle \sigma_0^+ \sigma_m^- \rangle - \langle \sigma_0^+ \sigma_m^- \sigma_n^z \rangle \right) + ig \left( \langle \sigma_0^- \sigma_n^+ \rangle + \langle \sigma_0^- \sigma_n^+ \sigma_m^z \rangle \right) - ig \langle \sigma_0^+ \sigma_n^z \sigma_m^- \rangle - ig \langle \sigma_0^- \sigma_n^+ \sigma_m^z \rangle \tag{E.4.14}$$

$$= 8g(N-2) \operatorname{Im}\left[ \langle \sigma_0^+ \sigma_n^- \rangle \langle \sigma_n^+ \sigma_m^- \rangle \right] + 2g \operatorname{Im}\left[ \langle \sigma_0^+ \sigma_n^- \rangle \right]. \tag{E.4.15}$$

The remaining terms from individual contributions can now be added to give in total

$$\partial_t \langle \sigma_0^z \sigma_n^+ \sigma_m^- \rangle = -(\kappa + \Gamma_T) \langle \sigma_0^z \sigma_n^+ \sigma_m^- \rangle - \kappa \langle \sigma_n^+ \sigma_m^- \rangle + 8g(N-2) \operatorname{Im}\left[ \langle \sigma_0^+ \sigma_n^- \rangle \right] \langle \sigma_n^+ \sigma_m^- \rangle. \tag{E.4.16}$$

## $\langle \sigma_0^+ \sigma_n^- \sigma_m^z \rangle$ ↑

From the interaction Hamiltonian only the terms $\sim \sigma_0^- \sigma_n^+$ contribute, since $\left(\sigma_0^+\right)^2 = 0$:

$$\partial_t \langle \sigma_0^+ \sigma_n^- \sigma_m^z \rangle \ni -ig \sum_{k=1}^{N} \Big[ \big\langle \sigma_0^+ \sigma_0^- \left( [\sigma_n^-, \sigma_k^+] \sigma_m^z + \sigma_n^- [\sigma_m^z, \sigma_k^+] \right) \big\rangle + \big\langle [\sigma_0^+, \sigma_0^-] \sigma_k^+ \sigma_n^- \sigma_m^z \big\rangle \Big] \tag{E.4.17}$$

$$= -i\frac{g}{2} \big\langle (1+\sigma_0^z) \left( -\sigma_n^z \sigma_m^z + 2\sigma_n^- \sigma_m^+ \right) \big\rangle - ig \sum_{k=1}^{N} \langle \sigma_0^z \sigma_k^+ \sigma_n^- \sigma_m^z \rangle . \tag{E.4.18}$$

$$= i\frac{g}{2} \left( \langle \sigma_n^z \sigma_m^z \rangle + \langle \sigma_0^z \sigma_n^z \sigma_m^z \rangle \right) - ig \left( \langle \sigma_n^+ \sigma_m^- \rangle + \langle \sigma_0^z \sigma_n^+ \sigma_m^- \rangle \right) - ig \sum_{k=1}^{N} \langle \sigma_0^z \sigma_k^+ \sigma_n^- \sigma_m^z \rangle . \tag{E.4.19}$$

To evaluate the remaining sum we consider the three cases:

$$k = n \Rightarrow \langle \sigma_0^z \sigma_k^+ \sigma_n^- \sigma_m^z \rangle = \frac{1}{2} \left( \langle \sigma_0^z \sigma_n^z \rangle + \langle \sigma_0^z \sigma_n^z \sigma_m^z \rangle \right) \tag{E.4.20}$$

$$k = m \Rightarrow \langle \sigma_0^z \sigma_k^+ \sigma_n^- \sigma_m^z \rangle = -\langle \sigma_0^z \sigma_n^- \sigma_m^+ \rangle \tag{E.4.21}$$

$$k \neq n, m \Rightarrow \langle \sigma_0^z \sigma_k^+ \sigma_n^- \sigma_m^z \rangle \approx \langle \sigma_0^z \sigma_n^+ \sigma_m^- \rangle \langle \sigma_n^z \rangle + \langle \sigma_0^z \rangle \langle \sigma_n^z \sigma_m^+ \sigma_k^- \rangle + \langle \sigma_0^z \sigma_n^z \rangle \langle \sigma_n^+ \sigma_m^- \rangle - 2 \langle \sigma_0^z \rangle \langle \sigma_n^z \rangle \langle \sigma_n^+ \sigma_m^- \rangle . \tag{E.4.22}$$

The last expression was the third-order cumulant expansion. Substituting into Eq. (E.4.19),

$$\partial_t \langle \sigma_0^+ \sigma_n^- \sigma_m^z \rangle \ni i\frac{g}{2} \left( \langle \sigma_n^z \sigma_m^z \rangle - \langle \sigma_0^z \sigma_n^z \rangle \right) - ig \langle \sigma_n^+ \sigma_m^- \rangle - ig(N-2) \Big( \langle \sigma_0^z \sigma_n^+ \sigma_m^- \rangle \langle \sigma_n^z \rangle + \langle \sigma_0^z \rangle \langle \sigma_n^z \sigma_m^+ \sigma_k^- \rangle + \langle \sigma_0^z \sigma_n^z \rangle \langle \sigma_n^+ \sigma_m^- \rangle - 2 \langle \sigma_0^z \rangle \langle \sigma_n^z \rangle \langle \sigma_n^+ \sigma_m^- \rangle \Big), \tag{E.4.23}$$

and then the full equation

$$\partial_t \langle \sigma_0^+ \sigma_n^- \sigma_m^z \rangle = \left( i(\omega - \epsilon) - \frac{\kappa + 3\Gamma_T}{2} \right) \langle \sigma_0^+ \sigma_n^- \sigma_m^z \rangle + \Gamma_\Delta \langle \sigma_0^+ \sigma_n^- \rangle + i\frac{g}{2} \left( \langle \sigma_n^z \sigma_m^z \rangle - \langle \sigma_0^z \sigma_n^z \rangle \right) - ig \langle \sigma_n^+ \sigma_m^- \rangle - ig(N-2) \Big( \langle \sigma_0^z \sigma_n^+ \sigma_m^- \rangle \langle \sigma_n^z \rangle + \langle \sigma_0^z \rangle \langle \sigma_n^z \sigma_m^+ \sigma_k^- \rangle + \langle \sigma_0^z \sigma_n^z \rangle \langle \sigma_n^+ \sigma_m^- \rangle - 2 \langle \sigma_0^z \rangle \langle \sigma_n^z \rangle \langle \sigma_n^+ \sigma_m^- \rangle \Big). \tag{E.4.24}$$

## $\langle \sigma_n^z \sigma_m^+ \sigma_k^- \rangle$ ↑

As the three spin indices are distinct, the contributions from the interaction are straightforward to determine:

$$\partial_t \langle \sigma_n^z \sigma_m^+ \sigma_k^- \rangle \ni -ig \left( \langle \sigma_0^+ [\sigma_n^z, \sigma_n^-] \sigma_m^+ \sigma_k^- \rangle + \langle \sigma_0^- [\sigma_n^z, \sigma_n^+] \sigma_m^+ \sigma_k^- \rangle \right) - ig \langle \sigma_0^+ \sigma_n^z [\sigma_m^+, \sigma_m^-] \sigma_k^- \rangle - ig \langle \sigma_0^- \sigma_n^z \sigma_m^+ [\sigma_k^-, \sigma_n^+] \rangle \tag{E.4.25}$$

$$= 2ig \langle \sigma_0^+ \sigma_n^- \sigma_m^+ \sigma_k^- \rangle - 2ig \langle \sigma_0^- \sigma_n^+ \sigma_m^+ \sigma_k^- \rangle - ig \langle \sigma_0^+ \sigma_n^z \sigma_m^z \sigma_k^- \rangle + ig \langle \sigma_0^- \sigma_n^z \sigma_m^+ \sigma_k^z \rangle \tag{E.4.26}$$

$$= -4g \,\mathrm{Im}\left[ \langle \sigma_0^+ \sigma_n^- \sigma_m^+ \sigma_k^- \rangle \right] + 2g \,\mathrm{Im}\left[ \langle \sigma_0^+ \sigma_n^z \sigma_m^z \sigma_k^- \rangle \right] . \tag{E.4.27}$$

From the cumulant expansions

$$\langle\sigma_0^+\sigma_n^-\sigma_m^+\sigma_k^-\rangle \approx \langle\sigma_0^+\sigma_n^-\rangle\langle\sigma_m^+\sigma_n^-\rangle\langle\sigma_0^+\sigma_k^-\rangle\langle\sigma_n^-\sigma_m^+\rangle \tag{E.4.28}$$
$$= 2\langle\sigma_0^+\sigma_n^-\rangle\langle\sigma_n^+\sigma_m^-\rangle \tag{E.4.29}$$

and

$$\langle\sigma_0^+\sigma_n^z\sigma_m^z\sigma_k^-\rangle = \langle\sigma_0^+\sigma_n^-\rangle\langle\sigma_n^z\sigma_m^z\rangle + \langle\sigma_0^+\sigma_n^-\sigma_m^z\rangle\langle\sigma_n^z\rangle - 2\langle\sigma_0^+\sigma_n^-\rangle\langle\sigma_n^z\rangle^2, \tag{E.4.30}$$

and then, introducing also the contributions from the non-interacting terms (note the unitary contributions $\sim\epsilon$ cancel to zero),

$$\partial_t\langle\sigma_n^z\sigma_m^+\sigma_k^-\rangle = -2\Gamma_T\langle\sigma_n^z\sigma_m^+\sigma_k^-\rangle + \Gamma_\Delta\langle\sigma_n^+\sigma_m^-\rangle - 8g\,\mathrm{Im}\left[\langle\sigma_0^+\sigma_n^-\rangle\right]\langle\sigma_n^+\sigma_m^-\rangle$$
$$+2g\Big(\mathrm{Im}\left[\langle\sigma_0^+\sigma_n^-\rangle\right]\langle\sigma_n^z\sigma_m^z\rangle - 2\,\mathrm{Im}\left[\langle\sigma_0^+\sigma_n^-\rangle\right]\langle\sigma_n^z\rangle^2 + 2\,\mathrm{Im}\left[\langle\sigma_0^+\sigma_n^-\sigma_m^z\rangle\right]\langle\sigma_n^z\rangle\Big). \tag{E.4.31}$$

### $\langle\sigma_n^z\sigma_m^z\rangle$ ⮥

This one's easy: no fourth-order moments are generated by the commutator $[\sigma_n^z\sigma_m^z, H]$ and the contributions from $\partial_t\langle\sigma_n^z\rangle$, $\partial_t\langle\sigma_m^z\rangle$ just 'double-up'!

$$\partial_t\langle\sigma_n^z\rangle = -\Gamma_T\langle\sigma_n^z\rangle + \Gamma_\Delta - 4g\,\mathrm{Im}\left[\langle\sigma_0^+\sigma_n^-\rangle\right] \tag{E.4.32}$$
$$\Rightarrow \partial_t\langle\sigma_n^z\sigma_m^z\rangle = -\Gamma_T\langle\sigma_n^z\sigma_m^z\rangle + \Gamma_\Delta - 4g\,\mathrm{Im}\left[\langle\sigma_0^+\sigma_n^-\sigma_m^z\rangle\right]$$
$$- \Gamma_T\langle\sigma_m^z\sigma_n^z\rangle + \Gamma_\Delta - 4g\,\mathrm{Im}\left[\langle\sigma_0^+\sigma_m^-\sigma_n^z\rangle\right] \tag{E.4.33}$$
$$= -2\Gamma_T\langle\sigma_n^z\sigma_m^z\rangle + 2\Gamma_\Delta\langle\sigma_n^z\rangle - 8g\,\mathrm{Im}\left[\langle\sigma_0^+\sigma_n^-\sigma_m^z\rangle\right]. \tag{E.4.34}$$

### $\langle\sigma_0^z\sigma_n^z\rangle$ ⮥

Similar to the equation for $\langle\sigma_n^z\sigma_m^z\rangle$,

$$\partial_t\langle\sigma_0^z\rangle = -\kappa\left(\langle\sigma_0^z\rangle + 1\right) + 4gN\,\mathrm{Im}\left[\langle\sigma_0^+\sigma_n^-\rangle\right] \tag{E.4.35}$$

and

$$\partial_t\langle\sigma_n^z\rangle = -\Gamma_T\langle\sigma_n^z\rangle + \Gamma_\Delta - 4g\,\mathrm{Im}\left[\langle\sigma_0^+\sigma_n^-\rangle\right] \tag{E.4.36}$$

imply that

$$\partial_t\langle\sigma_0^z\sigma_n^z\rangle = -(\kappa+\Gamma_T)\langle\sigma_0^z\sigma_n^z\rangle - \kappa\langle\sigma_n^z\rangle + \Gamma_\Delta\langle\sigma_0^z\rangle + 4g(N-1)\,\mathrm{Im}\left[\langle\sigma_0^+\sigma_n^-\sigma_m^z\rangle\right]. \tag{E.4.37}$$

That completes the derivation of Eqs. (5.2.16) to (5.2.24) from the main text (Section 5.2.4). The steady state of the equations was solved numerically using Mathematica's [464] `NSolve` function to produce the third-order data in Fig. 5.6a.

# Appendix F

# Transport

According to some, the modern world began in 1965 when J. Cooley and J. Tukey published their account of an efficient method for numerical computation of the Fourier transform. According to some others, the method was known to Gauss in the mid 1800s; the idea that lies at the heart of the algorithm is clearly present in an unpublished paper that appeared posthumously in 1866. Take your pick.

Brad Osgood

## F.1 Molecular basis and structure factors

### F.1.1 Basis

From the main text we recall a basis for the electronic and vibrational degrees of freedom $\mathcal{H}_{e\nu}$ is constructed from the $2N_\nu \times 2N_\nu$ ($2N_\nu = \dim\mathcal{H}_{e\nu}$) generalised Gell-Mann (GGM) matrices [417, 418] $\lambda^\alpha$ as

1. $\lambda^{i_0}$ the $2N_\nu^2 - 1$ GGM matrices that leave the electronic state unchanged
2. $N_\nu^2$ matrices $\lambda^{i_+}$ that add one electronic excitation, $\lambda^{i_+} = (1/\sqrt{2})\left(\lambda^S + i\lambda^A\right)$ where $\lambda^S$ is a symmetric ($\sigma^x$-like) GGM matrix with a single unit value in the upper right and lower left quadrants, and $\lambda^A$ the corresponding antisymmetric ($\sigma^y$-like) matrix
3. $N_\nu^2$ matrices $\lambda^{i_-} = (\lambda^{i_+})^\dagger$ which destroy an electronic excitation.

In this section we do not include the subscripts $n\mathrm{x}$ of the matrices (e.g. $\lambda^{i_0}_{n\mathrm{x}}$) since all expressions will be independent of the molecular ensemble and site.

The GGM matrices are traceless, Hermitian and satisfy the trace orthogonality condition

$$\mathrm{Tr}\left(\lambda^\alpha, \lambda^\beta\right) = 2\delta_{\alpha\beta}. \tag{F.1.1}$$

From these properties, we note $\lambda^{i_o}$ are traceless and Hermitian, $\lambda^{i_+}$ and $\lambda^{i_-}$ traceless and Hermitian conjugates, and the collection $\{\lambda^{i_o}, \lambda^{i_+}, \lambda^{i_-}\}$ satisfies

$$\mathrm{Tr}\big(\lambda^{i_\alpha}(\lambda^{j_\beta})^\dagger\big) = 2\delta_{i_\alpha j_\beta}\delta_{\alpha\beta}. \tag{F.1.2}$$

The decomposition of a general matrix $O$ representing an operator on a single molecule is

$$O = A_{i_0}\lambda^{i_o} + B_{i_+}\lambda^{i_+} + C_{i_-}\lambda^{i_-} + DI_{2N_\nu} \tag{F.1.3}$$

where $I_{2N_\nu}$ is the identity matrix and

$$A_{i_0} = \frac{1}{2}\mathrm{Tr}(O\lambda^{i_o}),\ B_{i_+} = \frac{1}{2}\mathrm{Tr}(O\lambda^{i_-}),\ C_{i_-} = \frac{1}{2}\mathrm{Tr}(O\lambda^{i_+}),\ D_{i_0} = \frac{1}{2N_\nu}\mathrm{Tr}(O). \tag{F.1.4}$$

Here we used $(\lambda^{i_o})^\dagger = \lambda^{i_o}$ and $(\lambda^{i_+})^\dagger = \lambda^{i_-}$.

### F.1.2 Structure tensors

To handle products of the matrices it is helpful to define structure tensors[1] $d$, $d^+$ as

$$d_{i_oj_op_o} = \frac{1}{4}\mathrm{Tr}(\{\lambda^{i_o}, \lambda^{j_o}\}\lambda^{p_o}), \quad d^+_{i_0i_+j_+} = \frac{1}{4}\mathrm{Tr}\big(\{\lambda^{i_o}, \lambda^{i_+}\}(\lambda^{j_+})^\dagger\big), \tag{F.1.5}$$

and $f$, $f^+$ via

$$f_{i_oj_op_o} = \frac{1}{4i}\mathrm{Tr}([\lambda^{i_o}, \lambda^{j_o}]\lambda^{p_o}), \quad f^+_{i_0i_+j_+} = \frac{1}{4i}\mathrm{Tr}\big([\lambda^{i_o}, \lambda^{i_+}](\lambda^{j_+})^\dagger\big), \tag{F.1.6}$$

so that

$$\begin{aligned}\{\lambda^{i_o}, \lambda^{j_o}\} = 2d_{i_oj_op_o}\lambda^{p_o} + \frac{2}{N_\nu}\delta_{i_oj_o}I_{2N_\nu}, \quad [\lambda^{i_o}, \lambda^{j_o}] = 2if_{i_oj_op_o}\lambda^{p_o}, \\ \{\lambda^{i_o}, \lambda^{i_+}\} = 2d^+_{i_0i_+j_+}\lambda^{j_+}, \quad [\lambda^{i_o}, \lambda^{i_+}] = 2if^+_{i_0i_+j_+}\lambda^{j_+},\end{aligned} \tag{F.1.7}$$

(F.1.8)

but also

$$\begin{aligned}\{\lambda^{i_o}, \lambda^{i_-}\} = 2\overline{d}^+_{i_0i_+j_+}\lambda^{j_-}, \quad [\lambda^{i_o}, \lambda^{i_-}] = 2i\overline{f}^+_{i_0i_+j_+}\lambda^{j_-}, \\ \{\lambda^{i_+}, \lambda^{j_-}\} = 2d^+_{i_0i_+j_+}\lambda^{i_o} + \frac{2}{N_\nu}\delta_{i_+j_-}I_{2N_\nu}, \quad [\lambda^{i_+}, \lambda^{j_-}] = 2if^+_{i_0i_+j_+}\lambda^{i_o}, \\ \{\lambda^{i_-}, \lambda^{j_+}\} = 2\overline{d}^+_{i_0i_+j_+}\lambda^{i_o} + \frac{2}{N_\nu}\delta_{i_-j_+}I_{2N_\nu}, \quad [\lambda^{i_-}, \lambda^{j_+}] = 2i\overline{f}^+_{i_0i_+j_+}\lambda^{i_o} = -2if^+_{i_0j_+i_+}\lambda^{i_o}.\end{aligned} \tag{F.1.9}$$

Rules for products follow as

$$\lambda^{i_o}\lambda^{j_o} = Z_{i_oj_op_o}\lambda^{p_o} + \frac{1}{N_\nu}\delta_{i_oj_o}I_{2N_\nu} \tag{F.1.10}$$

$$\lambda^{i_o}\lambda^{i_+} = Z^+_{i_0i_+j_+}\lambda^{j_+} \tag{F.1.11}$$

$$\lambda^{i_o}\lambda^{i_-} = Z^-_{i_0i_+j_+}\lambda^{j_-} \tag{F.1.12}$$

$$\lambda^{i_+}\lambda^{j_-} = Z^+_{i_0i_+j_+}\lambda^{i_o} + \frac{1}{N_\nu}\delta_{i_+j_-}I_{2N_\nu}. \tag{F.1.13}$$

[1]The point is it is useful to distinguish which type of matrix is produced by a given product or commutator (and so conserves electronic excitations or not), hence the division of the structure factors $d$ and $d^+$, $f$ and $f^+$.

with $Z_{i_0 j_0 p_0} = d_{i_0 j_0 p_0} + i f_{i_0 j_0 p_0}$, $Z^+_{i_0 i_+ j_+} = d^+_{i_0 i_+ j_+} + i f^+_{i_0 i_+ j_+}$ and $Z^-_{i_0 i_+ j_+} = \overline{d}^+_{i_0 i_+ j_+} + i \overline{f}^+_{i_0 i_+ j_+}$. Note in particular that $Z^-$ is not exactly the conjugate of $Z^+$.

The tensors $d$ and $f$ are, respectively, completely symmetric and antisymmetric. Further, they are real. The same cannot be said for $d^+$ and $f^+$ tensors for which index rearrangement is not generally simple due to the non-Herminicity of the $\pm$ matrices. However, it is useful to note one does have the conjugate (anti)symmetry $\overline{d}^+_{i_0 i_+ j_+} = d^+_{i_0 j_+ i_+}$, $\overline{f}^+_{i_0 i_+ j_+} = -f^+_{i_0 j_+ i_+}$. From this it follows that $\overline{Z}^+_{i_0 i_+ j_+} = Z^+_{i_0 j_+ i_+}$, for example.

## F.2 Second-order cumulant equations

The equations of motion to be derived are $\partial_t \langle a^\dagger_{k'} a_k \rangle$, $\partial_t \langle \lambda^{i_0}_{nx} \rangle$, $\partial_t \langle a_k \lambda^{i_+}_{nx} \rangle$, $\partial_t \langle \lambda^{i_+}_{nx} \lambda^{i_-}_{my} \rangle$. For each equation of motion to be derived we calculate contributions from the Hamiltonian part and any dissipative terms in the master equation separately, according to Eqs. (3.1.94) and (3.1.95):

$$\partial_t \langle X \rangle = -i \big\langle [X, H] \big\rangle - (1/2) \sum_i \big\langle [X, Y_i^\dagger]\, Y_i + Y_i^\dagger\, [Y_i, X] \big\rangle \tag{F.2.1}$$

$$= -i \, \langle [X, H] \rangle - \sum_i \mathrm{Re} \left\langle [X, Y_i^\dagger]\, Y_i \right\rangle \quad \text{for } X \text{ Hermitian} \tag{F.2.2}$$

For reference, we repeat the Hamiltonian Eq. (6.2.12) and master equation Eq. (6.2.13):

$$H = \sum_k \omega_k a^\dagger_k a_k + \sum_{nx} \left[ A_{i_0} \lambda^{i_0}_{nx} + \frac{1}{\sqrt{N_\mathrm{m}}} \sum_k \Big( B_{i_+} e^{-ikr_n} a_k \lambda^{i_+}_{nx} + \mathrm{H.c.} \Big) \right], \tag{F.2.3}$$

$$\partial_t \rho = -i \, [H, \rho] + \sum_k \kappa L[a_k] + \sum_{nx,\mu_0} L[\gamma^{\mu_0}_{i_0} \lambda^{i_0}_{nx}] + \sum_{nx} L[\gamma^+_{i_+} \lambda^{i_+}_{nx}] + \sum_{nx} L[\gamma^-_{i_-} \lambda^{i_-}_{nx}]. \tag{F.2.4}$$

### $\langle a^\dagger_{k'} a_k \rangle$

$$\begin{aligned}
-i \langle [a^\dagger_{k'} a_k, H] \rangle &= (1 - \delta_{kk'}) \Big( -i\omega_{k'} \langle [a^\dagger_{k'} a_k, a^\dagger_{k'} a_{k'}] \rangle - i\omega_k \langle [a^\dagger_{k'} a_k, a^\dagger_k a_k] \rangle \Big) \\
&\quad - i \frac{1}{\sqrt{N_\mathrm{m}}} \sum_{nx} \Big( B_{i_+} e^{-ik' r_n} \langle [a^\dagger_{k'} a_k, a_{k'} \lambda^{i_+}_{nx}] \rangle + B_{i_-} e^{ikr_n} \langle [a^\dagger_{k'} a_k, a^\dagger_k \lambda^{i_-}_{nx}] \rangle \Big) \\
&= i(\omega_{k'} - \omega_k) \langle a^\dagger_{k'} a_k \rangle + \frac{1}{\sqrt{N_\mathrm{m}}} \sum_{nx} \Big( i B_{i_+} e^{-ik' r_n} \langle a_k \lambda^{i_+}_{nx} \rangle - i B_{i_-} e^{ikr_n} \langle a^\dagger_{k'} \lambda^{i_-}_{nx} \rangle \Big),
\end{aligned}$$

$$\begin{aligned}
\sum_p \kappa \, \mathrm{Tr} \Big( a_{k'} a^\dagger_k L[a_p] \Big) &= -\frac{\kappa}{2} \sum_p \Big\langle [a^\dagger_{k'} a_k, a^\dagger_p] a_p + a^\dagger_p [a_p, a^\dagger_{k'} a_k] \Big\rangle \\
&= -\frac{\kappa}{2} \Big( \langle a^\dagger_{k'} [a_k, a^\dagger_k] a_k \rangle + \langle a^\dagger_{k'} [a_{k'}, a^\dagger_{k'}] a_k \rangle \Big) \\
&= -\kappa \langle a^\dagger_{k'} a_k \rangle.
\end{aligned}$$

Combining the above and introducing $\hat{\lambda}^{i_+}_{nx} = \sqrt{N_\mathrm{m}}\lambda^{i_+}_{nx}$,

$$\partial_t \langle a^\dagger_{k'} a_k\rangle = \Big[i(\omega_{k'} - \omega_k) - \kappa\Big] \langle a^\dagger_{k'} a_k\rangle + \frac{1}{N_\mathrm{m}} \sum_{nx} \left( iB_{i_+} e^{-ik' r_n} \langle a_k \hat{\lambda}^{i_+}_{nx}\rangle - iB_{i_-} e^{ikr_n} \langle a^\dagger_{k'} \hat{\lambda}^{i_-}_{nx}\rangle \right).$$

$\langle \lambda^{i_0}_{nx}\rangle$ ↩

Remembering that both molecular indices must be equal for a non-trivial commutator (operators on different sites commute),

$$-i\langle[\lambda^{i_0}_{nx}, H]\rangle = -iA_{j_0} \langle [\lambda^{i_0}_{nx}, \lambda^{j_0}_{nx}]\rangle - i\frac{1}{\sqrt{N_\mathrm{m}}} \sum_k B_{i_+} e^{-ikr_n} \langle a_k [\lambda^{i_0}_{nx}, \lambda^{i_+}_{nx}]\rangle - i\frac{1}{\sqrt{N_\mathrm{m}}} \sum_k B_{i_+} e^{ikr_n} \langle a_k [\lambda^{i_0}_{nx}, \lambda^{i_-}_{nx}]\rangle$$
$$= 2f_{i_0 j_0 p_0} A_{j_0} \langle \lambda^{p_0}_{nx}\rangle + \frac{1}{\sqrt{N_\mathrm{m}}} \sum_k \left(2B_{i_+} f^+_{i_0 i_+ j_+} e^{-ikr_n} \langle a_k \lambda^{j_+}_{nx}\rangle + \text{c.c.}\right).$$

Since $\lambda^{i_0}_{nx}$ is Hermitian for the dissipators we can use Eq. (F.2.2):

$$\begin{aligned}
\operatorname{Tr}\left(\lambda^{i_0}_{nx} L[\gamma^{\mu_0}_{j_0} \lambda^{j_0}_{nx}]\right) &= -\operatorname{Re}\langle [\lambda^{i_0}_{nx}, \overline{\gamma}^{\mu_0}_{j_0} \lambda^{j_0}_{nx}] \gamma^{\mu_0}_{p_0} \lambda^{p_0}_{nx}\rangle \\
&= 2f_{i_0 j_0 q_0} \operatorname{Im}\left[\overline{\gamma}^{\mu_0}_{j_0} \gamma^{\mu_0}_{p_0} \langle \lambda^{q_0}_{nx} \lambda^{p_0}_{nx}\rangle\right] \\
&= 2f_{i_0 j_0 q_0} \operatorname{Im}\left[\overline{\gamma}^{\mu_0}_{j_0} \gamma^{\mu_0}_{p_0} Z_{q_0 p_0 r_0}\right] \langle \lambda^{r_0}_{nx}\rangle + 2f_{i_0 j_0 q_0} \operatorname{Im}\left[\overline{\gamma}^{\mu_0}_{j_0} \gamma^{\mu_0}_{p_0} (1/N_\nu)\delta_{q_0 p_0}\right] \langle I_{2N_\nu}\rangle \\
&= 2f_{i_0 j_0 q_0} \operatorname{Im}\left[\overline{\gamma}^{\mu_0}_{j_0} \gamma^{\mu_0}_{p_0} Z_{q_0 p_0 r_0}\right] \langle \lambda^{r_0}_{nx}\rangle + \frac{2f_{i_0 j_0 p_0}}{N_\nu} \operatorname{Im}\left[\overline{\gamma}^{\mu_0}_{j_0} \gamma^{\mu_0}_{p_0}\right] \\
&= 2f_{i_0 r_0 q_0} \operatorname{Im}\left[\overline{\gamma}^{\mu_0}_{r_0} \gamma^{\mu_0}_{p_0} Z_{q_0 p_0 j_0}\right] \langle \lambda^{j_0}_{nx}\rangle + \frac{2f_{i_0 j_0 p_0}}{N_\nu} \operatorname{Im}\left[\overline{\gamma}^{\mu_0}_{j_0} \gamma^{\mu_0}_{p_0}\right],
\end{aligned}$$

for $\mu_0 \in \{1, 2, 3\}$, and (note $\gamma^\pm_{i_\pm}$ are real),

$$\begin{aligned}
\operatorname{Tr}\left(\lambda^{i_0}_{nx} L[\gamma^+_{i_+} \lambda^{i_+}_{nx}]\right) &= -\operatorname{Re}\left\langle [\lambda^{i_0}_{nx}, \gamma^+_{i_+} \lambda^{i_-}_{nx}] \gamma^+_{j_+} \lambda^{j_+}_{nx}\right\rangle \\
&= 2\operatorname{Im}\left[\gamma^+_{i_+} \gamma^+_{j_+} \overline{f}^+_{i_0 i_+ p_+} \left(Z^-_{j_0 p_+ j_+} \langle \lambda^{j_0}_{nx}\rangle + (1/N_\nu)\delta_{p_+ j_+}\right)\right] \\
&= 2\gamma^+_{i_+} \gamma^+_{j_+} \operatorname{Im}\left[\overline{f}^+_{i_0 i_+ p_+} Z^-_{j_0 p_+ j_+}\right] \langle \lambda^{j_0}_{nx}\rangle + \frac{2}{N_\nu} \operatorname{Im}\left[\gamma^+_{i_+} \gamma^+_{j_+} \overline{f}^+_{i_0 i_+ j_+}\right], \\
\operatorname{Tr}\left(\lambda^{i_0}_{nx} L[\gamma^-_{i_-} \lambda^{i_-}_{nx}]\right) &= -\operatorname{Re}\left\langle [\lambda^{i_0}_{nx}, \gamma^-_{i_+} \lambda^{i_+}_{nx}] \gamma^-_{j_-} \lambda^{j_-}_{nx}\right\rangle \\
&= 2\operatorname{Im}\left[\gamma^-_{i_+} \gamma^-_{j_+} f^+_{i_0 i_+ p_+} \left(Z^+_{j_0 p_+ j_+} \langle \lambda^{j_0}_{nx}\rangle + (1/N_\nu)\delta_{p_+ j_+}\right)\right] \\
&= 2\gamma^-_{i_+} \gamma^-_{j_+} \operatorname{Im}\left[f^+_{i_0 i_+ p_+} Z^+_{j_0 p_+ j_+}\right] \langle \lambda^{j_0}_{nx}\rangle + \frac{2}{N_\nu} \operatorname{Im}\left[\gamma^-_{i_+} \gamma^-_{j_+} f^+_{i_0 i_+ j_+}\right].
\end{aligned}$$

Defining

$$\begin{aligned}
\xi_{i_0 j_0} &= 2f_{i_0 p_0 j_0} A_{p_0} + 2f_{i_0 r_0 q_0} \sum_{\mu_0} \operatorname{Im}\left[\overline{\gamma}^{\mu_0}_{r_0} \gamma^{\mu_0}_{p_0} Z_{q_0 p_0 j_0}\right] \\
&\quad + 2\gamma^+_{i_+} \gamma^+_{j_+} \operatorname{Im}\left[\overline{f}^+_{i_0 i_+ p_+} Z^-_{j_0 p_+ j_+}\right] + 2\gamma^-_{i_+} \gamma^-_{j_+} \operatorname{Im}\left[f^+_{i_0 i_+ p_+} Z^+_{j_0 p_+ j_+}\right], \\
\phi_{i_0} &= \frac{2f_{i_0 j_0 p_0}}{N_\nu} \sum_{\mu_0} \operatorname{Im}\left[\overline{\gamma}^{\mu_0}_{j_0} \gamma^{\mu_0}_{p_0}\right] + \frac{2}{N_\nu} \operatorname{Im}\left[\gamma^+_{i_+} \gamma^+_{j_+} \overline{f}^+_{i_0 i_+ j_+} + \gamma^-_{i_+} \gamma^-_{j_+} f^+_{i_0 i_+ j_+}\right],
\end{aligned}$$

we have,

$$\partial_t \langle \lambda_{n\mathrm{x}}^{i_0} \rangle = \xi_{i_0 j_0} \langle \lambda_{n\mathrm{x}}^{j_0} \rangle + \phi_{i_0} + \frac{1}{N_\mathrm{m}} \sum_k \left( 2B_{i_+} f^+_{i_0 i_+ j_+} e^{-ikr_n} \langle a_k \hat{\lambda}_{n\mathrm{x}}^{j_+} \rangle + \text{c.c.} \right).$$

$\langle a_k \lambda_{n\mathrm{x}}^{i_+} \rangle$

$$\begin{aligned}
&- i\langle [a_k \lambda_{n\mathrm{x}}^{i_+}, H] \rangle \\
&= -i\omega_k \langle a_k \lambda_{n\mathrm{x}}^{i_+} \rangle - 2f^+_{i_0 i_+ j_+} A_{i_0} \langle a_k \lambda_{n\mathrm{x}}^{i_+} \rangle - i\frac{1}{\sqrt{N_\mathrm{m}}} \sum_{my} \sum_{k'} B_{j_+} e^{ik' r_m} [a_k \lambda_{n\mathrm{x}}^{i_+}, a^\dagger_{k'} \lambda_{m\mathrm{y}}^{j_-}] \\
&= -\left[ 2f^+_{i_0 i_+ j_+} A_{i_0} + i\omega_k \delta_{i_+ j_+} \right] \langle a_k \lambda_{n\mathrm{x}}^{i_+} \rangle - i\frac{1}{\sqrt{N_\mathrm{m}}} \sum_{my} \sum_{k'} B_{j_+} e^{ik' r_m} \left\langle \lambda_{n\mathrm{x}}^{i_+} \lambda_{m\mathrm{y}}^{j_-} [a_k, a^\dagger_{k'}] + [\lambda_{n\mathrm{x}}^{i_+}, \lambda_{m\mathrm{y}}^{j_-}] a^\dagger_{k'} a_k \right\rangle \\
&= -\left[ 2f^+_{i_0 i_+ j_+} A_{i_0} + i\omega_k \delta_{i_+ j_+} \right] \langle a_k \lambda_{n\mathrm{x}}^{i_+} \rangle - i\frac{1}{\sqrt{N_\mathrm{m}}} \sum_{my} B_{j_+} e^{ikr_m} \langle \lambda_{n\mathrm{x}}^{i_+} \lambda_{m\mathrm{y}}^{j_-} \rangle + \frac{2B_{j_+} f^+_{i_0 i_+ j_+}}{\sqrt{N_\mathrm{m}}} \sum_{k'} e^{ik' r_n} \langle a^\dagger_{k'} a_k \lambda_{n\mathrm{x}}^{i_0} \rangle \\
&\approx -\left[ 2f^+_{i_0 i_+ j_+} A_{i_0} + i\omega_k \delta_{i_+ j_+} \right] \langle a_k \lambda_{n\mathrm{x}}^{i_+} \rangle - i\frac{1}{\sqrt{N_\mathrm{m}}} \sum_{my} B_{j_+} e^{ikr_m} \langle \lambda_{n\mathrm{x}}^{i_+} \lambda_{m\mathrm{y}}^{j_-} \rangle + \frac{2B_{j_+} f^+_{i_0 i_+ j_+}}{\sqrt{N_\mathrm{m}}} \sum_{k'} e^{ik' r_n} \langle a^\dagger_{k'} a_k \rangle \langle \lambda_{n\mathrm{x}}^{i_0} \rangle,
\end{aligned}$$

where in the final line we set the third order cumulant $\langle\langle a^\dagger_{k'} a_k \lambda_{n\mathrm{x}}^{i_0} \rangle\rangle$ to zero and used the U(1) symmetry to write

$$\begin{aligned}
\langle a^\dagger_{k'} a_k \lambda_{n\mathrm{x}}^{i_0} \rangle &\approx \langle a^\dagger_{k'} a_k \rangle \langle \lambda_{n\mathrm{x}}^{i_0} \rangle + \langle a^\dagger_{k'} \lambda_{n\mathrm{x}}^{i_0} \rangle \langle a_k \rangle + \langle a_k \lambda_{n\mathrm{x}}^{i_0} \rangle \langle a^\dagger_{k'} \rangle - 2\langle a^\dagger_{k'} \rangle \langle a_k \rangle \langle \lambda_{n\mathrm{x}}^{i_0} \rangle \\
&= \langle a^\dagger_{k'} a_k \rangle \langle \lambda_{n\mathrm{x}}^{i_0} \rangle.
\end{aligned}$$

$$\begin{aligned}
\kappa \operatorname{Tr} \left( a_k \lambda_{n\mathrm{x}}^{i_+} L[a_k] \right) &= -\frac{\kappa}{2} \left\langle [a_k \lambda_{n\mathrm{x}}^{i_+}, a_k^\dagger] a_k \right\rangle \\
&= -\frac{\kappa}{2} \langle a_k \lambda_{n\mathrm{x}}^{i_+} \rangle,
\end{aligned}$$

$$\begin{aligned}
\operatorname{Tr} \left( a_k \lambda_{n\mathrm{x}}^{i_+} L[\gamma^{\mu_0}_{i_0} \lambda_{n\mathrm{x}}^{i_0}] \right) &= -\frac{1}{2} \left\langle \overline{\gamma}^{\mu_0}_{i_0} \gamma^{\mu_0}_{j_0} a_k \left( [\lambda_{n\mathrm{x}}^{i_+}, \lambda_{n\mathrm{x}}^{i_0}] \lambda_{n\mathrm{x}}^{j_0} + \lambda_{n\mathrm{x}}^{i_0} [\lambda_{n\mathrm{x}}^{j_0}, \lambda_{n\mathrm{x}}^{i_+}] \right) \right\rangle \\
&= i\overline{\gamma}^{\mu_0}_{i_0} \gamma^{\mu_0}_{j_0} \left\langle a_k \left( f^+_{i_0 i_+ j_+} \lambda_{n\mathrm{x}}^{j_+} \lambda_{n\mathrm{x}}^{j_0} - f^+_{j_0 i_+ j_+} \lambda_{n\mathrm{x}}^{i_0} \lambda_{n\mathrm{x}}^{j_+} \right) \right\rangle \\
&= i\overline{\gamma}^{\mu_0}_{i_0} \gamma^{\mu_0}_{j_0} \left\langle a_k \left( f^+_{i_0 i_+ j_+} \overline{Z}^-_{j_0 j_+ p_+} \lambda_{n\mathrm{x}}^{p_+} - f^+_{j_0 i_+ j_+} Z^+_{i_0 j_+ p_+} \lambda_{n\mathrm{x}}^{p_+} \right) \right\rangle \\
&= i\overline{\gamma}^{\mu_0}_{i_0} \gamma^{\mu_0}_{j_0} \left( f^+_{i_0 i_+ p_+} \overline{Z}^-_{j_0 p_+ j_+} - f^+_{j_0 i_+ p_+} Z^+_{i_0 p_+ j_+} \right) \langle a_k \lambda_{n\mathrm{x}}^{j_+} \rangle \\
&= i f^+_{i_0 i_+ p_+} \left( \overline{\gamma}^{\mu_0}_{i_0} \gamma^{\mu_0}_{j_0} \overline{Z}^-_{j_0 p_+ j_+} - \gamma^{\mu_0}_{i_0} \overline{\gamma}^{\mu_0}_{j_0} Z^+_{j_0 p_+ j_+} \right) \langle a_k \lambda_{n\mathrm{x}}^{j_+} \rangle,
\end{aligned}$$

$$\begin{aligned}
\operatorname{Tr} \left( a_k \lambda_{n\mathrm{x}}^{i_+} L[\gamma^+_{j_+} \lambda_{n\mathrm{x}}^{j_+}] \right) &= -\frac{1}{2} \left\langle \overline{\gamma}^+_{j_+} \gamma^+_{p_+} a_k [\lambda_{n\mathrm{x}}^{i_+}, \lambda_{n\mathrm{x}}^{j_-}] \lambda_{n\mathrm{x}}^{p_+} \right\rangle \\
&= -i f^+_{i_0 i_+ j_+} \overline{\gamma}^+_{j_+} \gamma^+_{p_+} \left\langle a_k \lambda_{n\mathrm{x}}^{i_0} \lambda_{n\mathrm{x}}^{p_+} \right\rangle \\
&= -i f^+_{i_0 i_+ j_+} \overline{\gamma}^+_{j_+} \gamma^+_{p_+} Z^+_{i_0 p_+ q_+} \langle a_k \lambda_{n\mathrm{x}}^{q_+} \rangle \\
&= -i f^+_{i_0 i_+ q_+} \overline{\gamma}^+_{q_+} \gamma^+_{p_+} Z^+_{i_0 p_+ j_+} \langle a_k \lambda_{n\mathrm{x}}^{j_+} \rangle \\
&= -i f^+_{i_0 i_+ p_+} \overline{\gamma}^+_{p_+} \gamma^+_{q_+} Z^+_{i_0 q_+ j_+} \langle a_k \lambda_{n\mathrm{x}}^{j_+} \rangle,
\end{aligned}$$

$$\begin{aligned}
\mathrm{Tr}\left(a_k\lambda^{i_+}_{n\mathrm{x}}L[\gamma^-_{j_-}\lambda^{j_-}_{n\mathrm{x}}]\right) &= -\frac{1}{2}\left\langle\overline{\gamma}^-_{j_+}\gamma^-_{p_+}a_k\lambda^{j_+}_{n\mathrm{x}}[\lambda^{p_-}_{n\mathrm{x}},\lambda^{i_+}_{n\mathrm{x}}]\right\rangle \\
&= -i\overline{f}^+_{i_0p_+i_+}\overline{\gamma}^-_{j_+}\gamma^-_{p_+}\left\langle a_k\lambda^{j_+}_{n\mathrm{x}}\lambda^{i_0}_{n\mathrm{x}}\right\rangle \\
&= -i\overline{f}^+_{i_0p_+i_+}\overline{\gamma}^-_{j_+}\gamma^-_{p_+}\overline{Z}^-_{i_0j_+q_+}\langle a_k\lambda^{q_+}_{n\mathrm{x}}\rangle \\
&= if^+_{i_0i_+p_+}\overline{\gamma}^-_{q_+}\gamma^-_{p_+}\overline{Z}^-_{i_0q_+j_+}\langle a_k\lambda^{j_+}_{n\mathrm{x}}\rangle.
\end{aligned}$$

To obtain the last line we used

$$\begin{aligned}
[\lambda^{p_-}_{n\mathrm{x}},\lambda^{i_+}_{n\mathrm{x}}] = -[\lambda^{i_+}_{n\mathrm{x}},\lambda^{p_-}_{n\mathrm{x}}] &= -2if^+_{i_0i_+p_+}\lambda^{i_0}_{n\mathrm{x}} \\
\Rightarrow \overline{f}^+_{i_0p_+i_+} &= -f^+_{i_0i_+p_+}.
\end{aligned}$$

and swapped the dummy indices $q_+$ and $j_+$. Note the $\lambda^{j_-}_{n\mathrm{x}}$ dissipator contribution is not simply the conjugate of that from $\lambda^{j_+}_{n\mathrm{x}}$, since $Y^\dagger[Y,X]$ is not the hermitian conjugate of $[X,Y^\dagger]Y$.

Combining the above,

$$\begin{aligned}
\partial_t\langle a_k\hat{\lambda}^{i_+}_{n\mathrm{x}}\rangle = \left[\xi^+_{i_+j_+} - (i\omega_k+\kappa/2)\delta_{i_+j_+}\right]\langle a_k\hat{\lambda}^{j_+}_{n\mathrm{x}}\rangle - \frac{1}{N_\mathrm{m}}\sum_{m\mathrm{y}} iB_{j_-}e^{ikr_m}\langle\hat{\lambda}^{i_+}_{n\mathrm{x}}\hat{\lambda}^{j_-}_{m\mathrm{y}}\rangle \\
+ 2B_{j_+}f^+_{i_0i_+j_+}\langle\lambda^{i_0}_{n\mathrm{x}}\rangle\sum_{k'}e^{ik'r_n}\langle a^\dagger_{k'}a_k\rangle,
\end{aligned}$$

with $\xi^+_{i_+j_+}$ as in the main text:

$$\begin{aligned}
\xi^+_{i_+j_+} = -2f^+_{i_0i_+j_+}A_{i_0} + if^+_{i_0i_+p_+}\sum_{\mu_0}\left(\overline{\gamma}^{\mu_0}_{i_0}\gamma^{\mu_0}_{j_0}\overline{Z}^-_{j_0p_+j_+} - \overline{\gamma}^{\mu_0}_{j_0}\gamma^{\mu_0}_{i_0}Z^+_{j_0p_+j_+}\right) \\
+ if^+_{i_0i_+p_+}\left(\gamma^-_{q_+}\gamma^-_{p_+}\overline{Z}^-_{i_0q_+j_+} - \gamma^+_{p_+}\gamma^+_{q_+}Z^+_{i_0q_+j_+}\right).
\end{aligned}$$

### $\langle\lambda^{i_+}_{n\mathrm{x}}\lambda^{j_-}_{m\mathrm{y}}\rangle$

Assuming $n\mathrm{y}\neq m\mathrm{y}$,

$$\begin{aligned}
-i\langle[\lambda^{i_+}_{n\mathrm{x}}\lambda^{j_-}_{m\mathrm{y}},H]\rangle &= -iA_{i_0}\langle[\lambda^{i_+}_{n\mathrm{x}},\lambda^{i_0}_{n\mathrm{x}}]\lambda^{j_-}_{m\mathrm{y}}\rangle - iA_{i_0}\langle\lambda^{i_+}_{n\mathrm{x}}[\lambda^{j_-}_{m\mathrm{y}},\lambda^{i_0}_{m\mathrm{y}}]\rangle \\
&\quad - i\sum_k\left[B_{p_+}e^{-ikr_m}\langle a_k\lambda^{i_+}_{n\mathrm{x}}[\lambda^{j_-}_{m\mathrm{y}},\lambda^{p_+}_{m\mathrm{y}}]\rangle + B_{p_+}e^{ikr_n}\langle a^\dagger_k[\lambda^{i_+}_{n\mathrm{x}},\lambda^{p_-}_{n\mathrm{x}}]\lambda^{j_-}_{m\mathrm{y}}\rangle\right] \\
&= -2f^+_{i_0i_+p_+}A_{i_0}\langle\lambda^{p_+}_{n\mathrm{x}}\lambda^{j_-}_{m\mathrm{y}}\rangle - 2\overline{f}^+_{i_0j_+p_+}A_{i_0}\langle\lambda^{i_+}_{n\mathrm{x}}\lambda^{p_-}_{m\mathrm{y}}\rangle \\
&\quad + \sum_k\left[2\overline{f}^+_{i_0j_+p_+}B_{p_+}e^{-ikr_m}\langle a_k\lambda^{i_+}_{n\mathrm{x}}\lambda^{i_0}_{m\mathrm{y}}\rangle + 2f^+_{i_0i_+p_+}B_{p_+}e^{ikr_n}\langle a^\dagger_k\lambda^{i_0}_{n\mathrm{x}}\lambda^{j_-}_{m\mathrm{y}}\rangle\right] \\
&\approx -2f^+_{i_0i_+p_+}A_{i_0}\langle\lambda^{p_+}_{n\mathrm{x}}\lambda^{j_-}_{m\mathrm{y}}\rangle - 2\overline{f}^+_{i_0j_+p_+}A_{i_0}\langle\lambda^{i_+}_{n\mathrm{x}}\lambda^{p_-}_{m\mathrm{y}}\rangle \\
&\quad + \sum_k\left[2\overline{f}^+_{i_0j_+p_+}B_{p_+}e^{-ikr_m}\langle a_k\lambda^{i_+}_{n\mathrm{x}}\rangle\langle\lambda^{i_0}_{m\mathrm{y}}\rangle + 2f^+_{i_0i_+p_+}B_{p_+}\overline{e^{-ikr_n}\langle a_k\lambda^{j_+}_{m\mathrm{y}}\rangle}\langle\lambda^{i_0}_{n\mathrm{x}}\rangle\right].
\end{aligned}$$

Here the third-order cumulants $\langle\langle a_k\lambda^{i_+}_{n\mathrm{x}}\lambda^{i_0}_{m\mathrm{y}}\rangle\rangle$ and $\langle\langle a^\dagger_k\lambda^{i_0}_{n\mathrm{x}}\lambda^{j_-}_{m\mathrm{y}}\rangle\rangle$ were set to zero to obtain the last approximation.

Since $\lambda^{i_+}_{n\mathrm{x}}$ and $\lambda^{j_-}_{m\mathrm{y}}$ commute for $n\mathrm{x}\neq m\mathrm{y}$, we can work out dissipator contributions for the operators one at a time. In fact, we already derived those for $\lambda^{i_+}_{n\mathrm{x}}$ in the previous part as contained in $\xi^{+}_{i_+j_+}$. Those for $\lambda^{j_-}_{my}$ are very similar:

$$\begin{aligned}
\mathrm{Tr}\left(\lambda^{i_+}_{nx}\lambda^{j_-}_{my}L[\gamma^{\mu_0}_{i_0}\lambda^{i_0}_{my}]\right) &= -\frac{1}{2}\left\langle\overline{\gamma}^{\mu_0}_{i_0}\gamma^{\mu_0}_{j_0}\lambda^{i_+}_{nx}\left(\left[\lambda^{j_-}_{my},\lambda^{i_0}_{my}\right]\lambda^{j_0}_{my}+\lambda^{i_0}_{my}\left[\lambda^{j_0}_{my},\lambda^{j_-}_{my}\right]\right)\right\rangle \\
&= -\frac{1}{2}\overline{\left\langle\lambda^{i_-}_{my}\overline{\gamma}^{\mu_0}_{j_0}\gamma^{\mu_0}_{i_0}\left(\left[\lambda^{j_+}_{my},\lambda^{j_0}_{my}\right]\lambda^{i_0}_{my}+\lambda^{j_0}_{my}\left[\lambda^{i_0}_{my},\lambda^{j_+}_{my}\right]\right)\right\rangle} \\
&= \overline{if^{+}_{j_0j_+p_+}\left(\overline{\gamma}^{\mu_0}_{j_0}\gamma^{\mu_0}_{i_0}\overline{Z}^{-}_{i_0p_+q_+}-\gamma^{\mu_0}_{j_0}\overline{\gamma}^{\mu_0}_{i_0}Z^{+}_{i_0p_+q_+}\right)\langle\lambda^{i_-}_{nx}\lambda^{q_+}_{my}\rangle} \\
&= -i\overline{f}^{+}_{i_0j_+p_+}\left(\overline{\gamma}^{\mu_0}_{j_0}\gamma^{\mu_0}_{i_0}Z^{-}_{j_0p_+q_+}-\gamma^{\mu_0}_{j_0}\overline{\gamma}^{\mu_0}_{i_0}\overline{Z}^{+}_{j_0p_+q_+}\right)\langle\lambda^{i_+}_{nx}\lambda^{q_-}_{my}\rangle,
\end{aligned}$$

$$\begin{aligned}
\mathrm{Tr}\left(\lambda^{i_+}_{nx}\lambda^{j_-}_{my}L[\gamma^{+}_{p_+}\lambda^{p_+}_{my}]\right) &= -\frac{1}{2}\left\langle\lambda^{i_+}_{nx}\overline{\gamma}^{+}_{p_+}\gamma^{+}_{q_+}\lambda^{p_-}_{my}[\lambda^{q_+}_{my},\lambda^{y_-}_{my}]\right\rangle \\
&= i\overline{f}^{+}_{i_0j_+q_+}\langle\overline{\gamma}^{+}_{p_+}\gamma^{+}_{q_+}\lambda^{i_+}_{nx}\lambda^{p_-}_{my}\lambda^{i_0}_{my}\rangle \\
&= i\overline{f}^{+}_{i_0j_+q_+}\overline{Z}^{+}_{i_0p_+r_+}\overline{\gamma}^{+}_{p_+}\gamma^{+}_{q_+}\langle\lambda^{i_+}_{nx}\lambda^{r_-}_{my}\rangle \\
&= i\overline{f}^{+}_{i_0j_+r_+}\overline{Z}^{+}_{i_0p_+q_+}\overline{\gamma}^{+}_{p_+}\gamma^{+}_{r_+}\langle\lambda^{i_+}_{nx}\lambda^{q_-}_{my}\rangle,
\end{aligned}$$

$$\begin{aligned}
\mathrm{Tr}\left(\lambda^{i_+}_{nx}\lambda^{j_-}_{my}L[\gamma^{-}_{p_-}\lambda^{p_-}_{my}]\right) &= -\frac{1}{2}\left\langle\lambda^{i_+}_{nx}\overline{\gamma}^{-}_{p_+}\gamma^{-}_{q_+}[\lambda^{y_-}_{my},\lambda^{p_+}_{my}]\lambda^{q_-}_{my}\right\rangle \\
&= -i\overline{f}^{+}_{i_0j_+p_+}\langle\overline{\gamma}^{-}_{p_+}\gamma^{-}_{q_+}\lambda^{i_+}_{nx}\lambda^{i_0}_{my}\lambda^{q_-}_{my}\rangle \\
&= -i\overline{f}^{+}_{i_0j_+p_+}Z^{-}_{i_0q_+r_+}\overline{\gamma}^{-}_{p_+}\gamma^{-}_{q_+}\langle\lambda^{i_+}_{nx}\lambda^{r_-}_{my}\rangle \\
&= -i\overline{f}^{+}_{i_0j_+p_+}Z^{-}_{i_0r_+q_+}\overline{\gamma}^{-}_{p_+}\gamma^{-}_{r_+}\langle\lambda^{i_+}_{nx}\lambda^{q_-}_{my}\rangle.
\end{aligned}$$

Taking everything together,

$$\begin{aligned}
\partial_t\langle\hat{\lambda}^{i_+}_{nx}\hat{\lambda}^{j_-}_{my}\rangle &= \xi^{+}_{i_+p_+}\langle\hat{\lambda}^{p_+}_{nx}\hat{\lambda}^{j_-}_{my}\rangle+\xi^{-}_{j_-p_-}\langle\hat{\lambda}^{i_+}_{nx}\hat{\lambda}^{p_-}_{my}\rangle \\
&\quad+\sum_k\left(2B_{p_+}\overline{f}^{+}_{i_0j_+p_+}e^{-ikr_m}\langle a_k\hat{\lambda}^{i_+}_{nx}\rangle\langle\lambda^{i_0}_{my}\rangle+2B_{p_+}f^{+}_{i_0i_+p_+}\overline{e^{-ikr_n}\langle a_k\hat{\lambda}^{j_+}_{my}\rangle}\langle\lambda^{i_0}_{nx}\rangle\right),
\end{aligned}$$

where, in addition to $\xi^{+}_{i_+p_+}$ defined previously,

$$\begin{aligned}
\xi^{-}_{i_-j_-} &= -2\overline{f}^{+}_{i_0i_-j_-}A_{i_0}+i\overline{f}^{+}_{i_0i_-p_+}\sum_{\mu_0}\left(\overline{\gamma}^{\mu_0}_{i_0}\gamma^{\mu_0}_{j_0}\overline{Z}^{+}_{j_0p_+j_-}-\overline{\gamma}^{\mu_0}_{j_0}\gamma^{\mu_0}_{i_0}Z^{-}_{j_0p_+j_-}\right) \\
&\quad+i\overline{f}^{+}_{i_0i_-p_+}\left(\gamma^{+}_{q_+}\gamma^{+}_{p_+}\overline{Z}^{+}_{i_0q_+j_-}-\gamma^{-}_{p_+}\gamma^{-}_{q_+}Z^{-}_{i_0q_+j_-}\right).
\end{aligned}$$

## F.3 Dark state coefficient identities

The relevant coefficients for this section are

$$C^{0}_{i_0}=\frac{1}{2}\mathrm{Tr}\left[\sigma^{+}\sigma^{-}\lambda^{i_0}\right],\ D^{0}=\frac{1}{2N_\nu}\mathrm{Tr}\left[\sigma^{+}\sigma^{-}\right],\ \varsigma_{i_+}=\frac{1}{2}\mathrm{Tr}\left[\sigma^{+}\lambda^{i_-}\right], \tag{F.3.1}$$

where $\sigma^{+}=\sigma^{+}\otimes I_{N_\nu}$, $\sigma^{+}\sigma^{-}=\sigma^{+}\sigma^{-}\otimes I_{N_\nu}$ are operators in the full exciton-vibron space. We intend to show

$$\varsigma_{i_+}\varsigma_{j_+}Z^{+}_{i_0i_+j_+}\equiv C^{0}_{i_0},\ \frac{1}{N_\nu}\varsigma_{i_+}\varsigma_{i_+}\equiv D^{0}. \tag{F.3.2}$$

In the following we draw matrices for $N_\nu = 2$ for illustration, but the results hold for any $N_\nu$.

First note that $\mathrm{Tr}_e[\sigma^+\sigma^-] = 1$, $\mathrm{Tr}_\nu[I_{N_\nu}] = N_\nu$, so $D^0 = 1/2$. Next, in the combined space $\sigma^+\sigma^-$ has $N_\nu$ non-zero entries (of 1) in the diagonal of the upper-right quadrant,

$$\sigma^+\sigma^- = \begin{pmatrix} 0 & 0 & 1 & 0 \\ 0 & 0 & 0 & 1 \\ 0 & 0 & 0 & 0 \\ 0 & 0 & 0 & 0 \end{pmatrix} \tag{F.3.3}$$

so only those $\lambda^{i_-}$ matrices with a non-zero entry in the corresponding entry of the lower-left quadrant will give a non-zero result under the trace,

$$\lambda^- = \begin{pmatrix} 0 & 0 & 0 & 0 \\ 0 & 0 & 0 & 0 \\ \sqrt{2} & 0 & 0 & 0 \\ 0 & 0 & 0 & 0 \end{pmatrix} \text{ or } \begin{pmatrix} 0 & 0 & 0 & 0 \\ 0 & 0 & 0 & 0 \\ 0 & 0 & 0 & 0 \\ 0 & \sqrt{2} & 0 & 0 \end{pmatrix}, \tag{F.3.4}$$

where the factor of $\sqrt{2}$ is due to the normalisation condition Eq. (6.2.11). As there are $N_\nu$ such matrices,

$$\begin{aligned} \varsigma_{i_+}\varsigma_{i_+} &= \frac{1}{2^2}\mathrm{Tr}\left[\sigma^+\lambda^{i_-}\right]\mathrm{Tr}\left[\sigma^+\lambda^{i_-}\right] && \text{(F.3.5)} \\ &= \frac{N_\nu}{4}\left(\sqrt{2}\right)^2 && \text{(F.3.6)} \\ &= \frac{1}{2} && \text{(F.3.7)} \\ &= D^0, && \text{(F.3.8)} \end{aligned}$$

as required. For the other identity, following the above we have that the product $\varsigma_{i_+}\varsigma_{j_+}$ is non-zero only if $i_+ = j_+$, in which case it equals $1/2$, so

$$\begin{aligned} \varsigma_{i_+}\varsigma_{j_+} Z^+_{i_0 i_+ j_+} &= \frac{1}{2}\delta_{i_+ j_+} Z^+_{i_0 i_+ j_+} && \text{(F.3.9)} \\ &= \frac{1}{2} Z^+_{i_0 i_+ i_+} && \text{(F.3.10)} \\ &= \frac{1}{2}\left(d^+_{i_0 i_+ i_+} + i f^+_{i_0 i_+ i_+}\right). && \text{(F.3.11)} \end{aligned}$$

Now, the anti-commutator of a $\lambda^{i_+}$ matrix with its Hermitian conjugate places two values (of 2) along the diagonal, e.g.,

$$\left\{\begin{pmatrix} 0 & 0 & \sqrt{2} & 0 \\ 0 & 0 & 0 & 0 \\ 0 & 0 & 0 & 0 \\ 0 & 0 & 0 & 0 \end{pmatrix}, \begin{pmatrix} 0 & 0 & 0 & 0 \\ 0 & 0 & 0 & 0 \\ \sqrt{2} & 0 & 0 & 0 \\ 0 & 0 & 0 & 0 \end{pmatrix}\right\} = \begin{pmatrix} 2 & 0 & 0 & 0 \\ 0 & 0 & 0 & 0 \\ 0 & 0 & 2 & 0 \\ 0 & 0 & 0 & 0 \end{pmatrix}. \tag{F.3.12}$$

Hence, summing over all $i_+$ under the trace as in $d^+_{i_0 i_+ i_+}$ gives (2 times) $\lambda^{i_0}$ multiplied by the identity, which vanishes, since $\lambda^{i_0}$ is traceless. On the other hand, the commutator $[\lambda^{i_+}, \lambda^{i_0}]$ gives values with opposite signs in the upper-left and lower-right quadrants,

$$\left[\begin{pmatrix} 0 & 0 & \sqrt{2} & 0 \\ 0 & 0 & 0 & 0 \\ 0 & 0 & 0 & 0 \\ 0 & 0 & 0 & 0 \end{pmatrix}, \begin{pmatrix} 0 & 0 & 0 & 0 \\ 0 & 0 & 0 & 0 \\ \sqrt{2} & 0 & 0 & 0 \\ 0 & 0 & 0 & 0 \end{pmatrix}\right] = \begin{pmatrix} 2 & 0 & 0 & 0 \\ 0 & 0 & 0 & 0 \\ 0 & 0 & -2 & 0 \\ 0 & 0 & 0 & 0 \end{pmatrix}, \tag{F.3.13}$$

hence in this case the sum under the trace gives $\lambda^{i_0}$ multiplied by a part propositional to $\sigma^+\sigma^-$:

$$
\begin{aligned}
if^{+}_{i_0 i_+ i_+} &= \frac{1}{4}\,\mathrm{Tr}\left[\lambda^{i_0}\left[\lambda^{i_+},\lambda^{i_0}\right]\right] && \text{(F.3.14)}\\
&= \frac{1}{4}\,\mathrm{Tr}\left[\lambda^{i_0}\begin{pmatrix}2&0&0&0\\0&2&0&0\\0&0&-2&0\\0&0&0&-2\end{pmatrix}\right] && \text{(F.3.15)}\\
&= \frac{1}{4}\,\mathrm{Tr}\left[\lambda^{i_0}\left(4\sigma^+\sigma^- - 2I_{N_\nu}\right)\right] && \text{(F.3.16)}\\
&= \mathrm{Tr}\left[\lambda^{i_0}\sigma^+\sigma^-\right]. && \text{(F.3.17)}
\end{aligned}
$$

Therefore

$$
\varsigma_{i_+}\varsigma_{j_+}Z^{+}_{i_0 i_+ j_+} = \frac{1}{2}Z^{+}_{i_0 i_+ i_+} = \frac{1}{2}\,\mathrm{Tr}\left[\lambda^{i_0}\sigma^+\sigma^-\right] = C^0_{i_0} \quad \square \qquad \text{(F.3.18)}
$$

# Index of names

Authors named or quoted in the text.

# Index